\documentclass[twoside,12pt]{Latex/Classes/PhDthesisPSnPDF}

\usepackage{lineno}
\usepackage{amsbsy}
\usepackage{xspace}
\usepackage{wtmmPkg}
\usepackage{natbib}
\usepackage{multirow}
\usepackage{paralist}
\usepackage{fancyhdr} 
\usepackage{lscape}

\newcommand{\aap}{    {\it Astronomy \& Astrophysics}}

\newcommand{\apj}{    {\it Astrophysical Journal}}

\newcommand{\solphys}{{\it Solar Physics}}

\newcommand{\flt}{
\hrule height 1mm
\vspace{0.5mm}
\hrule height 0.5mm 
\noindent 
\\}

\newcommand{\flb}{ \\
\hrule height 0.5mm
\vspace{0.5mm}
\hrule height 1mm }

\usepackage{titlesec}
\titlespacing*{\chapter}{0pt}{-50pt}{20pt}[20pt]
\titleformat{\chapter}[display]{\normalfont\huge\bfseries}{\chaptertitlename\ \thechapter}{20pt}{\fontsize{22.8pt}{20}\selectfont}

\ifpdf  
    \pdfcatalog { /PageMode (/UseOutlines)
                  /OpenAction (fitbh)  }
\fi

\title{The Evolution and Space Weather Effects of Solar Coronal Holes}

\ifpdf
  \author{\href{mailto:kristal@tcd.ie}{Larisza D. Krista}}
  \collegeordept{\href{http://www.physics.tcd.ie}{School of Physics}}
  \university{\href{http://www.tcd.ie}{University of Dublin, Trinity College}}

  \crest{
  \vspace{1cm}
  \includegraphics[width=0.3\textwidth]{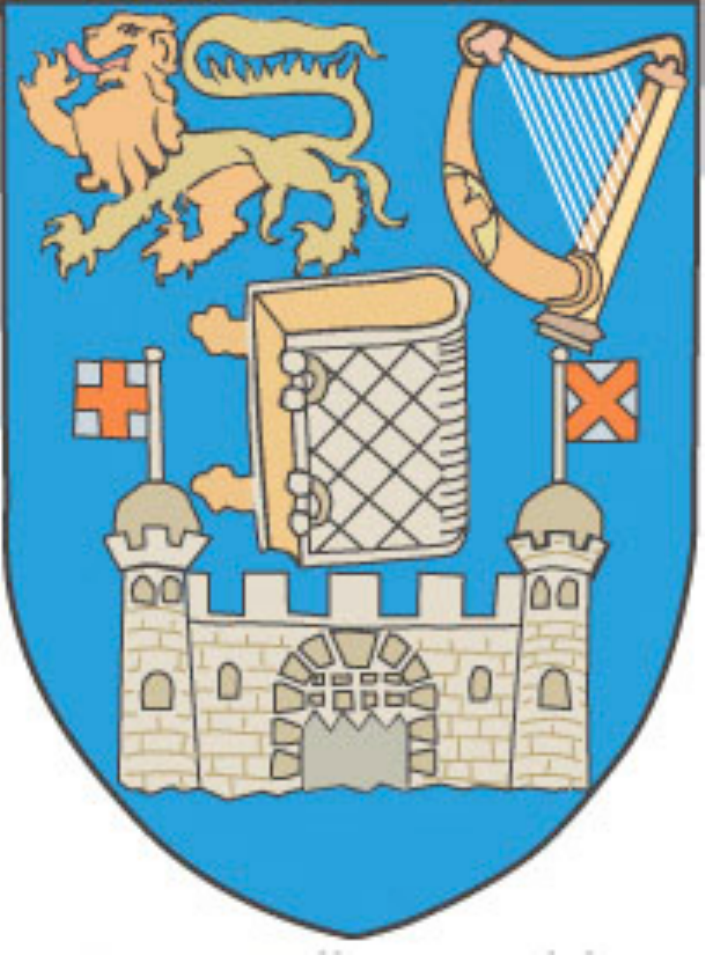}
   \vspace{1cm}
    }
  
\else
  \author{\\ Larisza D. Krista}
  \collegeordept{School of Physics}
  \university{University of Dublin, Trinity College}
  \crest{\includegraphics[width=6cm]{TCD_Crest.pdf}}
\fi
\degree{Philosophi\ae Doctor (PhD)}
\degreedate{2011 February}

\usepackage{setspace}
\begin{document}


\renewcommand\baselinestretch{1.2}
\baselineskip=18pt plus1pt


\maketitle  

\newpage
\thispagestyle{plain}

\noindent
This is an online version of this thesis and contains lower resolution images to reduce file size. The original version of the thesis and the CHARM website can be found at: http://charm-mist.appspot.com. 
\\
\\
\noindent
Larisza D. Krista
\\
\noindent {\it Boulder, Colorado, 15 October 2012.}

%









\frontmatter



\begin{declaration}        

I, Larisza D. Krista, hereby certify that I am the sole author of this thesis and that all the work presented in it, unless otherwise referenced, is entirely my own. I also declare that this work has not been submitted, in whole or in part, to any other university or college for any degree or other qualification. 

\textbf{Name:} Larisza D. Krista
 \begin{center}
 \textbf{Signature:}  ........................................		\textbf{Date:}  ..............
\end{center}
\vspace{5mm}

The thesis work was conducted from October 2007 to March 2011 under the supervision of Dr. Peter T. Gallagher at Trinity College, University of Dublin.

\vspace{5mm}
In submitting this thesis to the University of Dublin I agree that the University Library may lend or copy the thesis upon request. 

\textbf{Name:} Larisza D. Krista
 \begin{center}
 \textbf{Signature:}  ........................................		\textbf{Date:}  ..............
\end{center}

\end{declaration}



\begin{abstracts}        

In recent years the role of space weather forecasting has grown tremendously as our society increasingly relies on satellite dependent technologies. As solar activity is the foremost important aspect of space weather, the forecasting of flare and CME related transient geomagnetic storms has become a primary initiative. Minor magnetic storms caused by coronal holes (CHs) have also proven to be of high importance due to their long lasting and recurrent geomagnetic effects. 
In order to forecast CH related geomagnetic storms, the author developed an automated CH detection package (Coronal Hole Automated Recognition and Monitoring; {\it CHARM}) to replace the user-dependent CH boundary intensity thresholding methods used in previous studies. {\it CHARM} uses a local intensity thresholding method to identify low intensity regions in extreme ultraviolet or X-ray images. CHs are known to be regions of ``open" magnetic field and predominant polarity, which allowed the differentiation of CHs from other low intensity regions using magnetograms. 
An additional algorithm (Coronal Hole Evolution; {\it CHEVOL}) was developed and used in conjunction with {\it CHARM} to study individual CHs by tracking their boundary evolution. It is widely accepted that the short term changes in CH boundaries are due to the interchange reconnection between the CH open field lines and small loops. In order to test the interchange reconnection model, the magnetic reconnection rate and the diffusion coefficient at CH boundaries were determined using observed CH boundary displacement velocities. The results were found to be in agreement with those determined by the theory. Our results also indicate that the short-term CH boundary evolution is caused by random granular motions bringing open and small closed field lines close, and thus accommodating the magnetic reconnection and displacement of field lines.
The Minor Storm ({\it MIST}) algorithm was developed by the author to build on the {\it CHARM} package, providing a fast and consistent way to link CHs to high-speed solar wind (HSSW) periods detected at Earth. This allowed us to carry out a long-term analysis (2000--2009) to study the relationship between CHs, the corresponding HSSW properties, and the geomagnetic indices. The study found a strong correlation between the velocity and proton plasma temperature of the HSSW stream. This indicates that the heating and acceleration of the solar wind plasma in CHs is closely related, and perhaps are caused by the same mechanism. Many authors accept this mechanism to be Alfv\'en wave heating.
The research presented in this thesis includes the small scale analysis of individual CHs on time scales of days, which is complemented with large scale analysis of CH groups on time scales of years. This allowed us to further our understanding of CH evolution as a whole.

\end{abstracts}



\begin{dedication} 

Apunak \'es Anyunak, \\ Ti nyitott\'atok fel szemeim a Csillag\'aszat sz\'eps\'eg\'ere

For Dad and Mum, \\ who opened my eyes to the beauty of Astronomy

\end{dedication}


\begin{acknowledgements}      

First and foremost I would like to acknowledge the financial support of the Irish Research Council for Science, Engineering \& Technology (IRCSET) through the Embark Initiative Postgraduate Research Scholarship. I also acknowledge the financial support of the European Commission's Seventh Framework Programme (FP7; Project No. 238969) through the HELIO program.

I would like to thank Dr. Peter Gallagher, my supervisor, for the excellent guidance and advice he has given me from the day I set foot in Dublin. Thank you for letting me work on what I love and pursue my ideas and still keeping me on track. Your insight and experience was vital. 

I would also like to thank Dr. Shaun Bloomfield, for taking time and interest in my work. Chatting with you has given me many 'aha!' moments.

I would also like to thank the The Office (the old and the new members - Paul C, Claire, Jason, Dave, Shane, Joe, Paul H, Sophie, Eoin, Dan, Eamon, Peter B, Pietro and Aidan), the postdocs (Shaun, James, David) and Peter G, for all the chats, the cups of tea, the pints and the group dinners that so often led to the dance floor (to my pure delight). The time spent with you guys has been a true experience, and has so naturally worn down my ever-so-serious attitude towards things that should be laughed at. Thank you so much, you have all been great friends to me.

I want to thank my Mum for showing me that almost everything can be achieved if one puts their mind to it.
Thank you for always being there when I needed a break, and for so cleverly reminding me of my priorities and why it is so important to stick to one's dreams and goals. 

Dad, thank you for those childhood bed-time stories about planets, stars and galaxies - they have kept me fascinated ever since. That love and curiosity for Science you has given me so much happiness - it defines who I am. Also, thank you for always being so proud of me and being such a big fan of my work. What else could I possibly wish for.

Henry, my furry little friend, I know you cannot read, but I do want to thank you for keeping me company on many the hours spent writing this thesis. For occasionally distracting me by chewing on anything you could lay your paws on, for gazing at the screen possibly wondering what takes so long in writing a thesis. You are a dear little thing, and blessed is the moment when Tom chose you to be our cat.

Last but not least, I would like to thank Tom, my wonderful husband, for moving to Dublin with me so I could pursue my research. Thank you for listening to my relentless rant about my work through these years. On many occasions your listening was all I needed to clear my head and find a new perspective. Thank you for all the cheering on and advice on writing a thesis. I couldn't have done it without You and the excellent cups of tea you make.

\end{acknowledgements}


\chapter{List of Publications}
\label{chapter:publications}

\textbf{\Large{\underline{Refereed}}}
\begin{enumerate}

 \item \textbf{Krista, L. D.} and Gallagher, P. T. (2009)\\
 ``Automated Coronal Hole Detection Using Local Intensity Thresholding Techniques'' \\ \solphys, 256: 87-100
 
 \item \textbf{Krista, L. D.} and Gallagher, P. T. and Bloomfield, D. S. (2011)\\
 ``Short-Term Evolution of Coronal Hole Boundaries'' \\\apj, 731, L26

 \item \textbf{Krista, L. D.} and Gallagher, P. T. (2011)\\
 ``The Geoeffectiveness of Solar Coronal Holes'' \\\aap, in prep

\end{enumerate}

\newpage
\thispagestyle{plain}


\setcounter{secnumdepth}{3} 
\setcounter{tocdepth}{3}    
\tableofcontents            


\listoffigures	

\listoftables  







\mainmatter



\doublespacing
\pagestyle{fancy}


\chapter{Introduction}
\label{chapter:introduction}

\ifpdf
    \graphicspath{{1/figures/PNG/}{1/figures/PDF/}{1/figures/}}
\else
   \graphicspath{{1/figures/EPS/}{1/figures/}}
\fi


\flt
{\it This Chapter introduces the Sun - its interior structure, and the solar atmosphere that extends into the vast expanse of interplanetary space. The outer solar atmosphere, the corona, is described using a hydrostatic model in conjunction with the Heliospheric model of the solar wind. The origins of the solar wind are discussed with a more in-depth overview of coronal holes, the sources of high-speed solar wind and geomagnetic disturbances. Since coronal holes are the fundamental subject of the present thesis, their observation, morphology and physical properties are presented together with models describing their evolution. The Chapter ends with the overview of space weather and the  sources of geomagnetic storms.
 }
\flb

\newpage

\section{The structure of the Sun}

\subsection{The solar interior}

Our Sun is a main sequence star with a spectral type G2V, classified as a dwarf star with 5200--6000~K surface temperature and a yellow visible colour. Like other stars, the Sun is made out of high temperature mostly ionised plasma. At its present age of $\sim$4.5$\times$10$^{9}$ years the Sun still holds most of its fuel reserves consisting of 90\% hydrogen and $\sim$10\% helium, with traces of other heavier elements \citep{Priest84}. It might make one wonder how this extraordinary hydrogen burning furnace we call our Sun manages to stay stable instead of going out in an almighty explosion. A very simplified explanation lies in the equilibrium between the solar gas pressure and gravity, which is also responsible for the stratification of the solar interior. This stratification and hydrostatic equilibrium is what ensures that the solar fuel is burned gradually over thousands of millions of years. The temperature in the solar core is $\sim$1.5$\times$10$^{7}$~K which is hot enough to ignite thermonuclear reactions which produce 99\% of the total solar energy output. Nevertheless, it takes milions of years for this energy to be transferred out to the solar surface. In different layers of the solar interior different means of energy transfer processes dominate. The layer above the core is the highly opaque {\it radiative zone} (0.25--0.68~R$_{\odot}$) where the energy is constantly absorbed and emitted by photons, thus the transfer happens mainly by radiative diffusion. The rigid-rotating radiative zone is separated from the above lying differentially rotating layer by a largely sheared region called the {\it tachocline}. 

From 0.68--1~R$_{\odot}$ lies the {\it convection zone} where the energy is transferred by kinetic motions. 
\begin{figure}[!t]
\vspace{-0.05\textwidth}
\centerline{\includegraphics[scale=0.6]{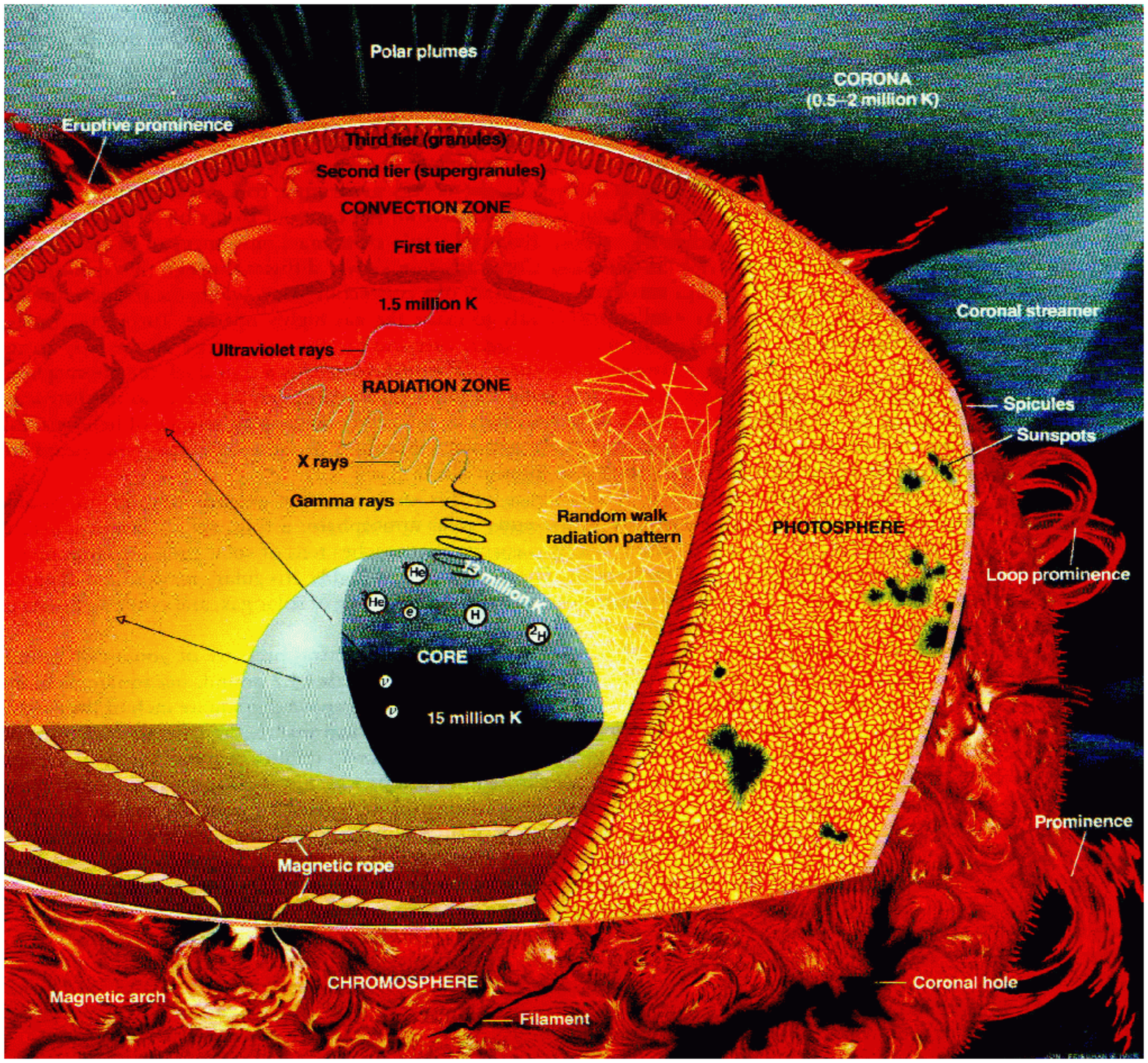}}
\caption{Schematic of the solar interior with structures of the solar atmosphere.}
\label{interior}
\end{figure}
Here, the temperature gradient is so high that the plasma cannot remain in static equilibrium and convective instability occurs allowing the plasma to rise and fall in a cellular circulation, creating the structures better known as {\it supergranular cells} (see Figure \ref{interior}). 

\subsection{The solar atmosphere}

\subsubsection{The photosphere}

\begin{figure}[!t]
\centerline{\includegraphics[scale=0.5, trim=0 150 0 150, clip]{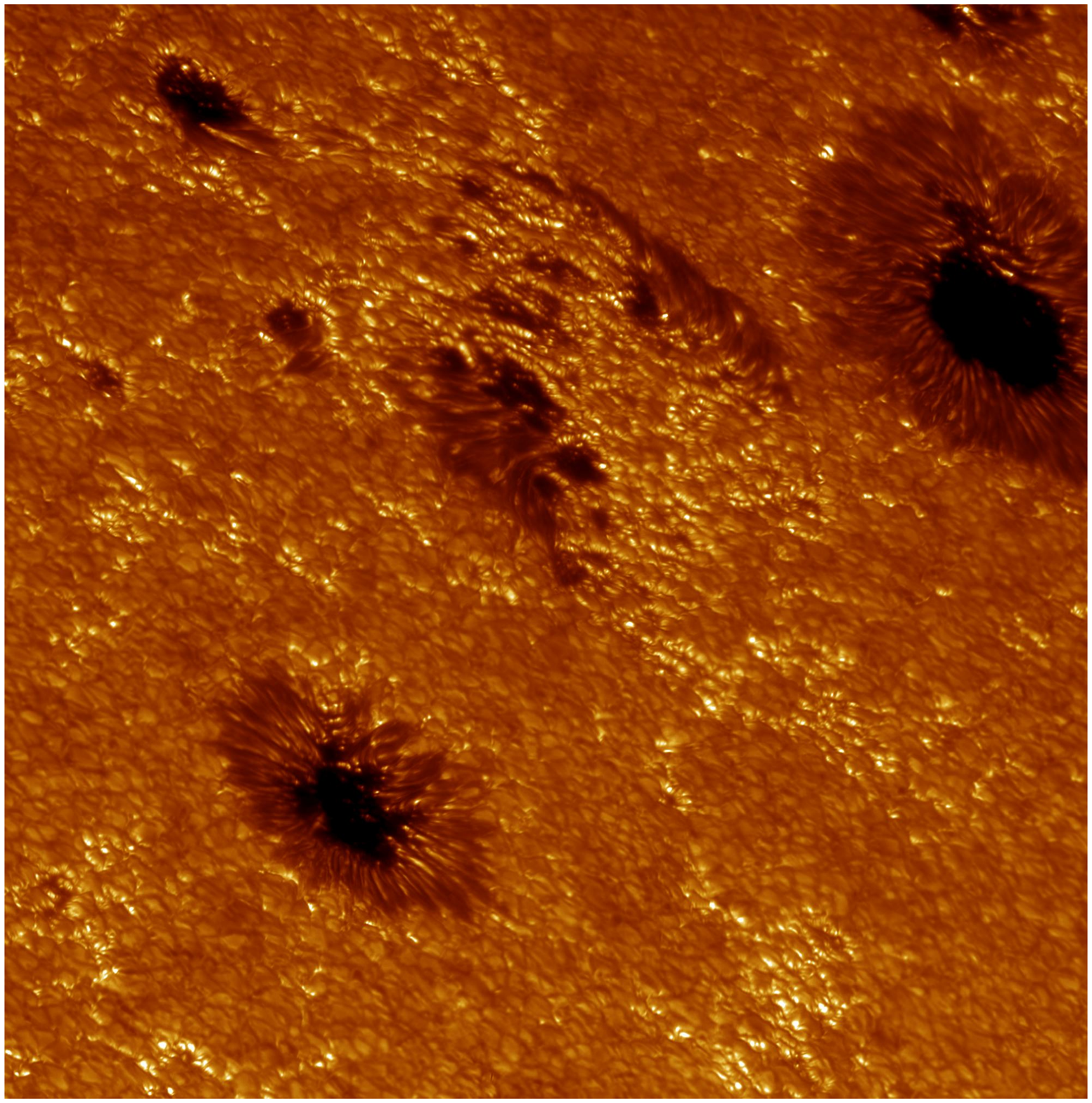}}
\caption{Sunspots and faculae observed by the Swedish solar Telescope.}
\label{faculae}
\end{figure}

At approximately 1 R$_{\odot}$ is the commonly accepted boundary between the solar interior and atmosphere. This region is the solar {\it photosphere}, also popularly referred to as the `surface' of the Sun since historically, it was the first part of the solar atmosphere to be observed with the most basic techniques (i.e., projection, or direct observation using a brightness decreasing filter). The observations showing the photosphere are the so-called white-light images that reveal sunspots (also known as {\it active regions} in corona images), which are the sites of solar magnetic activity. The magnetic field in the quiet Sun is $\pm5-10$~G while in sunspots it can be well over 1000~G. The average temperature of the photosphere is $\sim5800$~K while sunspots are relatively cooler with temperatures $\sim4500$~K. This lower temperature also means a lower intensity in sunspots and makes them appear darker than the rest of the Sun. Unlike sunspots, {\it faculae} are brighter than the rest of the solar photosphere (Figure~\ref{faculae}). 
This can be explained by reduced plasma density due to the strong magnetic fields conglomerated at the granular lanes. The low density plasma is nearly transparent, and hence allows us to see deeper layers, where the gas is hotter and brighter. At high spatial resolutions the {\it granulation} pattern caused by plasma convection can also be observed. In these features the brighter, middle part of a granule is the rising hot material, while the darker sides are the downward moving cooler plasma.

The majority of solar radiation comes from the photosphere creating a continuous spectrum with dark absorption lines, due to the absorption of radiation in the atmospheric layers above. Most of the absorption lines are formed in the upper photosphere, but one of the most prominent lines - the hydrogen~$\alpha$ (H$\alpha$) - is created in the lower chromosphere, and can be used to determine the temperature, magnetic field strength (based on the Zeemann-splitting of the absorption line) and plasma motions in the line-of-sight direction (using the Doppler broadening of the line).

\setstretch{2}
\subsubsection{The chromosphere}

The {\it chromosphere} was named after its colorful (mainly pink, due to the dominant red H$\alpha$ line) appearance during total solar eclipses. From the bottom to the top of the chromosphere the temperature rises from $\sim6000$~K to $\sim20~000$~K and the density drops by almost a factor of one million. This leads the magnetic field to dominate over gas motions, which can also be described by a parameter known as the {\it plasma beta} given by the equation,
\begin{equation}
\beta=\frac{P_{gas}}{P_{B}}=\frac{n_{e}k_{B}T}{B^{2}/8\pi},
\label{plasmabeta}
\end{equation}
where $P_{gas}$ is the gas pressure, $P_{B}$ is the magnetic pressure, $n_{e}$ is the electron density, $k_{B}$ is the Boltzmann constant, $T$ is the temperature and $B$ is the magnetic field strength.

\begin{figure}[!t]
\centerline{\includegraphics[scale=0.5, trim=0 150 0 150, clip]{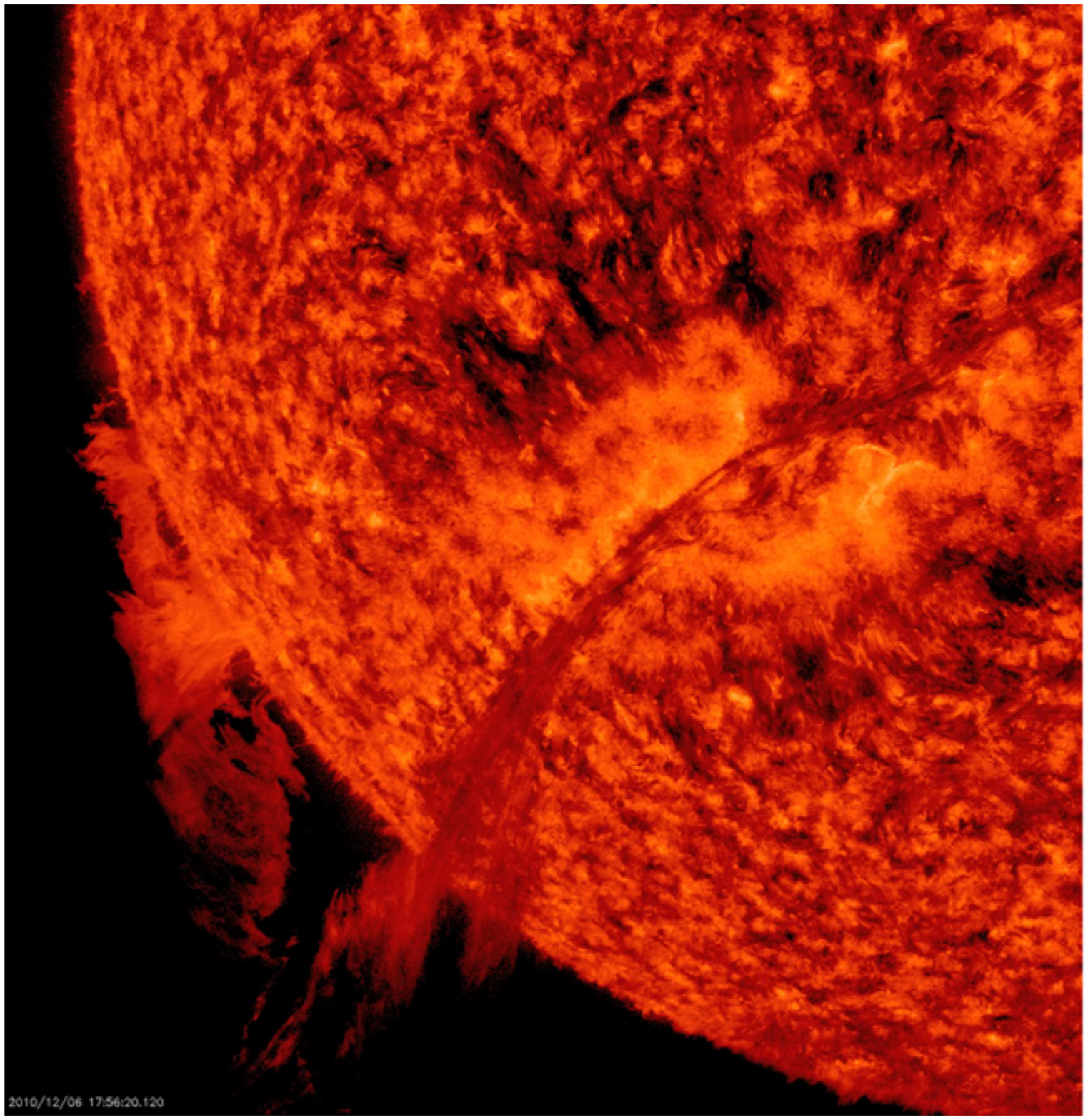}}
\caption{Filament eruption observed by the Solar Dynamic Observatory Atmospheric Imaging Assembly.}
\label{fil_erupt}
\end{figure}

Chromospheric phenomena are best observed at H$\alpha$ and Ca~{\sc ii}~K lines. The most prominent chromospheric features are prominences and filaments, which are essentially the same solar structures appearing as bright formations above the limb and as dark thin lines on-disk, respectively (Figure \ref{fil_erupt}). They are magnetically bound plasma ``clouds" that due to the motion and shearing of the magnetic field lines can build up magnetic stress to the point where magnetic reconnection occurs and the magnetic field restructures through a solar eruption known as a coronal mass ejection (CME). Further interesting chromospheric phenomena are the thinly distributed jets, the {\it spicules}, which are best visible at the solar limb. These jets have velocities of $\sim$25~km~s$^{-1}$ and they rise to an average of 11~000~km before the material falls back. The lifetime of spicules is in the range of minutes \citep{Suematsu08}. 

\subsubsection{The transition region}

The {\it transition region} is a thin and inhomogeneous atmospheric layer ($\sim$100~km) between the chromosphere and the corona, where the temperature rises rapidly from 10$^{5}$~K to over a million. It was previously modelled to be static, but high-resolution imaging has proved it to be a highly dynamic and finely structured region that can be easily disturbed \citep{Doyle06b, Gallagher99TR}. This is due to upward propagating waves and heated plasma flows, as well as cooling downward flows. These dynamic processes make the transition region behave as an energy source and sink with regards to the atmospheric layers above and below. The high temperature gradient in the transition region makes the plasma highly conductive which produces strong emission in the ultraviolet (UV) and extreme ultraviolet (EUV) range \citep{Dowdy86}. 
In a simplified model of the transition region the plasma is confined into filamentary strands of magnetic field with magnetic pressure dominating gas motions over gravity and gas pressure. It is important to note, however, that besides the steep temperature and density gradient, the modelling of the transition region is complicated due to the rapid time variations in the plasma $\beta$ parameter, and the transitions from optically thick to optically thin regimes as well as transitions from partial to full ionisation \citep{Aschwanden04}.

\subsubsection{The corona}

\begin{figure}[!t]
\centerline{\includegraphics[scale=0.5]{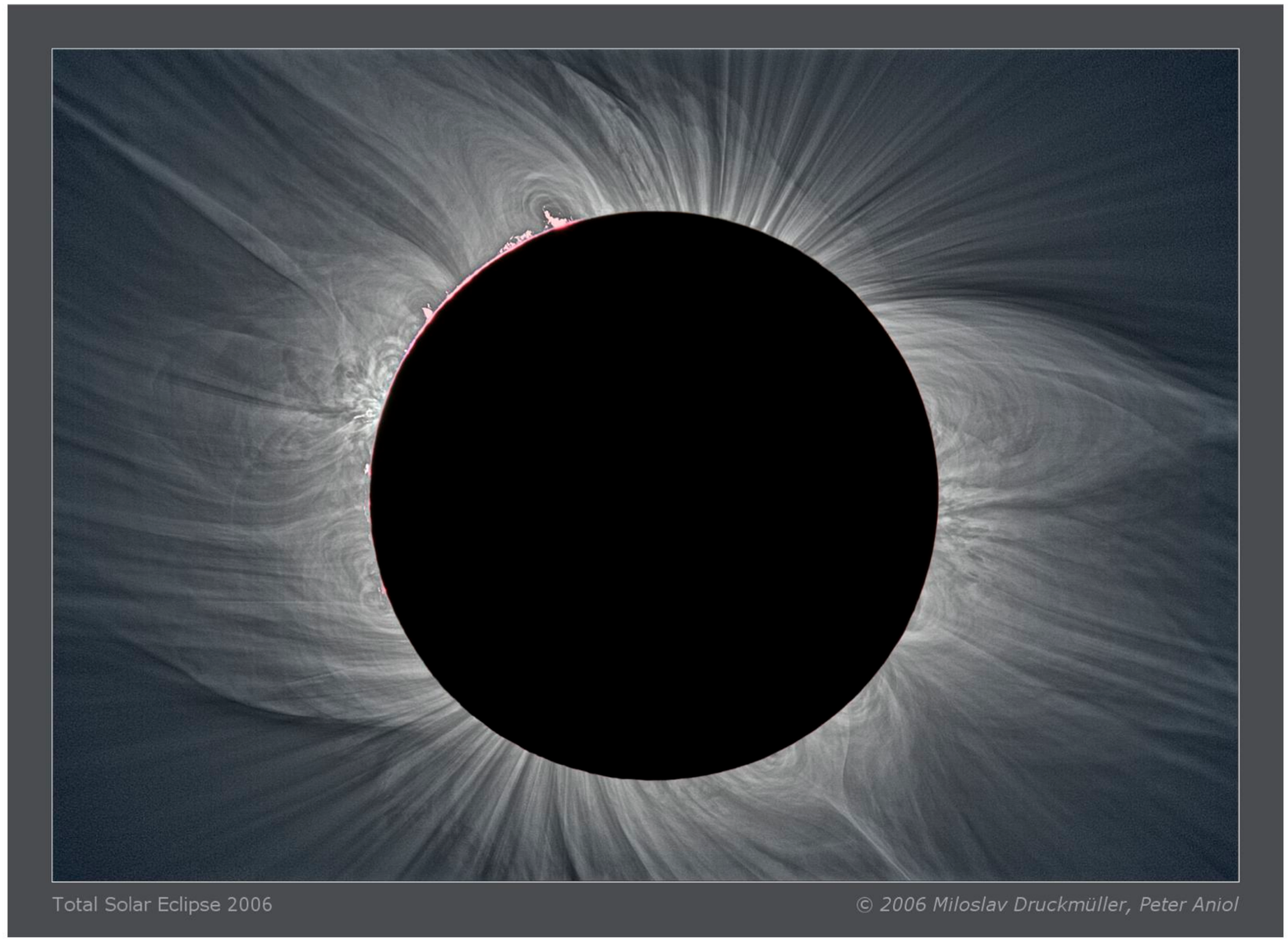}}
\caption{High-resolution image of the solar corona during a solar eclipse observed in 2006. Photo courtesy of Miloslav Druckm\"{u}ller, Brno University of Technology.}
\label{eclipse}
\end{figure}

The {\it corona} is the layer of the solar atmosphere that can only be seen by the naked eye during total solar eclipses or by using coronographs, instruments that block out the bright solar disk allowing the much fainter corona to be visible. The shape of the corona changes according to the time of the solar cycle, i.e. the global magnetic configuration of the Sun. At solar maximum the corona is visible all around the limb due to the shrunken polar coronal holes and the wide latitudinal range at which active regions are located (see Fig.~\ref{eclipse}). During the solar minimum the active regions appear closer to the solar equator and the large open polar magnetic fields (polar coronal holes) cause the bright coronal structures (e.g. streamers) to appear closer to the equator. The corona can also be observed on the solar disk at EUV or X-ray wavelengths using space-borne instruments. These observations reveal structures of million degree temperatures such as loops and active events like filament eruptions (flares, CMEs). Coronal holes can also be observed at these wavelengths due to their low density and reduced brightness compared to the other quiet regions of the Sun. 

\begin{figure}[!t]
\centerline{\includegraphics[scale=0.7, trim=0 410 0 70]{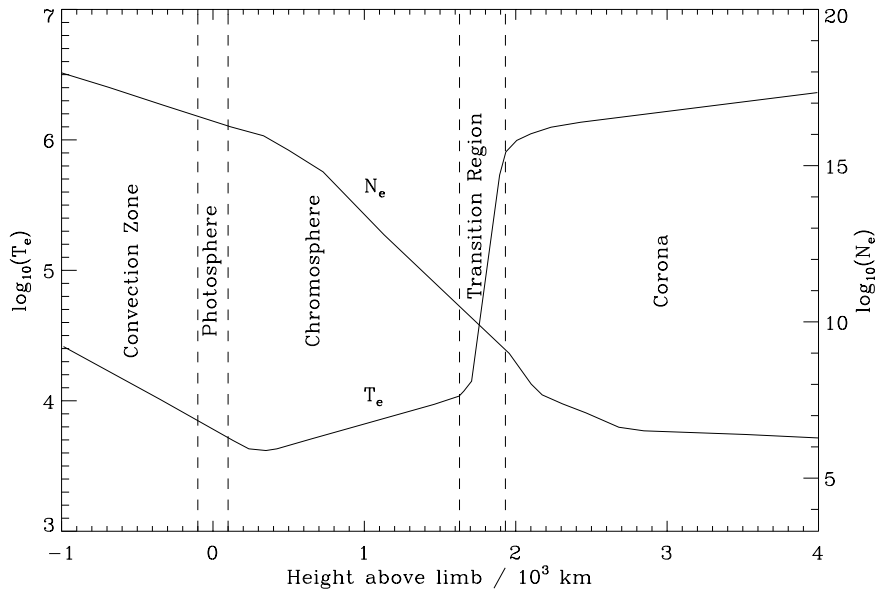}}
\caption{The electron density ($N_e$) and temperature ($T_e$) profile of the solar atmosphere \citep{Gallagher_thesis}.}
\label{atmosphere}
\end{figure}

 As demonstrated in Figure~\ref{atmosphere}, the density in the corona is at least 10$^{10}$ times lower than in the photosphere, while its temperature is vastly larger, reaching over a million degrees. If heating was only possible through thermal conduction, the temperature of the corona would gradually drop off with distance from the $\sim$5800~K solar surface (i.e. the photosphere). However, the corona reaches temperatures in excess of 10$^{6}$~K over a short distance of $\sim$100~km, through the previously mentioned transition region. In order to maintain such high temperatures the corona has to be continuously heated. Although it is not yet confirmed what mechanism(s) is responsible for the high temperature of the corona, there are several theoretical approaches. A number of models suggest that Alfv\'en waves are created during chromospheric magnetic footpoint motions, which deposit their energy by dissipating through magnetic reconnection, current cascade, magnetohydrodynamic (MHD) turbulence or phase mixing \citep{Wang91, Suzuki06, Verdini07, Cranmer07, Verdini09}.  
In these models the difference between slow and high-speed solar wind velocities are explained by differing rates of flux-tube expansion. From an observational point of view, the large number of impulsive small-scale phenomena such as nanoflares and microflares have been shown to produce energy comparable to the radiative loss of the quiet Sun (QS). This is in agreement with another group of heating models, which suggests that the primary source of the corona heating lies in the small-scale magnetic reconnection processes occurring between open and closed field lines in the transition region and lower corona. Here, the difference between slow and fast solar wind is explained by the differing rates of flux emergence, reconnection and coronal heating in different regions of the Sun \citep{Fisk03, Schwadron03, Woo04, Fisk06}. The high temperature of the corona also means that the corona plasma mainly consists of electrons, protons and highly ionised heavy elements (e.g. Fe~{\sc xii}, Ca~{\sc xv}, Mg~{\sc x} etc.), that continuously stream into interplanetary space due to the large pressure gradient as a function of distance from the Sun. This continuous flow of solar particles is the {\it solar wind}. 
 
In Sections~\ref{hydro} and \ref{sw} we discuss the theoretical models that describe the vast extent of the solar corona and the formation and heliospheric morphology of the solar wind. 

\section{Hydrostatic model of the corona}
\label{hydro}

There are many different temporal and spatial scales to solar activity (i.e. solar cycle, flares and CMEs), nevertheless the solar corona can be approximated as stable on a large scale which allows us to understand how the corona can extend into the heliosphere as predicted by Chapman and Parker in the 1950s.

In 1957 the basic model of a static solar corona was described by \cite{Chapman57} by assuming a static atmosphere where the energy transfer was purely by conduction. For simplicity, the lower boundary of the corona was set to be at $r_{0}\simeq 1R_{\odot}$ above the solar surface. As the corona was suggested to be highly thermally conductive, and hence the equation 
\begin{equation}
\kappa=\kappa_{0}{T_{e}}^{\frac{5}{2}}
\end{equation}
was used to describe thermal conduction \citep{Spitzer62}, where $T_{e}=10^6$~K is the plasma electron temperature and $\kappa_{0}=8\times10^{-7}$~W~K$^{-1}$~m$^{-1}$ is the thermal conductivity at the lower boundary. As there is no particle flow in the assumed hydrostatic equilibrium, the temperature gradient yields heat flux given by
\begin{equation}
q=-\kappa \frac{dT}{dr},
\end{equation}
where we also require $\nabla q=0$, as we assume that there are no energy sinks or sources. By assuming spherical symmetry, the divergence of the heat flux yields
\begin{equation}
\frac{1}{r^2}\frac{d}{dr}\left(r^{2}\kappa_{0}{T_{e}}^{\frac{5}{2}}\frac{dT_{e}}{dr}\right)=0.
\end{equation}
We require the coronal temperature to tend to zero at large distances, so the boundary condition will take the form 
\begin{equation}
T_{e}(r)=T_{0}\left(\frac{r_{0}}{r}\right)^{\frac{2}{7}},
\end{equation} 
where $r_{0}\sim7 \times 10^{5}$ km is the lower boundary at the base of the corona.
Chapman assumed the corona to be in hydrostatic equilibrium described by
\begin{equation}
\frac{dp}{dr}=-\rho \frac{GM_{\odot}}{r^{2}},
\label{hydro}
\end{equation}
where $p$ is the pressure, $G=6.67 \times 10^{-11}$ m$^3$~s$^{-2}$~kg$^{-1}$ is the gravitational constant, $M_{\odot}=2 \times 10^{30}$~kg is the mass of the Sun, and $\rho$ is the plasma density that can be given as
\begin{equation}
\rho=n(m_{e}+m_{p})\simeq nm_{p},
\end{equation}
where $\rho$ is the mass density, $n$ is the particle number density and $m_{e}$ and $m_{p}$ are the electron and proton mass. For simplicity a pure hydrogen atmosphere is assumed and thus $m_{e}$ is neglected. Assuming that the electrons and protons in the plasma have the same temperature, the coronal pressure can be written as
\begin{equation}
p=\frac{2nk_{B}T}{\mu}
\end{equation}
where $n\sim2\times10^{14}$~m$^{-3}$, $k_{B}$=1.38$\times10^{-23}$~m$^{2}$~kg~s$^{-2}$~K$^{-1}$ is the Boltzmann constant and $\mu$ is the mean atomic weight (equaling unity for pure hydrogen). As a final step $T_{e}$, $\rho$ and $p$ are substituted into the pressure-balance equation (Equation~\ref{hydro}), which gives
\begin{equation}
p(r)=p_{0}\exp \left\{ \frac{7}{5} \frac{GM_{\odot} m_{p}}{2k_{B}T_{0}r_{0}} \left[ \left(\frac{r_{0}}{r}\right)^{\frac{5}{7}} - 1\right] \right\}
\end{equation}
Note that as $r$ goes to infinity, the coronal pressure tends to a finite constant value ($p_{\infty}=10^{-6}-10^{-10}$~Pa; \citeauthor{Golub97},\citeyear{Golub97}). Although this solution demonstrates the vast extent of the corona, it fails by assuming the solar corona is at hydrostatic equilibrium at large distances. The interstellar gas pressure cannot contain the finite pressure given by the above equation. It was \cite{Parker58}, who pointed out a year later that stellar corona plasmas must go through continuous hydrodynamic expansion to allow for the pressure to drop off with distance and balance with the interstellar gas pressure (see Section~\ref{parker_model}).

\section{Solar wind}
\label{sw}

The solar wind is a continuous stream of electrons, protons and ionised particles originating from the Sun. solar wind flows emanating from QS regions travel at speeds of 300--400 km s$^{-1}$, while streams originating in coronal holes reach speeds up to 800~km~s$^{-1}$. Transient solar eruptions such as flares and CMEs can lead to plasma velocities over 2000~km~s$^{-1}$.

The existence of the continuous particle stream from the Sun was first suggested by \cite{Biermann51} based on his study on the direction of comet tails. After Chapman's model of the extended corona, it was \cite{Parker58} who gave the first physical model to describe the particle outflow from the Sun.

\subsection{Parker solar wind model}
\label{parker_model}

Due to the high density in the solar interior all plasma motions are dominated by gas pressure. In the solar corona the density is at least $10^{10}$ times lower and thus magnetic fields dominate all gas motions. However, at distances over $10^{3}$~Mm the plasma $\beta$ once again becomes larger than one, and hence plasma pressure becomes dominant over magnetic pressure (see Figure~\ref{plasmabeta}). Since the magnetic field lines are ``frozen-in" to the plasma (as explained later, in Section~\ref{spiral_structure}), the expanding plasma carries the magnetic field into interplanetary space and the rotation of the Sun subsequently deforms the magnetic field lines into an Archimedean spiral or {\it Parker spiral} (Section \ref{spiral_structure}). 

\begin{figure}[!t]
\centerline{\includegraphics[scale=0.6, trim=10 220 10 220 , clip]{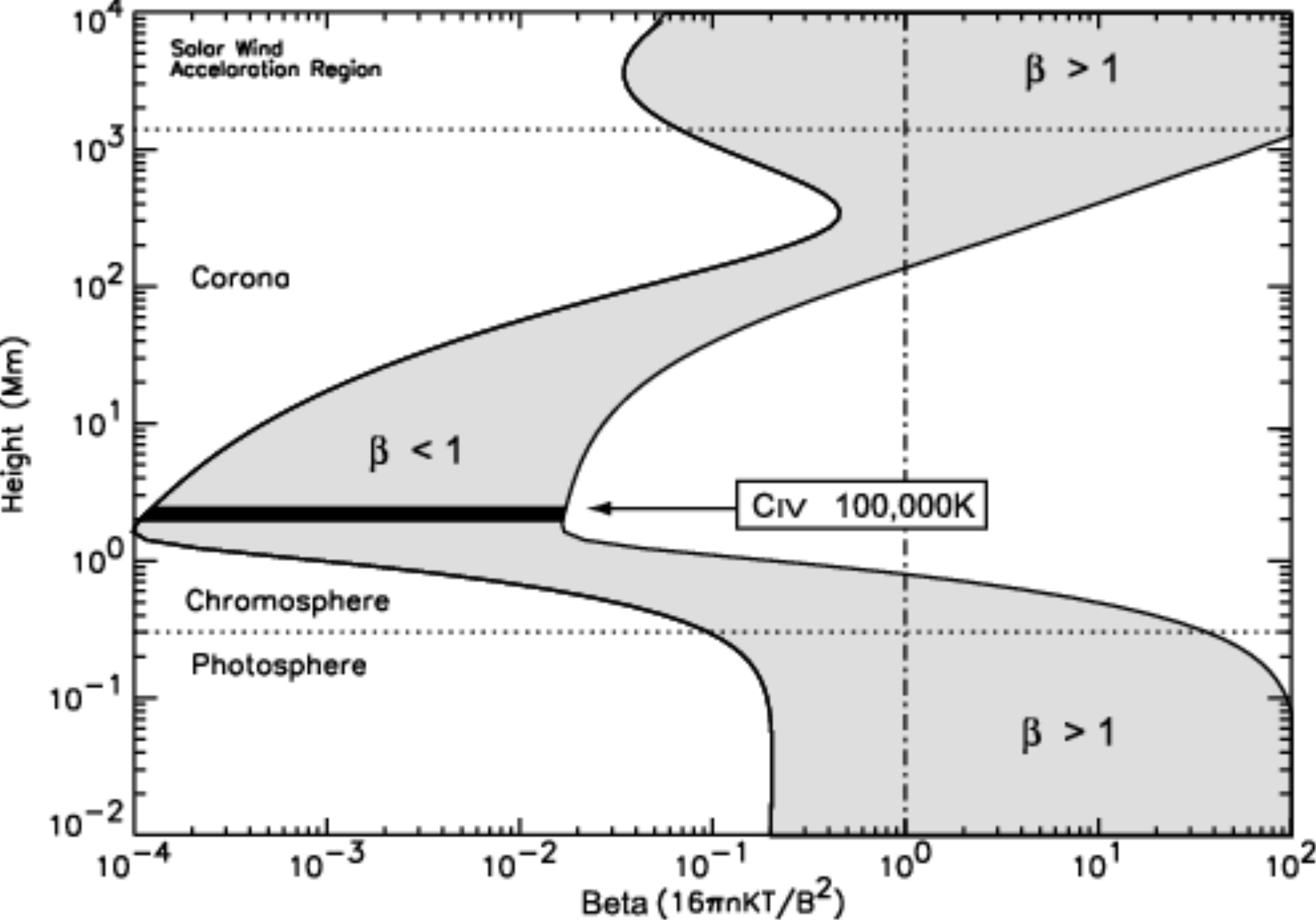}}
\caption{The variation of the plasma $\beta$ parameter as a function of distance in the solar atmosphere \citep{Gary01}.}
\label{plasmabeta}
\end{figure}
In the Parker solar wind model the atmosphere continuously expands and therefore we start with the {\it equation of mass conservation} ($\nabla \cdot(\rho u)=0$), written assuming a spherically symmetric radial outflow:
\begin{equation}
\frac{1}{r^{2}} \frac{d}{dr}(r^{2} \rho u)=0.
\label{mass_cons}
\end{equation}
Next, we consider the {\it momentum equation} ($\rho(u \cdot \nabla)u=-\nabla p + \rho g$, with $g=-GM_{\odot}/r^{2}$), which can be written as
\begin{equation}
\rho u \frac{du}{dr} = - \frac{dp}{dr} - \rho \frac{GM_{\odot}}{r^{2}},
\label{mom_cons}
\end{equation}
where $u$ is the radial solar wind velocity and the first and second term on the right-hand side refer to momentum exerted by gas pressure and gravity, respectively.
To relate the pressure and density, we use the {\it ideal gas law}: 
\begin{equation}
p=\rho RT.
\label{gaslaw}
\end{equation}
Here, we consider an isothermal corona by setting $T$ to be constant. Using the previous equations and the isothermal sound speed ($u_{c}=(p/\rho)^{1/2}$), we deduce the equation of motion,
\begin{equation}
\left(u-\frac{{u_{c}}^{2}}{u}\right)\frac{du}{dr}=2\frac{{u_{c}}^{2}}{r^{2}}(r-r_{c}),
\label{eq_mot}
\end{equation}
where the critical distance ($r_{c}$) is defined as
\begin{equation}
r_{c}=\frac{GM_{\odot}}{2{u_{c}}^2}.
\end{equation}
In Eq.~\ref{eq_mot} if $r=r_{c}$ then either $du/dr=0$ or $u=u_{c}$, and if $u=u_{c}$ then either $du/dr=\infty$ or $r=r_c$. From this it is clear that the critical distance is of great importance, since it is where the plasma velocity reaches the sound speed. At distances smaller than the critical distance the solar wind flow velocity remains subsonic, while at distances larger than the critical distance the flow is supersonic. Substituting typical lower corona temperatures ($T=1-2\times10^6$~K) into $u_{c}=(p/\rho)^{1/2}=($R$T)^{1/2}$ leads to a critical velocity of 90-130~km~s$^{-1}$. This leads to a critical distance of $\sim$5.7-11.5~R$_{\odot}$. Note the dependence of the critical values on the corona temperature.

\begin{figure}[!t]
\centerline{\includegraphics[scale=0.5, angle=-90, trim= 0 0 0 100, clip]{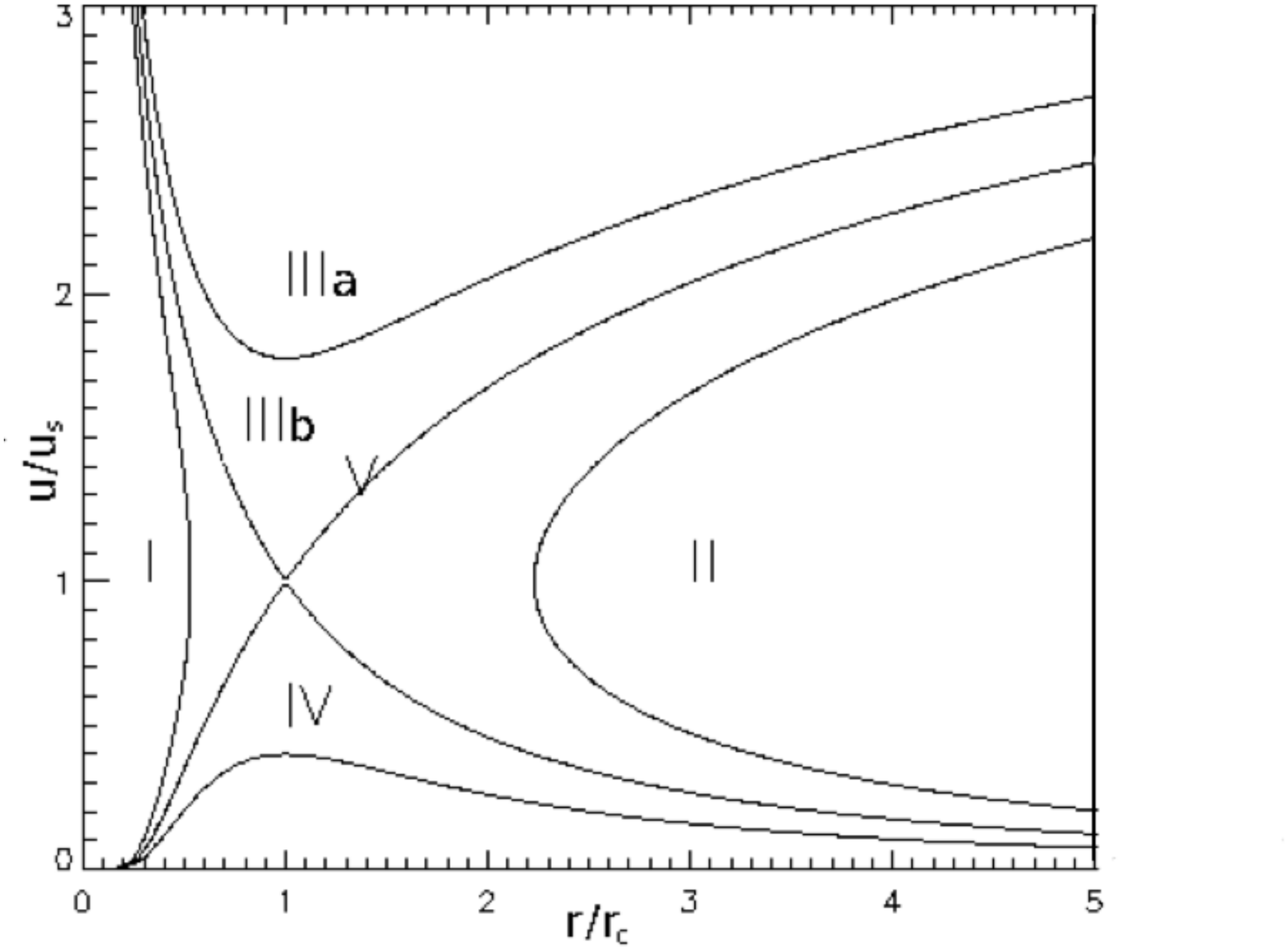}}
\caption{The five classes of solutions of the Parker solar wind model.}
\label{parkersolutions}
\end{figure}

To get the {\it Parker equation} we integrate Equation~\ref{eq_mot}:
\begin{equation}
\left(\frac{u}{u_{c}}\right)^{2}-log\left(\frac{u}{u_{c}}\right)^{2}=4log\left(\frac{r}{r_{c}}\right)+4\frac{r_{c}}{r}+C
\label{parker_eq}
\end{equation}
where C is a constant. There are five classes of solutions to the Parker equation (see Figure~\ref{parkersolutions}). Class I is double valued and indicates a solar plasma that does not extend into the heliosphere, while solution II does not connect to the base of the corona. Therefore, these solutions are considered unphysical. Solutions IIIa and IIIb can be ruled out as well, since they suggest supersonic flows at the base of the corona, which is not acceptable as such high velocity flows would cause shocks in the denser lower corona.
Classes IV and V are acceptable with their sub-sonic flows at the coronal base, but they predict very different flow velocities as $r$ goes to infinity. At large distances the Class IV solution gives a subsonic flow ($u<u_c$; aka "solar breeze solution"). Considering this solution, with $r\rightarrow \infty$ Equation \ref{parker_eq} can be reduced to the form 
\begin{equation}
-\log\left(\frac{u}{u_{c}}\right)^{2}\simeq 4\log \left(\frac{r}{r_{c}}\right),
\end{equation}
which can be approximated as 
\begin{equation}
\left(\frac{u}{u_{c}}\right)\simeq \left(\frac{r_{c}}{r}\right)^{2}
\end{equation}
This relation and Equation~\ref{mass_cons} ($\rho u=1/r^{2}$) yields $\rho\sim const/(r_{c}u_{c})$, and $p\sim const$ (Equation~\ref{gaslaw}) for $r\rightarrow \infty$. This shows that solution IV is unphysical as it gives a finite density and pressure at large $r$ values, which cannot be contained by the extremely small interstellar pressure ($\simeq10^{-13}$~Pa; \citeauthor{Suess90}, \citeyear{Suess90}).

Assuming  $u\gg u_c$ (Class V solution), the solar wind flow becomes supersonic beyond the critical point and with $r \rightarrow \infty$ Equation \ref{parker_eq} reduces to
\begin{equation}
\left(\frac{u}{u_{c}}\right)^{2}\simeq 4\log \left(\frac{r}{r_{c}}\right)
\end{equation}
which leads to
\begin{equation}
\left(\frac{u}{u_{c}}\right)\simeq 2\left(log\left(\frac{r}{r_{c}}\right)\right)^{1/2}
\end{equation}
Equation \ref{mass_cons} then yields $\rho\sim const/(r^{2}u)$, which leads to $\rho \rightarrow 0$ for $r\rightarrow \infty$. Consequently, $p \rightarrow 0$, which can be matched with the low pressure interstellar medium. For this reason we accept the Class V solution as a physically realistic model. 

Finally, let us calculate the solar wind velocity at the Earth's distance from the Sun. We substitute $r\simeq$~214~R$_{\odot}$, $r_c=10$~R$_{\odot}$ and $u_{c}\simeq90$~km~s$^{-1}$ into the Parker equation (Equation~\ref{parker_eq}) and use the Newton-Rapson method to solve the equation to get $\sim$270~km~s$^{-1}$ as the solar wind velocity at Earth, which is in good agreement with the observed slow solar wind velocities.

\subsection{Heliospheric structure of the solar wind}
\label{spiral_structure}

To construct a realistic geometry of solar wind streams, we have to understand the relationship between the solar wind and the interplanetary magnetic field lines that originate from the Sun. It is well known that the interplanetary magnetic field lines are ``frozen in" to the plasma due to high electrical conductivity. To demonstrate this we use the {\it induction equation}:

\begin{equation}
\frac{\partial B}{\partial t}= \nabla \times (v\times B)+\eta \nabla^{2}B,
\label{frozenin}
\end{equation}
where $v$ is the plasma velocity, $\eta$ is the magnetic diffusion, and the magnetic Reynolds number is defined as $R_{m}\sim 1/\eta$. In case $R_m \rightarrow \infty$ (perfectly conductive fluid), the magnetic diffusion is very small and therefore the induction equation can be written as
\begin{equation}
\frac{\partial B}{\partial t}= \nabla \times (v\times B).
\label{mod_ind}
\end{equation}
This equation is known as the {\it frozen-flux theorem}. Under these circumstances magnetic diffusion becomes very slow and the evolution of the magnetic field is completely determined by the plasma flow. Parker's model has shown that the corona continuously expands, and consequently magnetic field lines are dragged out by the expanding plasma. Therefore, if there was no solar rotation the solar magnetic field would take up a radial configuration, while solar rotation would wind the magnetic field lines up into a spiral. Here, we consider the latter scenario.

\begin{figure}[!t]
\centerline{\includegraphics[scale=0.5, trim=100 80 100 50, clip]{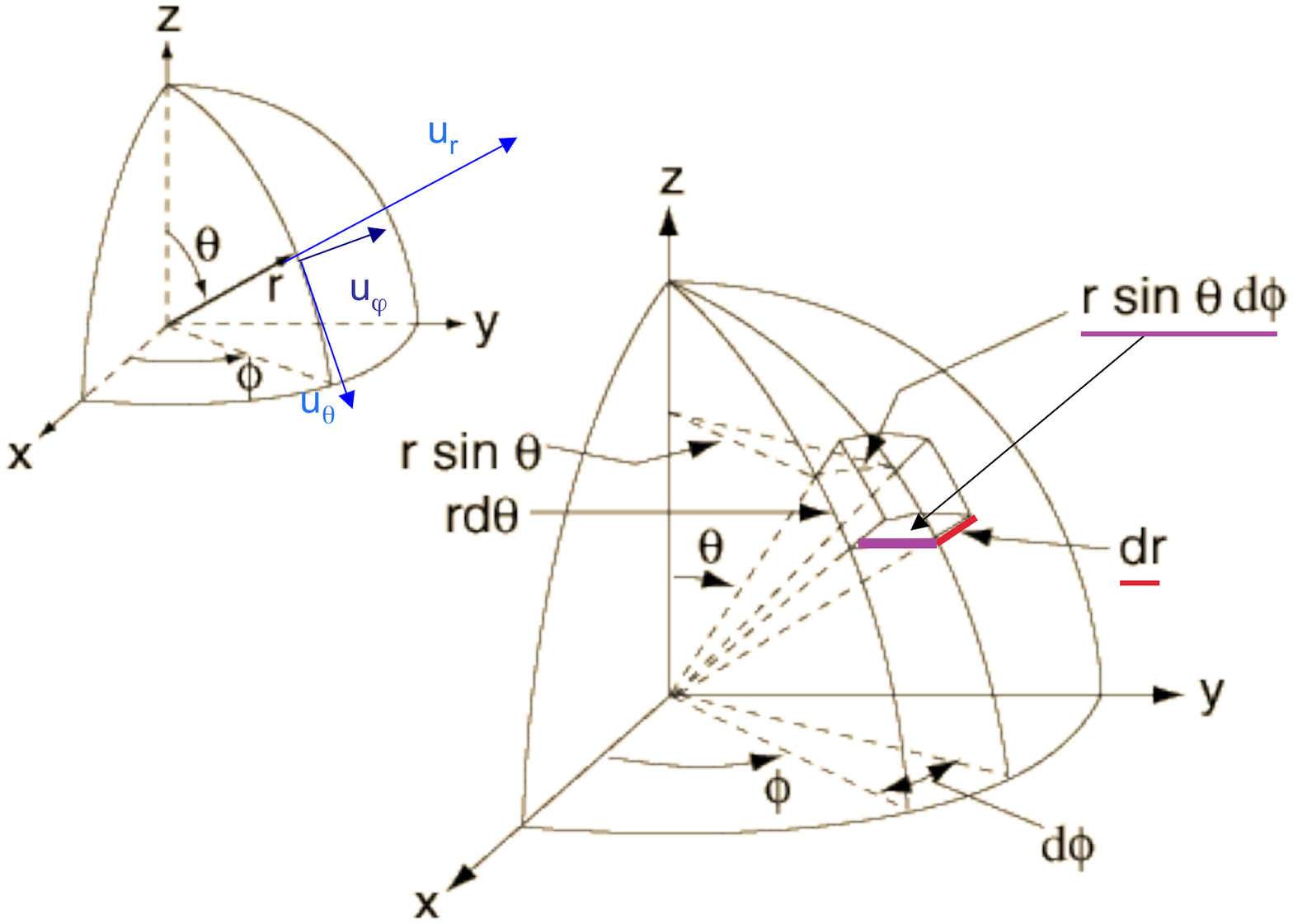}}
\caption{The spherical polar coordinate system with velocity components (left) and the plasma displacement (right).}
\label{spherical}
\end{figure}

Let us take a spherical polar coordinate system $(r, \theta, \phi)$ that rotates with the Sun. The velocity components of the solar wind can be written as $u_{r}=u, u_{\theta}=0,$ and $u_{\phi}=-\Omega r \sin \theta$, where $\Omega= 2.7 \times 10^{-6}$ rad s$^{-1}$ is the solar angular velocity, and $\theta$ is defined so it is 0$^\circ$ at the rotation axis and 90$^\circ$ at the solar equator (Figure \ref{spherical}).

\begin{figure}[!t]
\centerline{\includegraphics[scale=0.45, trim=0 30 0 40, clip]{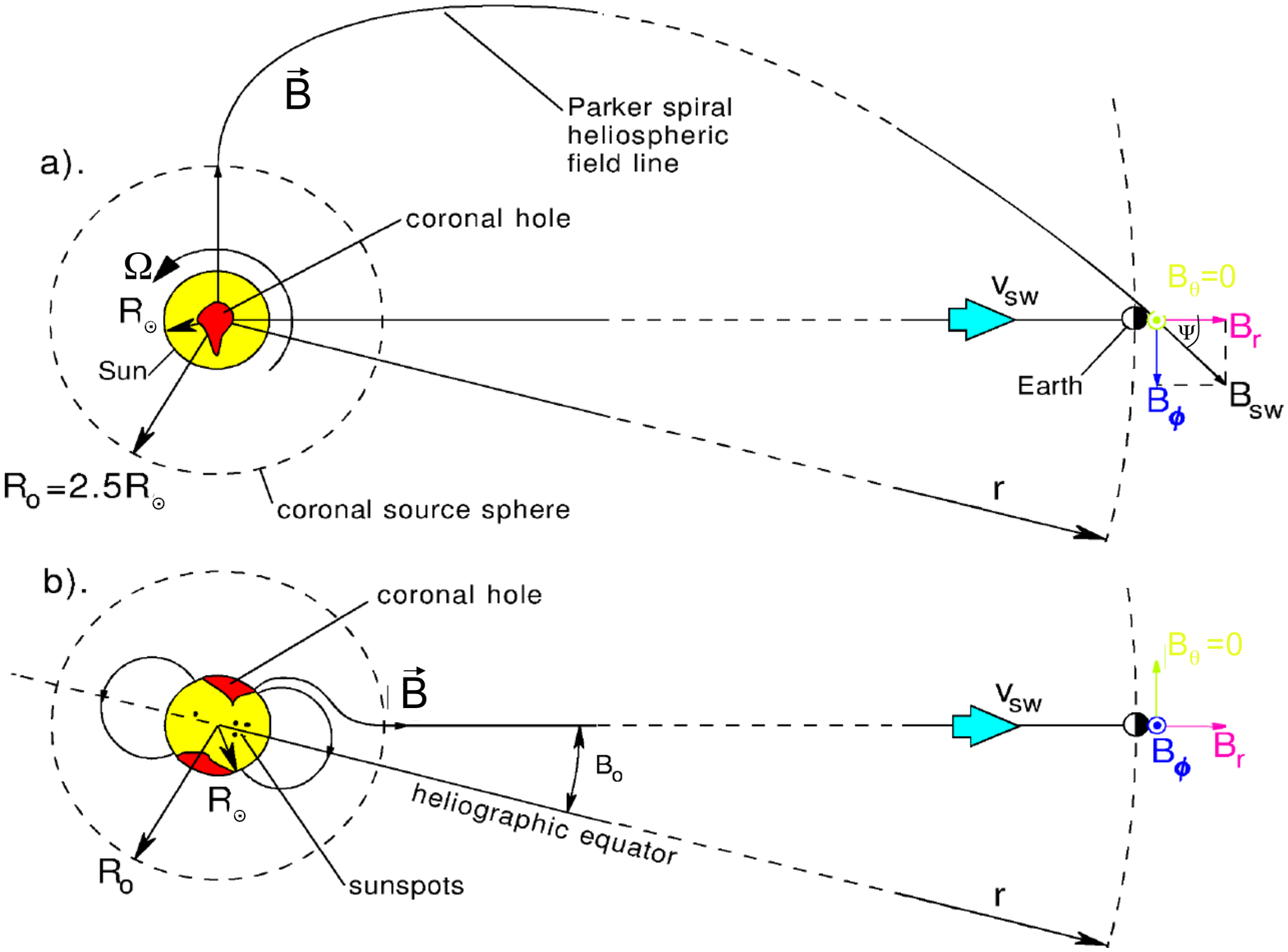}}
\caption{Top: the interplanetary magnetic field from a poleward view, bottom: as viewed from the ecliptic \citep{Lockwood99}.}
\label{IMF}
\end{figure}

From the radial plasma dislocation geometry shown in Figure~\ref{spherical}, we can establish that
\begin{equation}
\frac{r \sin \theta d\phi}{dr}=\frac{u_{\phi}}{u_{r}},
\label{rad1}
\end{equation}
and substituting $u_{\phi}=-\Omega r \sin \theta$ gives
\begin{equation}
\frac{r d\phi}{dr}=-\frac{r\Omega}{u_{r}}.
\label{rad}
\end{equation}
Because the solar wind plasma streams are equivalent to the magnetic field lines, the above equation describes the differential rotation of the solar wind streams in the heliosphere. Integrating Equation~\ref{rad} gives the equation for the Archimedean spiral geometry as described by \citeauthor{Parker58}:
\begin{equation}
r-r_{0}=-\frac{u_{r}}{\Omega}(\phi - \phi_{0}),	
\label{spiral}
\end{equation}
where $u_{r}=const$ for $r \rightarrow \infty$, since at large distances the solar wind speed tends to constant value. Equation \ref{spiral} will be later used in Chapter \ref{chapter:forecasting} to calculate the arrival of high-speed solar wind (HSSW) streams at Earth. To demonstrate the magnetic field configuration in the heliosphere, let us consider Figure~\ref{IMF}. It depicts a magnetic field line deformed into a spiral arm, with a coronal hole as the source of the continuous solar wind stream. As shown in Figure~\ref{IMF}a, the magnetic field is radial out to 2.5~R$_{\odot}$, and is gradually deformed with distance. The distance out to which all ``open" field lines are taken to be radial is commonly referred to as the {\it source surface}, and is used as a boundary condition for magnetic field extrapolations to model the large scale solar magnetic field. (Note, that this distance can be different in different models.) 

The figure shows the magnetic field line being Earth-connected, which is the case when the Earth passes through a high-speed solar wind stream originating from a CH. At Earth, the inclination of the magnetic field line relative to the radial direction is given by the angle $\psi$. Note, that $\psi$ indicates the inclination angle of the magnetic field to the radial direction. It is 0$^\circ$ at the solar surface where the solar wind stream is radial ($B_{sw}=B_{r}$), and becomes larger with distance, as the magnetic field lines become more deformed (see Figure~\ref{IMF}a for reference). 
Figure~\ref{IMF}b shows the same scenario from an ecliptic view, with the heliographic equator being tilted relative to the Earth's orbital plane (this angle is known as $\mathrm{B_{0}}$). Let us now calculate the magnetic field components ($B_{r}=B\cos\psi, B_{\theta}=0, B_{\phi}=-B\sin\psi$). We start with {\it Gauss's law of magnetism} ($\nabla \cdot {\bf B}=0$) written as
\begin{equation}
\frac{1}{r^2}\frac{d}{dr}(r^2 B_{r})=0
\end{equation}
Here, we get $r^2 B_{r}=const$ and we can express the constant as ${r_{0}}^2B_{0}$ so that
\begin{equation}
B_{r}=B_{0}\frac{{r_{0}}^2}{r^2}
\end{equation}
From this we can express $B_{\phi}$ using Equation~\ref{rad1} (where $\tan \psi=-r\Omega \sin \theta/u_{r}$):
\begin{equation}
B_{\phi}=-B_{r}\tan \psi=B_{0}\frac{{r_{0}}^2}{r^2}\frac{r\Omega \sin \theta}{u_{r}}.
\end{equation} 
Considering a single spiral arm at the equator ($\theta=0$ and $\sin \theta = 1$), from the previous equation we get
\begin{equation}
\tan \psi=-B_{\phi}/B_{r}=\frac{\Omega r}{u_{r}},
\end{equation}
\begin{figure}[!t]
\centerline{\includegraphics[scale=0.5, trim=0 50 0 270, clip]{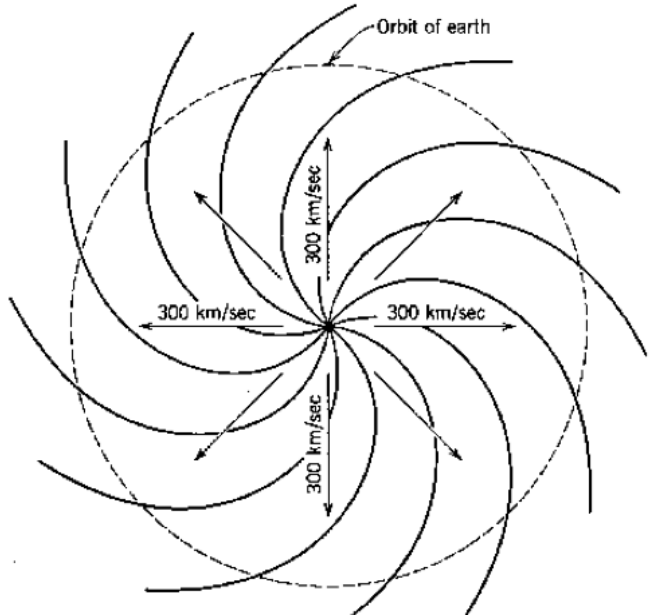}}
\caption{The 2D structure of the Parker spiral.}
\label{parker_sp}
\end{figure}
where substituting $\Omega$, $r$=1~AU$=150\times10^6$~km and $u_{r}\simeq300$~km~s$^{-1}$ (the average slow solar wind speed), at the distance of the Earth, we get $\psi\simeq45^{\circ}$ as the inclination angle of the magnetic field to the radial direction \citep{Goosens03}. This is in good agreement with observations. (However, note that solar wind originating from coronal holes are typically around 500-800~km~s$^{-1}$, and therefore the inclination angle $\psi$ would be less than the calculated 45$^{\circ}$ at Earth.) The spiraling slow solar wind streams are demonstrated in Figure~\ref{parker_sp}.

The spiral geometry changes further according to the location of the {\it current sheet} (the surface separating the positive and negative polarity global magnetic field), which in an idealised case of a perfectly dipole Sun would be a projection of the solar equator separating the northern and southern coronal hemisphere. However, the position of the equatorial current sheet can be modified by active regions (ARs) and coronal holes (CHs) located near the solar equator forcing it to move to higher or lower latitudes. The interaction of low and high-speed solar wind streams emanating from the QS and CHs can also change the spiral structure. The region where the different velocity streams meet produces compressed plasma (the high speed plasma catches up with slow plasma and typically forms a shock), also known as a corotating interactive region (CIR) (see Section \ref{cir}).

\begin{figure}[!t]
\centerline{\includegraphics[angle=-90, scale=0.37, trim=50 30 50 50, clip ]{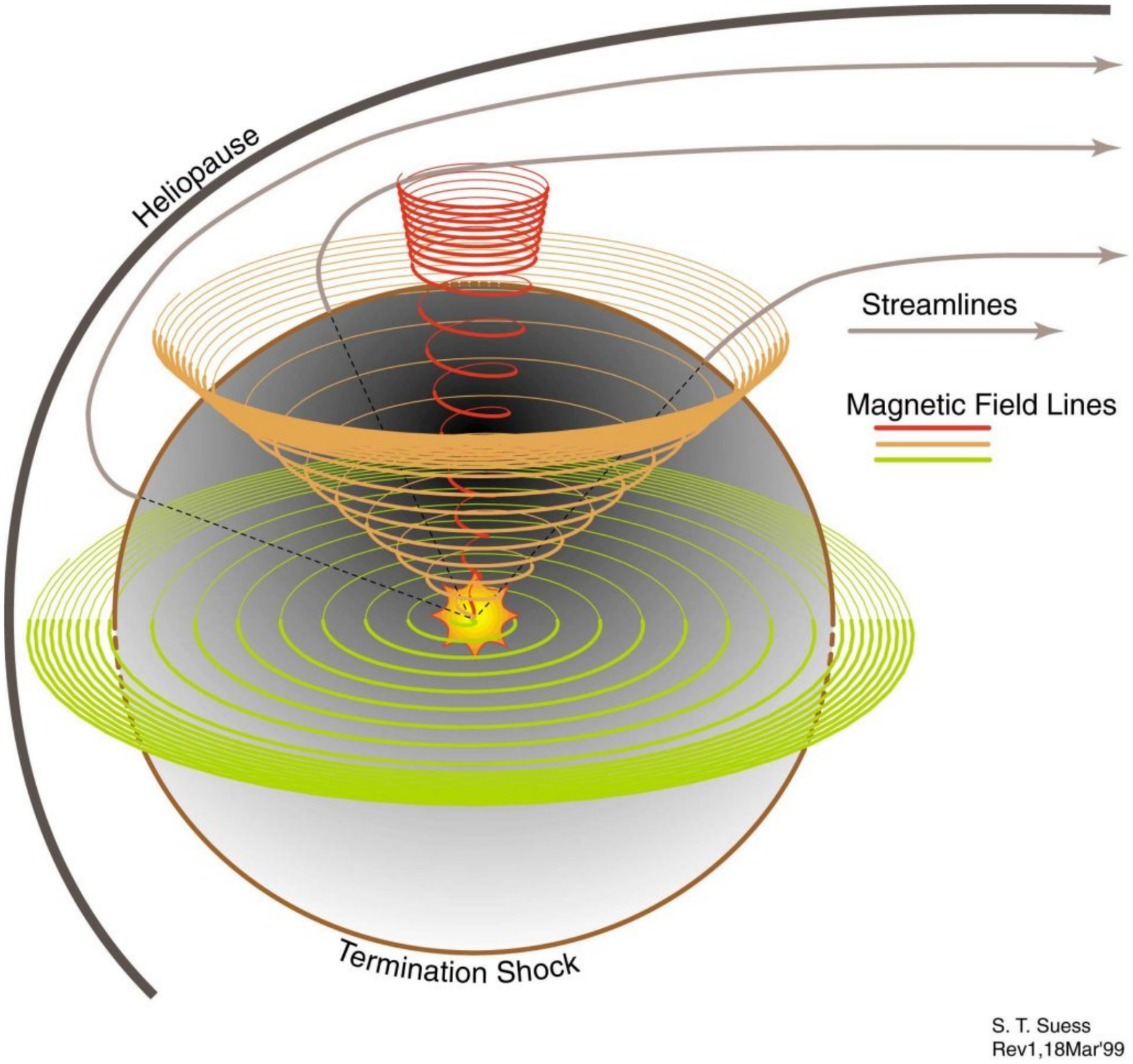}}
\vspace{0.1\textwidth}
\centerline{\includegraphics[scale=0.5, trim=0 150 0 120, clip]{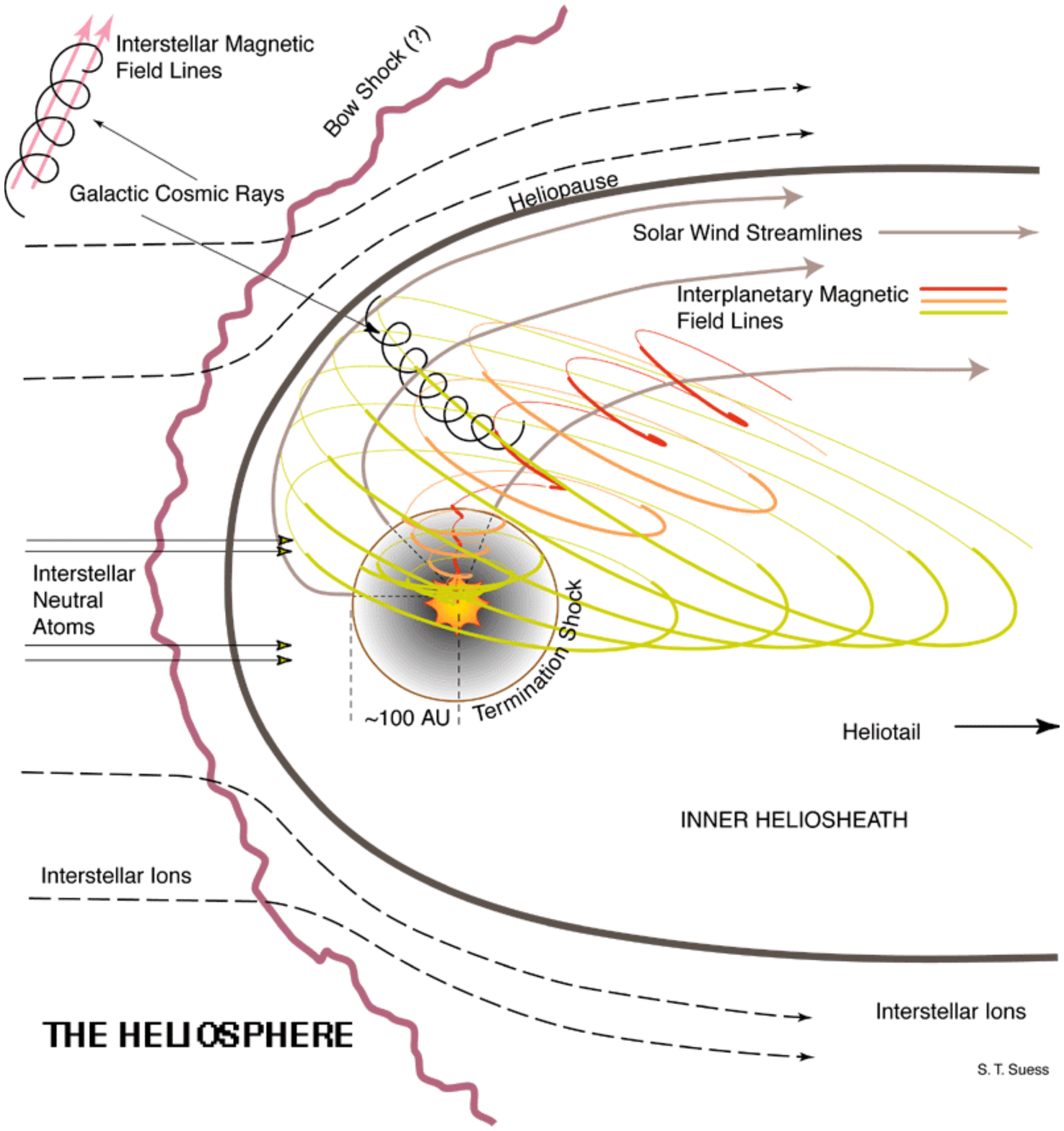}}
\caption{The 3D structure of the heliospheric magnetic field. Figure courtesy of Steve Suess, NASA/MSFC.}
\label{parkerspiral3D}
\end{figure}

The top schematic in Figure~\ref{parkerspiral3D} shows the spiral geometry in 3D, with the magnetic field lines originating at different latitudinal heights. It also shows the {\it termination shock} where the solar wind is met by the interstellar medium, which consequently slows the solar wind down to subsonic velocities. The streaming interstellar medium elongates the solar magnetospheric structure (and thus the interplanetary magnetic field lines), similar to how the solar wind deforms the Earth's magnetosphere. The {\it heliopause} is the region where the solar wind pressure and the interstellar medium pressure balance out. The bottom schematic in Figure~\ref{parkerspiral3D} shows an even larger-scale view of the interplanetary magnetic field (IMF). 
The IMF lines are shown to be further deformed and dragged out by the streaming interstellar medium, and the resulting ``tail" is known as the {\it heliosheath}. However, recent results by the Cassini and IBEX satellites indicate that the long accepted elongated tail-like structure of the heliosphere might be incorrect. It is suggested that the heliosphere is bubble-shaped with the interaction between the solar wind and the interstellar medium occurring in a narrow ribbon \citep{Krimigis09, Frisch10}. 

\begin{landscape}
\begin{figure}[!t]
\centerline{\includegraphics[angle=-90, scale=0.7, trim=50 0 50 0]{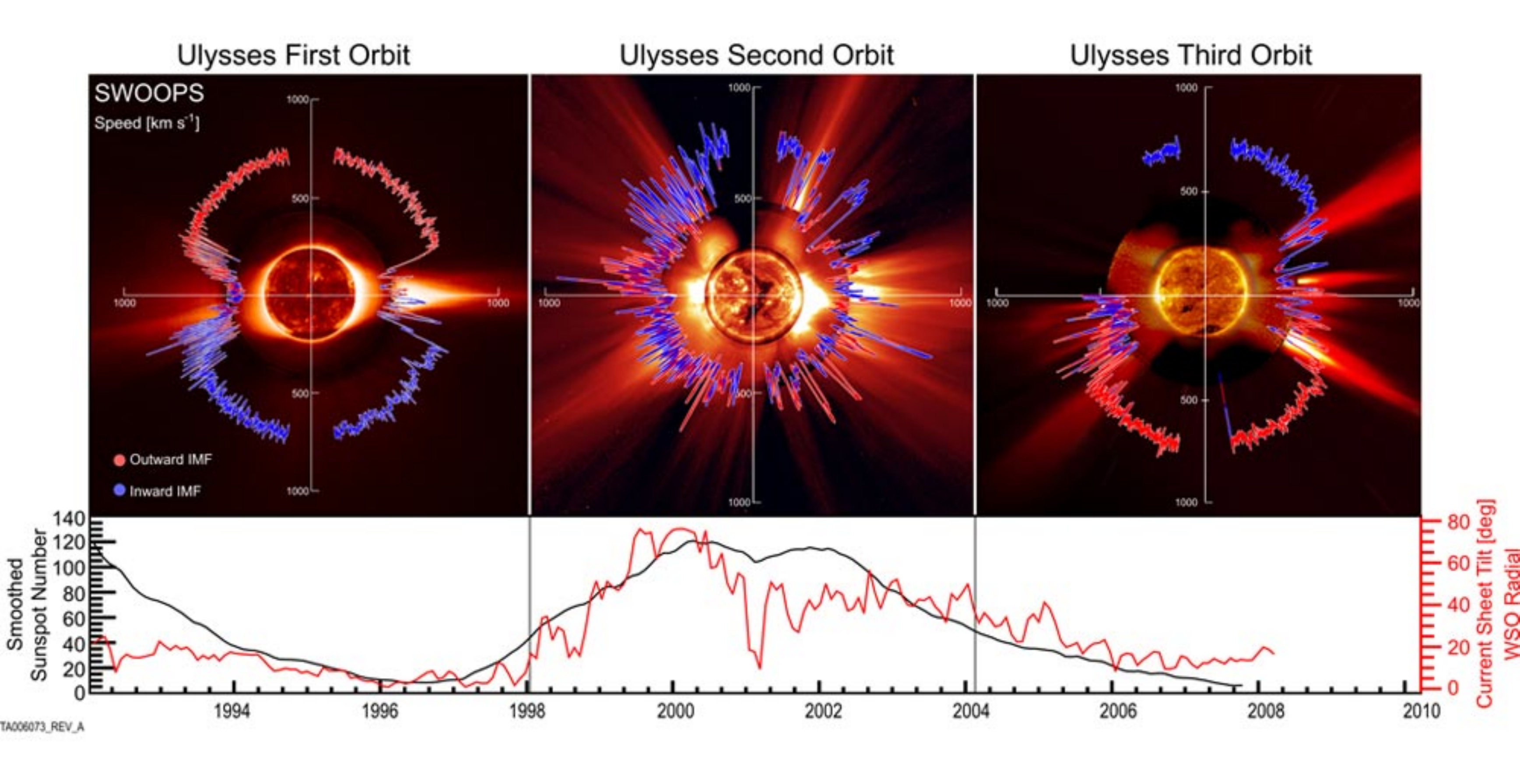}}
\caption{Solar wind speed and polarity variation seen at solar activity minimum (left and right) and maximum (middle) \citep{McComas08}.}
\label{swoops}
\end{figure}
\end{landscape}

Observations by the Solar Wind Observations Over the Poles of the Sun (SWOOPS) instrument aboard the Ulysses satellite have shown that the origin of the high-speed solar wind changes according to the time of the solar cycle. Figure~\ref{swoops} shows the solar wind velocity and the IMF polarity (blue: negative, red: positive) as a function of latitudinal location. The first and third images from the left show the high-speed velocity wind to dominate high latitudes and having distinctly separate polarities. 

This is in good agreement with the dipole structure of the global solar magnetic field during solar minima, a time when the solar poles are dominated by large polar CHs (sources of high-speed solar wind). During solar maximum , the polarities and the slow and high-speed solar wind sources are mixed, which can be explained by the multipolar structure of the solar magnetic field, the large number of low latitude CHs, and the generally enhanced activity of the Sun. The diagram at the bottom of Figure~\ref{swoops} shows the sunspot number (the indicator of solar activity) corresponding to the SWOOPS observation periods.

\section{Coronal Holes}

\begin{figure}[!t]
\centerline{\includegraphics[angle=-90, scale=0.45, trim=0 300 0 300]{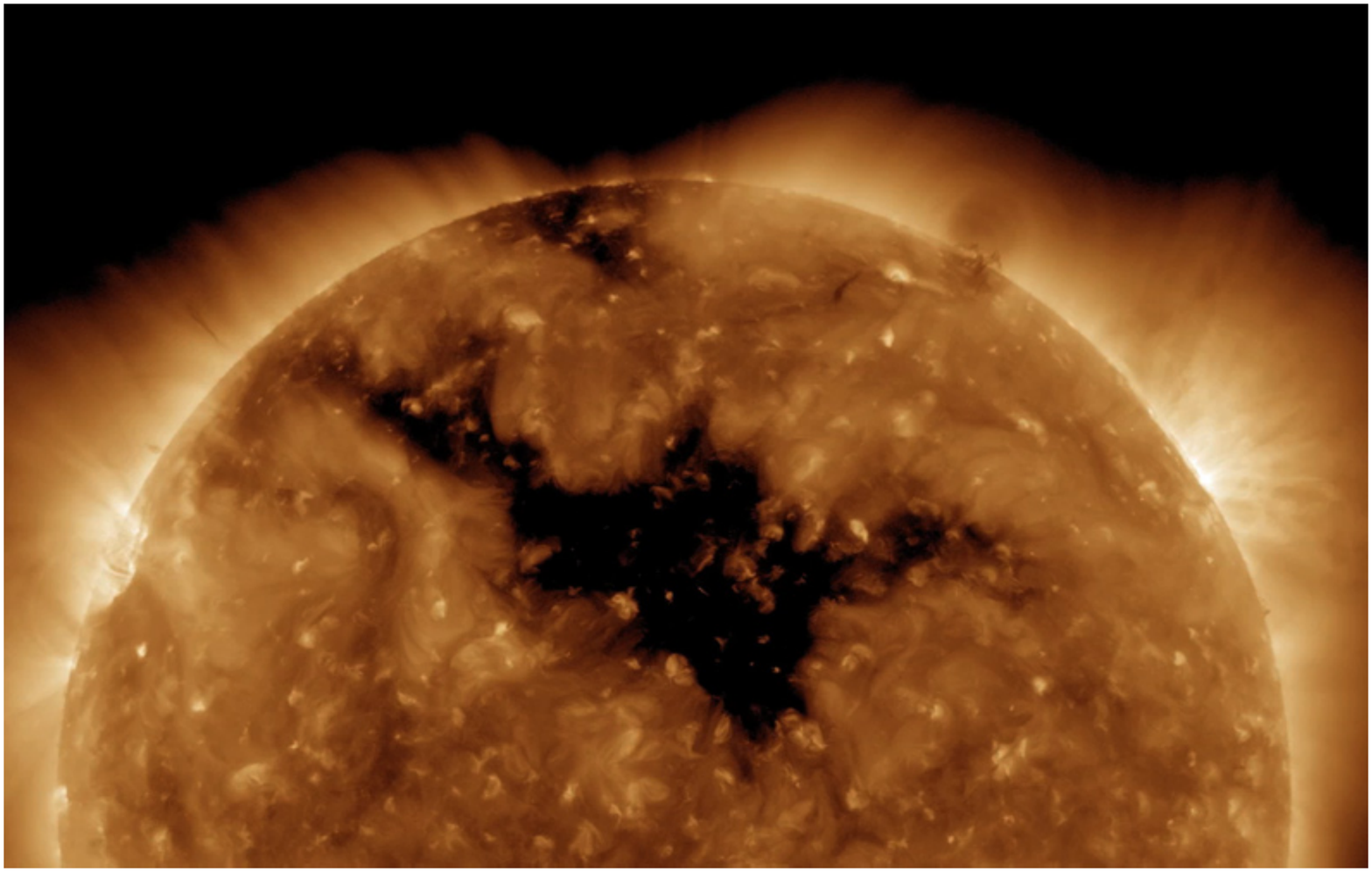}}
\caption{Coronal hole observed by the SDO/AIA instrument at 193~\AA.}
\label{ch}
\end{figure}

Due to their lower coronal temperature and density, CHs have reduced emission at X-ray and extreme ultraviolet wavelengths, and hence can be identified as dark regions \citep{Altschuler72, Vaiana76, delZanna97, Chapman02}. In Figure~\ref{ch} a CH is shown in a high spatial resolution image taken by the Solar Dynamic Observatory Atmospheric Imaging Assembly (SDO/AIA). 

\begin{figure}[!t]
\centerline{\includegraphics[scale=0.6, trim=0 300 0 300]{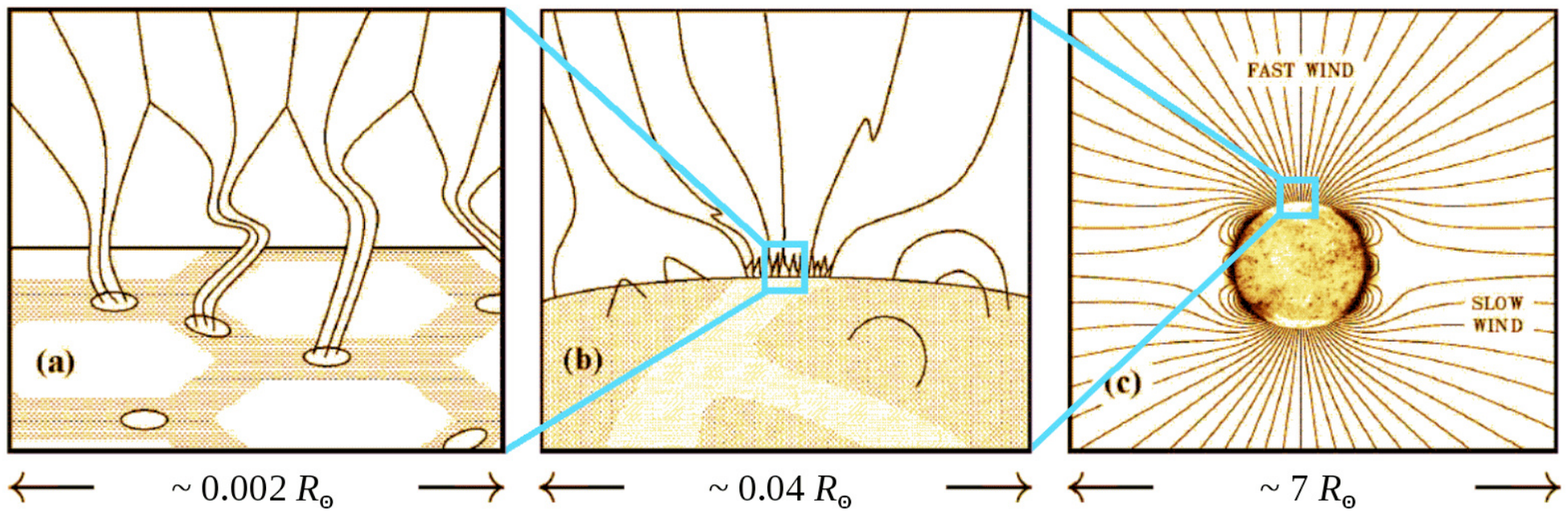}}
\caption{A schematic of the open flux structure in CHs \citep{Cranmer09}.}
\label{ch_funnel}
\end{figure}

Besides CH plasma properties being different from those of QS regions, their magnetic configuration is fundamentally different too. QS regions are dominated by small and large closed magnetic fields (aka {\it magnetic carpet}), while CHs are regions of conglomerated unipolar open magnetic field. The open field lines are brought up to the solar surface and are congregated by the convection occurring below the photosphere. When observing the origins of the CH open field lines at the photospheric level it is clear that they congregate at intergranular lanes as shown in the schematic in Figure~\ref{ch_funnel}. 

\begin{figure}[!ht]
\centerline{\includegraphics[ scale=0.45]{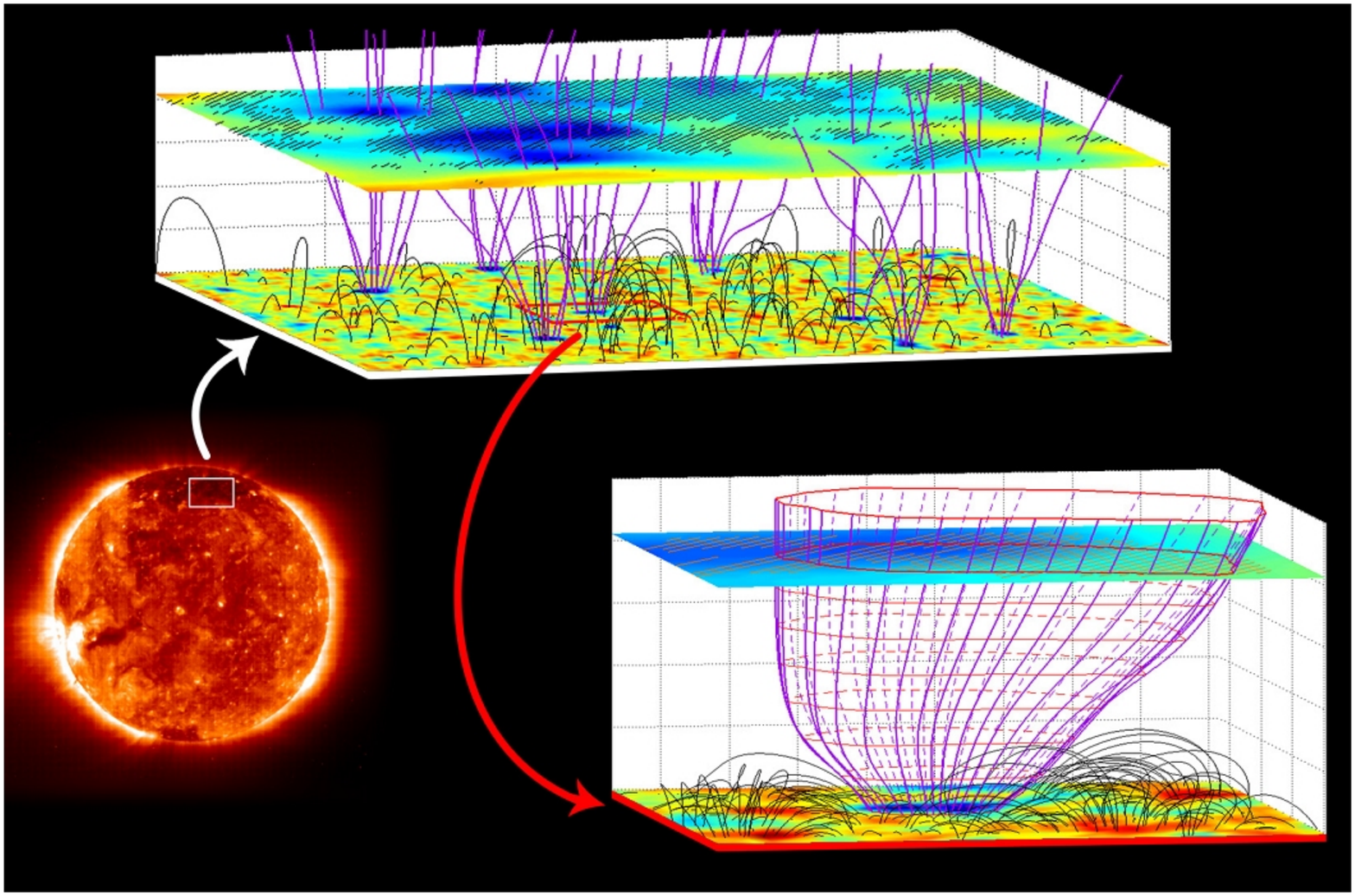}}
\caption{The closed magnetic field of the QS magnetic carpet and the expanding flux tubes of CH regions \citep{Tu05}.}
\label{magcarpet}
\end{figure}
As the density and pressure drops with height above the photosphere the open magnetic flux tubes expand and fill out to create what is observed as a CH at corona heights \citep{Cranmer05,Tu05}. Figure~\ref{magcarpet} shows the mentioned magnetic carpet as well as the open magnetic flux concentrations expanding with height to form a polar coronal hole. The CH related magnetic field concentrations in the photosphere have a predominant polarity, a characteristic often used to classify low-intensity regions as CHs.

It is debated whether the terminology of ``open" field is correct, since the field lines do connect back to the Sun despite their far reach into the heliosphere. Also, CHs are not entirely made up of open field lines (see Fig.~\ref{magcarpet}), at the transition region level lots of small loops (bipoles) can be observed \citep{Wiegelmann05}. These bipoles might have emerged inside the CH or have been admitted through magnetic reconnection occurring at the boundary of CHs. Whether the bipoles remain within the CH or leave the CH over time due to solar rotation is still unclear and requires a more detailed analysis (for more detail see Chapter~\ref{chapter:evolution}). Note the bright structures (bipoles) visible within the CH in Figure~\ref{ch}.

As previously discussed in Section \ref{parker_model} the corona plasma continuously expands into the heliosphere and the magnetic field lines follow the plasma, as explained by the frozen flux theorem. With CHs containing predominantly open field lines there is no constraint in their extension into the heliosphere. Due to the different solar wind acceleration mechanism (e.g. Alfv\'en wave acceleration discussed later in Section~\ref{theory}), the CHs solar wind velocity is higher than that of the QS solar wind. The high-speed solar wind velocities range between 500--800 km s$^{-1}$ \citep{Fujiki05,Schwadron05,Vrsnak07a,Obridko09}. The composition of the CH related HSSW is also very different from that of the slow solar wind and can be easily identified (see Section~\ref{plasma}). While both the slow and high-speed streams take up the spiral structure in the heliosphere, the higher velocity stream is more radial and thus the high-speed plasma runs into the slower stream creating a shock and a CIR (see Section~\ref{spaceweather}). Upon reaching Earth, the CIR gives rise to magnetic disturbances \citep{Schwadron03, Fujiki05, Choi09, He10}. Since CHs are long-lived features with lifetimes from days to over a year, these disturbances are recurrent and can be predicted (Chapter \ref{chapter:forecasting}).

\subsection{Observation of CHs}

\begin{figure}[!t]
\centerline{\includegraphics[scale=0.65, trim=20 34 20 34, clip]{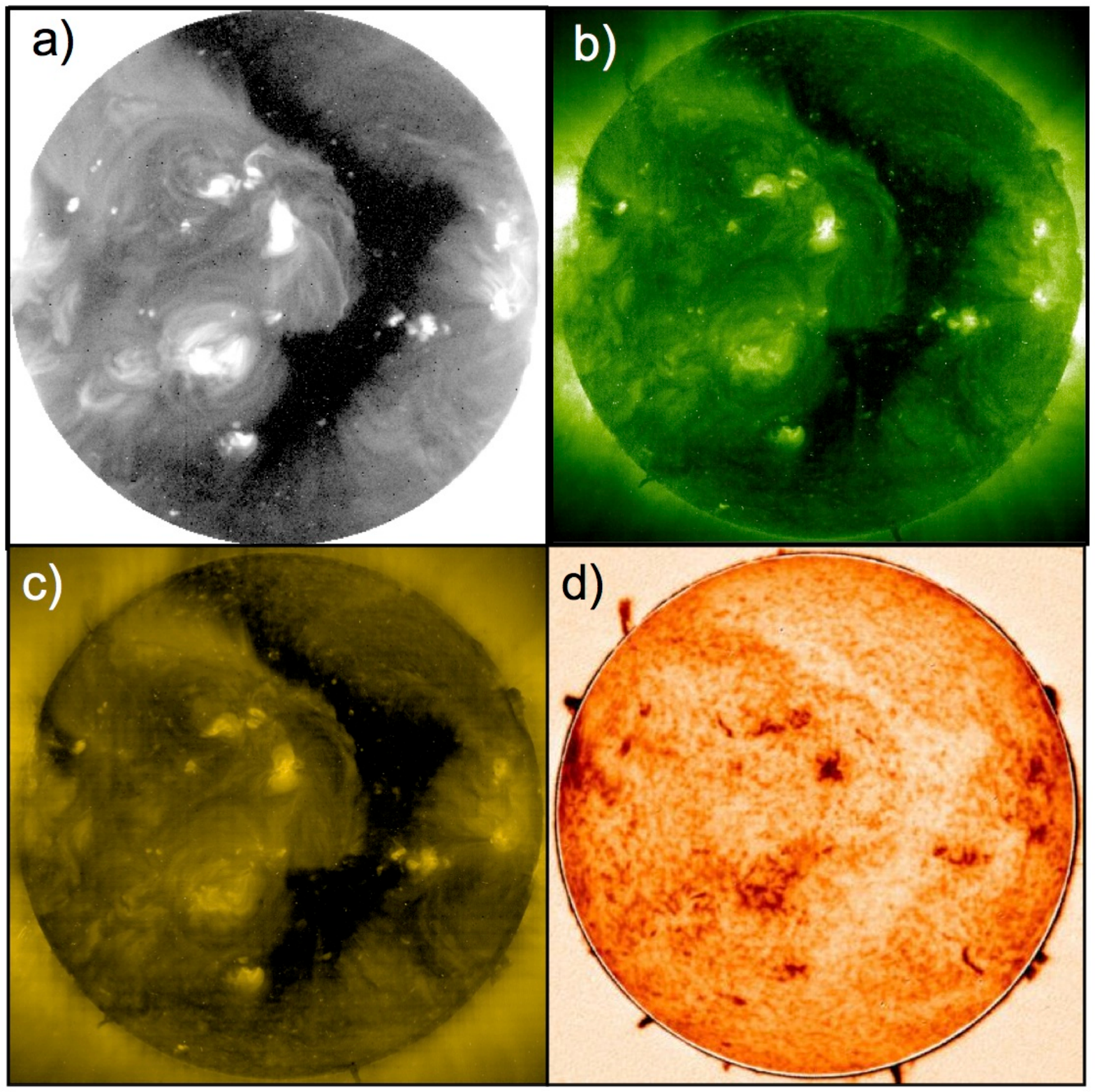}}
\caption{A coronal hole visible in four different wavelengths on 8 December 2000: a) 20~\AA\ (Yohkoh), b) 195~\AA\ (SOHO/EIT), c) 284~\AA\ (SOHO/EIT), d) He~{\sc i} 10830~\AA\ (KPVT).}
\label{4wave}
\end{figure}

The first evidence for CHs was found by E.W. Maunder in 1905, who identified geomagnetic disturbances recurring every 27 days. Knowing the solar rotation period he suggested that the disturbances could be caused by a solar region. In the 1930s using statistical methods on geomagnetic data Julius Bartels also suggested that the recurring geomagnetic disturbances were of solar origin and named them ``M-regions". The first observations to indirectly image CHs were made in the 1940s, but they were only identified in the 1950s when \cite{Waldmeier57} revisited the Fe~{\sc xiv} 5303~\AA\ coronograph images. Waldmeier identified the recurrence of dark regions near the solar limb and derived the shape of the region from limb-width measurements. Based on their low intensity and round shape, he named the regions ``l\"ocher" (the German word for ``holes"). The first on-disk observations of CHs were made in the 1960s and 1970s by X-ray and EUV sounding rockets and orbiting telescopes, and were confirmed to be low-emission regions. Using OSO-4 observations, \cite{Munro72} found CHs to also have a reduced temperature and density compared to the QS. The launch of Skylab in 1973 brought higher quality and quantity soft X-ray observations \citep{Huber74}, and CHs were discovered to be exempt from the differential rotation of the underlying photosphere. CHs can also be observed in He~{\sc i} 10830~\AA\ absorption line images as bright, blurry patches. In the QS, due to the radiation from the overlying corona, the He~{\sc i} 10830~\AA\ transition levels become overpopulated, which leads to increased absorption of radiation and hence QS regions appear darker. In contrast, since CHs have reduced density and hence coronal emission, the He~{\sc i} 10830~\AA\ transition levels are less populated, which leads to less absorption, making CHs appear brighter than the QS \citep{Zirin75, Andretta97}. Figure~\ref{4wave} shows a CH observed on 8 December 2000 at four different wavelengths: at 20~\AA\ using the Yohkoh Soft X-ray Telescope (Yohkoh/SXT), at 195~\AA\ and 284~\AA\ using the SOlar and Heliospheric Observatory Extreme ultraviolet Imaging Telescope (SOHO/EIT), and at 10830~\AA\ using the Kitt Peak Vacuum Telescope (KPVT).
The quality of corona observations improved significantly with the launch of SOHO in 1995, providing 12~minute cadence and 2.6~arcsec/pixel resolution EUV images and magnetograms. A further advance was made in observations with the Hinode X-ray telescope (Hinode/XRT), providing 1 arcsec/pixel resolution and 200~s cadence. EUV observations improved further with the Solar TErrestrial RElations Observatory Extreme Ultraviolet Imager (STEREO/EUVI), which has a 1.6~arcsec pixel resolution and 3~minute cadence. The latter two satellites were both launched in 2006. In 2010 SDO was launched providing the highest quality EUV images to date with a 1~arcsec/pixel spatial, and 10~second temporal resolution. All of the above mentioned instruments allow the visual detection of low intensity regions. However, these regions can only be categorised as CHs once a dominant magnetic polarity is found. 
\\ \indent
In the present thesis data from the SOHO/EIT and the Michelson Doppler Imager (MDI) were used as the former instrument allows the detection of low intensity regions, while the latter can be used to detect a predominant polarity within a region of interest. Furthermore, these instruments have been operational since 1995, providing a remarkable database spanning over a whole solar cycle, which is fundamental in the long-term CH study presented in the thesis.

\subsection{Coronal hole morphology}

The detection of CHs in soft X-ray images made by the Skylab Apollo Telescope Mount (ATM) and the Yohkoh Soft X-ray Telescope (SXT) was straightforward since they appeared as significantly darker regions in comparison to the QS. However, the low spatial resolution and the broad temperature range often lead to the boundary appearing ``blended" into the QS, making its identification difficult. The EUV observations by SOHO/EIT, STEREO/EUVI, SDO/AIA and the X-ray observations by Hinode/XRT have higher spatial resolution and thus provide a better CH boundary contrast. It is to be noted though, that due to the broad temperature range of the EUV and X-ray observations a large number of high temperature loops and other hot material become visible and can still cause line-of-sight obscuration effects at CH boundaries. This effect is reduced when observing CHs close to the solar disk centre, and increased near the limb. Also, during solar minimum this effect can be smaller due to the overall reduced activity of the Sun. He~{\sc i} 10830~\AA\ observations have a much narrower temperature range resulting in little or no obscuration effects at CH boundaries, however, the contrast between QS and CH is much lower making the CH boundary identification more challenging (see Figure~\ref{4wave}). 
\begin{figure}[!t]
\centerline{\includegraphics[ scale=0.50, trim=50 100 50 100, clip]{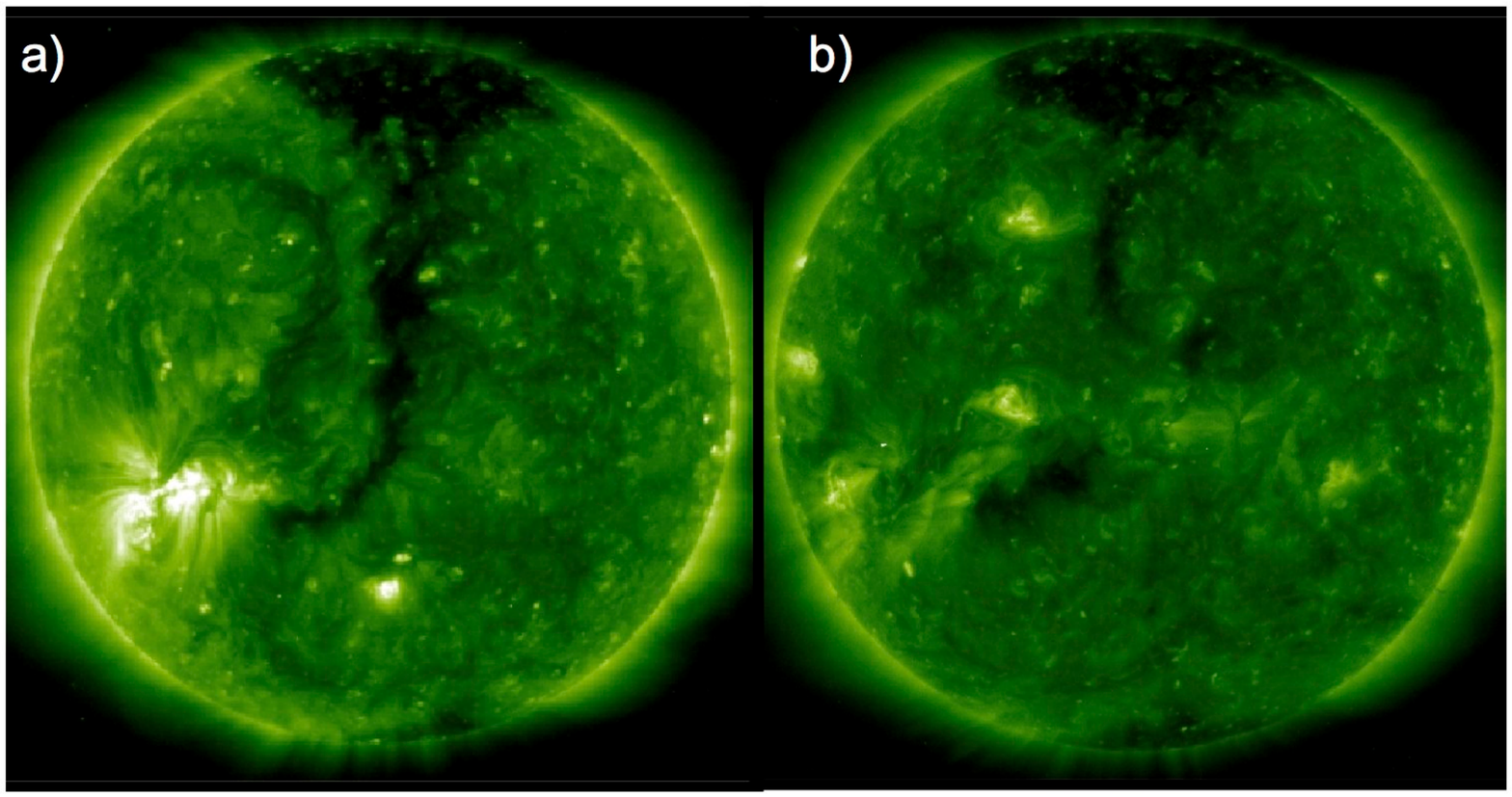}}
\caption{The ``Elephant's Trunk" transequatorial coronal hole, observed on 27 August 1996 (a), and the same region observed several rotations later on 1 Nov 1996, with equatorial CHs (b).}
\label{elephantstrunk}
\end{figure}
 Based on their location and lifetime, CHs are popularly divided into {\it polar}, {\it equatorial} and {\it transient coronal holes}. Some authors debate whether transient CHs is a valid group as they are essentially a by-product of CME eruptions when plasma vacates the region (also known as {\it dimming regions}). However, by definition CHs are regions of conglomerated open magnetic flux of similar polarity as well as reduced density, and as such, dimming regions qualify as transient CHs. 
\\ \indent The most long-lived CHs are the polar holes that start to develop near the time of the pole-reversal from pre-polar holes, that form at high-latitudes from the diffused magnetic flux of decaying active regions. These CHs then extend to form large polar holes once the pole-reversal is complete. At the start of their growing phase polar holes tend to be asymmetric and later develop equatorial extensions (e.g. the well known ``elephant's trunk", see Figure~\ref{elephantstrunk}a). Polar coronal holes areas reach maximum during the solar minimum \citep{Harvey02}. Equatorial CHs (Figure~\ref{elephantstrunk}b) are present throughout the solar cycle but are more frequent and extensive during the solar minimum when the number of ARs is lower \citep{Wang96}.

\subsection{Plasma properties}
\label{plasma}

Apart from low EUV and X-ray emission and a dominant polarity, CH regions can also be detected based on their elemental abundances \citep{Feldman98}. Unlike QS and ARs, CHs exhibit similar abundances in the photosphere, upper chromosphere, transition region and the lower corona, while QS regions and ARs show significant enhancements in the abundance of low first ionisation potential (FIP) elements. It is commonly accepted that in the upper chromosphere low-FIP elements become ionised and high-FIP elements remain more neutral. Since CHs are regions with temperatures typically lower than the QS and are dominated by open magnetic field, the FIP abundances found near the photosphere remain unchanged at higher atmospheric layers. The differing abundances of QS related slow solar wind and CH related HSSW can be detected at 1~AU when the HSSW streams sweep past Earth \citep{vonSteiger95, Zurbuchen02}. It is not yet agreed upon what processes cause this preferential ionisation in the chromosphere.  

\begin{figure}[!t]
\vspace{-3mm}
\centerline{\includegraphics[scale=0.4]{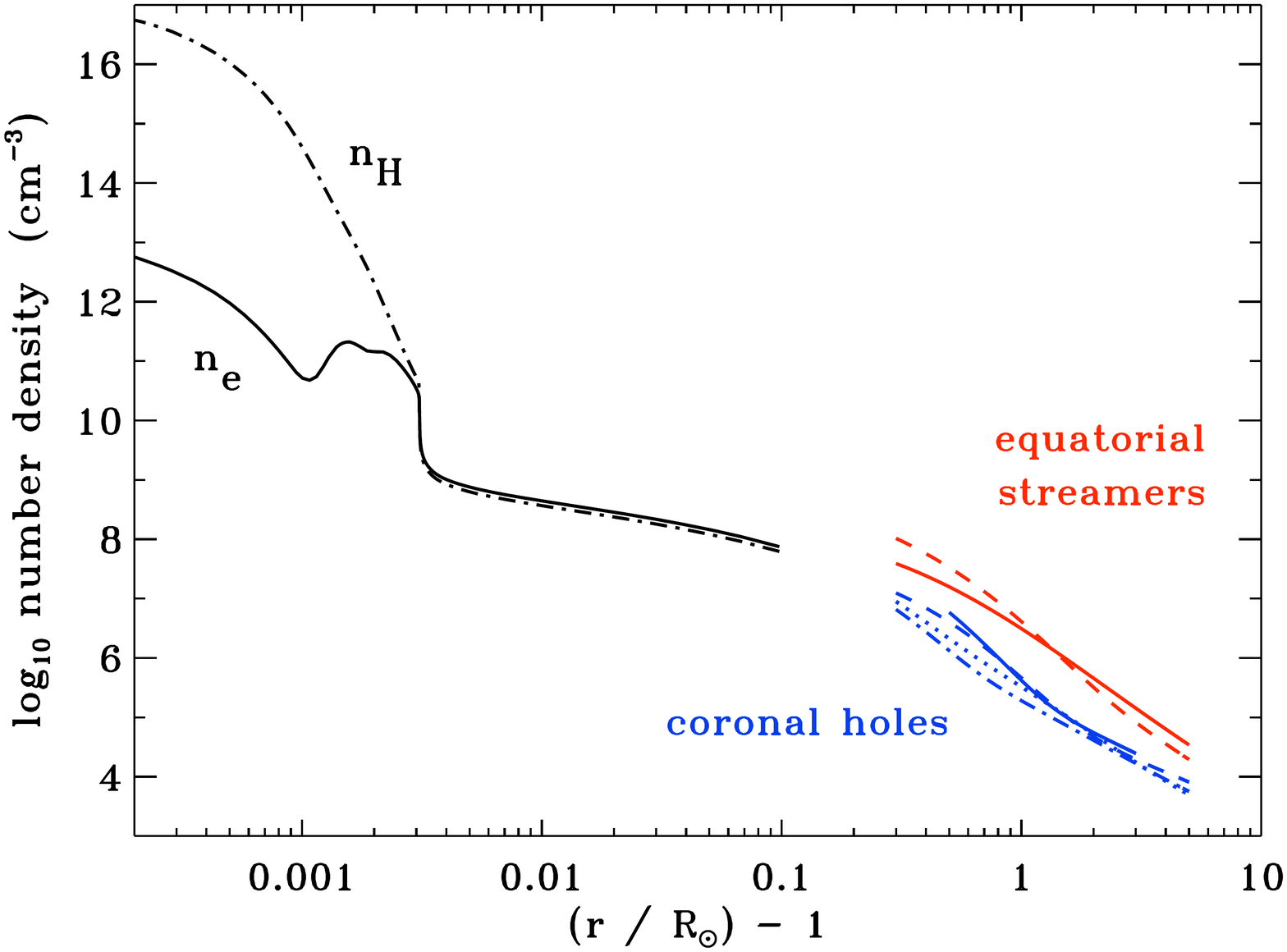}}
\caption{ The empirically acquired densities in the chromosphere, transition region, and low corona. The solid black line shows the electron number density, while the dot-dashed black line corresponds to the total hydrogen number density \citep{Avrett08}. Visible-light electron number densities for CHs and streamers (blue and red curves, respectively) are shown for comparison \citep{Cranmer09}.}
\label{ch_dens0}
\end{figure}

Figure~\ref{ch_dens0} shows a semi-empirical model of the electron density (solid black line) and total hydrogen number density (dot-dashed black curve) in a polar coronal hole (PCH) and streamers at the atmospheric heights of the chromosphere, transition region, and low corona \citep{Avrett08}. The blue curves are visible-light electron CH density measurements \citep{Fisher95, Guhathakurta99a, Doyle99, Cranmer99b}. The red curves correspond to visible-light measurements of equatorial helmet streamers \citep{Sittler99, Gibson99}. It must be noted that although the streamers were measured to be about a factor of ten denser, a factor of 2-3 variation is common due to the different number of polar plumes projected in the line-of-sight measurements \citep{Cranmer09}.
\begin{figure}[!t]
\vspace{-3mm}
\centerline{\includegraphics[scale=0.35, trim=0 150 0 150, clip]{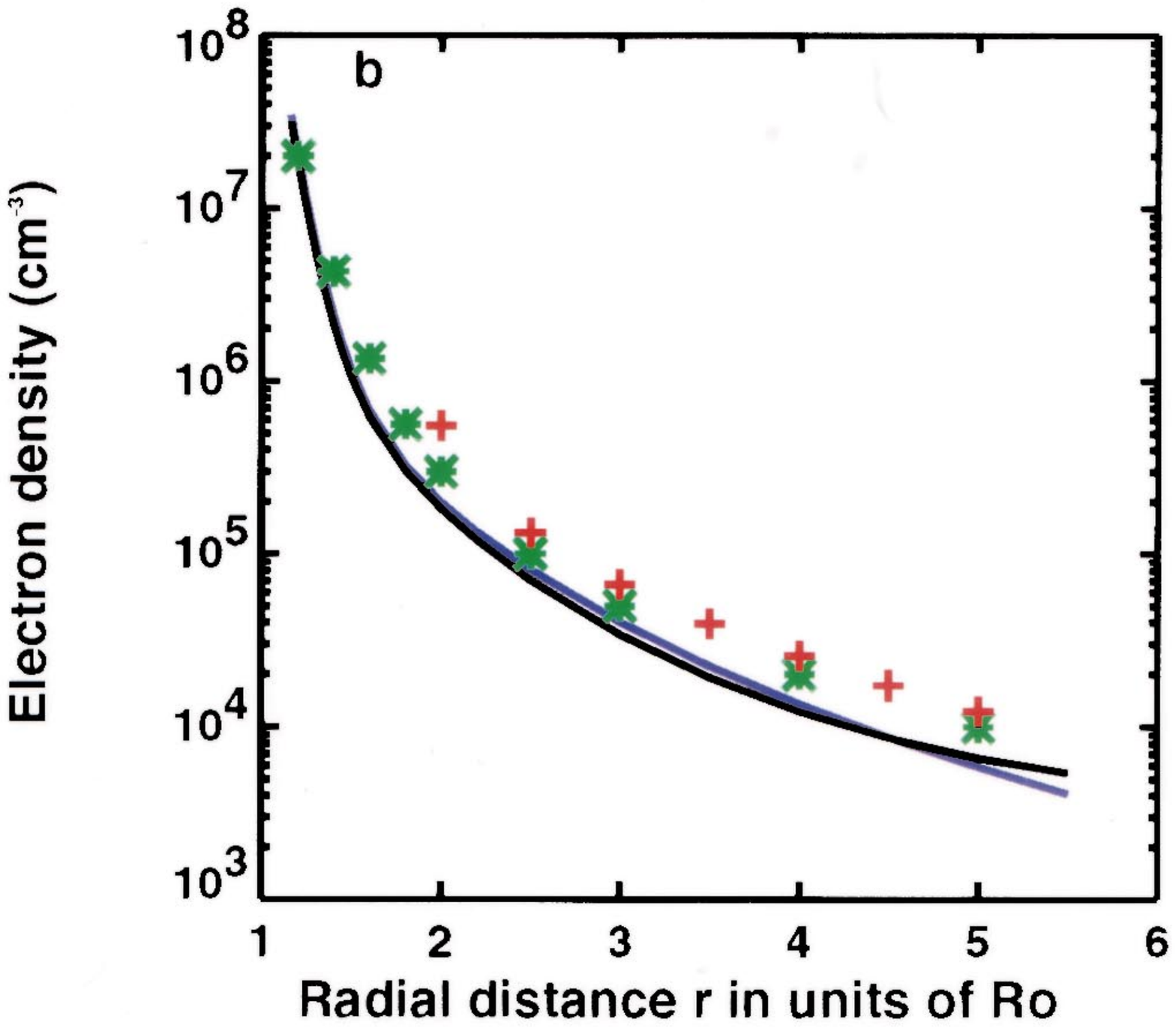}}
\caption{The north polar coronal hole plasma density as a function of the radial distance \citep{Guhathakurta98}. The black and blue lines represent density estimates for the northern PCH in 1995 and 1993, respectively. The crosses correspond to measurements by \citep{Munro77}, while the asterisks represent observations by \cite{Allen73}.}
\label{ch_dens1}
\end{figure}
In Figure~\ref{ch_dens1} the black and blue lines correspond to density estimates of the northern PCH in 1995 and 1993, respectively. The crosses correspond to measurements by \citep{Munro77} and the asterisks represent observations by \citep{Allen73}.

\begin{figure}[!t]
\vspace{-3mm}
\centerline{\includegraphics[scale=0.47, trim=0 70 0 70, clip]{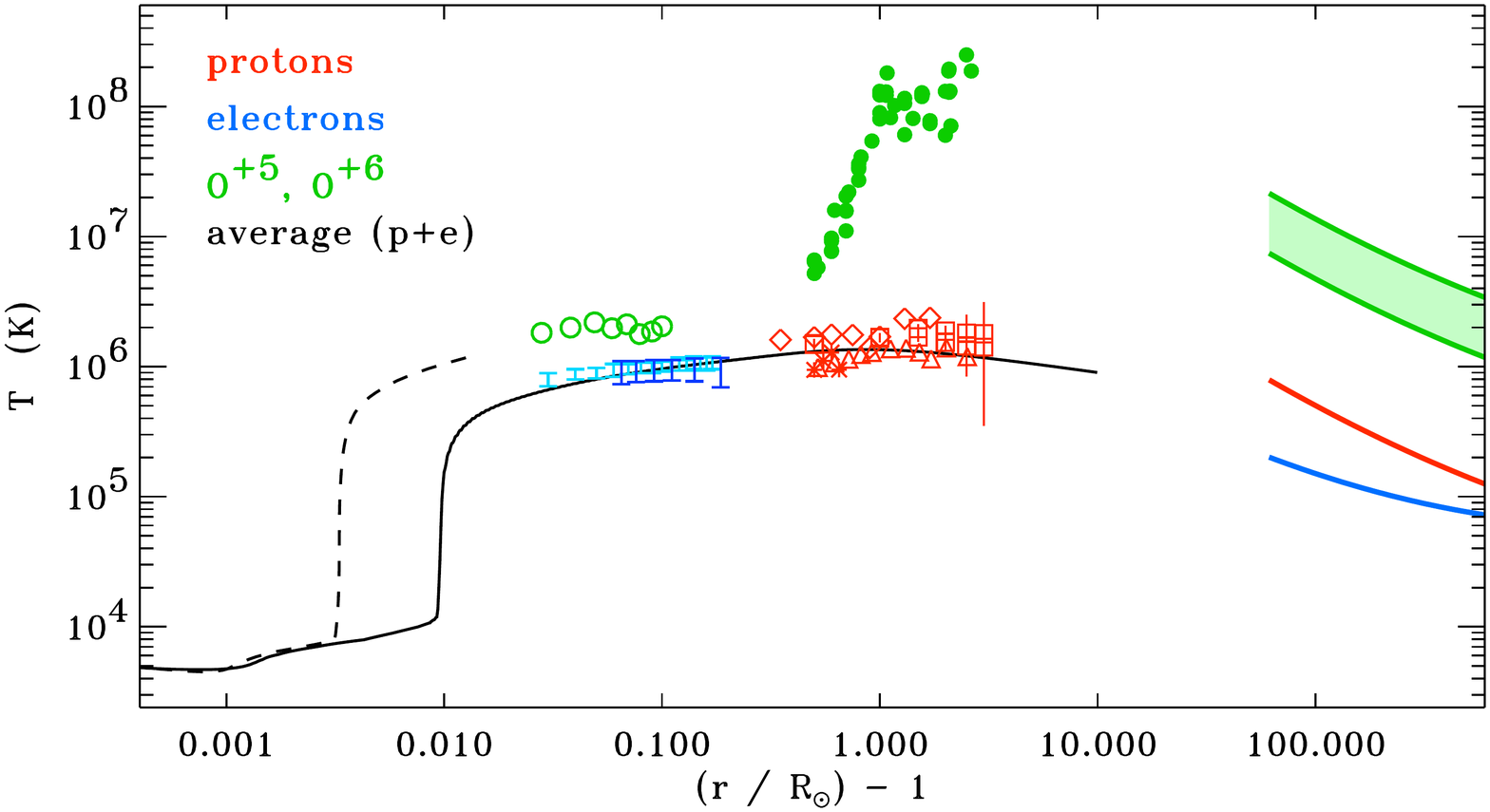}}
\caption{The CH plasma temperatures determined using a semi-empirical model (dashed line) and a turbulence-driven coronal heating model (solid line). The blue symbols correspond to electron temperatures (SOHO/SUMER), the red symbols correspond to proton temperatures (SOHO/UVCS), and the green symbols correspond to O$^{+5}$ ion temperature measurements \citep{Cranmer09}.}
\label{ch_temp}
\end{figure}

The empirical and model temperatures in PCHs and fast wind streams are shown in Figure~\ref{ch_temp}. The mean plasma temperatures from semi-empirical model of \cite{Avrett08} are demonstrated with the dashed black curve, while the turbulence-driven coronal heating model of \cite{Cranmer07} are shown with the solid black curve. The dark and light blue bars correspond to electron temperatures measured off-limb by the SOHO Solar Ultraviolet Measurements of Emitted Radiation (SUMER) instrument by \cite{Wilhelm06} and \cite{Landi08}, respectively. The proton temperatures were assembled by \cite{Cranmer04b} using the SOHO Ultraviolet Coronagraph Spectrometer (UVCS) instrument, and the O$^{+5}$ ion temperatures (based on the O~{\sc vi} 1032 and 1037~\AA\ spectral lines) were determined by \cite{Landi09} and \cite{Cranmer08} (open and filled green circles, respectively). It is clear from Figure~\ref{ch_temp} that the electron and proton temperatures are significantly lower than the heavy ion temperatures. 

Due to the high temperature of the corona, the hydrogen atom loses its electron and exist most of the time as a free proton. Hence, the measured properties of neutral hydrogen are considered as valid approximations of proton properties below 3~R$_{\odot}$ \citep{Allen00}. Observations by the UVCS instrument indicate that in comparison to protons, O$^{+5}$ ions are heated much more strongly, reaching temperatures similar or higher than the core of the Sun ($\>2\times$10$^{8}$~K). These extreme temperatures of the heavy ions measured in CHs can be explained by the preferential heating of different elements by processes, such as reconnection and turbulence driven heating, Alfv\'en wave heating, and heavy ion velocity filtration (for details see \citealt{Cranmer09}). 

\begin{figure}[!t]
\vspace{-3mm}
\centerline{\includegraphics[scale=0.4, trim=0 50 0 50, clip]{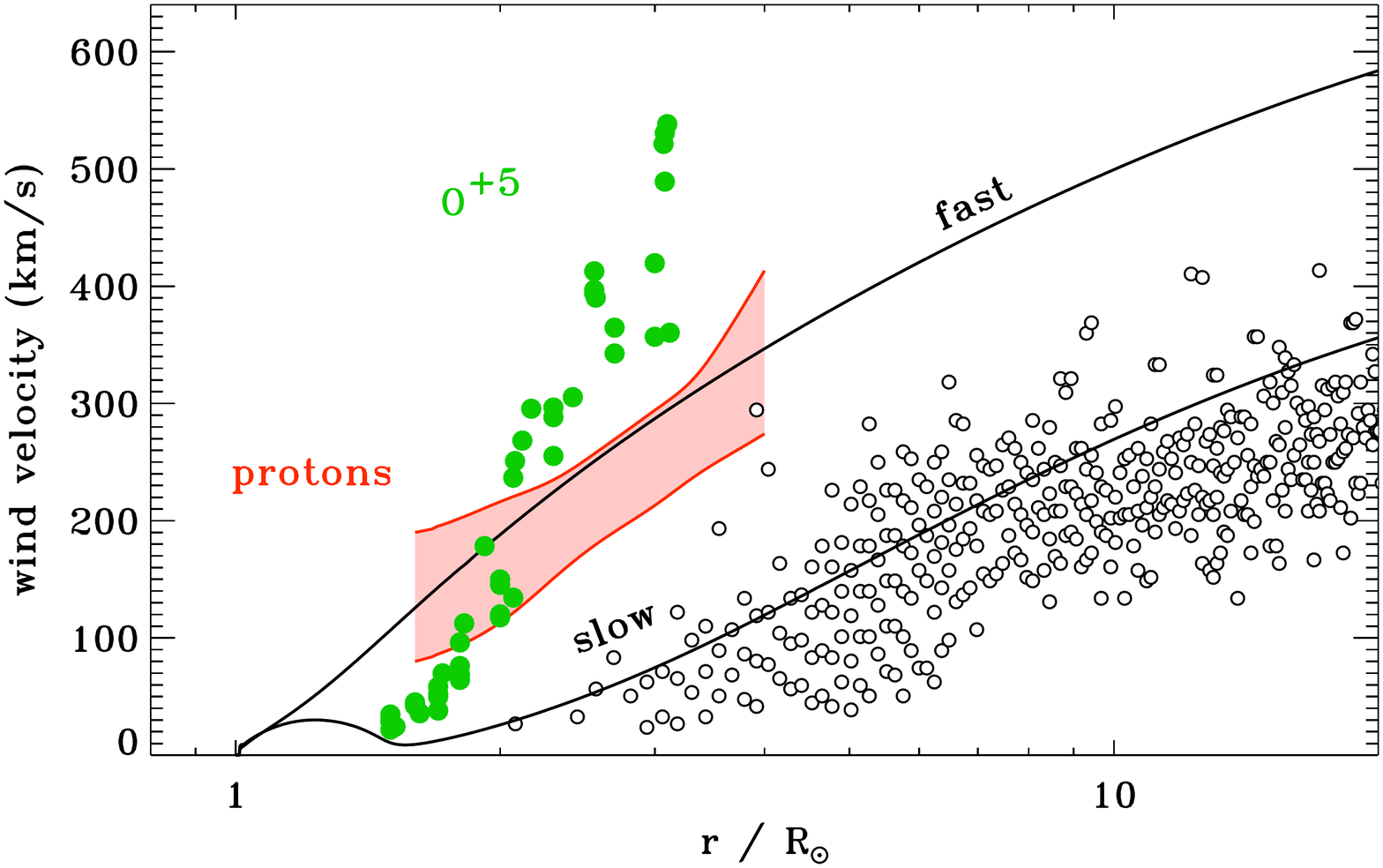}}
\caption{Slow and fast solar wind stream speeds are shown as a function of radial distance. Polar coronal hole proton and O$^{+5}$ ions velocities are shown in green. Theoretical models of the polar and equatorial solar wind are demonstrated with the black line \citep{Cranmer07} and the open circles correspond to the velocity measurements of ÒblobsÓ above equatorial streamers using the SOHO/LASCO instrument \citep{Cranmer09}.}
\label{ch_vel}
\end{figure}
\begin{figure}[!t]
\vspace{-3mm}
\centerline{\includegraphics[scale=0.51]{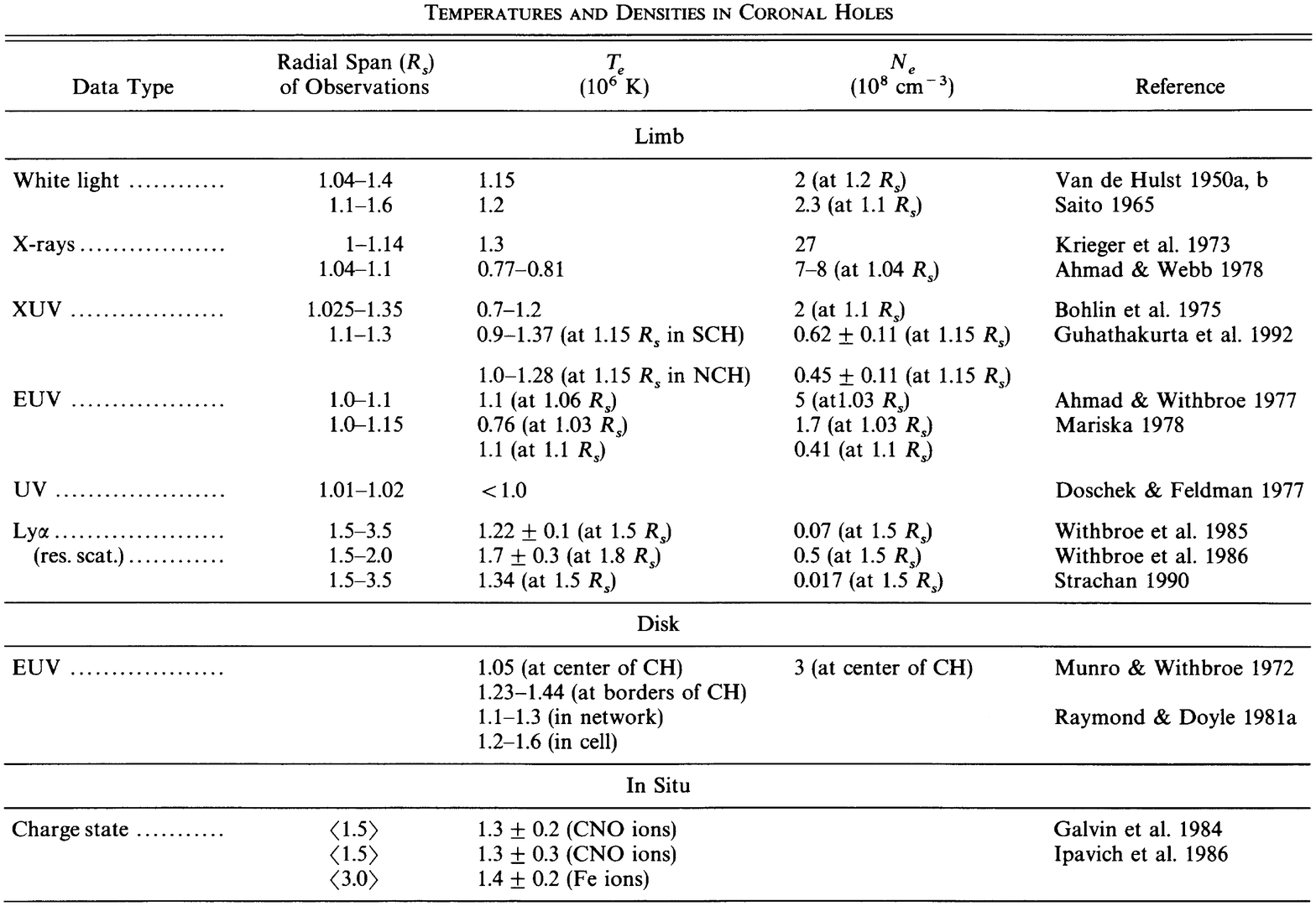}}
\caption{List of CH density and temperature measurements using on- and off-disk observations \citep{Habbal93}.}
\label{temp_dens_list}
\end{figure}
Figure~\ref{ch_vel} shows the solar wind speeds measured in a PCH using the UVCS instrument, together with the velocities derived from a theoretical model of the fast solar wind \citep{Cranmer07}. Polar coronal hole proton velocities are shown in red \citep{Kohl06} and O$^{+5}$ ions velocities are shown in green \citep{Cranmer08}. For comparison, the observed and theoretically derived velocities for slow solar wind streams (associated with equatorial helmet streamers) are shown at the solar minimum. For the latter, ``blobs" were tracked above helmet streamers using the SOHO Large Angle and Spectrometric Coronagraph (LASCO) instrument \citep{Sheeley97}.

For an extensive list of CH density and temperature measurements see \cite{Habbal93} and references therein (Figure~\ref{temp_dens_list}).

\setstretch{1.8}

\subsection{Solar wind acceleration theory}
\label{theory}

In order to recreate the measured properties of CHs and the corresponding HSSWs, a MHD model needs to be constructed, including an appropriate coronal heating source \citep{Tziotziou98}. We start with the hydrodynamic model that includes the continuity equations of mass (Equation~\ref{mass_cons}), momentum (Equation~\ref{mom_cons}), and energy. The latter can be written as:
\begin{equation}
\frac{1}{r^{2}}\frac{d}{dr}(r^{2}v(e+p))=-\rho vg+Q_{tot}+vF_{Alfv\acute{e}n},
\label{en_cons}
\end{equation}
where $F_{Alfv\acute{e}n}$ is the force due to Alfv\'en waves, $Q_{tot}$ is the sum of all external energy sources:
\begin{equation}
Q_{tot}=Q_{op}+Q_{rad}+Q_{cond}+Q_{mech},
\end{equation}
with $Q_{op}$ being the opacity heating of the gas due to the absorption of photospheric radiation, $Q_{rad}$ representing the radiation losses, $Q_{cond}$ is the heating through thermal conduction and $Q_{mech}$ is the mechanical heating. The $e$ in Equation~\ref{en_cons} is the total energy density defined as
\begin{equation}
e=\frac{1}{2}\rho v^{2}+\frac{p}{\gamma-1},
\end{equation}
where $\gamma$ is the ratio of specific heats (equals 5/3). We also include the ideal gas law (Eq.~\ref{gaslaw}). Additionally, it is assumed that Alfv\'en waves propagate along the magnetic field lines and exert a force on the gas flow. This force is the gradient of the Alfv\'en wave pressure. It is assumed that the Alfv\'en waves propagate in a magnetic field that varies with distance as
\begin{equation}
B=B_{0}\left(\frac{R_{\odot}}{r}\right)^3,
\end{equation}
where $B_{0}$ is the magnetic field at the surface of the Sun. The Alfv\'en wave force can then be written as
\begin{equation}
F_{Alfv\acute{e}n}=-\nabla \mathbf{P}_{w}=-\frac{1}{2}\frac{d\epsilon}{dr},
\end{equation}
where $\mathbf{P}_{w}$ is defined as the wave stress tensor and $\epsilon$ as the energy density of the Alfv\'en waves. For further details see \cite{Tziotziou98} and references therein. This model successfully reproduces the high speed velocities ($\sim$630~km~s$^{-1}$) typical of CH solar wind streams observed at Earth, while a simple hydrodynamic model would yield a velocity of $\sim$5~km~s$^{-1}$. For a comparison between the properties deduced from a model excluding and including the Alfv\'en wave acceleration, together with the observed properties, see Figure~\ref{CHmodel}.

\begin{figure}[!ht]
\centerline{\includegraphics[scale=0.9]{tziotziou_CHmodel.pdf}}
\caption{The CH and related HSSW properties derived from a model excluding and including Alfv\'en wave acceleration. The observed CH and solar wind properties are listed for comparison \citep{Tziotziou98}.}
\label{CHmodel}
\end{figure}

Regarding the interaction of open and closed magnetic field lines at the boundaries and inside CHs, one of the main theories is the interchange model \citep{Wang96, Fisk05, Lionello06}. It is based on the assumption that the dominant process in coronal open field evolution is the reconnection between open and closed magnetic field lines (described in detail in Chapter~\ref{chapter:evolution}). This view was initiated by observations showing streamer evolution (inflows and outflows) which could be explained with the closed flux opening into the solar wind and open flux closing back into the streamer \citep{Sheeley02}. Interchange reconnection can also explain the rigid rotation of the CH boundaries \citep{Timothy75}, by allowing closed magnetic fields rotating at the local QS differential rotation rate to continuously reconnect with the CH open field lines and effectively ``pass" through CHs. This would allow CHs to resist the shearing effect of the differential rotation at the photospheric level. At corona heights no such compensation mechanisms are necessary as the corona is well known to be rigidly rotating. The counter-argument to interchange reconnection is that the continuous opening and closing of coronal field lines would result in the injection of magnetic flux into the solar wind and cause a bi-directional heat flux \citep{Antiochos07}. However, the bi-directional plasma jets observed at CH boundaries by \cite{Madjarska04} does support the interchange reconnection model. 

The other main theory is the ``quasi-steady model" \citep{Antiochos07} in which the global solar magnetic field is determined by the instantaneous distribution of photospheric magnetic flux and the balance between magnetic and gas pressure in the corona. The disadvantage of the quasi-steady model is the assumption of a static coronal magnetic field  with topologically well separated regions of open and closed magnetic field. Since the corona does change in time due to photospheric motions and flux emergence, its evolution is modelled through a series of time stationary stages. The simplest quasi-steady model was the potential field source surface model \citep{Altschuler69}, where the gas pressure was neglected and the magnetic field was considered current free. The magnetic field was then extrapolated based on photospheric flux values, assuming the field to be radial from the given distance of the source surface \citep{Luhmann03,Schrijver03}. In recent years, the quasi-steady model has been improved by including the solution of the full MHD equations, which made the assumption of a fixed surface and a current-free corona unnecessary \citep{Pisanko97,Roussev03,Odstrcil03}. These models even allow the relaxation of the quasi-steady approach by including photospheric time-dependence. However, improvement is still needed since it makes the model computationally heavy and no flux emergence and cancellation can yet be included.

\section{Space weather}
\label{spaceweather}

The growing dependence of our society on advanced technologies has lead to an increasing vulnerability to space weather effects. While our understanding of the physical processes connecting solar activity and geomagnetic storms has advanced enormously in the past fifty years, forecasting the time, duration and scale of Earth-bound severe space weather events still remains a challenge. First and foremost, the solar regions which are likely to erupt in the form of a flare and/or a CME have to be identified and located. Secondly, the time of eruption needs to be predicted, which holds probably the largest challenge. Despite numerous models having been developed, the success rate of flare and CME forecasts is still relatively low. Nevertheless, if a region is forecasted to erupt, it still has to be determined whether the flare or CME is Earth-bound and whether it is likely to be geoeffective. Once a geomagnetic storm is forecasted, the estimated time of arrival (ETA) of the geomagnetic storm at Earth has to be provided together with the duration of the storm and the estimated end-time. This would allow public, commercial and military organisations to take precautionary measures in preparation of a geomagnetic storm. 

Programs that aim to develop space weather forecasting methods and provide space weather warnings publicly have only emerged in the past twenty years. To mention a few: the European Space Agency (ESA) Space Weather Web Server, the Solar Influences Data Analysis Center (SIDC) operated by the Royal Observatory of Belgium, the European Space Weather Portal (ESWeP), the U.S. National Space Weather Program (such as the National Oceanic and Atmospheric Administration Space Weather Prediction Center (NOAA/SWPC)), the Russian Space Weather Initiatives (such as ISMIRAN), the Australian Space Weather Agency, the Solar Activity Prediction Center (China), and the organisation connecting numerous space weather centres worldwide - the International Space Environment Service (ISES). 
Many of the above agencies provide space weather forecasts, real-time now-casts on solar and geomagnetic indices, observed halo-CMEs as well as mail-alerts. 

SWPC established three scales for the differentiation of space weather events based on their driving source. All three categories scale from 1 to 5, corresponding to ``minor" and ``extreme", respectively \citep{Biesecker08}. 
\\
1. The {\it Radio Blackout Scale} is based on X-ray flux measurements by Geostationary Satellites (GOES). Enhanced levels of X-ray flux cause heightened ionisation in the day-side ionosphere that can lead to sudden ionospheric disturbances (SIDs), which cause interruptions in telecommunications systems. Since many communication systems transmit signals using radio waves reflected off the ionosphere, geomagnetic storms can lead to some of the radio signals being absorbed in the disturbed ionosphere, while some still get reflected but at an altered angle. This results in rapidly fluctuating signals (scintillation), phase shifts and changed propagation paths. High-frequency radio bands are especially sensitive to geomagnetic activity, while TV and commercial radio stations are barely affected due to the use of transmission masts. 
\\
\begin{figure}[!t]
\centerline{\includegraphics[scale=0.4, trim=0 100 0 100, clip]{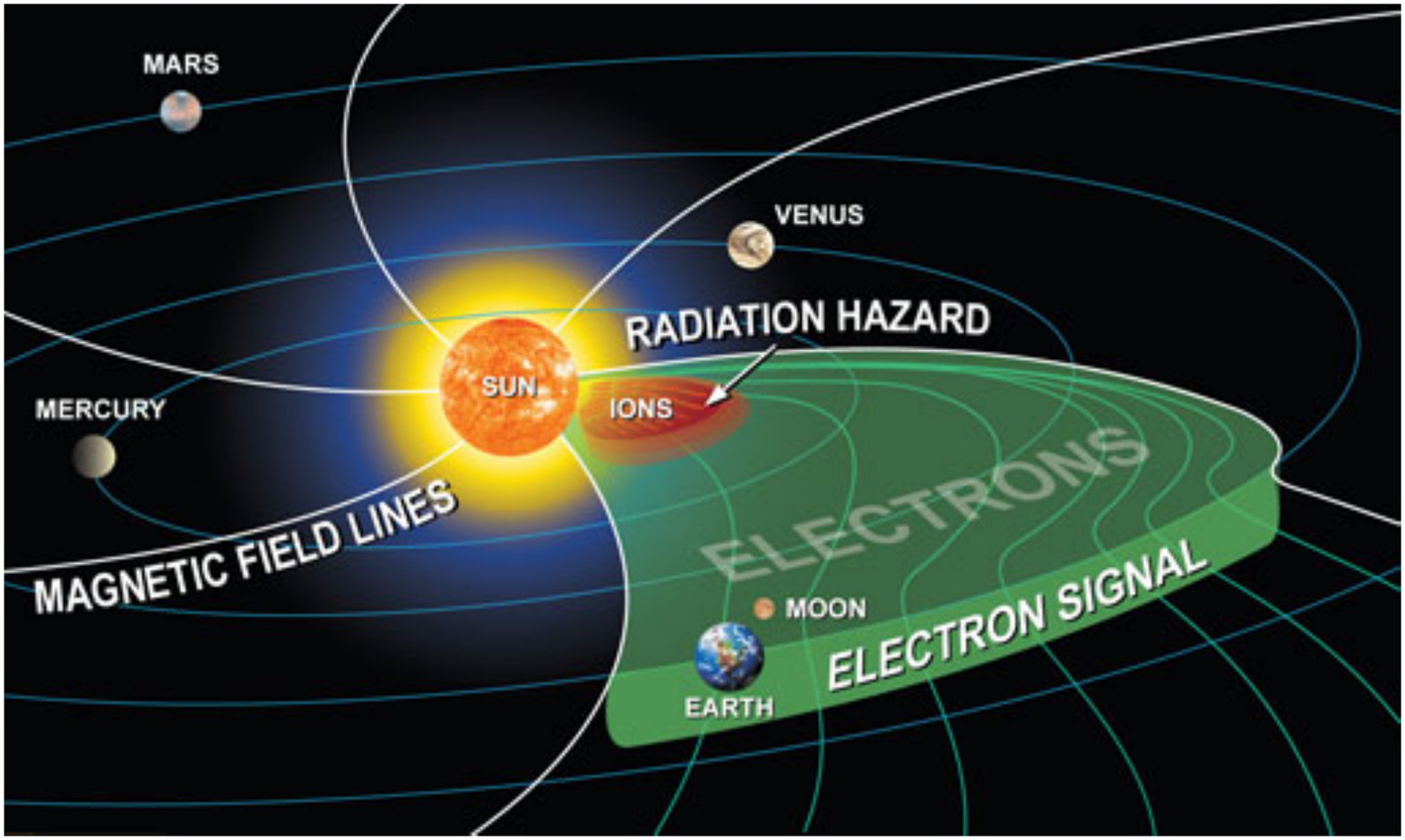}}
\centerline{\includegraphics[scale=0.4, trim=0 100 0 100, clip]{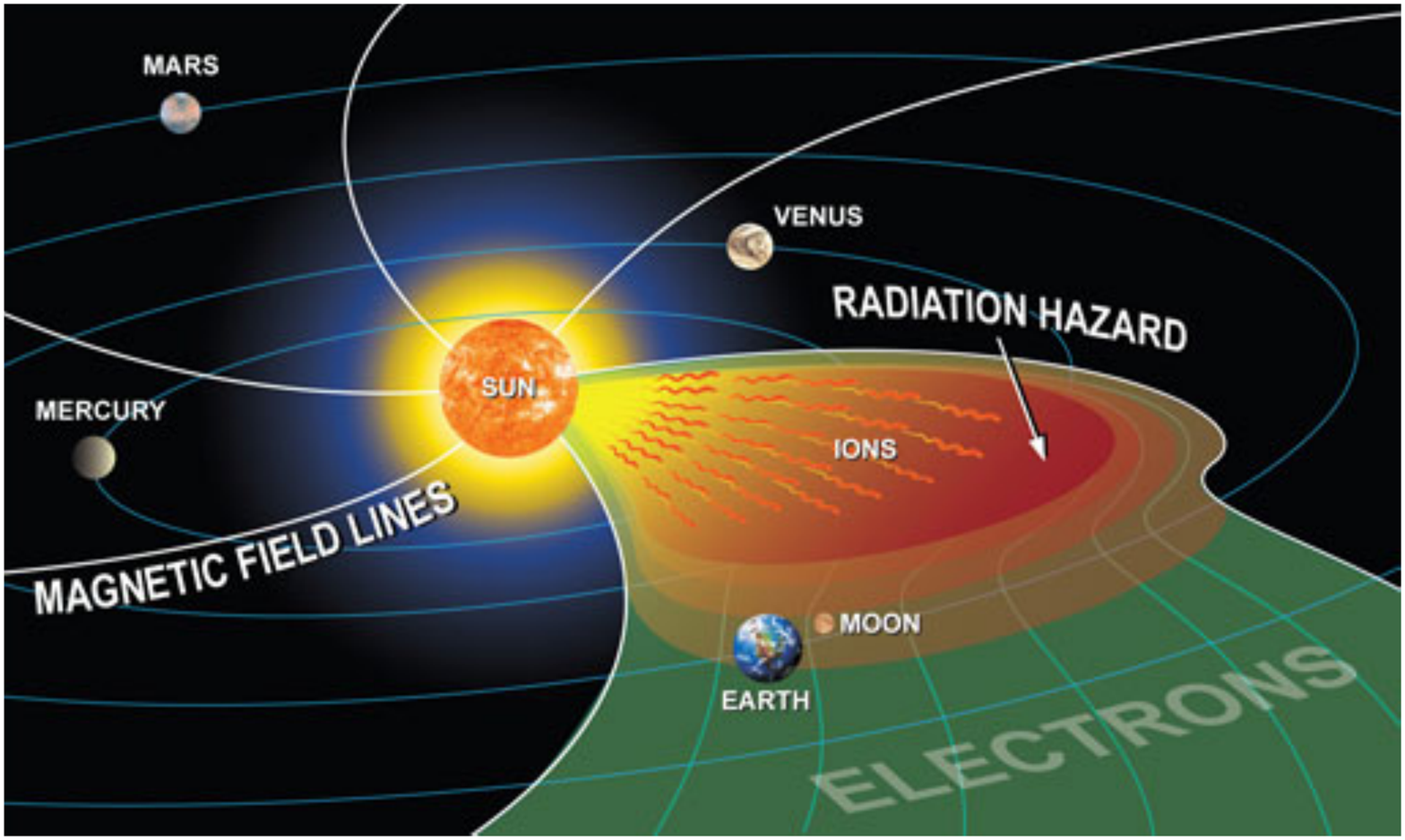}}
\caption{During a solar eruption electrons, protons and heavy ions are ejected, creating a hazardous environment for astronauts and electrical systems aboard satellites. Since electrons are much faster, they can be detected minutes before the ion radiation reaches Earth, providing a useful forecasting tool. Figure courtesy of the Southwest Research Institute.}
\label{geostorm}
\end{figure}
2. The {\it Solar Radiation Storm Scale} is measured from the solar energetic particle (SEP) flux. These energetic particles have energies 10-100~MeV, and sometimes in excess of 1~GeV. SEPs accelerated during solar eruptions take minutes to hours to reach Earth depending on their energies. The SEP events disturb the ionosphere which can lead to communication and navigation problems, while the elevated energetic proton and heavy ion flux levels can harm astronauts during extra-vehicular activities (EVAs) and damage electronic devices aboard satellites. The top cartoon in Figure~\ref{geostorm} shows a solar eruption during which electrons and ions are ejected from the Sun and are heading towards Earth. The first type of particles to arrive are the relativistic electrons, which are followed by the ions tens of minutes later. Hence, the electron signal can be used to predict the radiation hazard posed by energetic ions, allowing astronauts to seek shelter, and electrical devices on satellites to be put into safe-mode. Such ion radiation forecasts are made by the Comprehensive Suprathermal and Energetic Particle Analyzer (COSTEP) aboard SOHO \citep{Posner09}. Note, satellites, and spacecraft in low-Earth orbit are protected to a certain extent from energetic particles by the Earth's magnetosphere. However, radiation is still higher outside the protective shield of the Earth's atmosphere, and during severe solar storms astronauts on a space station may reach or even exceed a whole year's radiation limit. During extreme space weather conditions even high-altitude aircraft routes have to be modified due to the enhanced risk of radiation exposure to flight crews.
\\
\begin{figure}[!t]
\centerline{\includegraphics[scale=2]{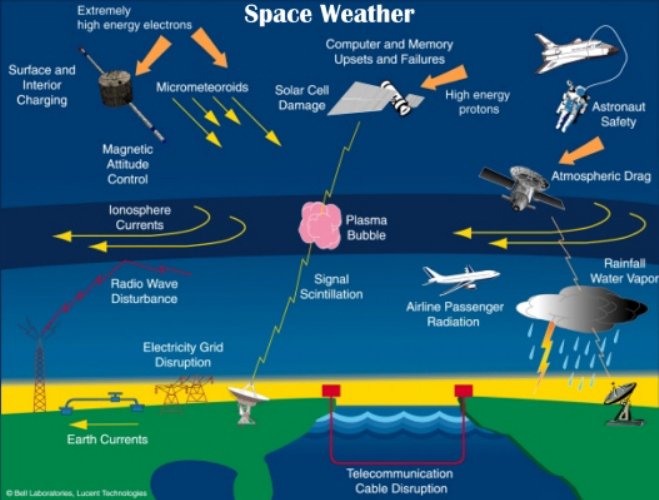}}
\caption{The effects of geomagnetic storms experienced in the Earth's atmosphere and on-ground.}
\label{spaceweather}
\end{figure}
3. {\it Geomagnetic Storm Scale} is based on energetic solar events, during which large amounts of plasma and energy departs the Sun in the form of CMEs. The ejected plasma contains imbedded magnetic structures, also known as a {\it magnetic cloud}. The magnetic fields in CMEs need to be of strong southward direction to reconnect with the mainly northward directed magnetic fields of the day-side magnetosphere. This reconnection gives rise to strong induced fields and electric currents in the magnetosphere, ionosphere and the solid body of Earth (Figure~\ref{spaceweather}). The latter can adversely affect large-scale power grids since the induced electric currents can enter long transmission lines and reduce the power by unbalancing the grid operation, and can even destroy transformers through heating and vibration. The most well-known example to a power-grid failure due to a severe geomagnetic storm was in Quebec on 13 March 1989, which caused a power-outage for 9 hours. 

A CME can be geoeffective even if it does not contain southward magnetic fields. The compressed plasma between the shock and the CME can contain a southward directed magnetic cloud, which can be geoeffective. Another geomagnetic effect of a CME is the compression of the magnetosphere on the day-side of the Earth. This may leave the high-altitude satellites exposed to harmful energetic solar wind particles.

\begin{figure}[!t]
\centerline{\includegraphics[scale=0.58, trim=0 200 0 150, clip]{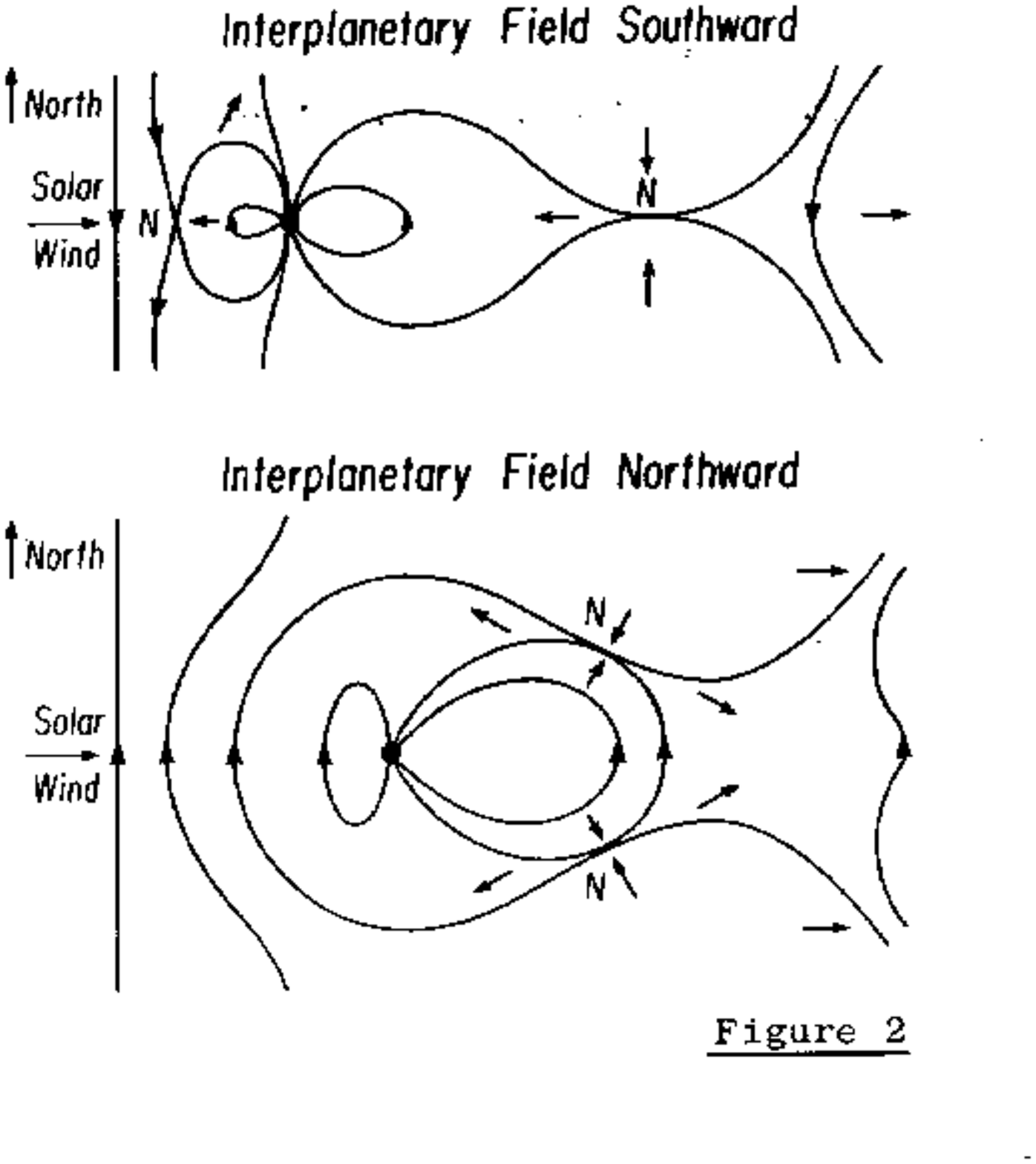}}
\caption{The reconnection process shown in case of a southward (top) and northward (bottom) directed IMF impacting the Earth's magnetosphere \citep{Dungey63}.}
\label{dungey}
\end{figure}

\begin{figure}[!t]
\centerline{\includegraphics[scale=0.6, trim=0 0 0 100, clip]{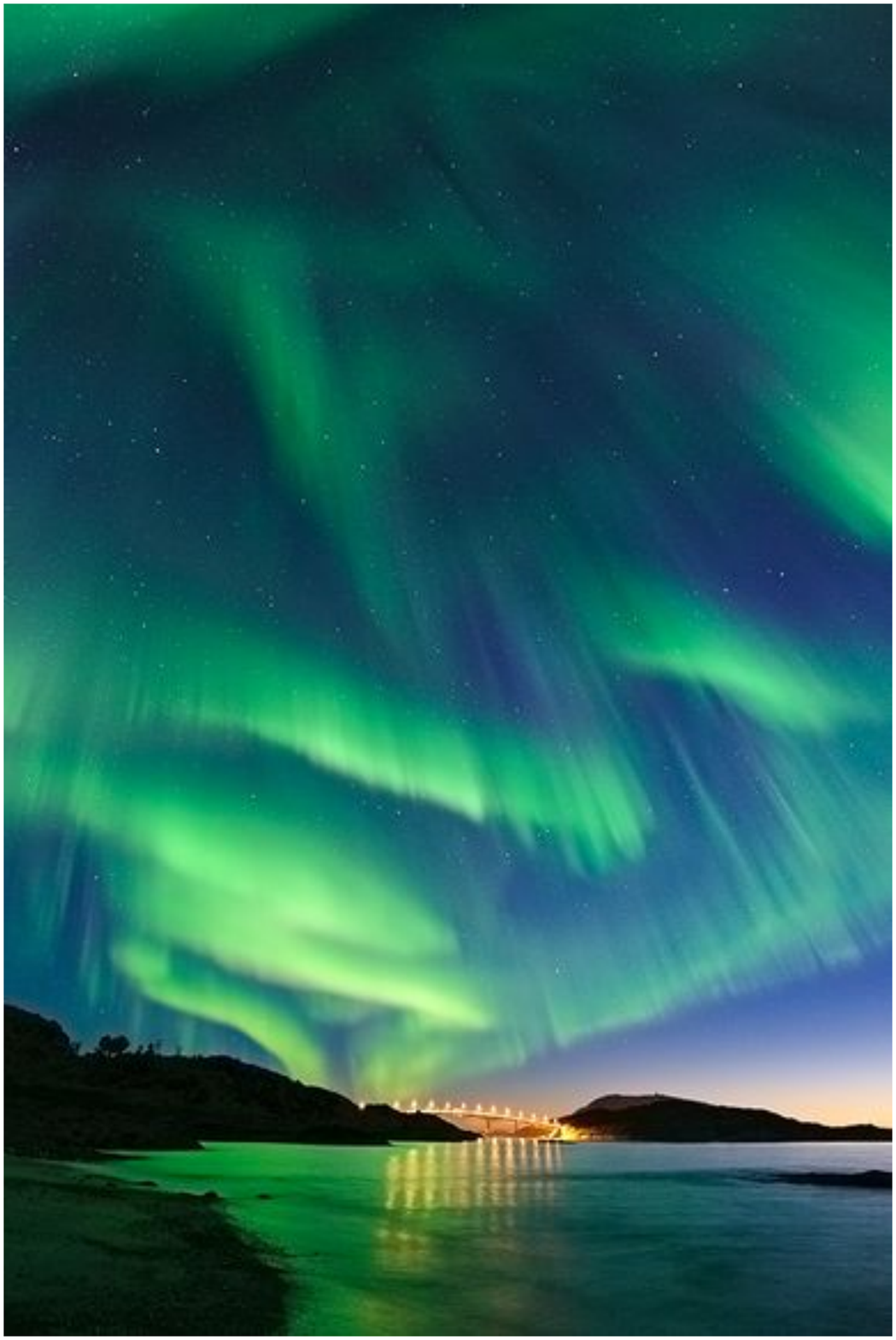}}
\caption{Aurora seen in Sommaroya, northern Norway on 8 September 2010. Photo courtesy of Thilo Bubek.}
\vspace{-0.02\textwidth}
\label{aurora}
\end{figure}

Figure~\ref{dungey} shows the classic models of the magnetospheric reconnection suggested by \cite{Dungey63}. The top diagram shows the change in the Earth's magnetospheric topology as an incoming south-directed IMF reconnects with the typically north-directed magnetic field of the Earth's day-side magnetosphere. Reconnection first occurs at the Sun-facing side of the magnetosphere, where the oppositely directed magnetic field lines meet. Then the reconnected field lines travel with the solar wind, which extends them to become part of the magnetotail where reconnection occurs again where oppositely oriented magnetic field lines meet. During the second reconnection two separate magnetic structures are created - one that connects the North and South Pole of the Earth and a disconnected field line that is carried away by the solar wind. During the second reconnection solar energetic particles are channelled towards the poles of the Earth where the atmospheric particles get excited and create the well known phenomenon - the {\it aurorae borealis} (Figure~\ref{aurora}). At the same time geomagnetic disturbances may be experienced at high latitudes based on the severity of the storm. 

\setstretch{2}

The second, geomagnetically less effective scenario is shown in the bottom of Figure~\ref{dungey}. Here, the incoming magnetic field is north-directed, similar to the Earth's magnetic field. In this case no reconnection occurs at the day-side of the Earth's magnetosphere, however the incoming magnetic field gets deformed around the magnetosphere. At the tail the incoming magnetic field encounters the night-side closed magnetic field of the Earth which is oppositely directed to it and causes and double reconnection as shown in the figure. Through this double reconnection a close field is created that becomes part of the day-side magnetosphere, while a disconnected field is also created which is carried away by the streaming solar wind. 
\setstretch{2}

It is important to note that flux is added to the geomagnetic tail during the reconnection process involving a southward interplanetary magnetic field, while flux is removed from the tail in the case of an incoming northward interplanetary field.

\subsection{Corotating Interactive Regions}
\label{cir}

\begin{figure}[!ht]
\vspace{0.18\textwidth} 
\centerline{\includegraphics[trim=0 120 0 250, scale=0.45]{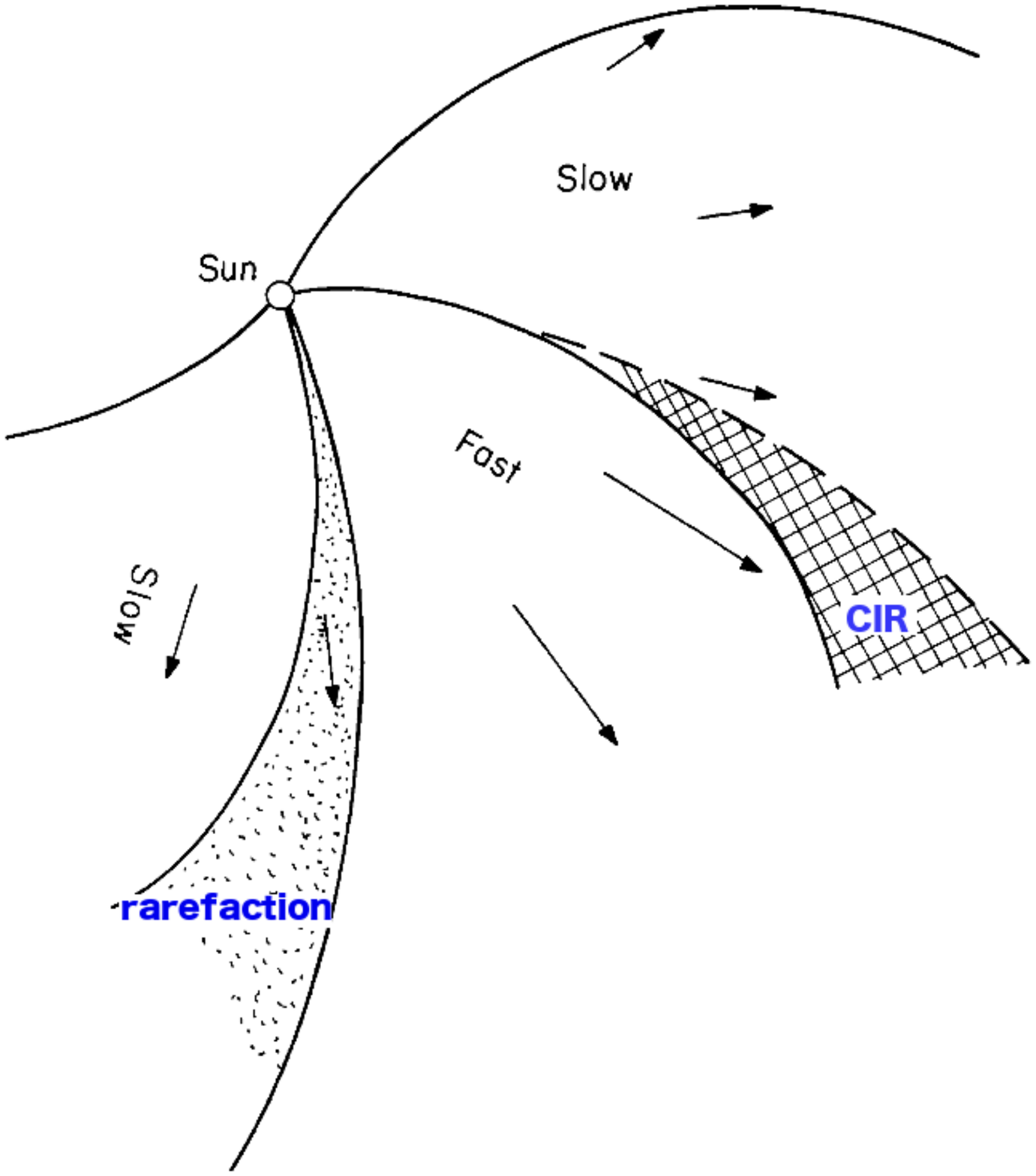}}
\caption{The CIR and rarefaction caused in the heliosphere by the interaction of the slow and high-speed solar wind plasma. Modified from \cite{Parker65}.}
\label{spiralcir}
\end{figure}

Although on a smaller scale, CHs can also cause geomagnetic disturbances. When the HSSW encounters the slower solar wind, the solar wind plasma becomes compressed (creating a CIR), while at the trailing part of the HSSW a rarefaction region is created (Figure~\ref{spiralcir}). Most CIRs are bounded by shocks, but mostly at distances of 2~AU or more (in general, waves have to go through several nonlinear steepening phases before a shock is formed). Since the interaction region is of high pressure, it expands into the plasma ahead and behind at high speed creating two shocks on the two sides of the pressure enhancement. This phenomenon is called the ``forward-reverse shock pair", where one shock travels towards and another away from the Sun \citep{Gosling96}. The propagation and the pressure gradient of the shock waves lead to the acceleration of the slow solar wind and the deceleration of the high-speed solar wind.  
This interaction limits the steepening of the stream and transfers energy and momentum from the high-speed solar wind to the slow solar wind. At large distances the stream pressure enhancement gradually reduces and the compression regions merge as they expand.

An extensive statistical study carried out by \cite{Choi09} has shown that approximately half of the CH related CIRs caused geomagnetic disturbances with D$_{st}\leq$-50~nT (-5$\times$10$^{-4}$ G). D$_{st}$ index indicates geomagnetic storm severity, with high values e.g. 100~nT (10$^{-3}$ G) corresponding to a calm magnetosphere, while low values e.g. -300~nT (-3$\times$10$^{-3}$ G) indicate a large storm. It has also been shown that 70$\%$ of the CHs were located between $\pm$30$^{\circ}$ latitude, and the area of geoeffective CHs were found to be larger than 0.12$\%$ of the visible solar hemisphere. \cite{Choi09} used KPVT CH maps and Advanced Composition Explorer (ACE) and Wind solar wind data to link CHs to CIRs. 

With real-time data available, the arrival of CH related CIRs at Earth can also be forecasted. This is possible since CHs are recurrent and visible for long periods of time, thus making the corresponding geomagnetic storms easier to forecast than disturbances caused by flares and CMEs, where the eruption time is challenging to predict. Knowing the range of typical CH solar wind stream velocities, the path of the streaming plasma can be calculated together with its arrival time at Earth. In Chapter~\ref{chapter:forecasting} we discuss how the CH detection algorithm presented in this thesis can be useful in predicting minor geomagnetic storms based on the early detection of CHs.

\section{Thesis outline}

The CH detection method presented in the thesis provides a fast, subjective and automated method to identify CHs and improve the determination of their physical properties throughout the solar cycle. The automated nature of the algorithm allows the detection of CHs on short and long timescales which can further our understanding of the magnetic and morphological evolution of CHs. This reflects on the small scale processes governing the CH boundary evolution as well as the mechanism of solar wind acceleration in CHs. Linking CH observations to HSSW streams using in-situ solar wind data provided a practical aspect to the presented research by enabling the real-time forecasting of minor storms and the investigation of correlation between CH and solar wind properties. The geomagnetic effects of CH related HSSW streams were also compared to changes observed in the Earth's radiation belt. Furthermore, the observations were used in conjunction with the solar cycle to derive the role of CHs in the solar magnetic pole reversal mechanism.
 
Chapter~\ref{chapter:instrumentation} presents the instruments that are used in this study to detect CHs. The most optimal instrument pair with the largest database, SOHO/EIT and SOHO/MDI, were fundamental for developing the CH detection algorithm and the research presented in Chapter~\ref{chapter:evolution} and Chapter~\ref{chapter:forecasting}. Chapter~\ref{chapter:detection} describes previous methods of CH detection as well as detailing the detection algorithm developed by the author. The algorithm is further extended to study the small-scale evolution of CH boundaries in conjunction with the interchange reconnection model, as presented in Chapter~\ref{chapter:evolution}. The physical properties of the studied CHs are determined together with the diffusion coefficient and the rate of magnetic reconnection occurring at CH boundaries. The large-scale and long-term changes in CHs are discussed in Chapter~\ref{chapter:forecasting} where the results involving the analysis of over nine years of EUV data are presented. The observed properties of CHs in terms of the solar cycle are discussed with respect to the geomagnetic effects experienced at Earth over the same period. In Chapter~\ref{chapter:conclusions} the theoretical implications of the short and long-term evolution of CHs is discussed, and the minor geomagnetic storm forecasting method is evaluated with suggestions for further development and improvements.


\chapter{Instrumentation} 
\label{chapter:instrumentation}

\ifpdf
    \graphicspath{{2/figures/PNG/}{2/figures/PDF/}{2/figures/}}
\else
    \graphicspath{{2/figures/EPS/}{2/figures/}}
\fi

\flt
{\it In this Chapter, we discuss the EUV and X-ray imagers that were used in the present research to detect low-intensity regions. These include the SOHO/EIT, STEREO/EUVI and Hinode/XRT instruments. In order to determine whether a low intensity region is a CH, magnetic field analysis was carried out using full-disk magnetograms provided by SOHO/MDI. For forecasting purposes the detected CHs were linked to high-speed solar wind streams using in-situ solar wind data from the ACE/SWEPAM and STEREO/PLASTIC solar particle monitors.
 }
\flb

\newpage

\section{Introduction}

Until the 1970s, imaging the solar corona was only possible from ground-based observatories. Although ground based solar telescopes are still used world-wide, space-based telescopes are now the dominating source of data. The main problem ground-based observations face is the atmospheric distortions due to air turbulence. An additional distortion is caused by the optics expanding due to heat from the environment or the instrument itself. Although there are now advanced adaptive optics that correct distortions due to atmospheric turbulence within split seconds, airborne telescopes still have the advantage of consistently providing high resolution images independent of weather conditions. Furthermore, EUV and X-ray imaging can only be done from space due to atmospheric absorption at these wavelengths. The disadvantage of airborne telescopes is maintenance and repairs, which is extremely expensive and sometimes even impossible. The size of the telescope is also an issue as the larger the instrument, the more it costs to put it into orbit. 

One of the solar features that could only be imaged from space were CHs. Despite this, the effects of CHs were detected long before X-ray and EUV imaging was carried out from space. In 1905, British astronomer E.W. Maunder found recurring magnetic disturbances caused by sources invisible on white-light solar images \citep{Cliver94}. These were later called M-regions (M for `magnetic') by Julius Bartels. The existence of CHs was first confirmed by rocket-borne X-ray telescopes in the 1960s \citep{Cranmer09}. In 1973 Skylab was launched, which amongst others allowed the continuous observation of CHs. In 1991 came Yohkoh with the Soft X-ray Telescope and in 1995 SOHO was launched enabling the EUV imaging of CHs and providing data for over a decade now. In recent years Hinode, STEREO A \& B and SDO have been launched, providing ever more impressive X-ray and EUV images of the Sun with the highest resolution and time cadence to date.

Although magnetograms do not reveal CHs like EUV or X-ray images, they have proved crucial in differentiating CH from filaments and other features that might appear as low-intensity regions at the mentioned wavelengths \citep{Wang96}. CHs can also be detected indirectly by analysing the solar wind properties, as the particle composition of CH related solar wind streams differs from that originating elsewhere.

\section{SOHO}

SOHO is a joint mission of the European Space Agency (ESA) and National Aeronautics and Space Administration (NASA). The main goal of the mission is threefold - to study the solar interior using helioseismology techniques, to reveal the mechanisms that heat the solar corona and to explain how the solar wind is created and what causes its acceleration.

The spacecraft was launched from the Cape Canaveral Air Station (Florida, United States) on 2 December 1995. SOHO is located at the first Lagrangian point (L1) approximately 1.5 million kilometres from Earth (four times the distance between the Earth and the Moon). It is always facing the Sun, providing uninterrupted observation. The mission had a two-year lifetime, but has now been in operation for almost 15 years.

\begin{figure}[!t]
\centerline{\hspace*{0.015\textwidth} 
\includegraphics[clip=,scale=0.45]{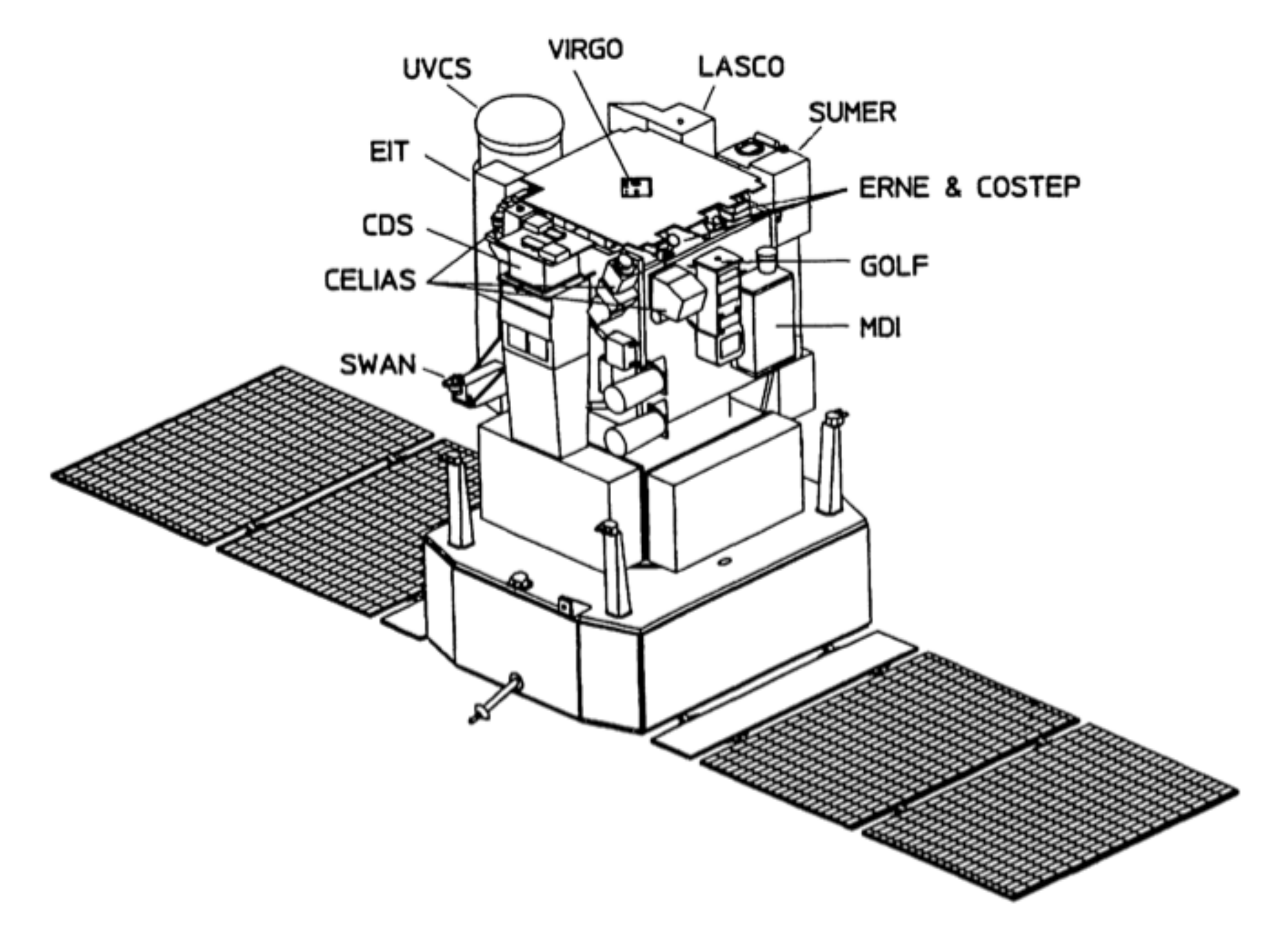} }
\caption{A schematic of SOHO and its instruments \citep{Domingo95}.} 
\label{soho} 
\end{figure}

There are twelve instruments aboard SOHO \citep{Domingo95}: three spectrometers that study the plasma of the solar atmosphere (CDS, SUMER, UVCS), three instruments detect particles of solar, interplanetary or galactic origin (CELIAS, COSTEP, ERNE, SWAN), two instruments that measure the oscillations in the solar irradiance and gravity, and global velocity (VIRGO and GOLF, respectively), a coronograph that studies the off-disk solar corona (LASCO), an extreme-ultraviolet imaging telescope (EIT), and a magnetograph that measures the vertical motions on the solar surface, the acoustic waves in the solar interior and the longitudinal magnetic field of the Sun (MDI). The instruments aboard SOHO are shown in Figure \ref{soho} 

Data from the EIT and MDI instruments were used extensively in this thesis work and are given a detailed review in this section.

\subsection{EIT}

\begin{figure}[!t]
\includegraphics[clip=,scale=0.48]{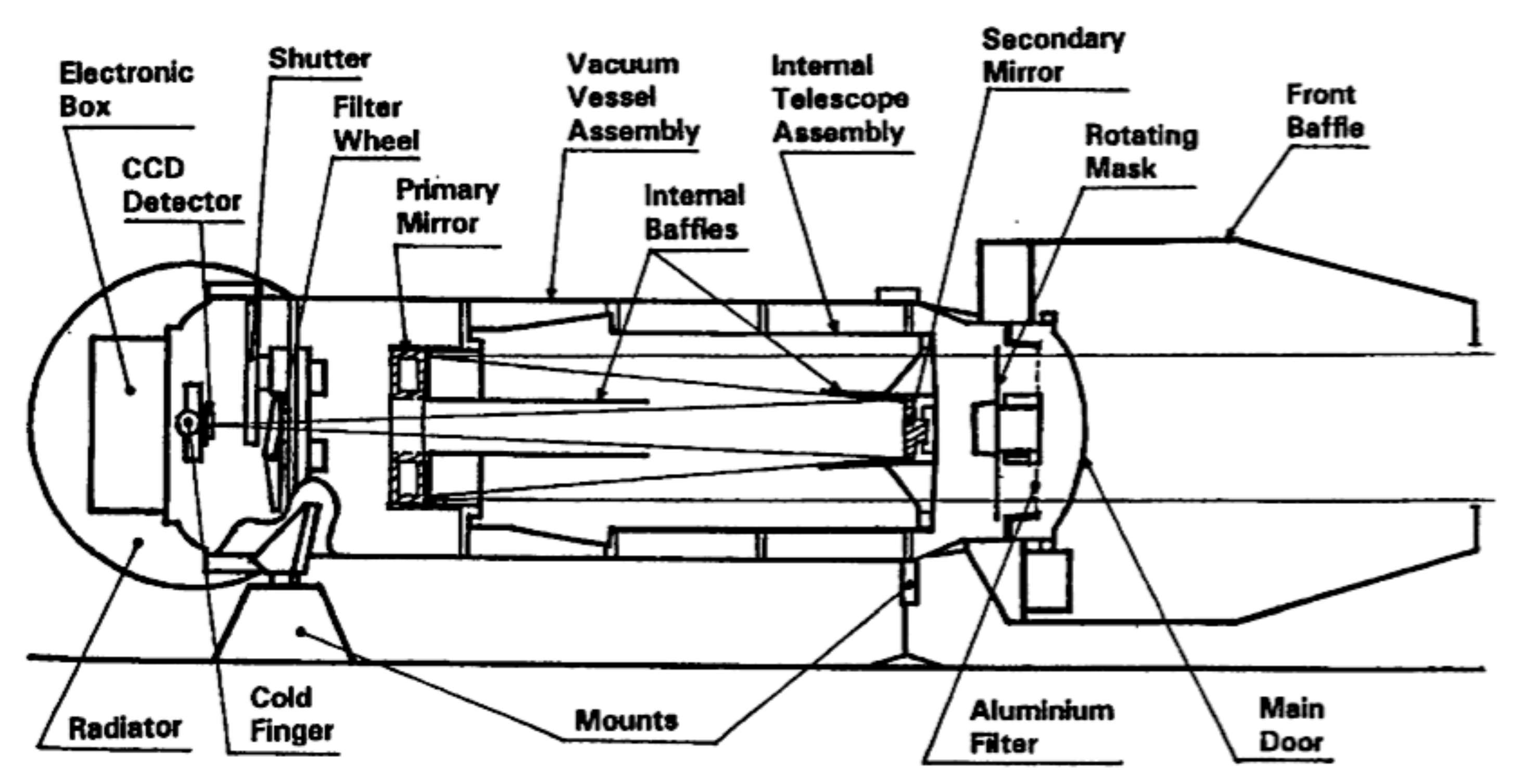}
\caption{The schematic of the EIT instrument \citep{Delaboudiniere95}.} 
\label{eit} 
\end{figure}

The EIT instrument \citep{Delaboudiniere95} provides full-disk solar images of the transition region and the inner corona out to 1.5 solar radii above the limb. The telescope is a Ritchey-Chr\'etien design (Figure \ref{eit}). The primary and secondary mirrors consist of four quadrants, each of which have multilayer coatings that define four spectral bandpasses by interference effects. At any time of observation, a rotating mask allows only one quadrant of the mirrors to be illuminated by the Sun. The multilayers consist of alternating layers of molybdenum and silicon. The telescope field of view is 45 $\times$ 45~arcmin with 2.6~arcsec pixels and a maximum spatial resolution of 5~arsecs. The detector is a 1024 $\times$ 1024~pixel backside-illuminated charge coupled device (CCD).  

\begin{figure}[!t]
\includegraphics[clip=,scale=0.9]{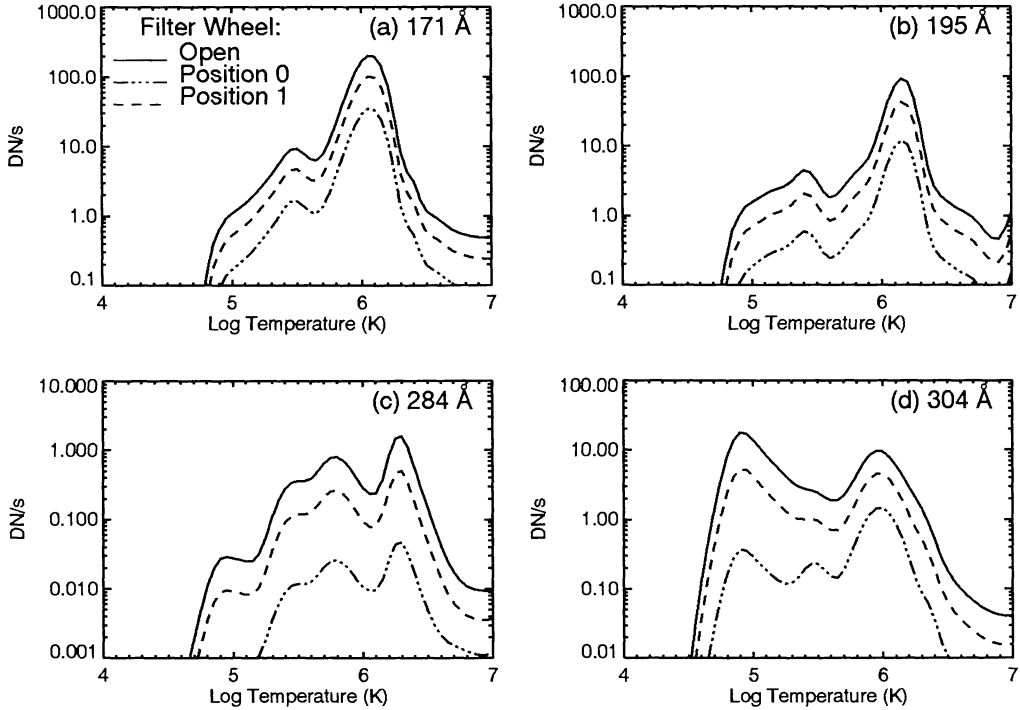}
\caption{The temperature response of the 171~\AA, 195~\AA, 284~\AA\ and 304~\AA\ EIT filters \citep{Delaboudiniere95}.} 
\label{eit_tempresp} 
\end{figure}

\begin{figure}[!t]
\vspace{-0.05\textwidth}
\includegraphics[clip=,scale=0.54, trim=60 50 60 58, clip]{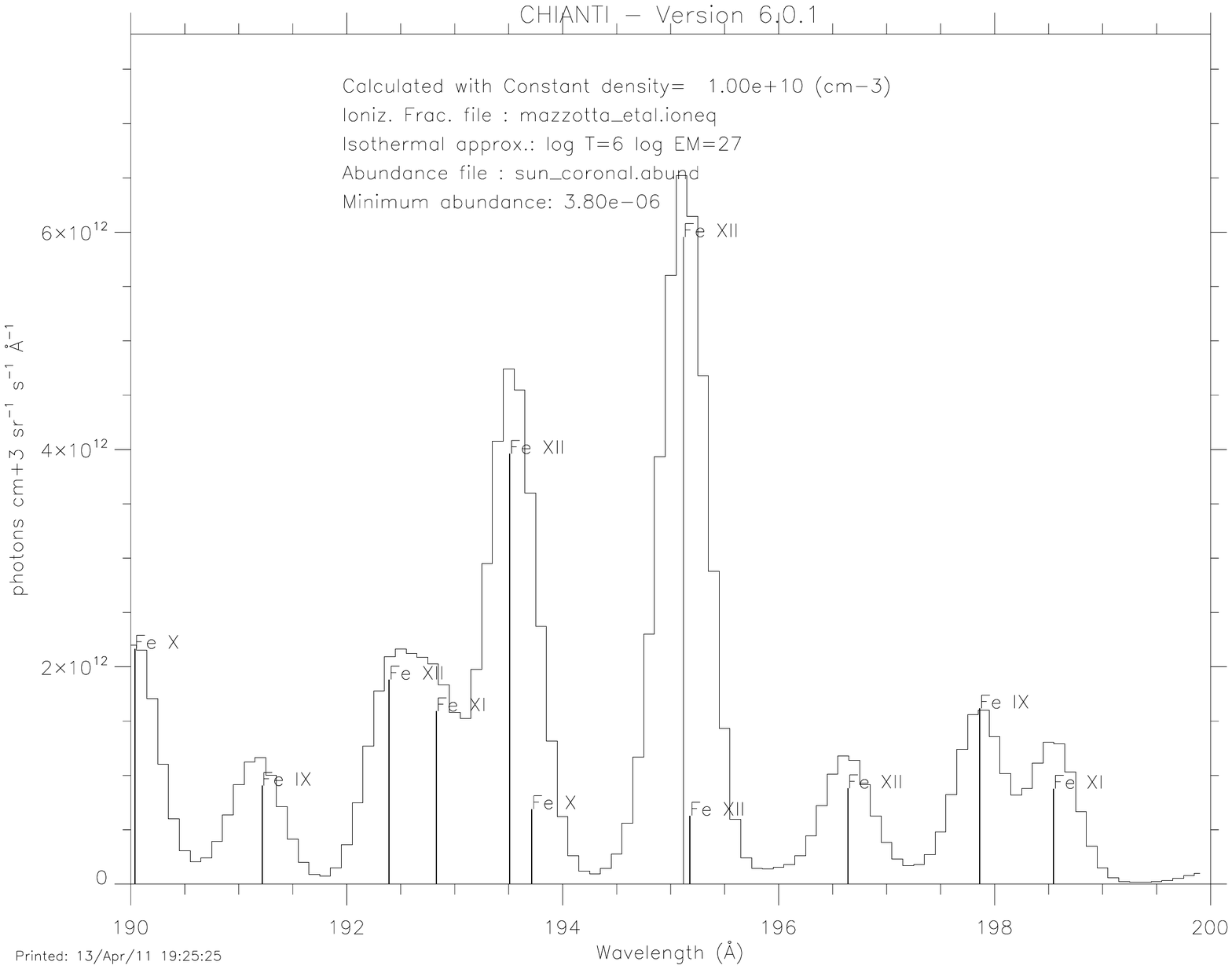}
\includegraphics[clip=,scale=0.54, trim=60 50 60 58, clip]{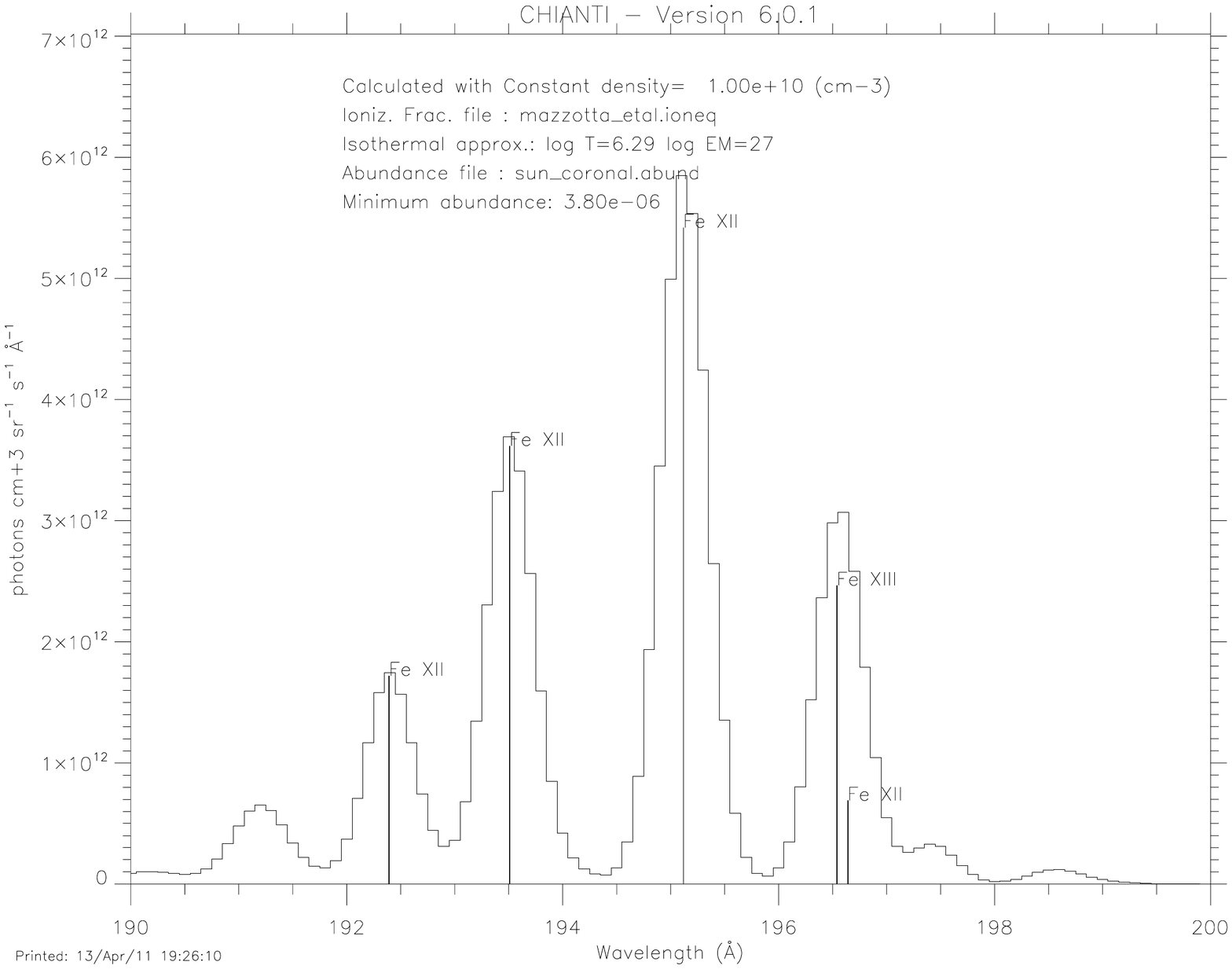}
\caption{Spectral line intensities between 190-200~\AA, at 1MK and 2MK (top and bottom, respectively).} 
\label{spectra1} 
\end{figure}

\begin{figure}[!t]
\vspace{-0.05\textwidth}
\includegraphics[clip=,scale=0.54, trim=60 50 60 58, clip]{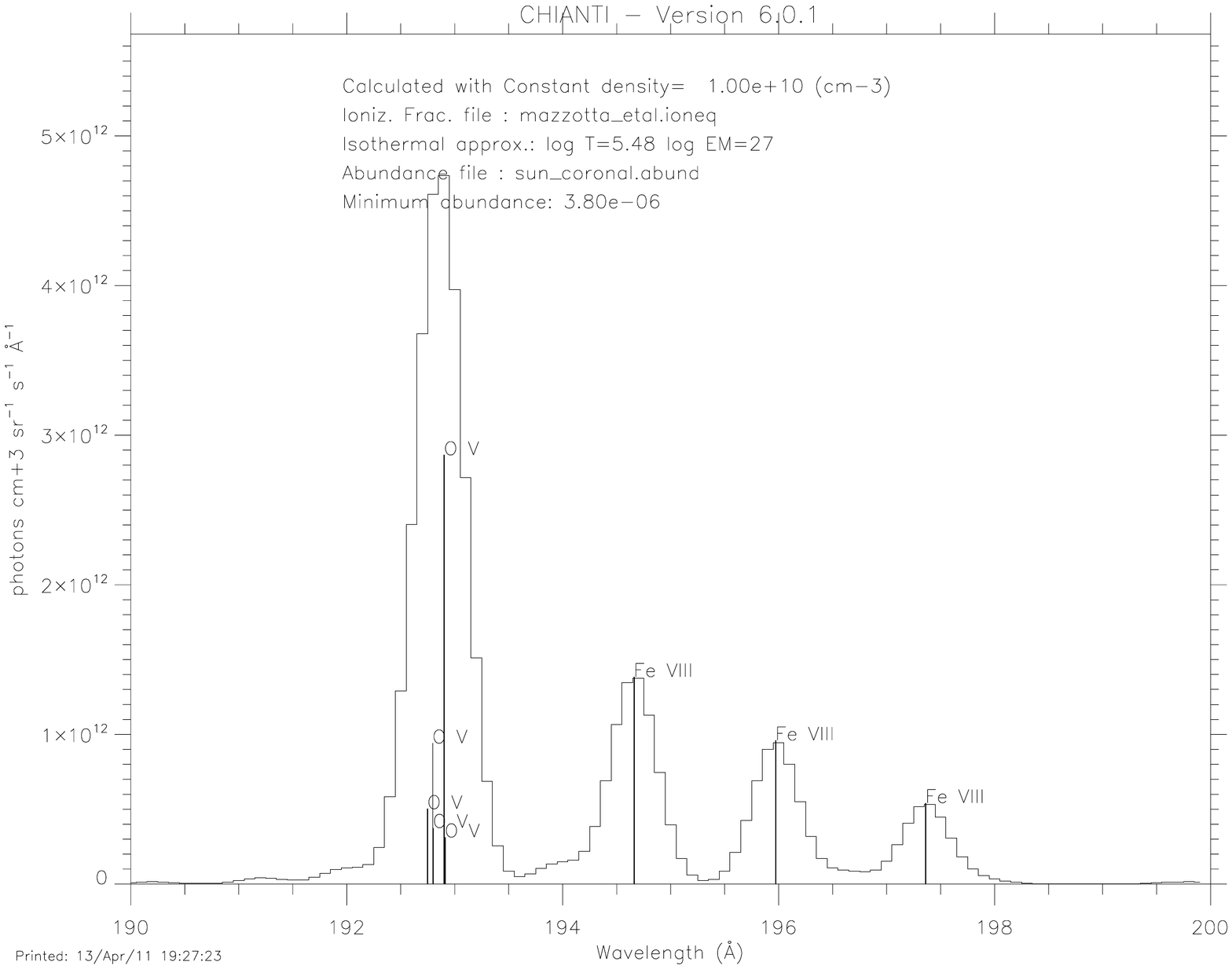}
\includegraphics[clip=,scale=0.54, trim=60 50 60 58, clip]{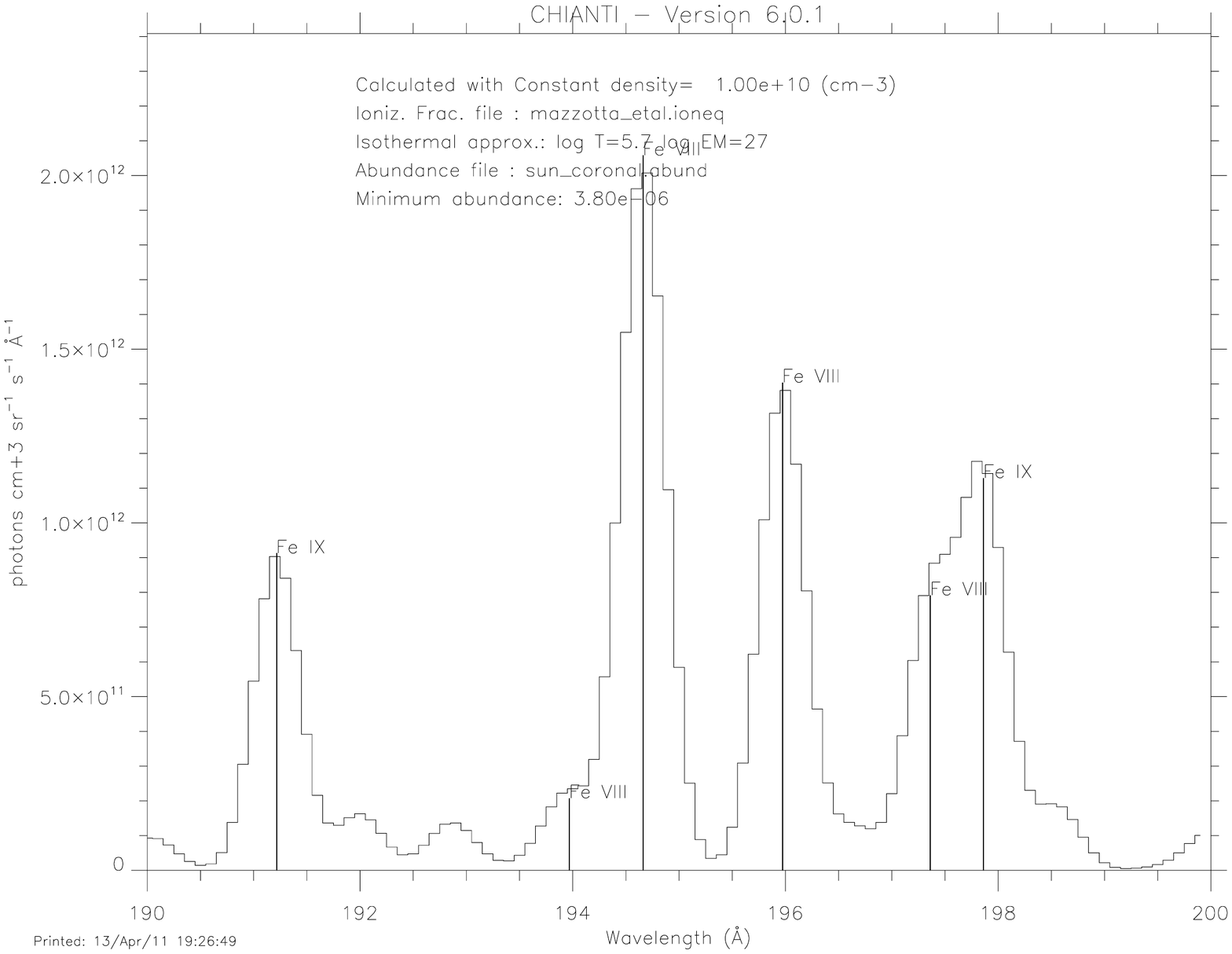}
\caption{Spectral line intensities between 190-200~\AA, at 0.3MK, 0.5MK (top and bottom, respectively).} 
\label{spectra2} 
\end{figure}

The four narrow bandpasses are centered at 171~\AA\ (Fe~{\sc ix/x}), 195~\AA\ (Fe~{\sc xii}), 284~\AA\ (Fe~{\sc xv}) and 304~\AA\ (He~{\sc ii}) are used to study solar features visible at corresponding temperatures as listed in Table \ref{eit_passbands}. The temperature response of the EIT filters are shown in Figure~\ref{eit_tempresp}. In the present work we use 195~\AA\ images. Note, that the peak temperature for the 195~\AA\ filter is at 1.6~MK, and at lower temperatures the signal decreases significantly. This allows the lower temperature CHs to appear dark in comparison to the higher temperature QS and ARs. Figure~\ref{spectra1} also shows that for temperatures of $\sim$1-2~MK the Fe~{\sc xii} line intensity is significant at 195~\AA\, while at lower temperatures of 0.3-0.5~MK there is no contribution from Fe~{\sc xii} (Figure~\ref{spectra2}). The synthetic spectra shown in Figure~\ref{spectra1} and \ref{spectra2} were created using the CHIANTI atomic database and software (for more information see http://www.ukssdc.ac.uk/solar/chianti/applications/index.html).

In the present work we used data from the 195~\AA\ wavelength band as it provides the best contrast between the brighter QS and the darker CHs. During the EIT data analysis, the data was calibrated using the standard techniques provided by the $\it{eit\_prep}$ algorithm available in the SolarSoft software package \citep{Freeland98}. The calibration procedures are detailed in Section~\ref{dataprep}.

\begin{table}[!t]
\begin{center}
\caption[The EIT bandpasses, peak temperatures and observables]{The EIT bandpasses, peak temperatures and observables \citep{Delaboudiniere95}.}
\label{eit_passbands}
\begin{tabular}{llll}
\\
\hline
\\
Wavelength & Ion           & Peak Temperature      & Observational Objective \\
\\
\hline
\\
304 \AA\   & He~{\sc ii}   & 8.0 $\times$ 10$^4$~K & chromospheric network, coronal holes \\
171 \AA\   & Fe~{\sc ix/x} & 1.3 $\times$ 10$^6$~K & corona/transition region boundary,\\
& & & structures in coronal holes\\
195 \AA\   & Fe~{\sc xii}  & 1.6 $\times$ 10$^6$~K & quiet corona outside coronal holes\\
284 \AA\   & Fe~{\sc xv}   & 2.0 $\times$ 10$^6$~K & active regions\\
\\ 
\hline 
\end{tabular}
\end{center}
\end{table}


\subsection{MDI}

The MDI instrument \citep{Scherrer95} was primarily built to probe the solar interior using helioseismological techniques and measure the line-of-sight magnetic field, line and continuum brightness at photospheric levels. The detector is a 1024 $\times$ 1024 pixel CCD, which can operate at 2 or 0.6 arcsec pixel resolution. A Lyot filter and two Michelson interferometers are used to create sets of filtergrams  obtained at five wavelengths which are combined onboard to produce the continuum intensity maps (white-light images), velocity maps (Dopplergrams) and magnetograms. The line-of-sight velocity is determined from the Doppler shift of the Ni~{\sc i} 6768~\AA\ solar absorption line, while the line-of-sight magnetic field is determined from the circular polarisations of the $v_{||}$ component of the velocity. The latter allows the magnetic field intensity maps (magnetograms) to be created from the filtergrams based on the physical process known as the {\it Zeeman effect}. This is explained in the following section.

\subsubsection{Zeeman effect}
\setstretch{1.8}

\begin{figure}[!t]
\includegraphics[clip=,scale=0.55, trim=40 0 40 0]{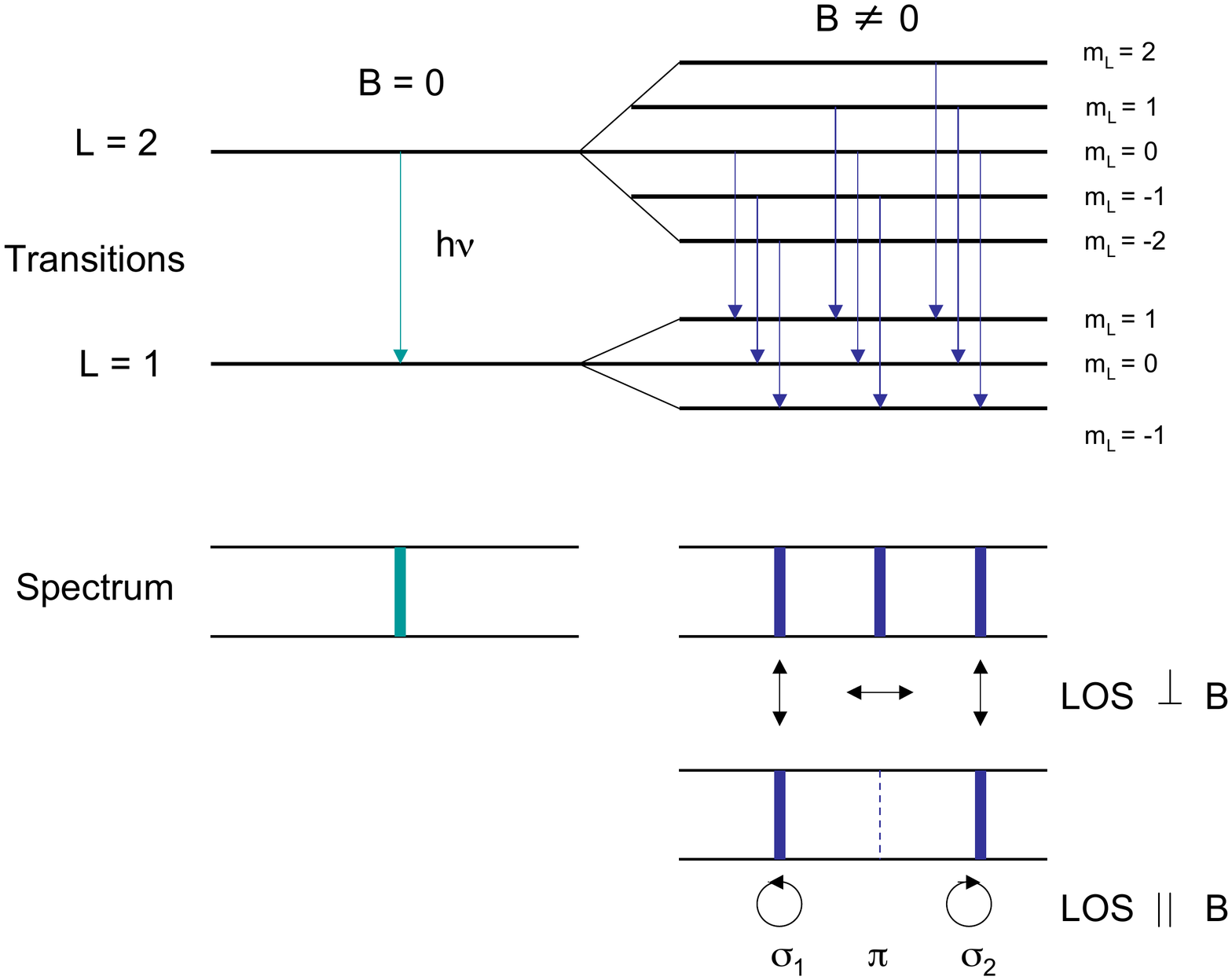}
\caption{A schematic of the normal Zeeman line-splitting. When no magnetic field (B) is present in the plasma, the electron transitions produce one line (shown on the left), while multiple lines are produced with magnetic field present (as shown in the right). In the normal Zeeman-effect three linearly polarised lines appear, two $\sigma$ and the middle $\pi$ when the magnetic field is perpendicular to the line-of-sight (LOS). Only the two $\sigma$ lines are seen with opposite circular polarisation when the magnetic field is along the LOS.} 
\label{zeeman} 
\end{figure}

The effect occurs when external magnetic field is applied to a plasma, which results in sharp spectral lines splitting into multiple closely spaced lines due to the interaction between the magnetic field and the magnetic dipole moment of electrons. The left side of Figure \ref{zeeman} shows the single hydrogen line produced due to the electron transfer from the higher energy level to the lower when no magnetic field is present. This transition marked with a green arrow from $L=2$ to $L=1$, where $L$ is the angular momentum of the electron. The right side of the figure shows three lines produced due to the transitions from the energy levels created by the different magnetic dipole moments ($m$) of the electron when magnetic field is present. 

\begin{figure}[!t]
\centerline{
\includegraphics[clip=,scale=1]{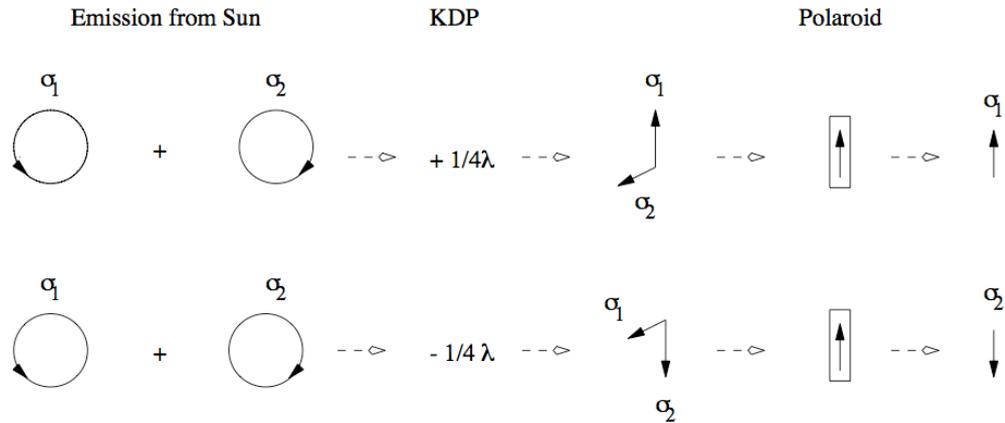}}
\caption{The schematic of the Babcock type magnetograph operation. The circularly polarised light is converted into plane-polarised light by an electrooptic crystal, and is shifted by 90$^{\circ}$ based on the sign of the voltage applied. The Polaroid then selects the components \citep{Gallagher_thesis}.} 
\label{babcock} 
\end{figure}

The number of lines produced depends on the alignment of the magnetic field to the line-of-sight. If the magnetic field is transverse, a line is produced at the original line position ($\pi$, perpendicular to the magnetic field) and two other offset lines are created which are linearly polarised ($\sigma$, parallel with the magnetic field). If the magnetic field is longitudinal, two offset lines are produced which are circularly polarised ($\sigma$) in opposite directions. Longitudinal magnetic fields can be shown by the circular polarisation of the produced $\sigma$ lines. The wavelength difference between the $\pi$ and a $\sigma$ spectral line is proportional to the magnetic field present:
\begin{equation}
\bigtriangleup\lambda_{B}=4.7\times10^{-13}g\lambda^{2}B.
\end{equation} 
Here, $B$ is the magnetic field strength and $g$ is the Land\'e factor of the spectral line. The detection of the Zeeman splitting can be difficult or even impossible for weak magnetic fields. For this reason the polarisation properties are used to determine the magnetic field strength. The MDI instrument operates on the principle of the standard Babcock magnetograph (Figure~\ref{babcock}). First, the $\sigma$ line components are separated using an electrooptic cell, which converts the $\sigma$ components into linearly polarised beams at right-angles. Next, a linear polariser is used to let through one beam at a time, using different sign voltages.

The intensity difference between the signals in each state is directly proportional to the line-of-sight magnetic field strength, and hence can be used to create magnetograms.

\section{STEREO}

The STEREO A \& B twin satellites were launched on the 25 October 2006 from Cape Canaveral Air Force Station in Florida. The mission includes two identical satellites - the STEREO-A spacecraft that orbits ahead and slightly inside of Earth's orbit, and the STEREO-B that orbits behind and slightly outside the Earth's orbit. For this reason the STEREO satellites provide a unique opportunity to carry out stereoscopic observations of the Sun. The spacecrafts drift away in opposite directions from Earth by 22 degrees per year due to the mentioned small differences in their distance from the Sun. When the spacecrafts are furthest from each other (being at the opposite sides of the Sun, like at the time of writing) they provide a nearly 360$^\circ$ angle view of the Sun. This has been achieved for the very first time in the history of solar observation.
\\ \indent
The STEREO mission goals are to study the initiating mechanisms of CMEs and their propagation through the heliosphere, to determine the location of energetic particle acceleration in the low corona, and to determine the structure of the solar wind. There are four instrument packages aboard STEREO, the names of which speak for themselves: the Sun Earth Connection Coronal and Heliospheric Investigation (SECCHI), the In-situ Measurements of Particles and CME Transients (IMPACT), the Plasma and Suprathermal Ion Composition (PLASTIC) instrument, and the interplanetary radio disturbance tracker STEREO/WAVES (SWAVES). The SECCHI package contains two coronagraphs (COR1 \& 2), a heliospheric imager (HI), and an EUV imager (EUVI).
\\ \indent
In this thesis we used data from the EUVI and PLASTIC instruments, which are discussed in detail in the sections below.

\subsection{EUVI}
The STEREO/EUVI instrument \citep{Wuelser04} works on the same principle as the SOHO/EIT instrument in creating images at extreme ultraviolet wavelengths (171~\AA, 195~\AA, 284~\AA, and 304~\AA). Also a Ritchey-Chr\'etien telescope in its optical design, it images the Sun at the same four wavelengths as SOHO/EIT, but at higher image cadence and spatial resolution.

\begin{figure}[!ht] 
\centerline{\hspace*{0.015\textwidth} 
\includegraphics[clip=,scale=0.8]{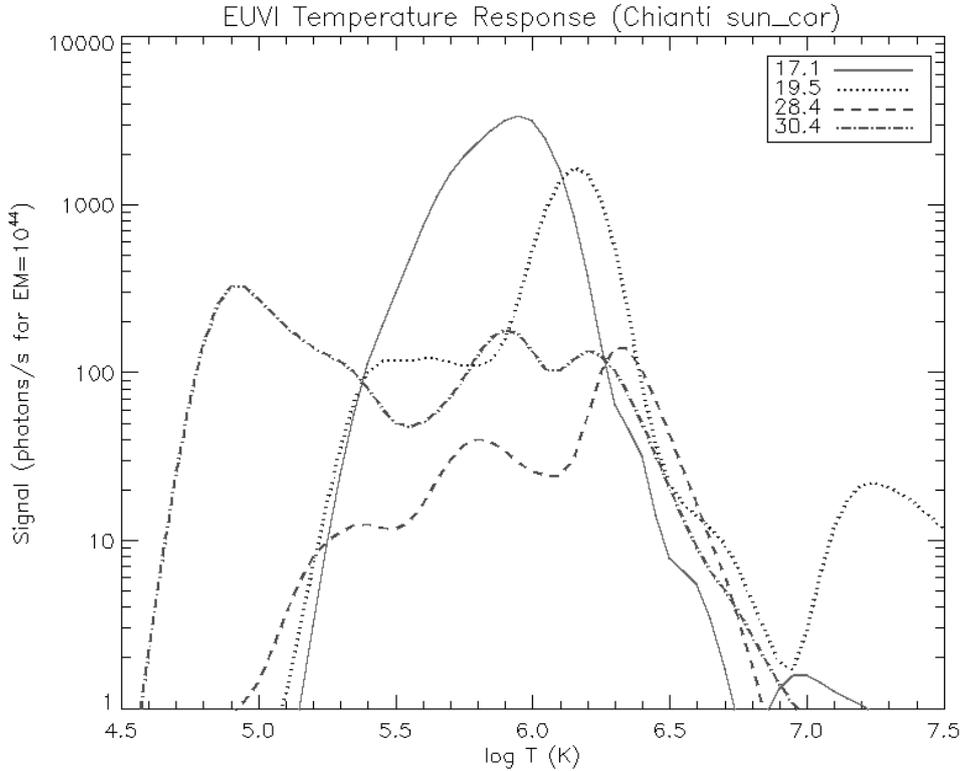}}
\caption{The temperature response of the EUVI instrument \citep{Wuelser04}.} 
\label{euvi_tempresp} 
\end{figure}

The primary and secondary mirrors consist of four quadrants, all of which are coated with a narrow-band reflective coating optimised for one of each four wavelengths mentioned. The coating material is MoSi. Two metal film filters are used at both the entrance aperture and the focal plane of the telescope to reduce solar heat, and filter out the undesired optical, EUV and IR radiation. For the short wavelength (171~\AA, 195~\AA) quadrants this filter is an aluminium-on-polyamid foil on a coarse nickel grid, while on the longer wavelength quadrants a single layer of aluminium foil was used on a fine nickel mesh. The temperature response of the different filters is shown in Figure \ref{euvi_tempresp}. The EUVI full field of view displays the Sun out to $\pm1.7$ solar radii, with a $1.6$ arcsec pixel resolution. The detector is a backside illuminated 2048 $\times$ 2048 pixel CCD. 

Although the above description refers to one instrument, two identical EUVI instruments were built for the STEREO mission. One is located on STEREO-A and the other on STEREO-B, allowing the 3D imaging of the solar corona.

\subsection{PLASTIC}

The Plasma and Suprathermal Ion Composition (PLASTIC; \citeauthor{Galvin08} \citeyear{Galvin08}) aboard STEREO makes in-situ measurements of the solar wind elemental composition, ionic charge states, ion bulk flow parameters, and heliospheric suprathermal ions in the energy range from $\sim$0.3 to 80 keV/e (including ions created in local shock acceleration processes). These measurements were designed to investigate the link between coronal and solar wind ion processes, and the link between solar wind and heliospheric processes.

PLASTIC has three fields-of-view and four instrument sections to observe both plasma and suprathermal particles. The PLASTIC Solar Wind Sector (SWS) provides a 45$^\circ \times $20$^\circ$ Sun-Earth line centered field-of-view and observes in the 0.3--10.6 and 0.3--80 keV energy range. The SWS consists of two parts: the Small Channel measures the density, velocity, thermal speed and the alpha/proton ratios of the solar wind protons, while the Main Channel measures the composition, charge state, and the bulk and thermal speeds of the more abundant solar wind minor ions (e.g., C, O, Mg, Si, and Fe).

The two channels of the Suprathermal Ions Wide-Angle Partition Sector (WAP) cover a 50$^\circ \times $6$^\circ$ and a 175$^\circ \times $6$^\circ$ field-of-view in the off-Sun azimuth direction. 
Both channels measure the composition, flux and direction of suprathermal ions in the 0.3--80 keV energy range. Similarly to all STEREO instruments, identical PLASTIC instruments were built for STEREO-A and B.

\section{Hinode}

The Hinode (formerly known as Solar-B) spacecraft was launched on 23 September 2006 from the Uchinoura Space Center in Kagoshima, Japan. The Hinode mission objectives are to study the magnetic field generation and transport, the processes responsible for the heating and structuring of the chromosphere and corona, and the mechanisms triggering flares and CMEs. The three instruments aboard Hinode are the Solar Optical Telescope (SOT) (including a Narrow and Broadband Filter Instrument, and a Spectropolarimeter), the Extreme Ultraviolet Imaging Spectrometer (EIS) and the X-Ray Telescope (XRT; \citeauthor{Golub07} \citeyear{Golub07}). In the present research data from the XRT instrument was used.

\subsection{Hinode/XRT}

\begin{figure}[!t] 
\centerline{\hspace*{0.015\textwidth} 
\includegraphics[scale=0.5, trim=0 150 0 130, clip ]{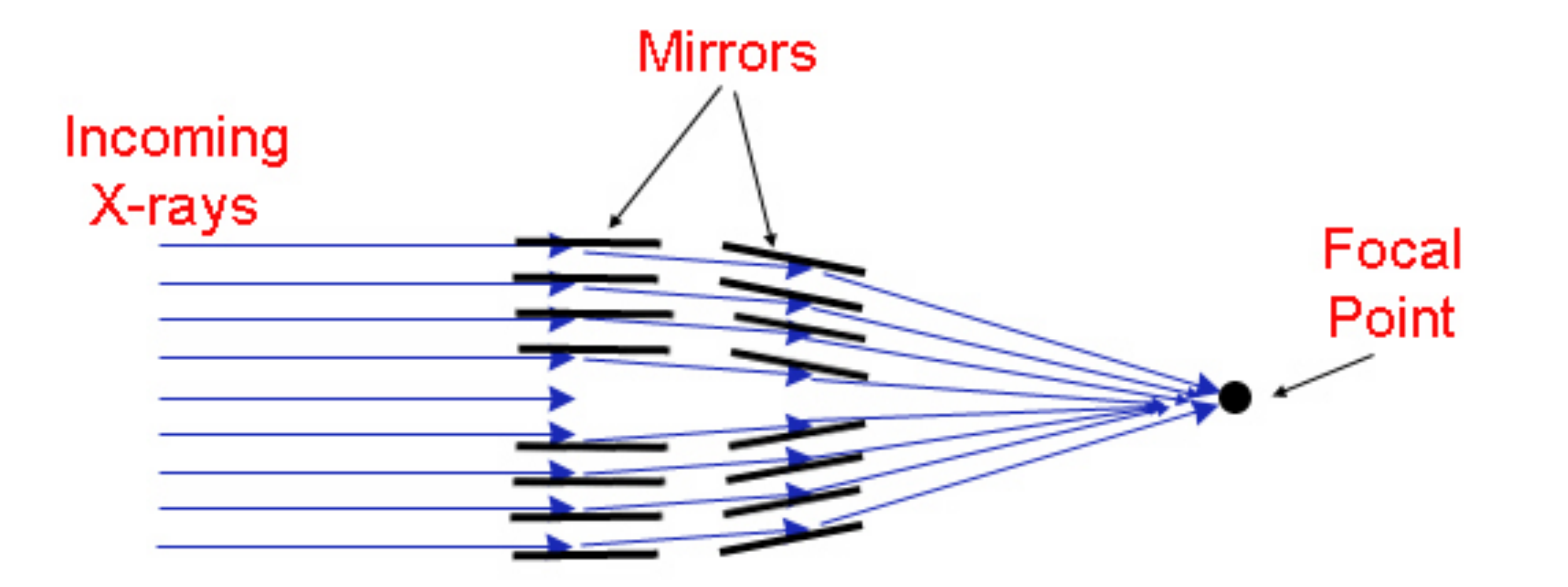}}
\caption{Schematic of the grazing incidence technique, which allows X-ray imaging.} 
\label{grazinginc} 
\end{figure}

The design of X-ray telescopes is radically different from classic reflectors since the X-ray photons are of such high-energy, they would pass through a standard reflector mirror. To capture the photons, they must be bounced off at very low angles. This technique is called {\it grazing incidence}, where the mirrors are mounted slightly off the parallel line of the incoming X-rays (see Figure~\ref{grazinginc}). The X-rays are then focused and recorded with the CCD camera. On XRT this is carried out with a modified Wolter I grazing incidence telescope, which has a 30~cm aperture. 

\begin{figure}[!t] 
\centerline{\hspace*{0.015\textwidth} 
\includegraphics[clip=,scale=1.2]{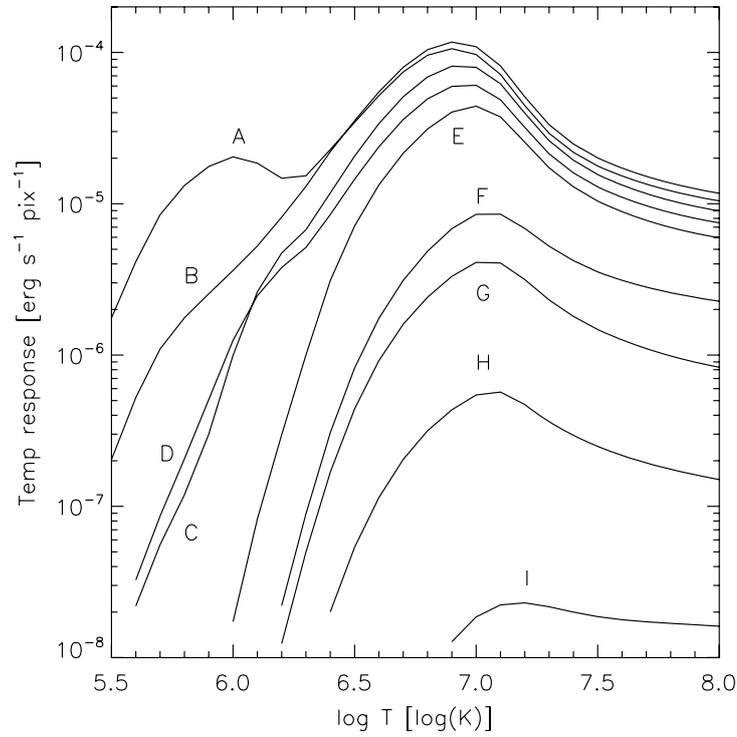}}
\caption{The temperature response of the XRT instrument for all the focal-plane filters (A = Al-mesh, B = Al-poly, C = C-poly, D = Ti-poly, E = Be-thin, F = Be-med, G = Al-med, H = Al-thick, and I = Be-thick). Figure courtesy of \cite{Golub07}.} 
\label{xrt_tempresp} 
\end{figure}

To filter out visible light and reduce heat, nine different aluminium based pre-filters are used made of aluminium, carbon, titanium and beryllium. The filters provide five passbands in the 6--60~\AA\ range allowing the observation of solar features in the temperature range $\sim$1--30 MK. The temperature responses of the different filters is shown in Figure \ref{xrt_tempresp}. This shows that apart from the first two filters (Al-mech and Al-poly), all the filters are sensitive at very high temperatures ($>$~1~MK), while this sensitivity drops considerably at lower temperatures ($\sim$~0.8~MK - typical of CHs), which provides a high contrast between CHs and the QS. The instrument has a 35 arcmin field of view with a 1~arcsec pixel spatial resolution. The detector is a back-illuminated 2048$\times$2048 pixel CCD.

\section{ACE}

ACE \citep{Stone98} was launched on 25 August 1997 from the Kennedy Space Center in Florida. Similarly to SOHO, ACE orbits in the L1 Lagrangian point observing the Sun without interruption from a distance of 148.5$\times$10$^{6}$~km and should be able to maintain its orbit until 2024, based on its fuel reserves. Due to its continuous ability to sample the solar wind, ACE can provide space weather warnings shortly ($\sim$1~hour) before a magnetic storm hits Earth. Often it is difficult to provide earlier warnings in the case of large geomagnetic storms originating from flares and CMEs, however minor storm warnings can be improved by detecting the features responsible (e.g. CHs) on solar images several days in advance. The HSSW forecasting method discussed in this thesis relies on the data from the Solar Wind Electron Proton Alpha Monitor (SWEPAM) instrument detailed in the following section.

ACE has three monitoring instruments and six high-resolution sensors which detect particles of solar and galactic origin, measure electron and ion fluxes, and determine the direction and the strength of the interplanetary magnetic field. These instruments are the Electron, Proton, and Alpha Monitor (EPAM), the Solar Wind Electron Proton Alpha Monitor (SWEPAM), the Magnetic Field Experiment (MAG), the Solar Energetic Particle Ionic Charge Analyzer (SEPICA), the Solar Isotope Spectrometer (SIS), the Cosmic Ray Isotope Spectrometer (CRIS), the Solar Wind Ion Composition Spectrometer (SWICS), the Solar Wind Ions Mass Spectrometer (SWIMS), and the Ultra Low Energy Isotope Spectrometer (ULEIS).

\subsection{SWEPAM}

SWEPAM \citep{McComas98} was designed to study the bulk flow and kinetic properties of the solar wind, and the suprathermal electrons and ions. Its primary goals are to provide solar wind measurements to further our understanding of the corona and solar wind formation, as well as the particle acceleration.

\begin{figure}[!t]
\includegraphics[width=\textwidth]{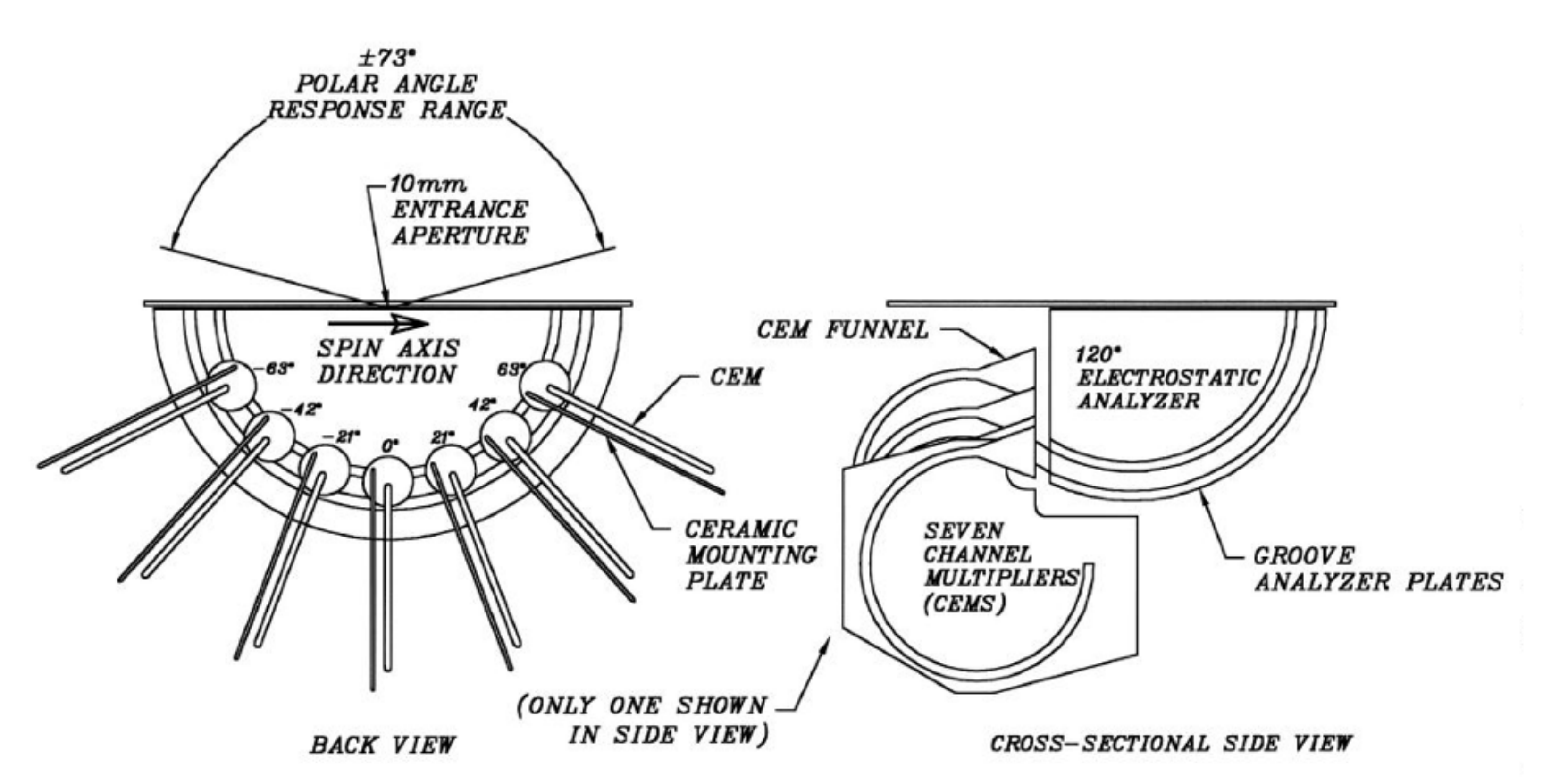}
\caption{Schematic of the SWEPAM-E instrument \citep{McComas98}.} 
\label{swepam} 
\end{figure}

It consists of two main parts, the ion and the electron detector: SWEPAM-I and SWEPAM-E, respectively. As ions enter the SWEPAM-I aperture, the electrostatic analyser allows only ions of a specific narrow energy and angular range to pass through (Figure~\ref{swepam}). The ions are then detected by channel electron multipliers. Similarly to SWEPAM-I, the SWEPAM-E instrument uses an electrostatic analyser and channel electron multipliers, but while SWEPAM-I separates ions by applying a negative voltage, SWEPAM-E applies a positive voltage to filter out electrons of a narrow energy and angle range. The field of view is fan-shaped in both instruments with a wide azimuthal angle. The incidence measured in the separate electron multipliers determine the angle at which the ions or the electrons entered the instrument. The electrons are detected in the energy range 1.6 eV -- 1.35 keV. The fan shaped field of view covers 95\% of 4$\pi$ steradian due to the spinning of the spacecraft. 

\chapter{Coronal hole detection} 
\label{chapter:detection}

\ifpdf
    \graphicspath{{3/figures/PNG/}{3/figures/PDF/}{3/figures/}}
\else
    \graphicspath{{3/figures/EPS/}{3/figures/}}
\fi

\flt
{\it Since the discovery of coronal holes, many different methods have been devised to improve their detection. From manual drawing to advanced intensity thresholding techniques, the results can differ considerably. The inconsistency between the coronal hole maps is sometimes due to the difference in the detection approaches, while in other cases it is an artifact of using different wavelengths that represent a different height and temperature range at which the coronal holes are observed. In this Chapter we discuss different previous methods developed to detect coronal holes, and compare the resulting coronal hole maps to those obtained using the method developed by the author. The foundation of this thesis research, the Coronal Hole Automated Recognition and Monitoring ({\it CHARM}) algorithms, are discussed in detail. The diverse use of the detection method is demonstrated by the use of several instruments and wavelengths. The research detailed in this Chapter has been published in the Solar Physics Journal \citep{Krista09}.
}
\flb 

\newpage

\section{Introduction}
In order to provide reliable information on their physical properties, CHs must be detected in a consistent manner. Initial CH detection methods involved hand-drawn maps, such as the method by \cite{Harvey02} based on the two-day average of ground-based He~{\sc i} (10830~\AA) images and magnetograms. However in recent years, a number of automated detection methods have been developed: \citet{Henney05} used He~{\sc i} (10830~\AA) images, magnetograms and the above-mentioned hand-drawn CH maps; \citet{Malanushenko05} used He~{\sc i} imaging spectroscopy to determine CH boundaries; while \citet{deToma05} have combined space-based EUV images with ground based He~{\sc i} (10830~\AA) and H$\alpha$ observations and magnetograms to identify coronal hole regions. \citet{Chapman02} and \citet{Chapman07} constructed synoptic CH maps using SOHO Coronal Diagnostic Spectrometer (CDS; \cite{Harrison97}) measurements. \citet{Barra07} used a multichannel segmentation method on EUV images to differentiate CHs, QS and ARs. More recently, \citet{Scholl08} applied an image segmentation technique based on a combination of region and edge-based methods using 171~\AA\ (Fe~{\sc ix/x}), 195~\AA\ (Fe~{\sc xii}), 304~\AA\ (He~{\sc ii}) images and Michelson Doppler Imager (MDI; \cite{Scherrer95}) magnetograms to determine CH boundaries.
\\ \indent
The first aspect to be discussed is the visibility of CHs at different wavelengths and using different instruments. Some of the above mentioned methods used ground-based images which are limited by environmental effects (atmospheric motion can lower the resolution, while clouds disturb the continuity in the observations). The analysis and results presented in this thesis are based solely on images produced by space-based instruments that are free of environmental effects. Some authors used He~{\sc i} images that provide low contrast between CHs and the QS. It should be noted though that He~{\sc i} images do have an advantage in having a relatively narrow temperature range which can reduce line-of-sight obscuration effects. However, EUV and X-ray images have a larger contrast between CHs and QS, but are produced with filters of broad temperature response that resolve high-temperature solar structures. This can lead to the obscuration of CH boundaries, and the effect becomes worse near the solar limb where the line-of-sight angle is smaller.
\\ \indent
The second aspect that should be considered is the run-time of the data analysis. The algorithm described in this chapter processes a single high-resolution 1024 $\times$ 1024 pixel EUV or X-ray image and a magnetogram under two minutes using a computer with a 2.2 GHz Intel Core 2 Duo processor and 2GB 667 MHz memory. The detection of CHs in near real-time manner proves a useful application for space weather forecasting purposes. 
\\ \indent
The third and most important aspect is the intensity threshold determination for CHs. Although determining the intensity threshold is of fundamental importance to accurately locate CH boundaries (and thus determine their physical properties), many of the methods discussed in the following section rely on arbitrary choices. The method developed by the author determines the CH boundary threshold based on an intensity thresholding technique that is also dependent on the mean QS intensity which ensures that the method works reliably at any time of the solar cycle.

\section{Previous methods of CH detection}

In this section we discuss the methods of CH detection developed by other authors. As there are numerous approaches, only those detecting CHs on the whole visible surface are included here (e.g we exclude studies that only concern polar CH identification). The method descriptions concentrate on what instrument(s) and wavelength(s) were used, the method of identification, what period was covered in the study and whether the CH maps are publicly available. Where it is possible, an arbitrarily chosen CH map from the study is compared with a corresponding map obtained using the author's method. 
 
\subsection{The K. Harvey and Recely method}
\label{harvey_method} 

\cite{Harvey02} used the ground-based National Solar Observatory/Kitt Peak (NSO/KP) synoptic rotation magnetograms and He~{\sc I} 10830~\AA\ spectroheliograms to identify coronal holes. The method includes the magnetograms and spectroheliograms being remapped into sine-latitude and Carrington longitude, but the exact details of the identification process is not discussed. A relevant later paper \citep{Henney05} indicates that the \citeauthor{Harvey02} CH maps were drawn by hand using computer assistance. 

CHs are known to appear as regions of enhanced emission in He~{\sc I} 10830~\AA\ images \citep{Andretta97}. This is due to the overpopulation of the He~{\sc I} 10830~\AA\ transition levels from coronal radiation which leads to increased absorption of radiation originating in QS regions, while CHs have reduced coronal emission and thus appear as bright regions at 10830~\AA . To aid the differentiation of CHs from other QS features \cite{Harvey02} defined CHs as having less than 20\% quiet-Sun network contrast, a predominately unipolar magnetic field containing 75--100\% of one polarity, and having the area of at least 2 supergranules ($\sim$109~km$^{2}$).


The advantages of the method are the low line-of-sight obscuration in the He~{\sc I} 10830~\AA\ images, and the CHs are observed at their ``base" (the chromosphere). Also, magnetograms were used to confirm the predominant polarity of the CHs detected. The disadvantages of the method are the low intensity contrast between the QS and CH regions in the He~{\sc I} images and the hand-drawn quality of the CH maps.

The comparison between the \cite{Harvey02} CH map and the corresponding map created by the author is shown in the next section together with results of the next method. An extensive database (1989--2002) of the \cite{Harvey02} CH maps is available online\footnote[1]{\href{ftp://nsokp.nso.edu/kpvt/synoptic/choles/}{ftp://nsokp.nso.edu/kpvt/synoptic/choles/}}.

\subsection{The Henney and J. Harvey method} 
\label{henney_method}

\cite{Henney05} developed an automated CH detection algorithm using the above mentioned hand-drawn maps, SOLIS Vector SpectroMagnetograph (VSM) helium spectroheliograms and photospheric magnetograms. As a first step, consecutive pairs of spectroheliograms were derotated to the time of the later observation and remapped into sine-latitude and longitude. This was followed by taking two-day averages of the image pairs, after which the previously mentioned hand-drawn CH contours were overlaid on the average magnetogram and helium spectroheliogram in order to determine the normalized He~{\sc I} 10830~\AA\ equivalent width and mean percent unipolarity. 

The next step involved taking only values above 1/10 of the median of the positive values in the average helium image. This was followed by removing the negative values, smoothing the image, and all values were set to a fixed number. Next, a square kernel function was used as the shape-operator to fill gaps and connect nearby regions while small regions were removed (the same 2 supergranular size threshold was applied as in \cite{Harvey02}). The produced image was then used as a mask to set coinciding pixels in the median-selected image to a high value. The final image was then obtained by smoothing, and removing small CHs and those of low unipolarity percentage.

The advantages of the method are the automated quality of the detection, the use of a wavelength that provides low line-of-sight obscuration, and using magnetograms to confirm the predominant polarity of the CHs detected. The disadvantages are the low contrast between the CH and QS regions that could impair the detection of the CH boundary, and the intensity threshold, which was based on the hand-drawn maps.

It must also be noted that the publicly available SOLIS CH maps are Carrington rotation maps which are created from stacked central meridian plots over $\sim$27 days. ``Time-delayed" Carrington plots are only an approximation of the global CH locations and morphology. (I.e. by the time the Carrington map is created one end of the image is almost a month ``older" than the other end of the image.)

The \cite{Henney05} Carrington rotation CH maps are available from 2004, and are continuously updated. The database is available online\footnote[2]{\href{http://solis.nso.edu/vsm/vsm_maps.php}{http://solis.nso.edu/vsm/vsm$\_$maps.php}, ``Archive of solar wind source synoptic maps"}.
	
\begin{figure}[!ht]
\centerline{\hspace*{0.015\textwidth} 
\includegraphics[]{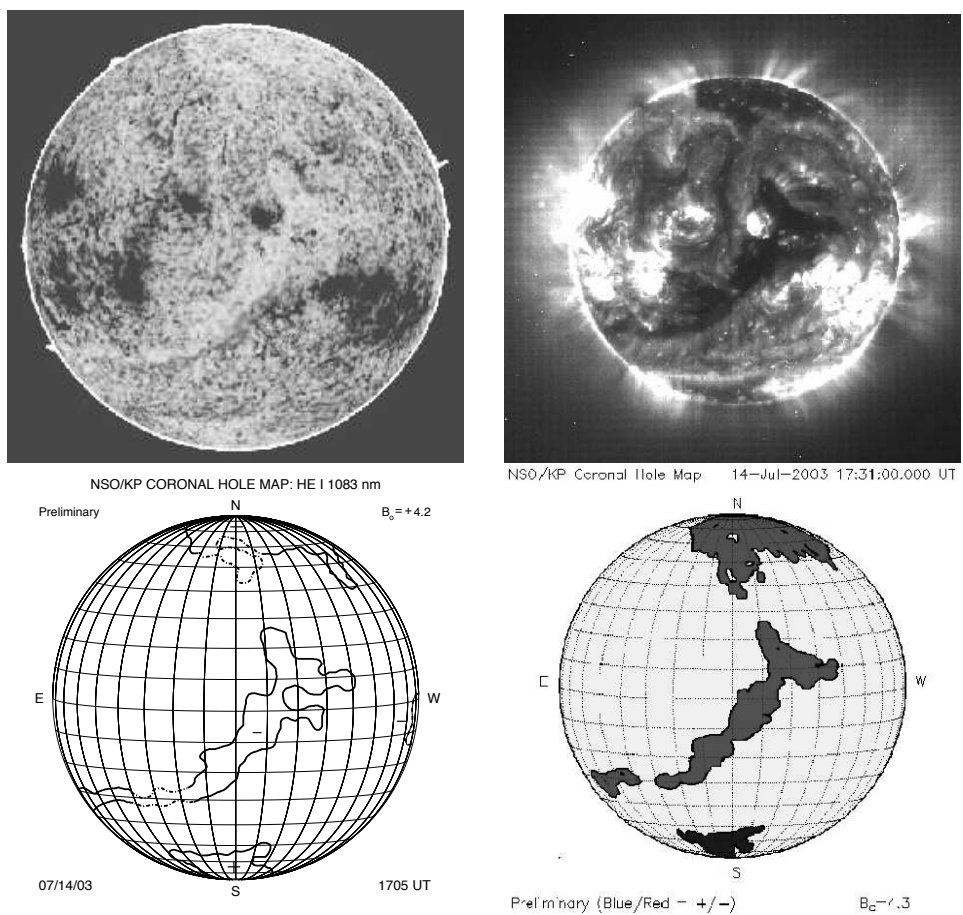} }
\caption{The top left and right images are a NSO/KP He~{\sc I} 10830~\AA\ spectroheliogram and a SOHO/EIT 195~\AA\ image taken on 14 July 2003, respectively. The bottom left and right images are the CH maps created with the computer assisted hand-drawn method devised by \cite{Harvey02} and the corresponding map from the \cite{Henney05} automated CH detection method. Figure courtesy of \citep{Henney05}.} 
\label{he_chmap} 
\end{figure}

\begin{figure}[!ht]
\includegraphics[scale=0.51, trim=30 0 30 0, clip]{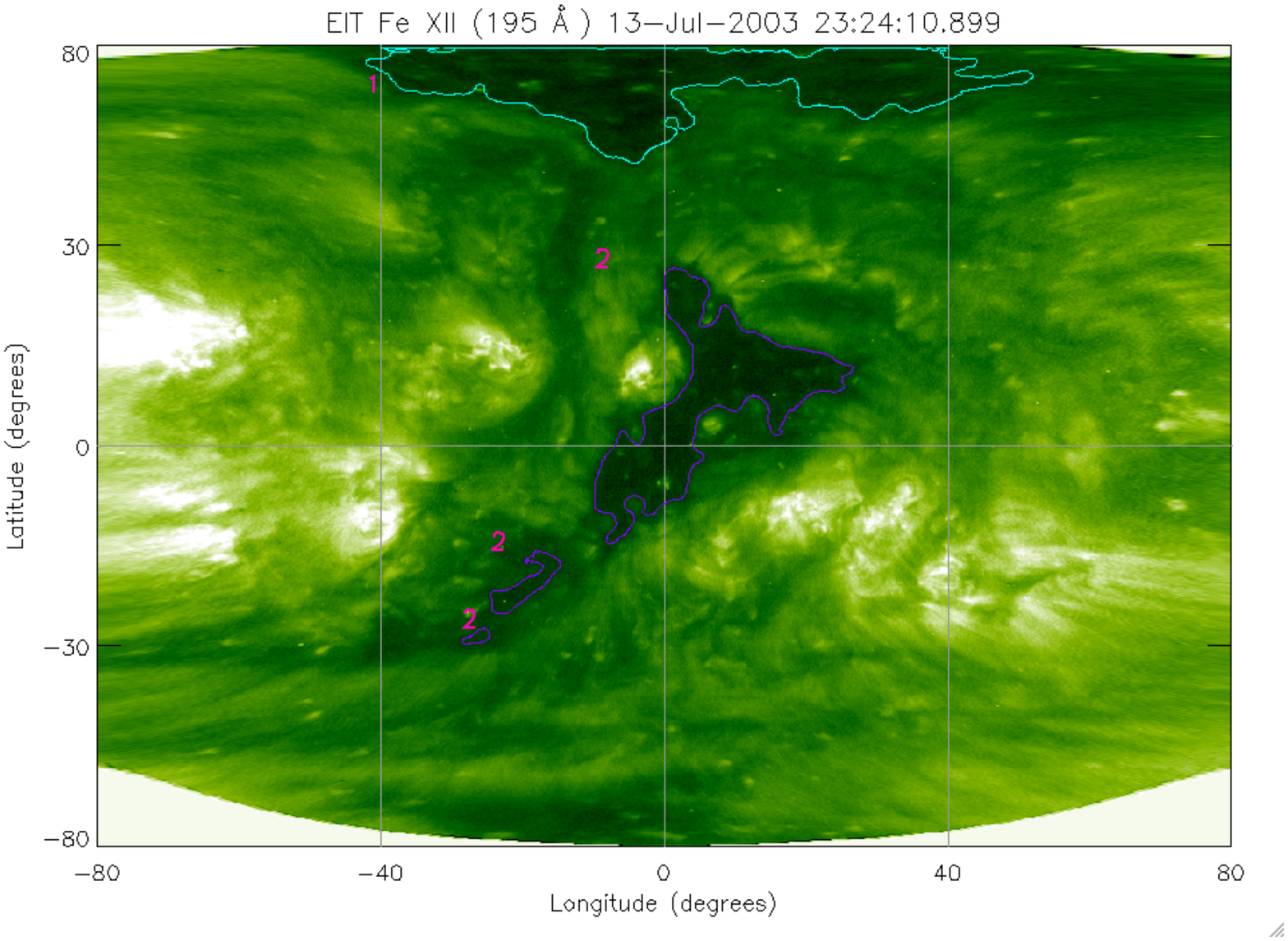}  
\caption{The SOHO/EIT CH map created by the author to compare with the results of the \citeauthor{Harvey02} and \citeauthor{Henney05} methods as seen in Figure \ref{he_chmap}. An EIT 195~\AA\ observation taken on 13 July 2003 and a corresponding magnetogram was used in the detection. (The numbers represent different groups of CHs, as discussed later in the chapter.) } 
\label{he_compare} 
\end{figure}

Figure \ref{he_chmap} shows a NSO/KP He~{\sc I} 10830~\AA\ spectroheliogram and a SOHO/EIT 195~\AA\ image taken on 14 July 2003 (top left and right, respectively), and the computer assisted hand-drawn maps by \cite{Harvey02} and the corresponding multi-step automated CH detection by \cite{Henney05} (bottom left and right, respectively). For comparison, the CH contours obtained using the authors' method is shown in Figure \ref{he_compare} on a Lambert equal area projection map.

\subsection{The Chapman and Bromage method}

\cite{Chapman02} used synodic data from the SOHO/CDS Normal Incidence Spectrometer (NIS) to develop a partially automated CH detection method. Nine 4$\times$4 arcmin rasters were used to create an image-strip along the solar meridian. The strips were obtained daily to produce a Carrington map. Among the available NIS wavelengths the Mg~{\sc ix} line observations were used for identification, since in these images CHs are distinguishable as dark regions. After the low emission regions were extracted, the CHs were manually differentiated from other low intensity regions using He~{\sc i} 10830~\AA\ images. Carrington rotation CH maps were then generated by joining raster images from consecutive days, removing overlapping areas. The method was applied to solar images in the period July 1997 -- Sep 2001. \cite{Chapman02} found that the CH area cycle was anti-correlated with the sunspot cycle.
\begin{figure}[!t]
\centerline{
\includegraphics[scale=0.55, trim=50 100 50 100, clip]{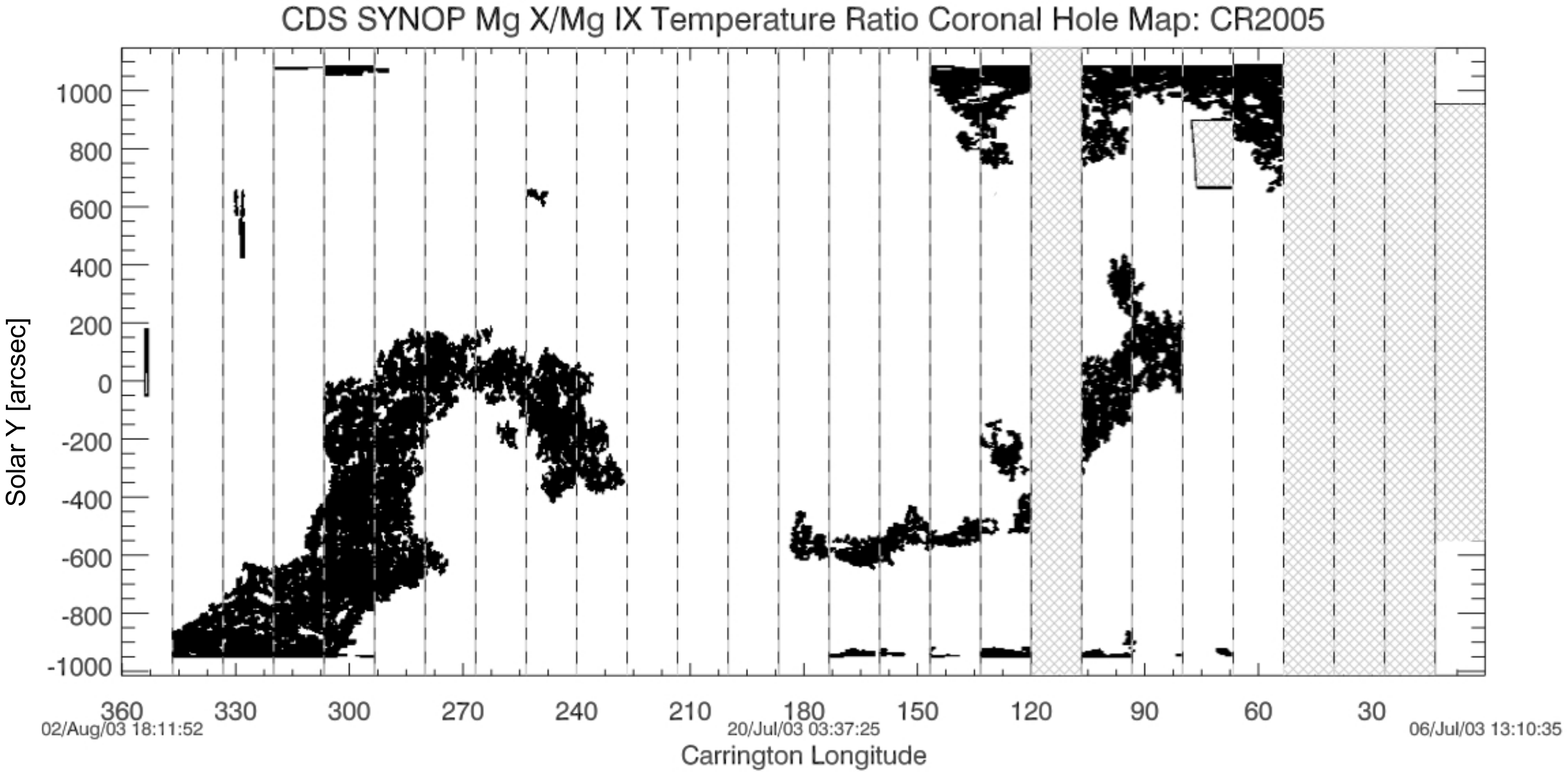} }
\caption{The CH map created by \cite{Chapman07} for Carrington rotation 2005.} 
\label{chapman} 
\end{figure}

The method was further developed in the following years. In the thesis research of \cite{Chapman07}, Mg~{\sc x}/Mg~{\sc ix} temperature ratio maps were used in conjunction with Mg~{\sc x} and Mg~{\sc ix} intensity images. Upper and lower boundaries were set to specify which low temperature and low intensity regions classified as CHs.  An example CH map by \citeauthor{Chapman07} showing the 13 July 2003 large transequatorial CH (at Carrington longitude 180--90) is shown in Figure~\ref{chapman} (for comparison see Figures~\ref{he_chmap} and \ref{he_compare} ).

The advantages of the method are the automated quality of the detection, the ability to investigate the temperature of the low intensity regions and the narrow bandpass imaging which limits line-of-sight obscuration effects. The shortcomings are the time-consuming analysis, the time-delayed raster-stacked Carrington maps and the absence of magnetic field analysis within the low temperature and intensity regions. 

The analysis was completed for the period 1998--2007, and the CH Carrington rotation maps are available online\footnote[3]{\href{http://solis.nso.edu/vsm/vsm_maps.php}{http://www.star.uclan.ac.uk/$\sim$sac/coronal$\_$holes/images.html}}.

\subsection{The Scholl and Habbal method}

\cite{Scholl08} used SOHO/EIT 304~\AA, 171~\AA, 195~\AA\ and 284~\AA\ observations and MDI magnetograms to identify CHs. A combination of image processing techniques were used to differentiate CHs and filaments. Although the steps are not detailed in the corresponding article, a list of image processing operations were given. Due to their lower contrast, the 304~\AA\ images were processed in a different way from the EUV images. From the numerous image processing steps, only the main ones are listed here. After calibration and normalisation, limb-darkening was removed from the 304~\AA\ images, and the contrast was enhanced in the images of all four wavelengths. This was followed by smoothing to connect dark regions, image-segmentation based on histogram-equalisation, and region-growing to isolate dark regions. A contouring method was used to extract regions of interest and remove small features. The candidate regions were then obtained separately at the four wavelengths. Finally, the regions were classified as CHs or filaments based on their magnetic field profile and the threshold values specified by the authors.

\begin{figure}[!t]
\includegraphics[scale=0.6, trim=80 0 80 0, clip]{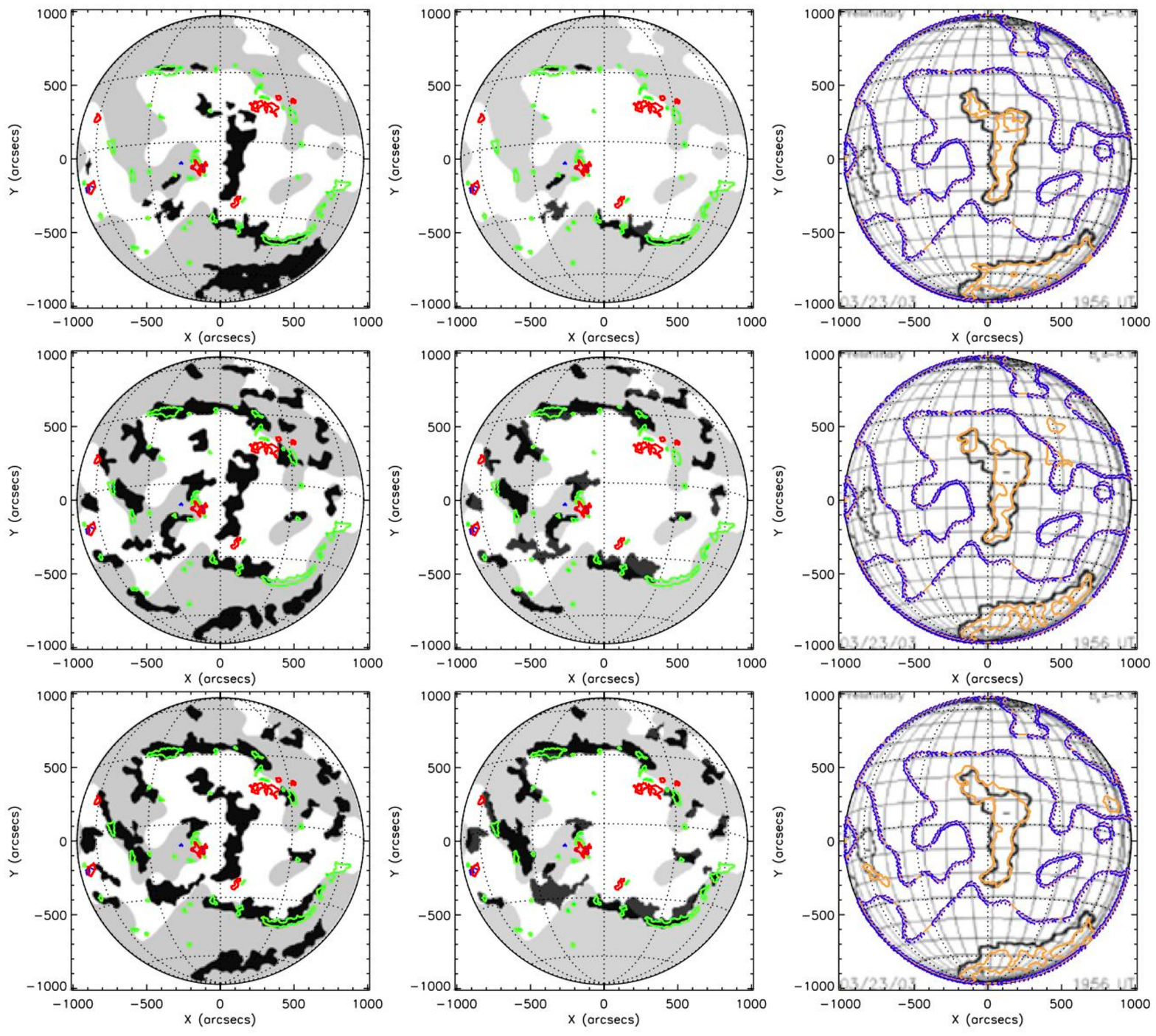}  
\caption{The feature classification map by \cite{Scholl08} showing CHs (orange, right panels), filaments (black, middle panels). SFC filaments (green) and active regions (red) are shown for comparison \citep{Bentley04}. The top, middle and bottom row maps are based on 304~\AA, 171~\AA, and 195~\AA\ SOHO/EIT observations, respectively. Observations were made on 27 May 1997. NSO/KP coronal holes are overplotted as black contours for comparison. The blue line separates the opposite-polarity magnetic-field areas shown as gray and white in the left and middle panels.} 
\label{scholl} 
\end{figure}

\begin{figure}[!t]
\includegraphics[scale=0.42, trim=30 0 30 0, clip]{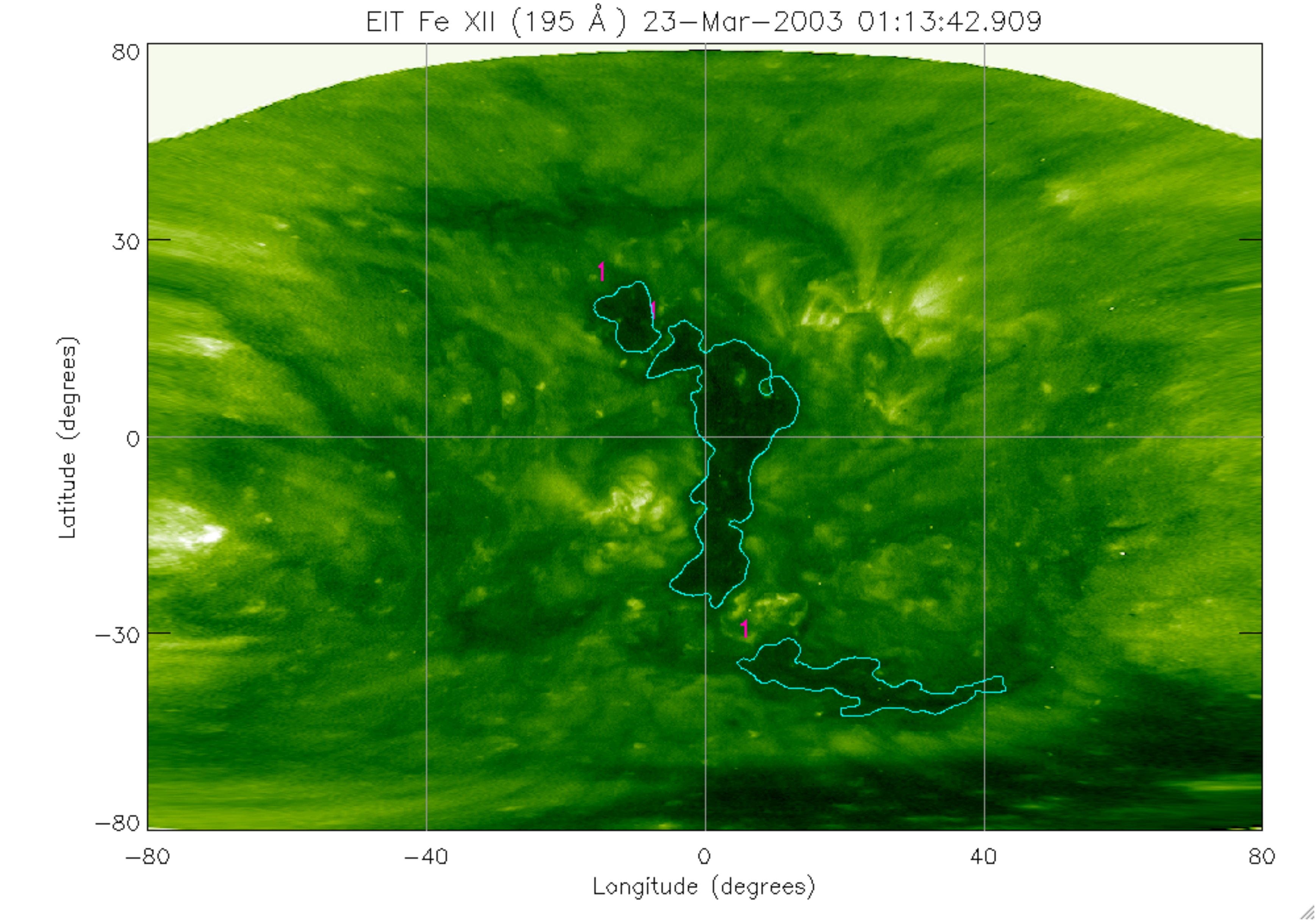}  
\caption{The SOHO/EIT CH map created by the author to compare with the results of the \citeauthor{Scholl08}. An EIT 195~\AA\ observation taken on 23 March 2003 and a corresponding magnetogram was used in the detection.} 
\label{scholl_compare} 
\end{figure}

An example for the \cite{Scholl08} method is shown in Figure~\ref{scholl}, with CHs (orange, right panels) and filaments contoured (black, middle panels). Filaments (green) and ARs (red) from the Solar Feature Catalogue (SFC) are shown for comparison \citep{Bentley04}. The top, middle and bottom row maps are based on 304~\AA, 171~\AA, and 195~\AA\ SOHO/EIT observations of 27 May 1997. In the right panels the NSO/KP CHs are overplotted in black contours for comparison. The corresponding CH map determined by our algorithm is shown in Figure~\ref{scholl_compare}.

The advantage of this method is the automated approach, the use of several wavelengths which map CHs from different perspectives of temperature and atmospheric height. Also, magnetograms were used to differentiate between CHs and filaments. The disadvantages are the time-consuming analysis of the numerous images, and the inconsistent success rate of the method during the solar cycle (it was shown to be more successful during the solar minimum). It is unclear whether an online database is available and what period was processed.


\section{The {\it CHARM} algorithm}

The {\it CHARM} algorithm presented in this section is a fast and automated method which allows the user to determine CH boundary thresholds in a consistent manner throughout the solar cycle. The detection tool can also be used at different wavelengths and with different instruments. Additional packages allow the user to study the short-term changes in CH boundaries, and to determine the arrival time of high-speed solar wind streams at Earth based on the CH boundary locations. These are discussed in Chapter~\ref{chapter:evolution} and Chapter~\ref{chapter:forecasting}, respectively.

\subsection{Data preparation}
\label{dataprep}

The detection tool requires the user to specify the wavelength and instrument. The use of EUV 195~\AA\ and X-ray wavelengths are advised for CH detection as they provide better CH boundary contrast than e.g. 171~\AA, 284~\AA\, 304~\AA\ or He~{\sc I} 10830~\AA\ images. The start time and duration of the analysis period are also determined by the user. This is followed by standard {\it SolarSoft} procedures \cite{Freeland98} to calibrate the EUV and X-ray images. These include:
\\
1. {\it Dark current subtraction}. Taking a time exposure with the shutter closed produces the thermal signature known as "dark count", caused by thermal electron excitation due to the temperature of the device. This is the zero flux CCD response, which is subtracted from the raw image. 
\\
2. {\it Instrument grid subtraction}. The aluminum filter near the focal plane of the instrument casts a grid-like shadow on the CCD detector, which has to be subtracted from the image.
\\
3. {\it CCD response correction}, also known as ``flat fielding". Each CCD pixel has a different sensitivity to incoming photons. This can be corrected for by dividing the data by flat-field exposures taken before the observation.
\\
4. {\it Exposure time normalisation} ensures that each observation is normalised according to the exposure time of the observation. The resulting image intensity unit is data numbers per second (DNs; the measure of photon incidence in the CCD camera).
\\
5. {\it Filter normalisation} corrects for the variable transmittivity of the clear and aluminium filters. 

The output is an index structure containing the observation details (stored in the header) and an image data cube. In this thesis work the calibration routines ``eit\_prep", ``secchi\_prep" and ``xrt\_prep" were used to prepare level-1 data. The MDI magnetograms and the PLASTIC and ACE solar wind data needed no further calibration.

\subsection{Data selection}
\label{dataselect}

The EUV images are used to detect low intensity regions, which are complemented by magnetograms to determine whether a region is of a predominant polarity (typical of CHs). SOHO provides high resolution (1024$\times$1024 pixel) full-disk EUV images and magnetograms, and has the largest database including over 10 years of data. For this reason, the present thesis research is mainly based on data from the SOHO EIT and MDI instruments. Hinode/XRT and STEREO/EUVI data are used for comparison and confirmation. 

The magnetograms analysed can be chosen to be one-minute or five-minute averaged full-disk magnetograms. In this work we used the five-minute images, as these are the lowest noise images produced daily. The per pixel noise in the Level~1.8 calibrated magnetograms is $\sim$30~G for a 1-minute magnetogram and $\sim$15~G for a 5-minute magnetogram. The Level~1.8 magnetograms are the definitively calibrated MDI images of the solar line-of-sight magnetic field.

In order to detect CHs at a given time, a magnetogram is found closest to the required time. Next, the closest EUV image is selected within $\pm$30~minutes (this time range can of course be changed). The reason for selecting the MDI magnetogram before the EUV image is the smaller number of MDI observations per day compared to EIT 195~\AA\ observations. There are fifteen MDI magnetograms created in a day and often over a hundred EUV images. Next, the intensity threshold is determined for the low intensity regions (as explained in Section~\ref{thresholding}), which is followed by the magnetogram being rigid-rotated to the time of the EUV observation to match the corresponding locations on the solar disk. The magnetograms are then used to determine whether the low intensity regions are CHs. 

\subsection{The Lambert cylindrical equal-area projection}
\label{Lambert}

The Lambert cylindrical equal-area projection allows a spherical surface to be projected onto a cylindrical surface whilst preserving the area and producing parallel latitudinal (and longitudinal) lines within approximately -60 and +60 degrees latitude and longitude \cite{McAteer05}. This projection was chosen to aid the local intensity thresholding procedure, which uses ``window-stepping" across the image (Section~\ref{thresholding}). It also provides a straightforward determination of the CH area and the boundary positions. 

\begin{figure}[!t]
\centerline{
\includegraphics[scale=0.7, trim=0 620 0 30, clip]{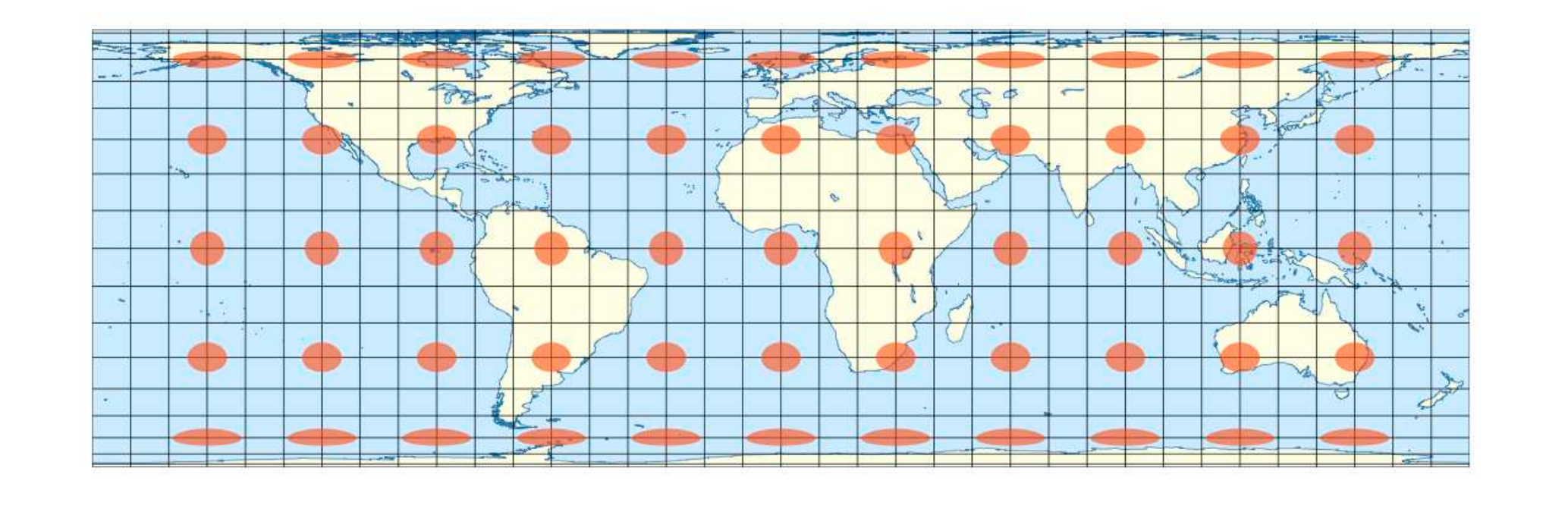}}  
\caption{The Tissot indicatrix showing the longitudinal and latitudinal line positions and the shape distortions after the Lambert cylindrical equal-area projection of the world map. Figure courtesy of Eric Gaba and U.S. NGDC World Coast Line.} 
\label{tissot_indicatrix} 
\end{figure}

The cylindrical equal-area projection was the fourth of seven projections invented by Johann Heinrich Lambert in 1772. The equations that describe the projection are
\begin{equation}
x = R (\lambda - \lambda_{0}) \cos \phi_{S}
\end{equation}
\begin{equation}
y = R \sin \phi / \cos \phi_{S},
\end{equation}
where $R$ is the sphere radius, $\lambda_{0}$ is the central meridian location (set to $0^{\circ}$ in our projection), $\lambda$ is the longitude, $\phi$ is the latitude, and $\phi_{s}$ is the standard parallel (set to $0^{\circ}$ in the equator centered projection), as described by \cite{Snyder87}. As an example, the Lambert projection of the whole Earth's surface is shown in Figure~\ref{tissot_indicatrix}. Note that the longitudinal lines are equally spaced, while the latitudinal lines are sine spaced.

\begin{figure}[!t] 
\centerline{
\includegraphics[scale=0.35, trim=20 0 20 0, clip]{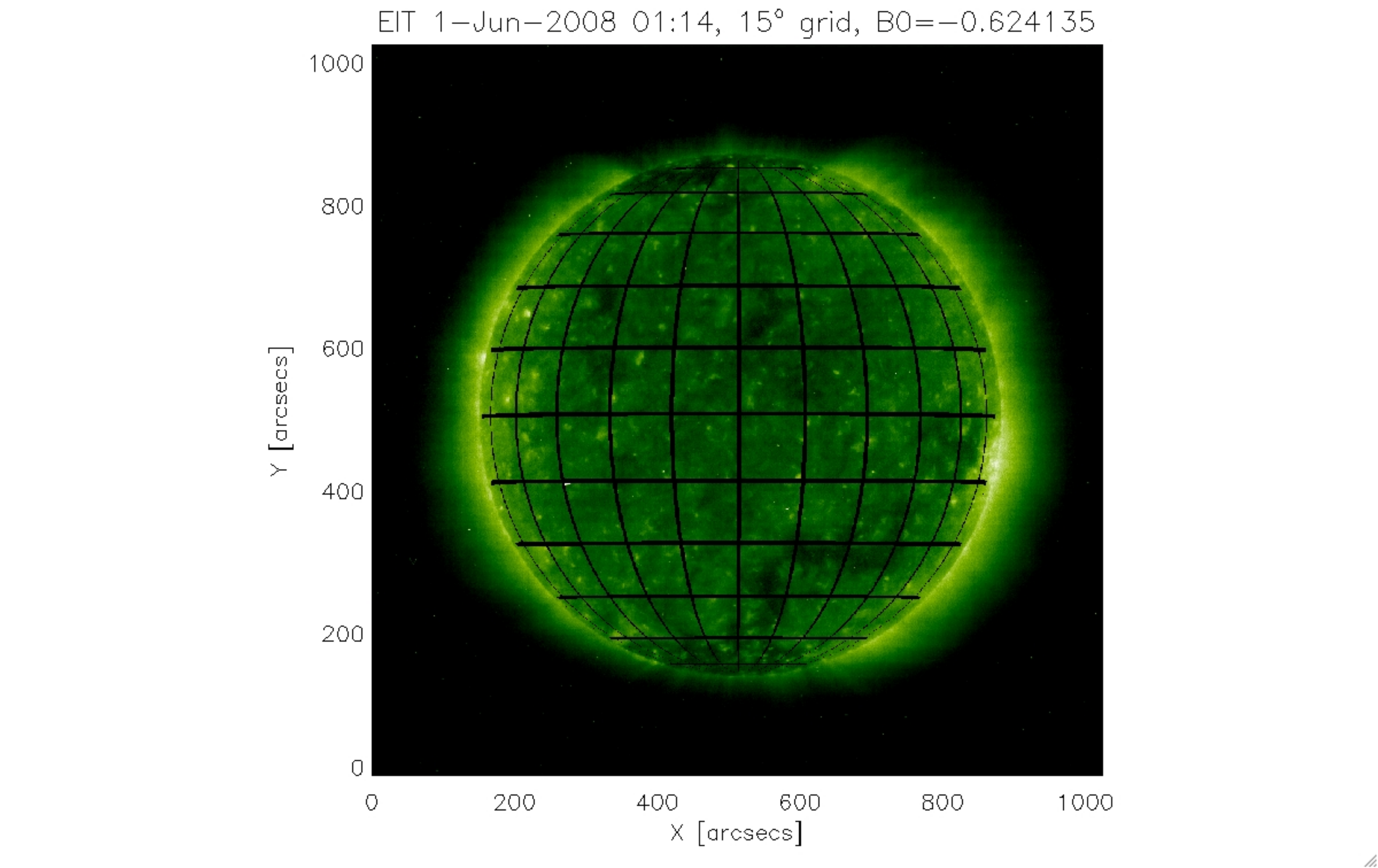} } 
\centerline{ 
\includegraphics[scale=0.3, trim=20 0 20 0, clip]{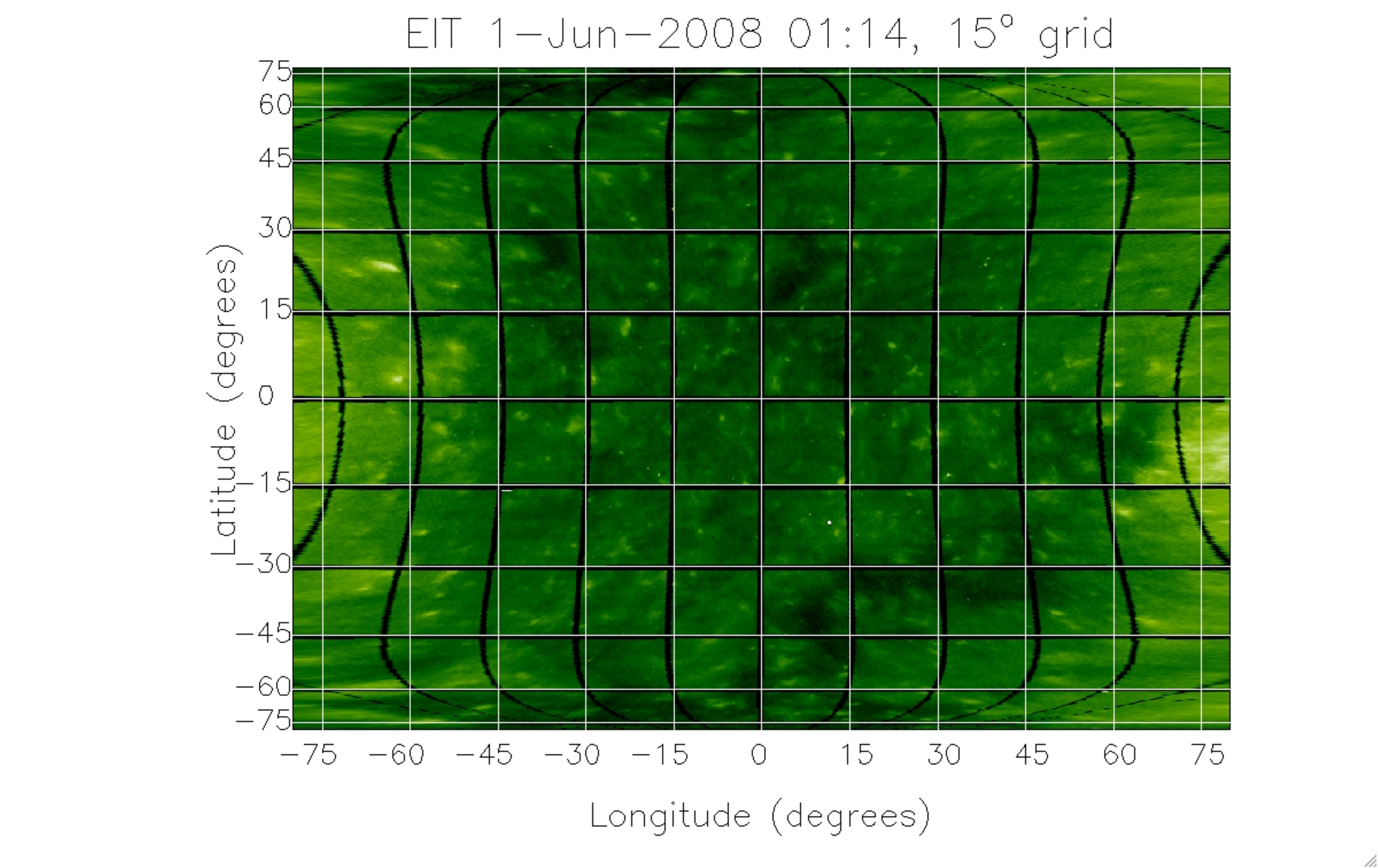} } 
\caption{The EIT disk image (top) and the Lambert cylindrical equal area projection of the visible solar surface with 15$^{\circ}$ grid spacing in latitude and longitude (bottom). (Observation: SOHO/EIT 195~\AA, 1 Jan 2008).} 
\label{proj_map} 
\end{figure}

The algorithm used for the projection is ${\it lambert\_project}$ (a modified version of ${\it map\_carrington3}$). The Lambert cylindrical equal-area projection is better known as Carrington projection in heliographic work. However, in the majority of cases Carrington maps are stacked maps of solar central meridian daily strip images. The Lambert projection procedure projects the visible solar surface into a cylindrical map, however, uncertainties are introduced as the surface visibility drops near the solar limb. For this reason the projection is not as reliable above $\sim\pm60^{\circ}$ latitude and longitude. Figure~\ref{proj_map} shows the projection deformation in latitudinal and longitudinal lines, which are drawn at 15$^{\circ}$ intervals. By default, the algorithm limits the projection to $\sim\pm80^{\circ}$ latitude and longitude due to lack of information near the limb. In addition to this, off-limb regions are visible in the Lambert map corners with varying scale due to the yearly variation is the solar B$_{0}$ angle. This is caused by the inclination of the Earth's orbit to the ecliptic giving a larger view of the northern or southern solar pole during an orbital year of Earth. This variation is corrected for in the algorithm. As a result, the solar equator is located in the image center in all Lambert maps. The off-limb corners are excluded from all analysis.

\subsection{Detecting low intensity regions}
\label{thresholding}

\begin{figure}[!t]
\vspace{-0.5\textwidth} 
\centerline{\hspace*{0.015\textwidth} 
\includegraphics[angle=90,clip=, width=1\textwidth]{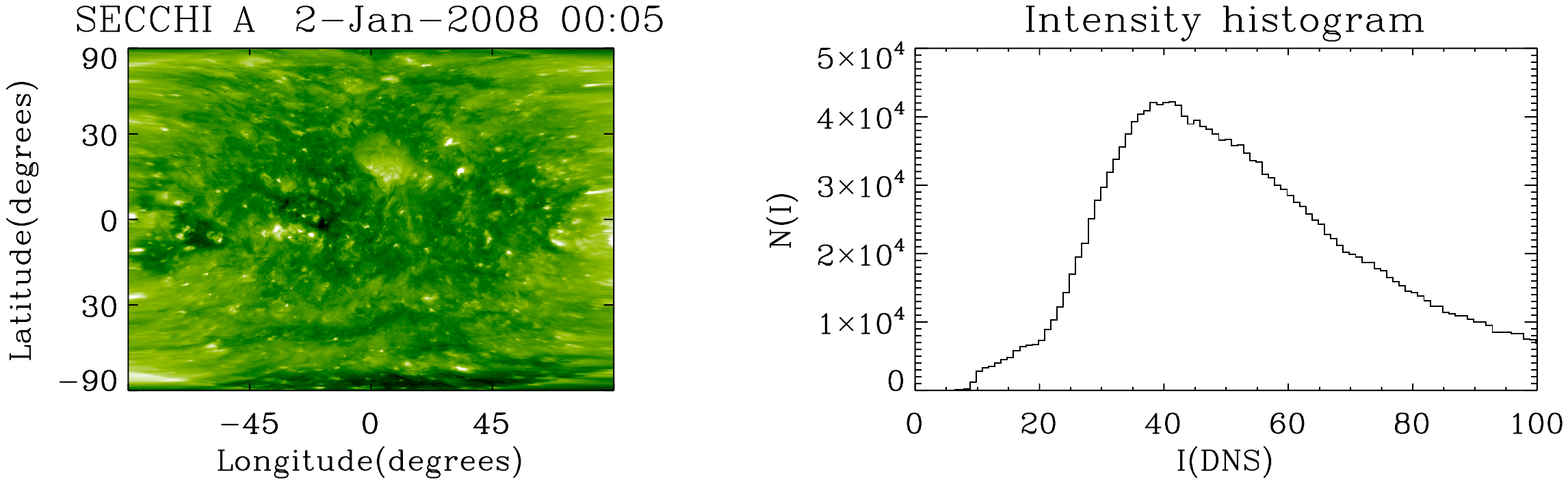} } 
\vspace{-0.5\textwidth} 
\centerline{\hspace*{0.015\textwidth} 
\includegraphics[angle=90,clip=, width=1\textwidth]{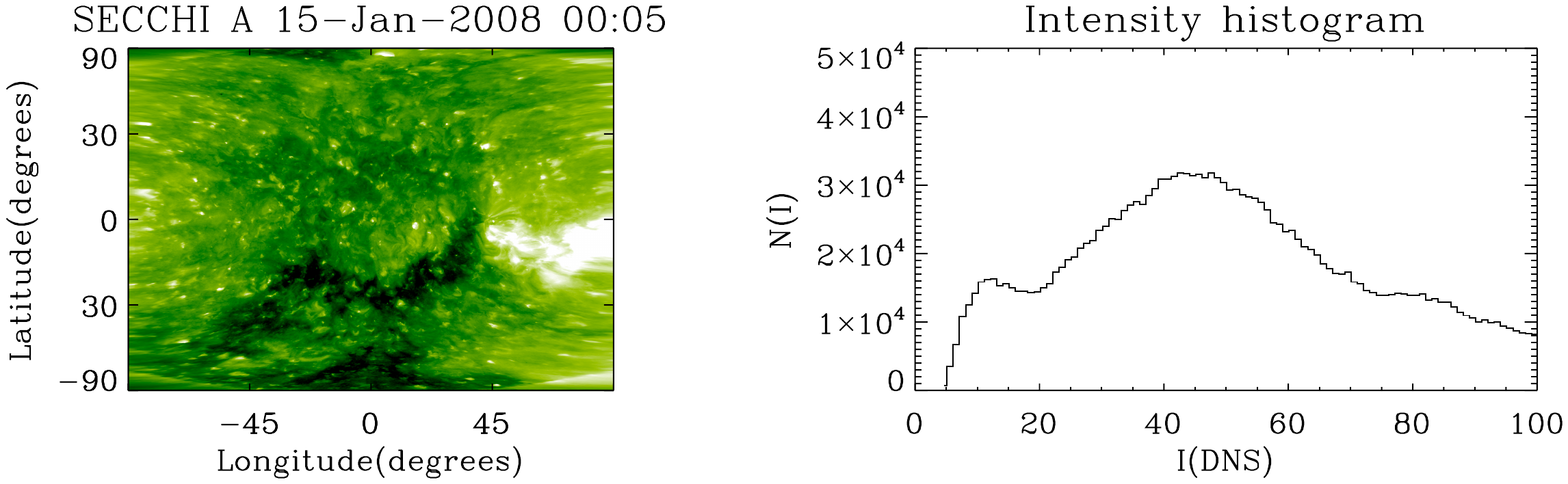} } 
\caption{{\it Top}: EUVI 195~\AA\ Lambert map showing no CHs, and the corresponding intensity histogram with a unimodal distribution. {\it Bottom:} Lambert map showing a large CH, and the corresponding intensity histogram with a bimodal distribution.} 
\label{carr_maps} 
\end{figure}

Intensity thresholding is a powerful method to distinguish regions of different intensities in an image (e.g. CHs and QS regions). The intensity histogram of an image gives a multimodal distribution where each frequency distribution corresponds to a feature in the image \cite{Gonzalez02,Gallagher98,Gallagher99,Belkasim03}. In the case of 195~\AA\ images, each major feature (ARs, low intensity regions - LIRs, and QS) corresponds to a distribution in the intensity histogram. When an image is dominated by QS, the corresponding intensity histogram shows a unimodal distribution (top panel of Figure~\ref{carr_maps}). When a relatively large part of the image is covered by LIRs, the resulting intensity histogram shows a bimodal distribution (bottom panel of Figure~\ref{carr_maps}), where the local minimum corresponds to the intensity threshold of the LIR boundary. Figure~\ref{carr_maps} gives the intensity measured in DNs as a function of CCD pixel number frequency.

\begin{figure}[!t]
\centerline{\hspace*{0.015\textwidth} 
\fbox{ 
\includegraphics[clip=, scale=0.4, trim=10 0 15 280]{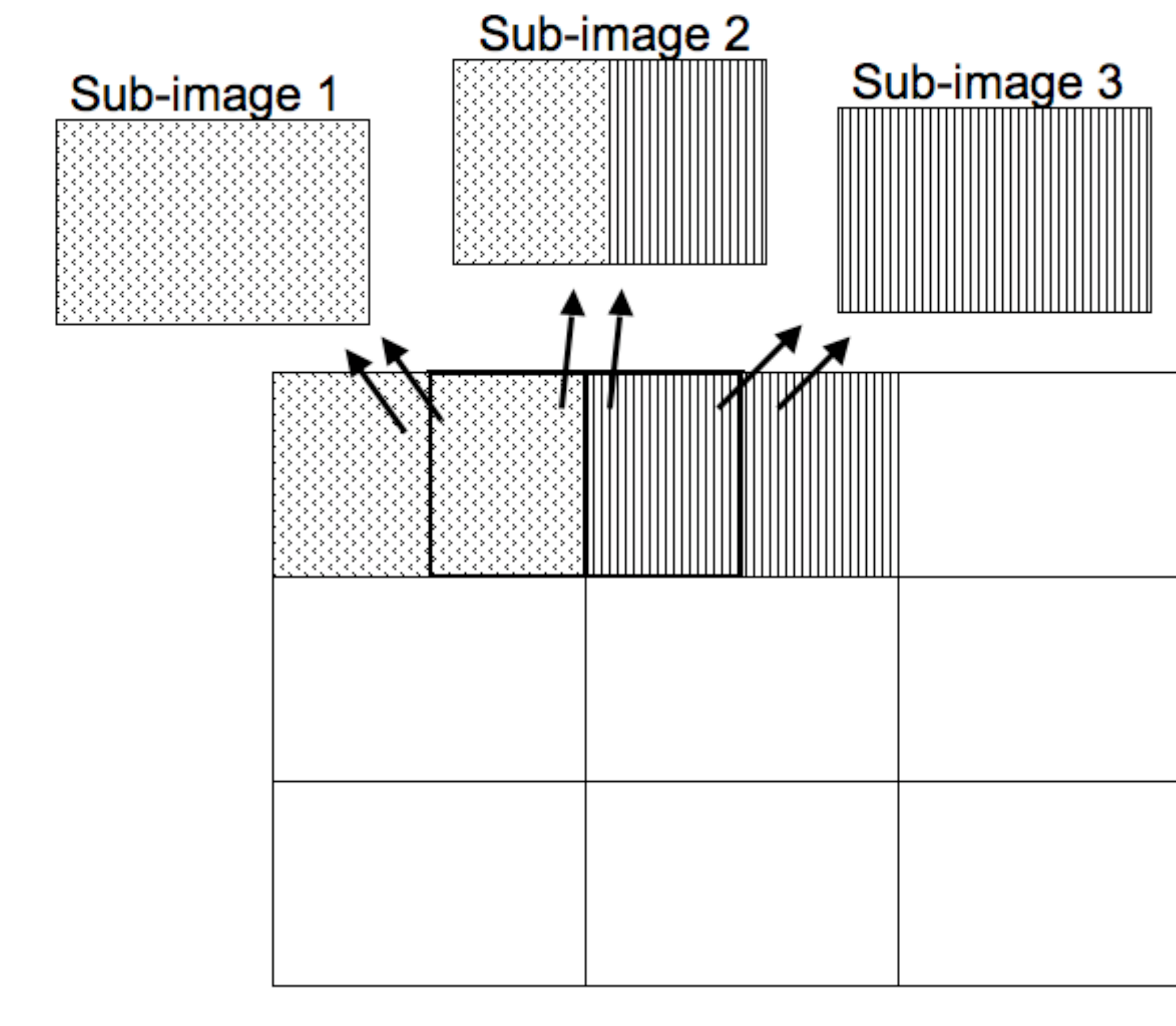} } }
\caption{A schematic of the Lambert map being partitioned into overlapping sub-images.} 
\label{windowing} 
\end{figure}

The minimum between the LIR and QS distribution was enhanced using a partitioning operation. By partitioning the solar image to sub-images, we obtained local histograms with more defined minima which aided the determination of thresholds. The sub-images used were 1/2, 1/3, 1/4 and 1/5 of the original Lambert map size. Each of the sub-image sizes were used to step across the map and obtain local intensity histograms from overlapping sub-images (Figure~\ref{windowing}). This process provided more clearly discernible local minima for thresholding. This can be explained by different size sub-images having different proportions of QS and LIR and hence frequency distributions. The threshold minima are defined as local minima that are located within 30--70\% of the QS intensity and are wider than 6 DNs. The threshold range was set based on CH boundary intensity tests through the solar cycle. The range is needed to reduce run time and to exclude occasional low intensity features inside CHs, or high intensity features in the QS.

\begin{figure}[!ht]
\vspace{-0.05\textwidth} 
\centerline{
\includegraphics[scale=0.54, trim=90 0 80 80, clip]{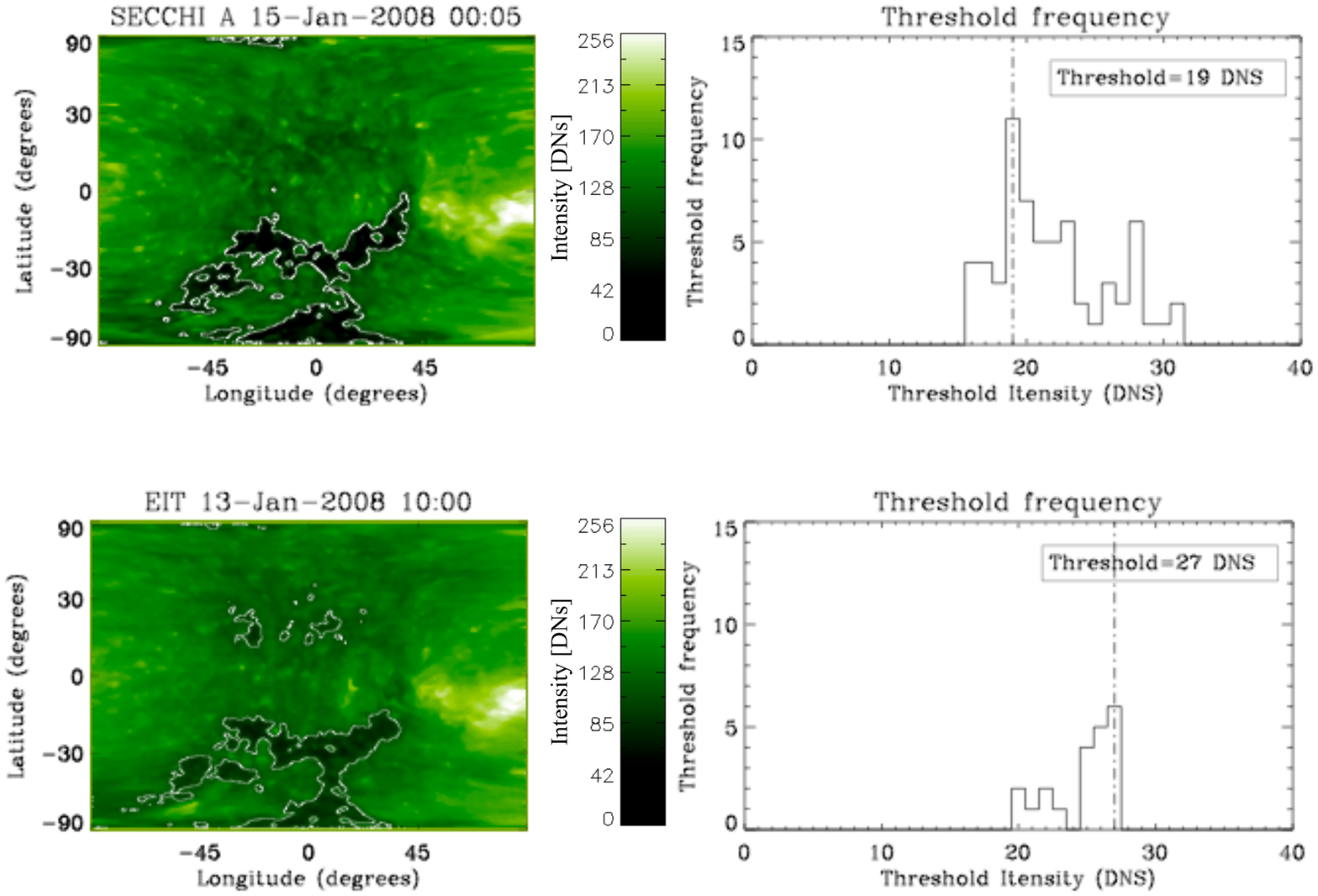} } 
\vspace{-0.1\textwidth} 
\centerline{
\includegraphics[scale=0.54, trim=90 80 80 0, clip]{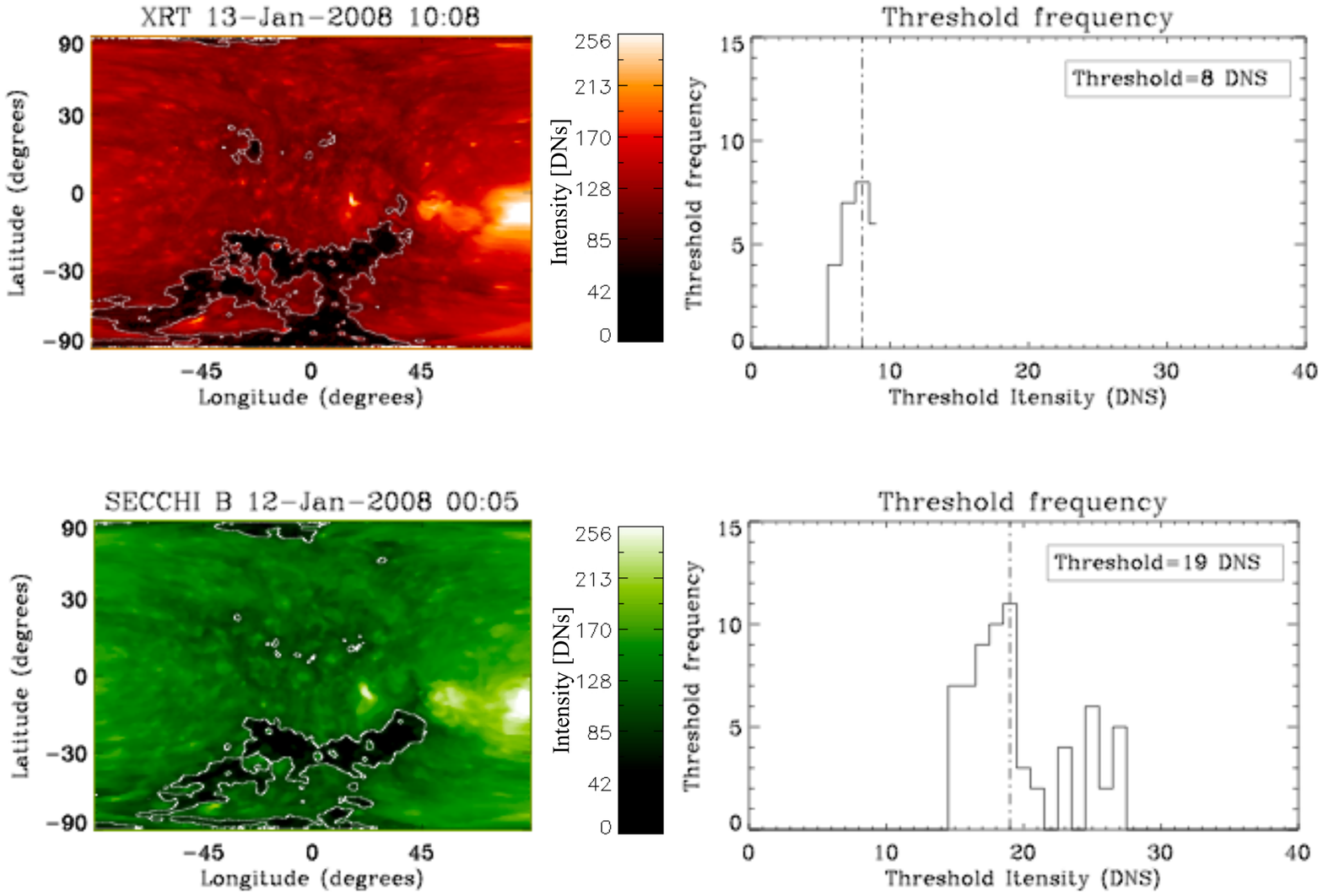} } 
\caption{{\it Left}: STEREO A/EUVI, SOHO/EIT, Hinode/XRT and STEREO B/EUVI Lambert maps with LIR contours. The images show a complex CH structure approximately at the same location on the map. {\it Right}: Threshold histogram, where the peak value is the optimal threshold.} 
\label{lambertmaps} 
\end{figure}

As mentioned above, the intensity range where the thresholds were identified depends on the QS intensity. For this reason the method is reliable at any time of the solar cycle regardless of the change in the overall intensity. It should be noted however, that with increasing solar activity, the acquired CH boundaries are less accurate due to bright coronal loops intercepting the line-of-sight and obscuring parts of the CH boundary.
\\ \indent
Once the intensity thresholds are obtained from the sub-images the final intensity threshold is determined by taking the most frequent value. The threshold can then be used to contour LIRs on the same Lambert map. In our studies we chose to determine the threshold for five images in a day and set the median value of the thresholds as the final threshold. This is done to overcome small intensity variations in time and secure a more reliable threshold. An example is shown in Figure~\ref{lambertmaps} for a CH complex observed with four different instruments in two wavebands (195~\AA\ and X-ray). The observation times are different to ensure that the CH is seen at the same central location observing with different satellites. The threshold histograms (right column of Figure~\ref{lambertmaps}) show distinct threshold peaks, which are used to contour LIRs. We would like to draw attention to the large-scale similarity in CH morphology despite the use of different instruments and observation times. This similarity is aided by the CH being close to the disk center, where line-of-sight obscuration is minimal. However, it is important to note that in other cases differences may appear between EUV and X-ray CH observations due to the difference in the temperature range they observe at. Observations made at different angles from the Sun (e.g. SOHO, STEREO A and B are at different orbital locations) may also lead to differences in the CH area and boundary location due possible line-of-sight obscuration.

\subsection{Differentiating coronal holes from filaments}

CHs were distinguished from filaments using MDI magnetograms. As previously mentioned, the magnetograms were chosen first in the analysis based on the time of interest. The magnetogram was then differentially rotated to the EUV image observation time. The LIR contours obtained from the EUV image was then overplotted on the magnetogram and the flux imbalance was determined for the contoured regions. CHs are well known to have a dominant polarity, while filaments have a bi-polar distribution of polarities \cite{Scholl08}. 
\begin{figure}[!t]
\vspace{-0.03\textwidth}
\centerline{
\includegraphics[angle=90, scale=0.6, trim=20 60 390 20, clip]{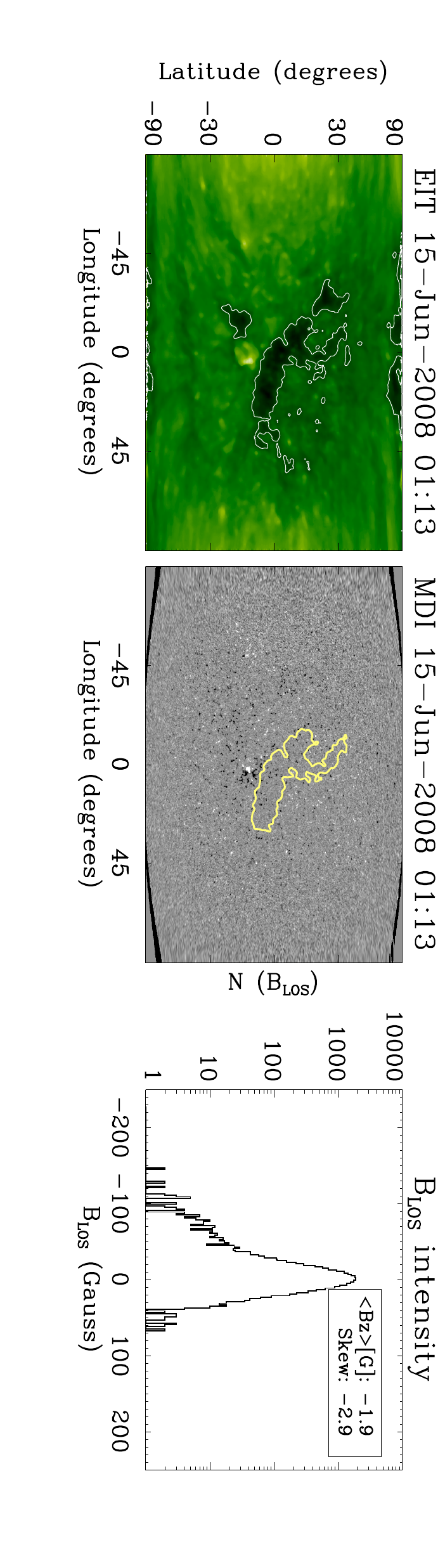} }
\centerline{
\includegraphics[angle=90, scale=0.6, trim=20 60 390 20, clip]{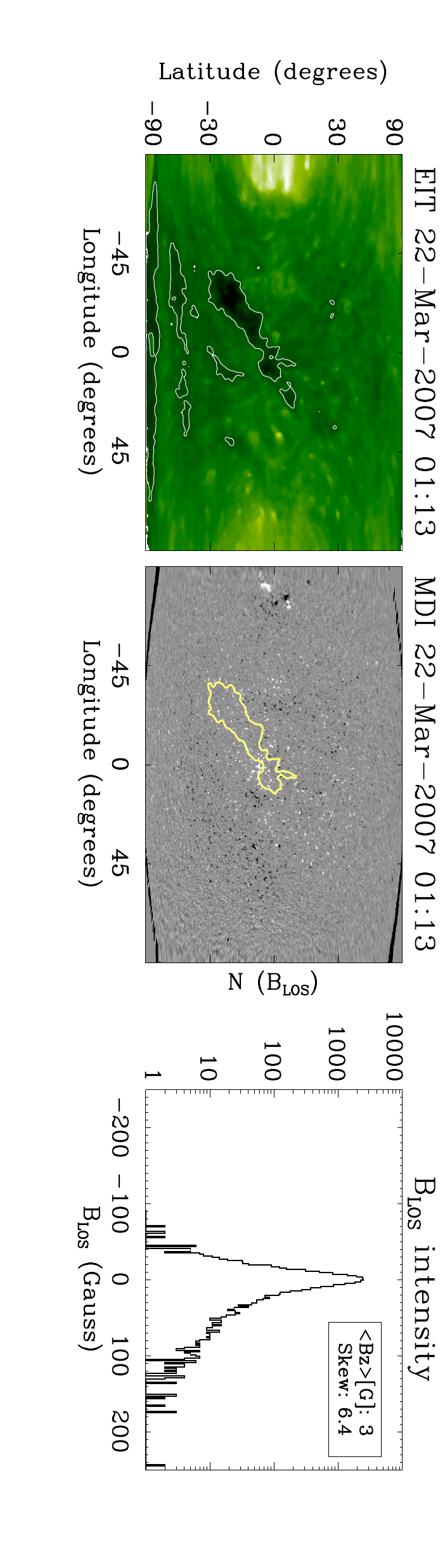} }
\centerline{
\includegraphics[angle=90, scale=0.6, trim=20 60 390 20, clip]{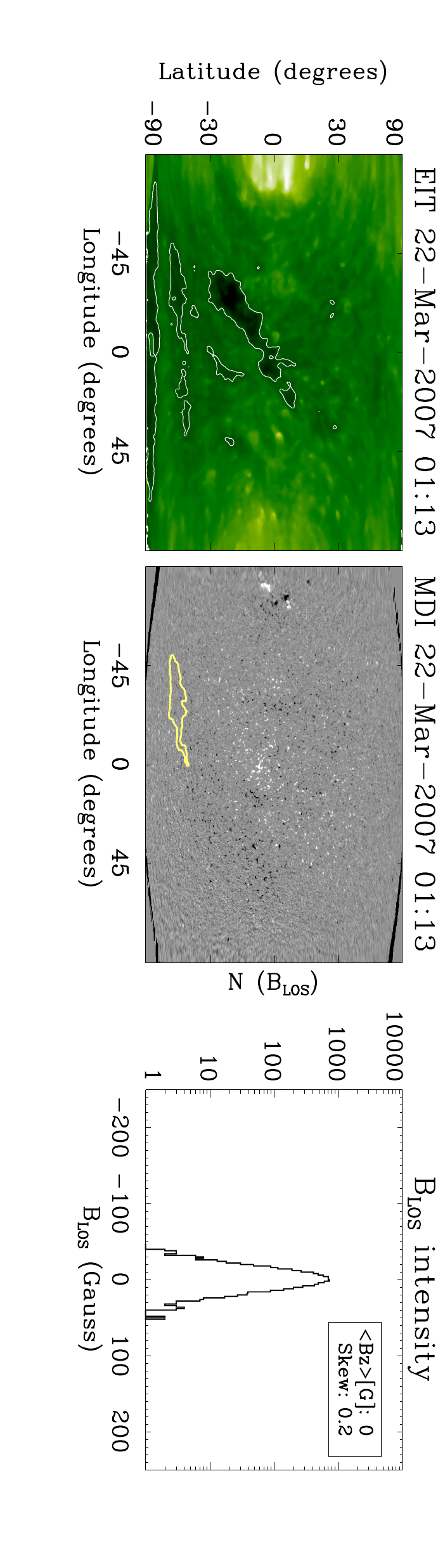} }
\vspace{-0.03\textwidth} 
\caption{{\it Left}: EIT Lambert maps with LIR contours. Middle: MDI Lambert maps with a LIR contours overlaid. {\it Right}: $B_{LOS}$ intensity histogram of the contoured regions. Regions highlighted in the first and second middle panels are CHs (B$_{LOS}$ histograms skewed), the third middle panel shows a filament ($B_{LOS}$ histogram not skewed).} 
\vspace{-0.01\textwidth}
\label{mdi} 
\end{figure}
Figure~\ref{mdi} shows examples of LIRs observed on 15 June 2008 and 22 March 2007, respectively. The 15 June 2008 LIR is an example of a CH with a negative dominant polarity, while the 22 March 2007 LIR is a CH with a positive dominant polarity (first and second row in Figure~\ref{mdi}). The skewness of the field strength distribution defines the scale of the flux imbalance in the observed region of interest. For the two CHs the skewness was found to be considerable with values $-2.9$ and $6.4$, respectively. The LIR highlighted on the bottom panel magnetogram in Figure~\ref{mdi} was found to be a filament based on the very small value of skewness ($0.2$) in its line-of-sight magnetic field strength distribution.
\vspace{-0.015\textwidth}

\subsection{Coronal hole grouping}

The grouping of CHs was done mainly for high-speed forecasting purposes (Chapter~\ref{chapter:forecasting}). However, it also simplifies the tracking and referencing of complex CH groups. The group association is not founded on physical concepts; it is simply a measure of distance between CHs. 
\\ \indent
Each CH is treated as a polygon located on the cylindrical Lambert map. The pixel coordinates of the polygons are converted into heliographic coordinates after which the Euclidean distance is determined between all coordinate points. The smallest distance between the two polygons' coordinate points is the minimum distance which determines whether two CHs belong to the same group. The distance threshold is set to 10$^{\circ}$, above which CHs are not grouped.
\begin{figure}[!t]
\vspace{-0.05\textwidth} 
\centerline{
\includegraphics[scale=0.37, trim=40 70 30 10, clip]{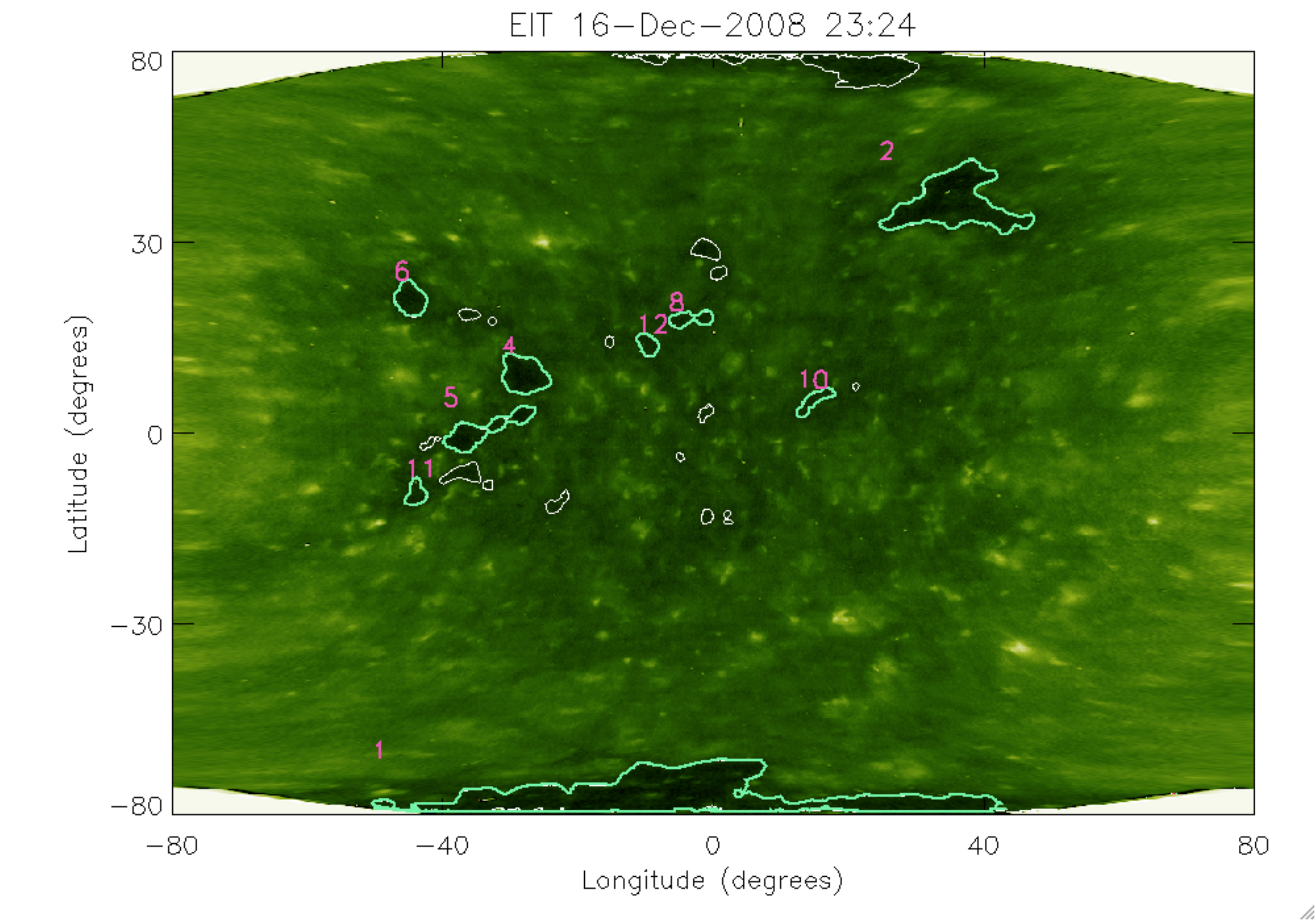} }
\centerline{
\includegraphics[scale=0.37, trim=40 20 30 38, clip]{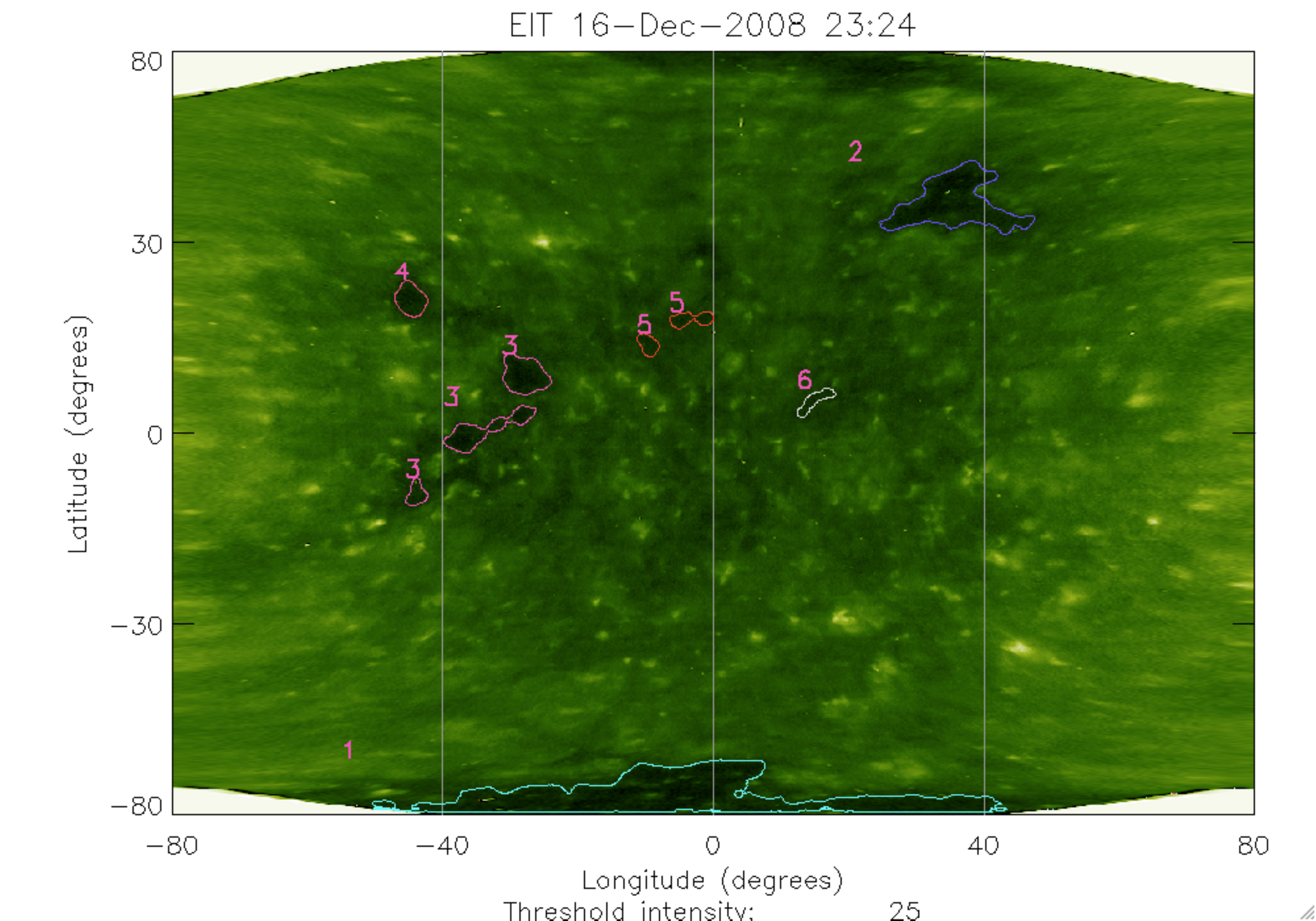} }
\caption{Top: The LIRs numbered are CHs, the white contours with no numbers are LIRs that did not meet the requirements to qualify as CHs (i.e. due to too small area or not large enough flux imbalance). Bottom: CHs after the grouping procedure. CHs of the same group have the same number and colour.} 
\vspace{-0.015\textwidth}
\label{ch_members} 
\end{figure}

\begin{figure}[!t]
\centerline{
\includegraphics[scale=0.5, trim=0 120 0 100, clip]{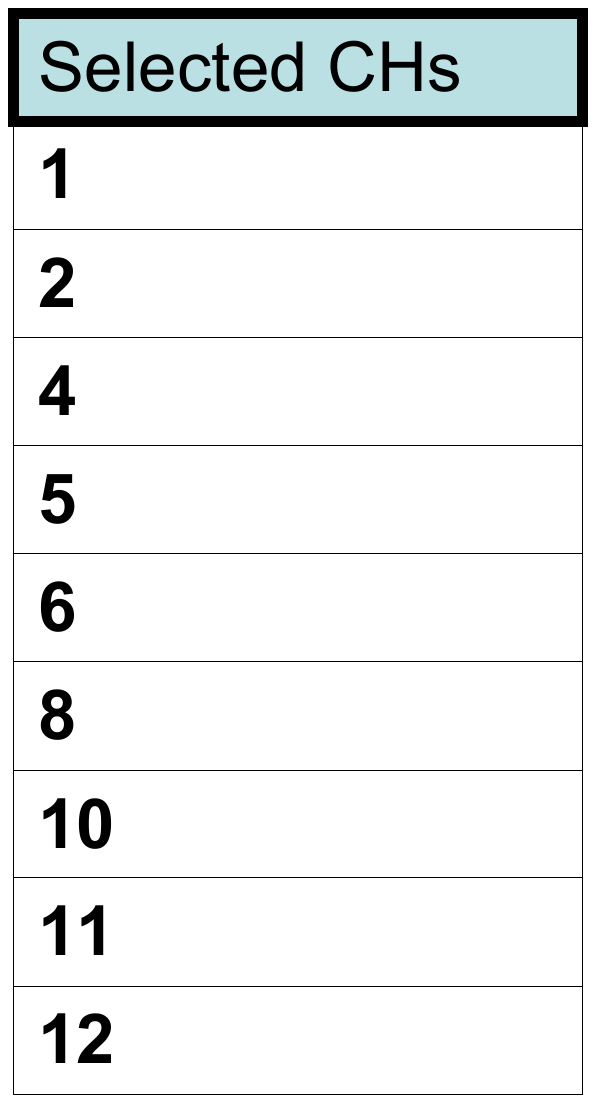} }
\caption{The CHs selected from the LIRs.} 
\label{ch_select} 
\end{figure}

Let us consider the LIRs detected in the EUV image shown in the Figure \ref{ch_members} top panel. These are numbered from 1 to 12 in the descending order of their areas (i.e. the largest CH is numbered 1). During the magnetic field analysis nine of the LIRs, numbered 1, 2, 4, 5, 6, 8, 10, 11, 12, were classified as CHs. The list of these CHs is shown in Figure~\ref{ch_select}. These CH numbers are non-consecutive due to the discarded LIRs. 

\begin{figure}[!t]
\centerline{
\includegraphics[scale=0.5, trim=0 20 0 20, clip]{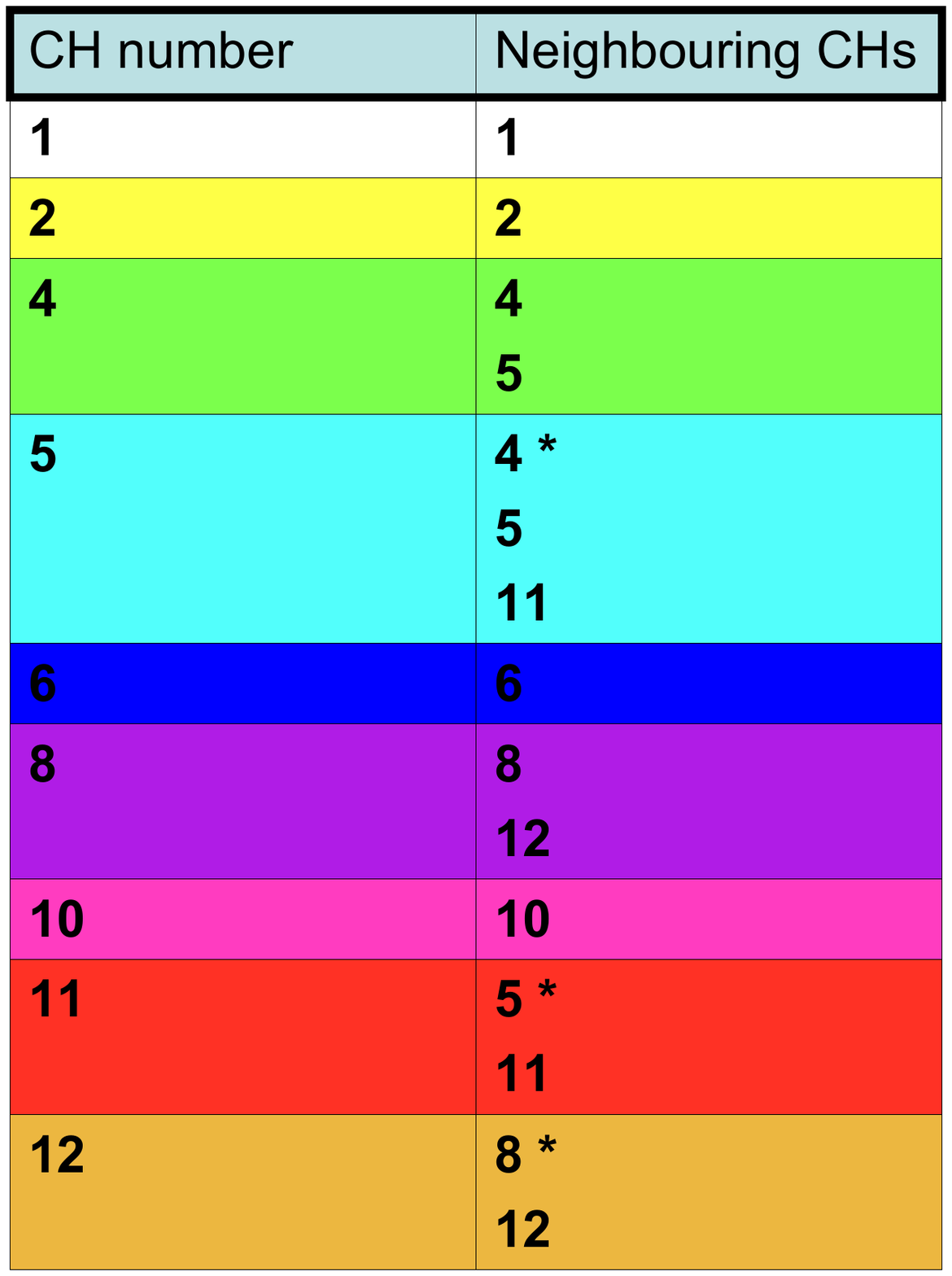} }
\caption{The CHs and their associated pairs. Each CH has beed given a different colour to highlight the pairs. The repetitions (crossed out with a red line) are removed from the list. } 
\label{ch_pair_cleanup} 
\end{figure}

First, the CHs are grouped with each other based on their distances as described before. The first and second column in Figure~\ref{ch_pair_cleanup} are the CH numbers in order, and the neighbouring CHs found at less than 10$^{\circ}$ distance, respectively. Repetitions, such as CH~4 paired with CH~5, and CH~5 paired with CH~4, have to be removed from the list. Note that the repeats with the higher CH number (thus the smaller area) are the ones to be removed. This ensures that the CH with the largest area becomes the {\it Group Leader}, and the smaller neighbouring CHs become the {\it Group Members}. 

The problem becomes complicated when a CH (e.g. CH~11) is not directly associated with a Group Leader (i.e. CH~4), but with a Group Member (e.g. CH~5) which is associated with a Group Leader (CH~4). Because all three CHs ( CH 4, 5 and 11) are close to each other, they become grouped, and the largest CH's number will be the Group Number (4 in this case). 

\begin{figure}[!t]
\centerline{
\includegraphics[scale=0.5, trim=0 130 0 100, clip]{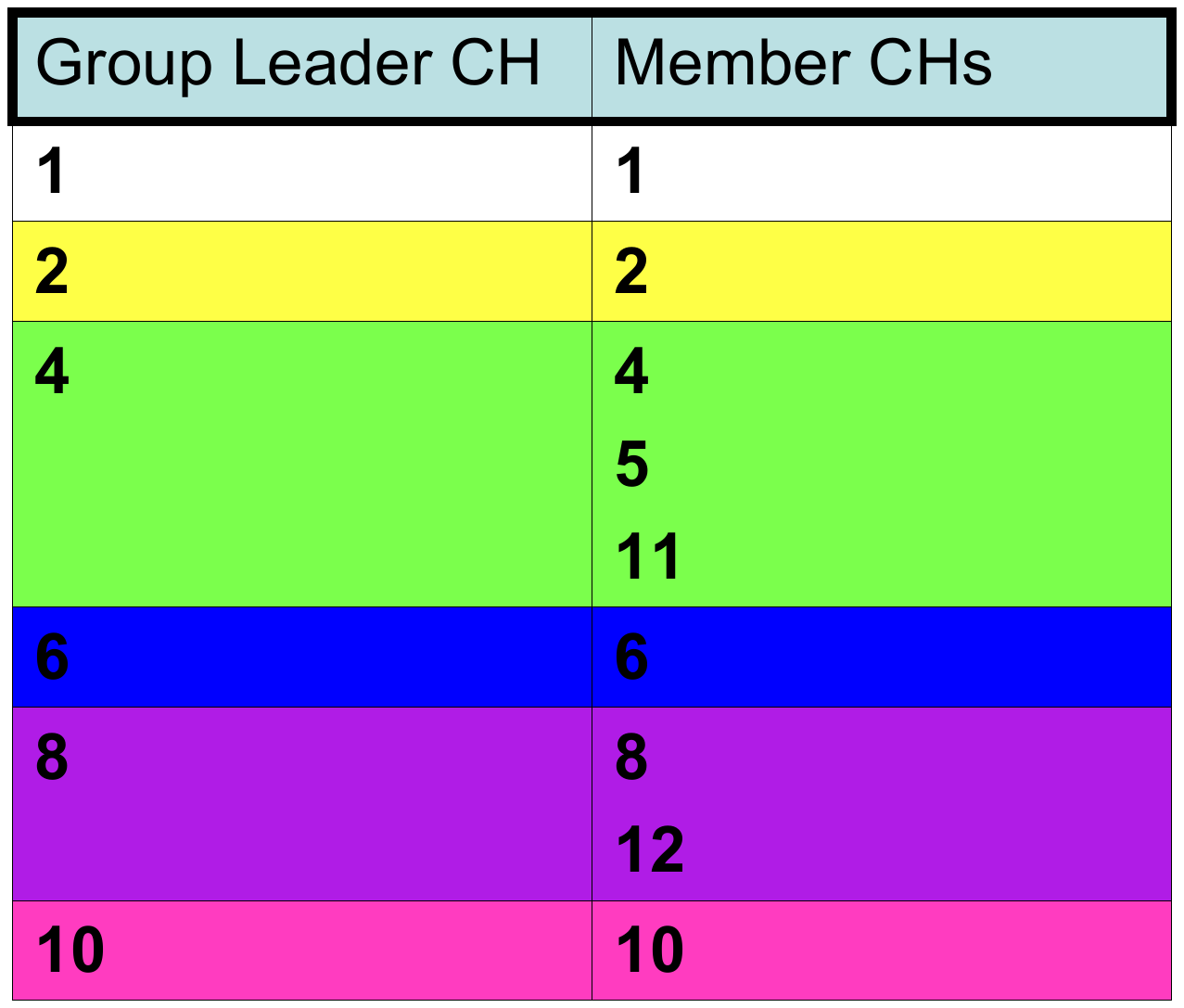} }
\caption{The CHs and their associated pairs.} 
\label{ch_pairs} 
\end{figure}

\begin{figure}[!t]
\centerline{
\includegraphics[scale=0.5, trim=0 150 0 80, clip]{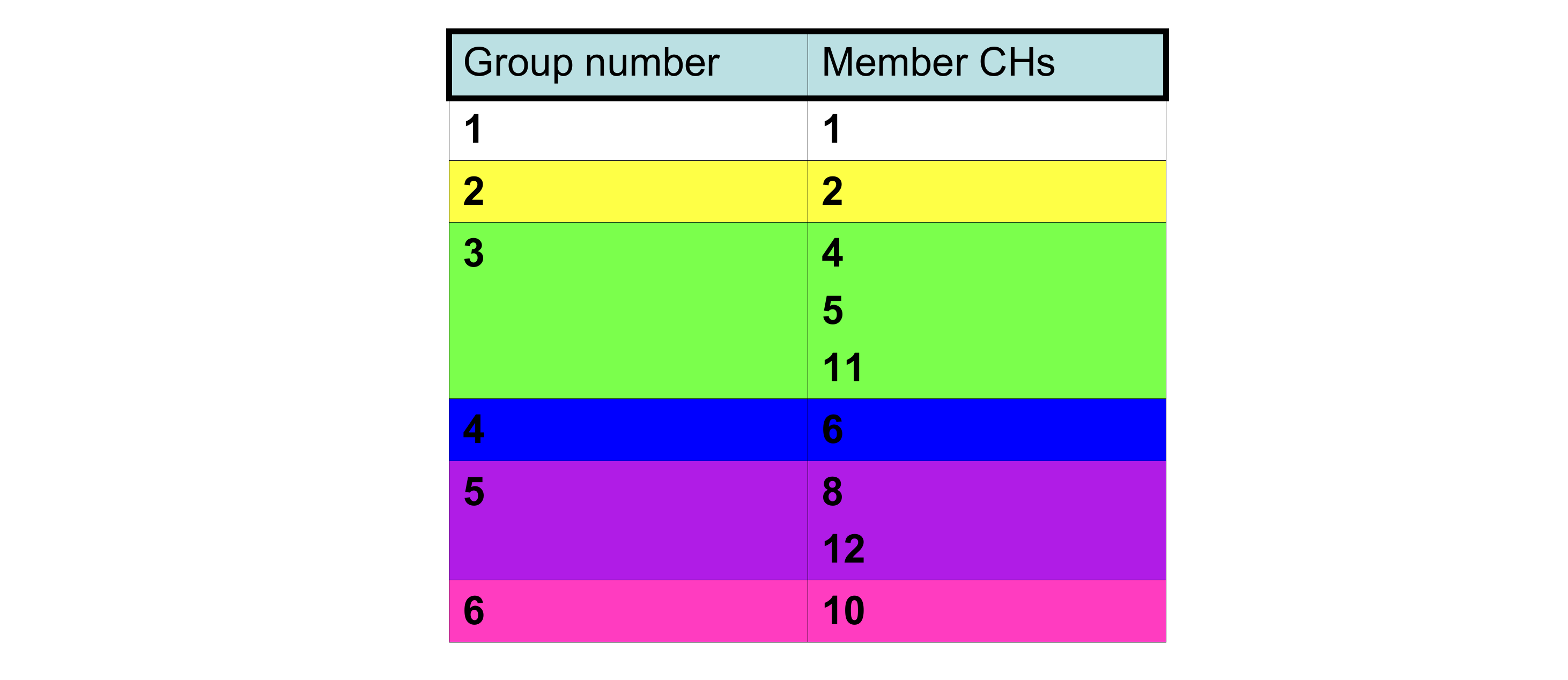} }
\caption{The CHs and their associated pairs were renumbered to produce consecutive Final Group numbers.} 
\label{ch_groups} 
\end{figure}

Figure~\ref{ch_pairs}) shows the previous table, but with repetitions excluded and further associations included (e.g. CH~11 is a member of Group~4 through being close to CH~5). Next, the Group Leaders are given consecutive group numbers ({\it Final Groups}), as shown in the table in Figure~\ref{ch_groups} (also see the corresponding CHs in Figure \ref{ch_members} bottom panel).

\section{Conclusions}

This Chapter introduced the {\it CHARM} algorithm, which was developed to identify CHs in fast and automated manner. It allows the use of different instruments and wavelengths (X-ray and EUV), and gives consistent CH boundary thresholds throughout the solar cycle. 

Previous methods have used arbitrary thresholds or a computationally intensive approach, while the {\it CHARM} method only uses a single EUV image and a magnetogram to detect CHs at the time of interest. The segmentation algorithm first identifies LIRs using local intensity histograms of either 195~\AA\ or X-ray observations, and then exploits MDI magnetograms to differentiate CHs from filaments by the flux imbalance determined for the identified regions. This was achieved assuming that CHs have a predominant polarity as opposed to filaments. Once the CH regions were determined, the position, area and mean $B_{\mathrm{LOS}}$ can be calculated for multiple CHs. Finally, the CHs are grouped mainly for forecasting purposes. However, this can also aid the tracking and referencing of CHs.

The main limitation of the method results from near-limb effects, which include: line-of-sight obscuration in EUV and X-ray images, increase of noise in $B_{\mathrm{LOS}}$ measurements, and interpolation effects caused by the Lambert equal-area projection. These near-limb effects can cause incorrect boundary determination, the inability to determine a dominant polarity, and incorrect area determination. Also, some CH boundaries located near the limb were found to have different threshold intensities which depended on the degree to which they were obscured. For this reason, only CHs within -60 and +60 degrees latitude and longitude were included in this analysis.

Due to the fast and automated approach of the present method, it allows the analysis of large databases (e.g. SOHO database). This provides an opportunity to study the long-term evolution of CHs (Chapter~\ref{chapter:forecasting}). On a smaller scale, the method can also be applied to analyse the small scale changes in CH boundaries, which may provide information on the processes governing CH growth and decay (Chapter~\ref{chapter:evolution}).



\chapter{Short-term evolution of coronal holes} 
\label{chapter:evolution}

\ifpdf
    \graphicspath{{4/figures/PNG/}{4/figures/PDF/}{4/figures/}}
\else
    \graphicspath{{4/figures/EPS/}{4/figures/}}
\fi

\flt
{\it The interaction of open and closed field lines at CH boundaries is widely accepted to be due to interchange magnetic reconnection. To date, it is unclear how CH boundaries vary on short timescales and what velocities this occurs at. The author developed an automated boundary tracking method to study short-term boundary displacements, which were found to be isotropic and have typical speeds of 0--2~km~s$^{-1}$. The magnetic reconnection rates at CH boundaries were calculated to be 0--3$\times$10$^{-3}$, and using the interchange reconnection model the diffusion coefficients were found to be 0--3$\times$10$^{13}$~cm$^{2}$~s$^{-1}$. These results are consistent with interchange reconnection in the corona, driven by the random granular motion of open and closed fields in the photosphere. This work has been accepted for publication in Astrophysical Journal Letters \citep*{Krista11}.}
\flb

\newpage

\section{Short-term changes in CH boundaries}
\label{chevol_intro}
\setstretch{2}

Small-scale changes in the CH morphology are thought to result from magnetic reconnection occurring between CH open fields and small closed fields found in the QS as well as in CHs \citep{Madjarska04}. Using the Coronal Hole Evolution {\it CHEVOL} algorithm we investigate the driving force behind the open and closed field interaction and determine the magnetic reconnection rate and the diffusion coefficient based on the observed displacements in CH boundaries.

A number of authors have previously studied the large-scale and long-term evolution of CHs and established the changes in the total CH area, flux, the latitudinal distribution over the solar cycle and the key role CHs play in the dipole-switch of the Sun \citep{Bravo97, Chapman02, Harvey02, Wang96, Wang09}. Although the small-scale changes in CH boundaries have only been studied in detail in recent years using high-resolution observations \citep{Madjarska09, Subramanian10}, it was already suggested in the late 1970s that the small-scale and short-term evolution of CH boundaries is strongly connected to bright point occurrences \citep{Nolte78, Davis85} and that reconnection could be a means of flux-transportation and ``diffusion'' of the open flux \citep{Nolte78}. A more in-depth discussion on reconnection occurring at CH boundaries was provided by \cite{Mullan82}, who proposed that reconnection was also a driver of the solar wind emanating from CHs. 

\section{Interchange reconnection}
\setstretch{2}

The process of continuous reconnection between open and closed field lines is known as ``interchange reconnection'' \citep{Wang93, Wang96, Wang04, Fisk05, Raju05, Edmondson10}. As suggested by \cite{Fisk01} and \cite{Fisk05}, interchange reconnection allows the transfer of magnetic flux from one point to another, which can be considered as a diffusion-convection process resulting from convective motions and random motions caused by non-stationary supergranular flows. The equation for this diffusion process is based on the induction equation which we derive using the Maxwell equations:
\begin{equation}
\nabla\times \mathbf{B} = \mu \mathbf{j} + \frac{1}{c^{2}} \frac{\partial \mathbf{E}}{\partial t} \ ,
\label{ampere}
\end{equation}
\begin{equation}
\nabla\times \mathbf{E} = - \frac{\partial \mathbf{B}}{\partial t}\ ,
\label{faraday}
\end{equation}
\begin{equation}
\nabla \cdot \mathbf{E} = \frac{\rho}{\varepsilon_{0}} \ ,
\label{gauss}
\end{equation}
\begin{equation}
\nabla \cdot \mathbf{B} = 0 \ ,
\label{nomonopoles}
\end{equation}
where $\mathbf{B}$ is the magnetic field strength, $\mathbf{E}$ is the electric field strength, $\mu$ is the magnetic permeability, $\mathbf{j}$ is the current density, $c$ is the speed of light, $\rho$ is the charge density and $\varepsilon_{0}$ is the electric permittivity. We also include the Ohm equation:
\begin{equation}
\mathbf{j} = \sigma(\mathbf{E}+\mathbf{v}\times \mathbf{B}) \ ,
\label{ohm}
\end{equation}
where $\sigma$ is the electrical conductivity, and $\mathbf{v}$ is the plasma velocity. In the MHD approximation the plasma velocity is much smaller than the speed of light ($v<<c$), thus $\nabla\times\mathbf{B}>>{1}/{c^{2}}\;{\partial \mathbf{E}}/{\partial t}$, and hence Equation~\ref{ampere} reduces to
\begin{equation}
\nabla\times \mathbf{B} = \mu \mathbf{j}\ ,
\label{ampere_mhd}
\end{equation}
which we rewrite as $\mathbf{j}=1/\mu \; (\nabla\times \mathbf{B})$ and substitute into Equation~\ref{ohm} to get
\begin{equation}
\mathbf{E} = -\mathbf{v}\times \mathbf{B}+\frac{1}{\mu\sigma}(\nabla \times \mathbf{B})
\label{ampere_mhd}
\end{equation}
This equation is substituted into Equation~\ref{faraday} with $\eta={1}/{\mu\sigma}$ to get
\begin{equation}
\frac{\partial \mathbf{B}}{\partial t}= \nabla \times (\mathbf{v}\times \mathbf{B})- \nabla \times (\eta\nabla \times \mathbf{B})\
\end{equation}
which reduces to 
\begin{equation}
\frac{\partial \mathbf{B}}{\partial t}= \nabla \times (\mathbf{v}\times \mathbf{B})- \eta \nabla^{2} \mathbf{B}\ ,
\label{diffeq}
\end{equation}
also known as the {\it induction equation}, where the first term on the right is the convection term, and the second is the diffusion term. Here, we assume that the magnetic field lines diffuse due to random motions in the medium. The resistive diffusion is ignored ($\nabla \times (\mathbf{v}\times \mathbf{B})=0$) as we only consider the field line motions due to the velocity of the medium. If only the radial motions along the solar surface are considered, Equation~\ref{diffeq} can be written as
\begin{equation}
\frac{\partial B_{z}}{\partial t}= -\nabla_{s} \cdot (\mathbf{v_{rand}} B_{z})\ ,
\end{equation} 
where $\mathbf{v_{rand}}$ is the plasma velocity along the surface due to random motions, and $B_{z}$ is the radial component of the magnetic field. Following \cite{Fisk01} it can be shown that
\begin{equation}
\frac{\partial \overline{B}_{z}}{\partial t}= -\nabla^{2}_{s} (\mathbf{\kappa} \overline{B}_{z})\ ,
\end{equation}
where $\overline{B}_{z}$ is the mean radial magnetic field and $\mathbf{\kappa}$ is the diffusion coefficient tensor, which in a simplified form can be written as
\begin{equation}
\kappa=\frac {(\delta r)^{2}}{2 \delta t} \ ,
\label{kappa}
\end{equation}
where $\delta r$ is the open magnetic field displacement distance along the surface and $\delta t$ is the time elapsed.

Despite all of the flux transfer processes involved being convective in nature, the non-stationary random supergranular convection is included in the diffusion term while those that are uniform (such as differential rotation and meridional flow) are included in the convection term. As supergranular motions are too slow for flux transportation, interchange reconnection was suggested as a faster mechanism for field-line diffusion. Interchange reconnection allows for the transfer of magnetic flux from one point to another through magnetic reconnection between open and closed field lines, which \cite{Fisk01} modelled using a diffusion-convection equation,
\begin{equation}
\frac{\partial B_{z}}{\partial t} = \kappa \nabla^{2}_{s} B_{z} - \nabla \cdot (v_{conv}B_{z}) \ ,
\label{eq_fisk}
\end{equation}
where $v_{conv}$ is the uniform convective flow velocity. It is assumed that the radial component of the magnetic field is equivalent of a surface density and is subject to diffusion and convection along the solar surface. The diffusion term corresponds to random supergranular motions, while all convective motions including the differential rotation and meridional flows are included in $v_{conv}$. Because supergranular motions are considered too slow to transport open magnetic flux on the Sun. Instead, a faster process of interchange reconnections between numerous open and closed field lines is considered as a process for diffusion.

\begin{figure}[!t]
\centerline{\hspace*{0.015\textwidth} 
\includegraphics[scale=0.45, trim=75 100 80 110, clip]{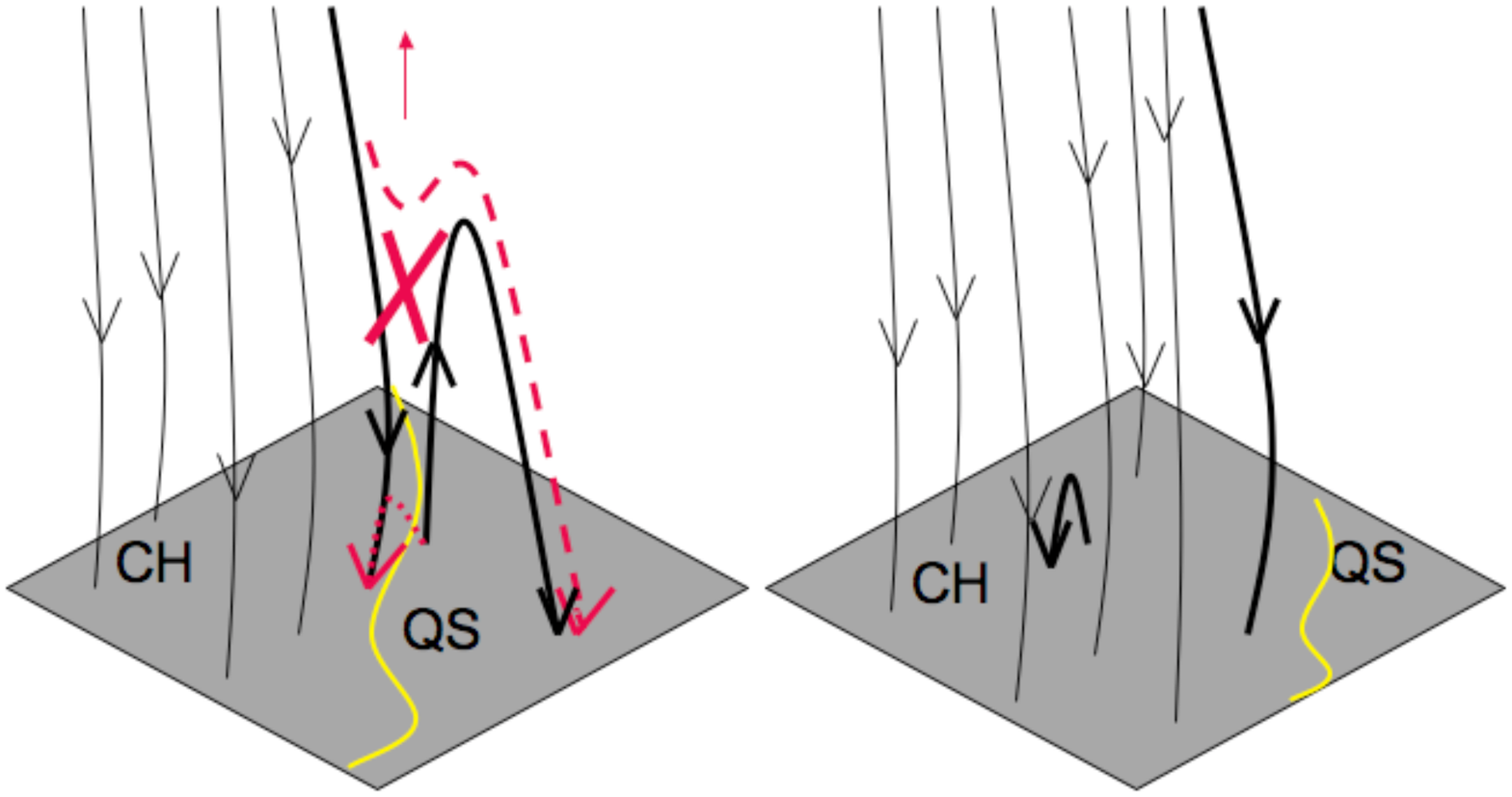} } 
\vspace{-0.01\textwidth} 
\caption{Interchange reconnection: The location of the CH and the QS are indicated on the cartoon with a line representing the boundary between the two regions.The scenario on the left shows a QS loop located near a CH open field which yields magnetic reconnection due to the proximity of the opposite polarity loop footpoint to the open field. The result of the reconnection is shown on the right - a small loop is created inside the CH with opposite magnetic orientation and the open field line ``relocated" from within the CH to the further foot-point of the previously larger loop. This process can explain the rigid rotation as well as the growth of CHs \citep[modified from][]{Fisk05}.}
\label{fig1}
\end{figure}

A graphical representation of interchange reconnection is given in Figure~\ref{fig1}, where a QS loop has moved close to a CH boundary due to random granular motions. Magnetic reconnection between the open and closed field leads to the displacement of the open field and thus the CH boundary moves further out. During this process the larger loop is replaced by a smaller loop with oppositely oriented footpoint polarities imbedded inside the CH, while the open field line is ``relocated" to the former foot-point of the larger loop.  

Interchange reconnection at CH boundaries is also supported by observational results, e.g., the prevalence of very small loops inside CHs with heights $<$3~Mm and larger loops with heights $>$3~Mm outside CHs \citep{Wiegelmann04}, the presence of bidirectional jets \citep{Madjarska04, Doyle06, Kamio09}, and the boundary displacements observed due to the emergence and disappearance of bright small-scale loops in the form of bright points \citep{Madjarska09, Subramanian10}. 

In this Chapter, we study the short-term evolution of CH boundaries by objectively determining CH boundaries based on intensity thresholding methods, as well as introducing a new automated boundary tracking tool ({\it CHEVOL}) that allows us to follow the changes that occur (detailed in Section \ref{chevol_methods}). In Section \ref{chevol_results} the magnetic reconnection rate and diffusion coefficient are determined using the interchange reconnection model, and the measured boundary displacement velocities. The processes that influence the CH boundary evolution are discussed in Section \ref{chevol_conclusions}.

\section{Observations and Data Analysis}
\label{chevol_methods}

Finding objective CH boundary intensity thresholds is fundamental in deriving consistent physical properties (e.g., area, magnetic field strength) throughout the CH evolution. For this reason we used the {\it CHARM} algorithm (see Chapter~\ref{chapter:detection}) on a localised small region of interest. The size of the studied region was chosen to be a 31$^{\circ}$~$\times$~35$^{\circ}$ ``box", which was centered on each analysed CH. 

In this study we used data most suitable for CH detection: 12~minute cadence 195~\AA\ images from SOHO/EIT to locate low intensity regions, and 96~minute cadence magnetograms from SOHO/MDI to detect predominant polarities and hence identify CHs. The magnetograms were also used to determine the magnetic properties of the CHs and the surrounding QS. The study includes eight relatively small CHs, which were chosen arbitrarily between 2008 September 8 and 2009 October 17, and tracked for two to six days. The CHs studied included both predominantly positive and negative polarity holes, with equal numbers located in the northern and southern hemispheres. All of the CHs studied were non-fragmenting and of relatively simple morphology throughout their evolution.

The {\it CHEVOL} method only requires the start-date and duration of the analysis, and the location of the region of interest to be given. The start-date should preferably be given for when the studied CH is in the eastern hemisphere. This allows the CH to be studied for a longer period of time before it rotates off-limb. To avoid near-limb effects, the CH should not be further than 50$^{\circ}$ from the central meridian. The location of the CH is provided by the user by clicking on the EIT Lambert map that appears after the program is executed. The user is requested to click south-east of the studied CH, as this determines the lower-left corner of the box. The box location is saved together with the start-date, thus the region specification only needs to be provided once for each CH studied. The size of the box can be altered if needed. From this point on the analysis is completely automated, and the specified box ``follows" the CH in consecutive EIT images in the time period requested. 
\begin{figure}[!t]
\vspace{-0.02\textwidth}
\centerline{ 
\includegraphics[scale=0.85, trim=0 0 250 190, clip]{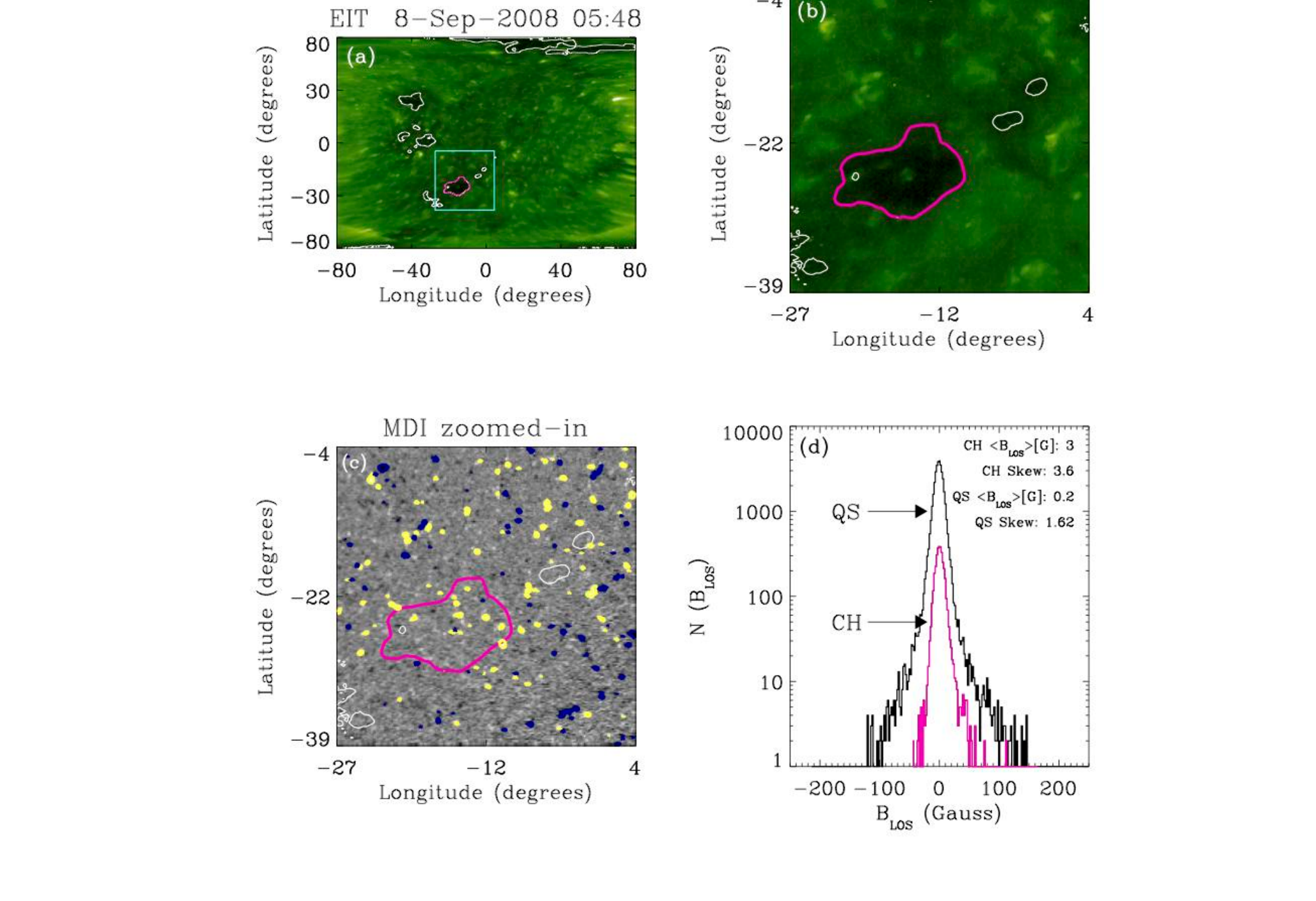} }
\vspace{-0.05\textwidth} 
\caption{CH contours in an EIT 195~\AA\  Lambert projection map (a) and the corresponding sub-image (b). MDI sub-image with the CH contours (c) ($>$50~G fields: yellow, $<$-50~G fields: blue) and the corresponding $B_{\mathrm{LOS}}$ histogram (d) (magenta: CH, black: QS).}
\label{fig2}
\end{figure}

All EIT images are initially transformed into Lambert equal-area projection maps to aid the intensity thresholding and to simplify the tracking of the region of interest. Figure~\ref{fig2}a shows the EIT Lambert projection map with the CH contours and the box around one of the CHs studied (only the largest CH in the box was a candidate in our study). Figures~\ref{fig2}b and \ref{fig2}c show the EIT and MDI close-up images of the box with the CH contours, while Figure~\ref{fig2}d shows the histogram of the line-of-sight magnetic field ($B_{\mathrm{LOS}}$) from the largest CH (magenta line) and the surrounding QS (black line), excluding small satellite CHs. The skewness of the $B_{\mathrm{LOS}}$ histogram is used to establish whether a predominant polarity is found, thus allowing the differentiation of CHs (skewness $>$~0.5) from other low intensity regions (skewness $<$~0.5). The agreement between the visual and the analytically-obtained CH boundary, together with the predominant polarity found in the contoured regions, validate the appropriateness of the boundary positions and justifies the CH status of the detected regions. In principle, the intensity of a CH boundary can change due to the line-of-sight obscuration at higher longitudes and latitudes. However, all CH candidates in this study lie within $\pm50^{\circ}$ in longitude and latitude, which lowers both the obscuration effect in EUV images and the errors in line-of-sight magnetic field measurements.

\begin{figure}[!t]
\centerline{\hspace*{0.015\textwidth} 
\includegraphics[scale=0.8, trim=0 45 400 20]{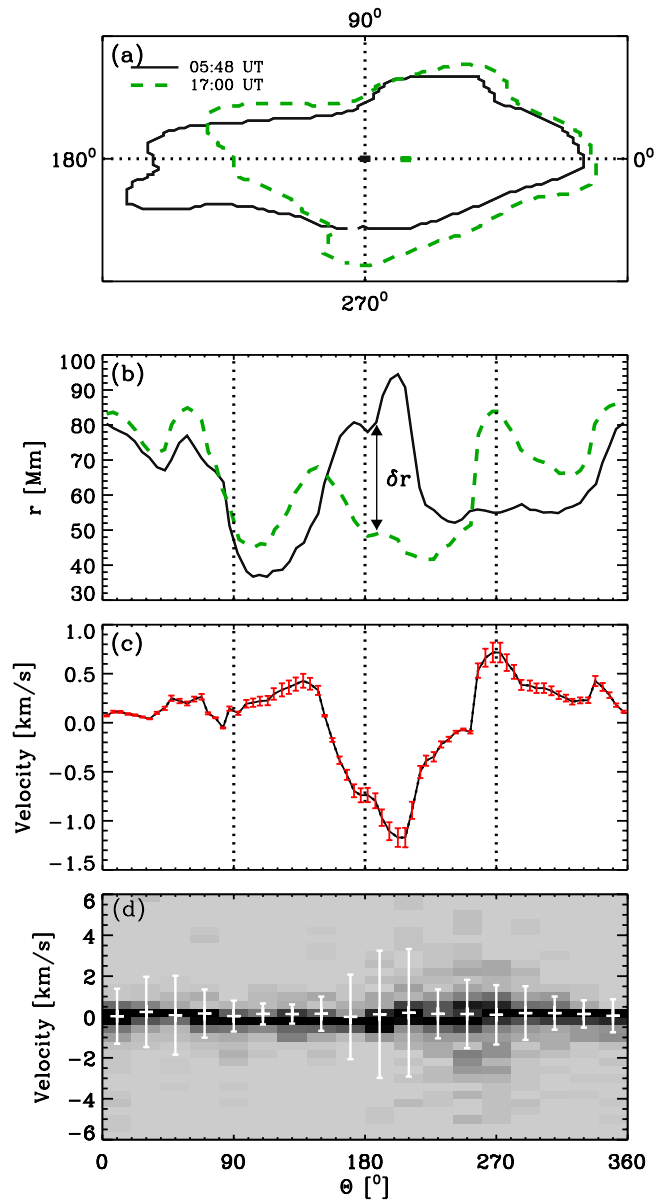} }
\vspace{-0.01\textwidth} 
\caption{(a) The CH contour observed at 05:48 UT (solid line) and 17:00 UT (dashed line) on 2008 September 8. The frame of reference is the earlier contour centroid and angle is measured anti-clockwise from the right direction of the earlier contour centroid. (b) The radial location of the boundary plotted as a function of position angle (solid: earlier; dashed: later). (c) Boundary displacement velocities as a function of position angle. (d) Boundary displacement velocity distributions as a function of position angle for all observations of the CH. The dotted lines in (a), (b), and (c) mark the boundary locations at $0^{\circ}$, $90^{\circ}$, $180^{\circ}$ and $270^{\circ}$. The error bars in (d) correspond to $\pm 1~\sigma$ for each $20^{\circ}$ bin.}
\label{fig3}
\end{figure}

Once a CH was identified, consecutive boundary observations were compared. Considering an example of contour pairs as shown in Figure~\ref{fig3}a, the centroid of the earlier observed contour (solid line, observed on 2008 September 8 at 05:48 UT) was considered as the origin of the frame of reference for both contours. The later contour (dashed line, observed on 2008 September 8 at 17:00 UT) will have moved with the synodic solar rotation rate compared to the earlier contour, due to solar rotation. For this reason a correction was applied to the second contour coordinates. The locations of both contours were measured radially from the earlier centroid in 5$^{\circ}$ segments and anti-clockwise from west of the earlier contour's centroid (Figure~\ref{fig3}b). Each boundary segment location was determined as the median of the contour point distances from the reference centroid. The displacement of a segment ($\delta r$) and the time difference between the two observations yielded the displacement velocity of the segment. This calculation was repeated to get all the boundary displacement velocities over 360$^{\circ}$ (Figure~\ref{fig3}c). (Note that the velocity mentioned here depends of the temporal resolution of the used instrument and the time-scale at which the boundary displacement happens.) 
The displacement velocities measured for the whole observation period of a single CH were also determined, as shown in the 2D histogram of Figure~\ref{fig3}d. Here, dark regions correspond to more frequent occurrences of the boundary velocity values, where the observations were binned in 20$^{\circ}$ and 0.4~km~s$^{-1}$ intervals. These calculations were further extended to include every boundary displacement velocity of each CH in the study, as detailed in the following Section.

\section{Results}
\label{chevol_results}

\subsection{Coronal hole boundary displacement velocities}

In the case of the discussed example CH, similar average velocities were observed over all angle space (Figure~\ref{fig3}d). This indicates that there is no preferred direction in the boundary evolution of the CH, and hence the boundary evolution is isotropic. The value of the average signed velocities, $\langle v \rangle$, is $\sim$0~km~s$^{-1}$ in all angle bins, which indicates no consistent expansion or contraction in the CH during the time of observation. Furthermore, the majority of $\pm 1~\sigma$ values for all angle bins are below 2~km~s$^{-1}$ (as is the case for all eight CHs studied here), which is in good agreement with the average velocities of random granular motions. 

In order to extract characteristic velocities, all CH observations were combined over all angles to obtain the unsigned boundary velocity distribution (Figure~\ref{fig4}a). From this distribution typical unsigned velocities were found to be 0--2~km~s$^{-1}$.

\subsection{Magnetic reconnection rate}

\begin{figure}[!t]
\centerline{\hspace*{0.015\textwidth}  
\includegraphics[scale=0.6, trim=100 130 70 50, clip]{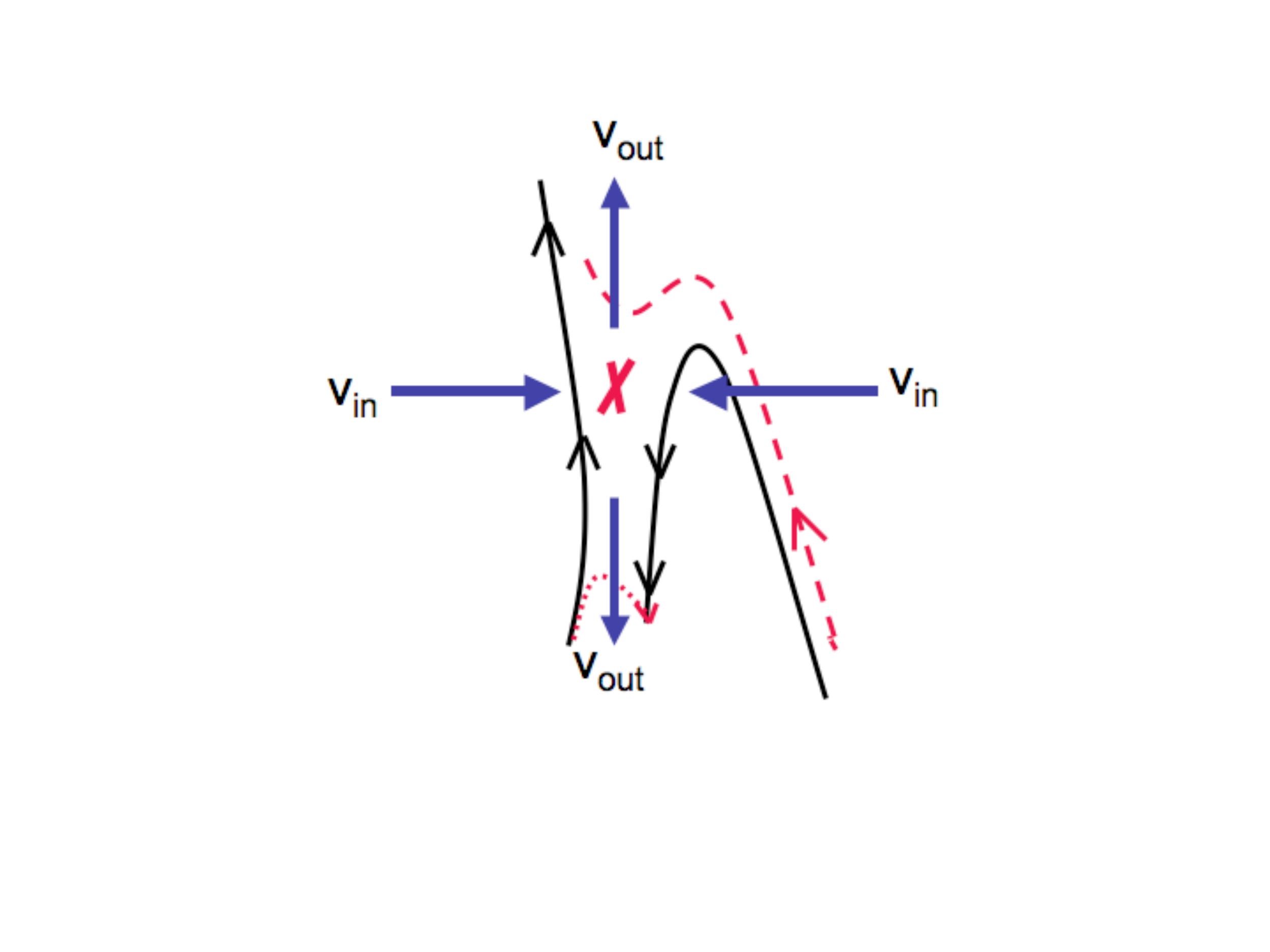} }
\vspace{-0.001\textwidth} 
\caption{Magnetic reconnection between a closed and an open field line. The solid lines show the field lines pre-reconnection, and the dashed lines demonstrate the field lines after the reconnection. The arrows indicate the direction of the inflow with velocity $v_{in}$, and the outflow with velocity $v_{out}$.}
\label{fig1a}
\end{figure}

Magnetic reconnection between CH open fields and small-scale QS loops (with heights of $<3$~Mm; \citeauthor{Wiegelmann04} \citeyear{Wiegelmann04}) is expected to occur at the transition region level. However, the CH boundary displacements are obtained from 195~\AA\ images that show transition region and the inner coronal heights. Here we can make the reasonable assumption that as a small-scale loop reconnects with the open magnetic field in the transition region it will consequently cause a displacement between open and closed field lines in the inner corona. In fact, the displacement at the transition region should be roughly proportional to the displacement detected in the EUV images if the surrounding QS has a relatively stable magnetic configuration in the corona and the mentioned reconnection does not fundamentally change that. We use velocities obtained from EIT 195~\AA\ images as we have no information on the displacement velocity at the reconnection site. The Alfv\'en speed, $v_{\mathrm{A}}$, was calculated using $v_{\mathrm{A}}=B/\sqrt{4\pi m_{p}n_{e}}$, where $B$ is the magnetic field strength and $n_{e}$ is the electron number density. Typical values for the magnetic field ($B=2$~G) and density ($n_{e}=10^{8}$~cm$^{-3}$) in the lower corona were used from \cite{Schrijver05} to calculate the Alfv\'en speed, which was found to be $\sim$600~km~s$^{-1}$. The magnetic reconnection rate was determined from
\begin{equation}
M_{0}=v_{\mathrm{in}}/v_{\mathrm{A}},
\label{asch}
\end{equation}
where $v_{\mathrm{in}}$ is the inflow speed (see Figure~\ref{fig1a}), the relative speed of the opposite polarity magnetic field lines being brought together prior to reconnection. The distribution of displacement velocities was used for $v_{\mathrm{in}}$ and substituted together with the Alfv\'en speed into Equation~\ref{asch} to obtain the reconnection rate distribution. This distribution is shown in Figure~\ref{fig4}b, where typical reconnection rates are in the range 0--3~$\times$~10$^{-3}$. 

\subsection{Magnetic diffusion coefficient}
\begin{figure}[!t]
\centerline{\hspace*{0.015\textwidth}  
\includegraphics[scale=0.8, trim=0 0 350 0, clip]{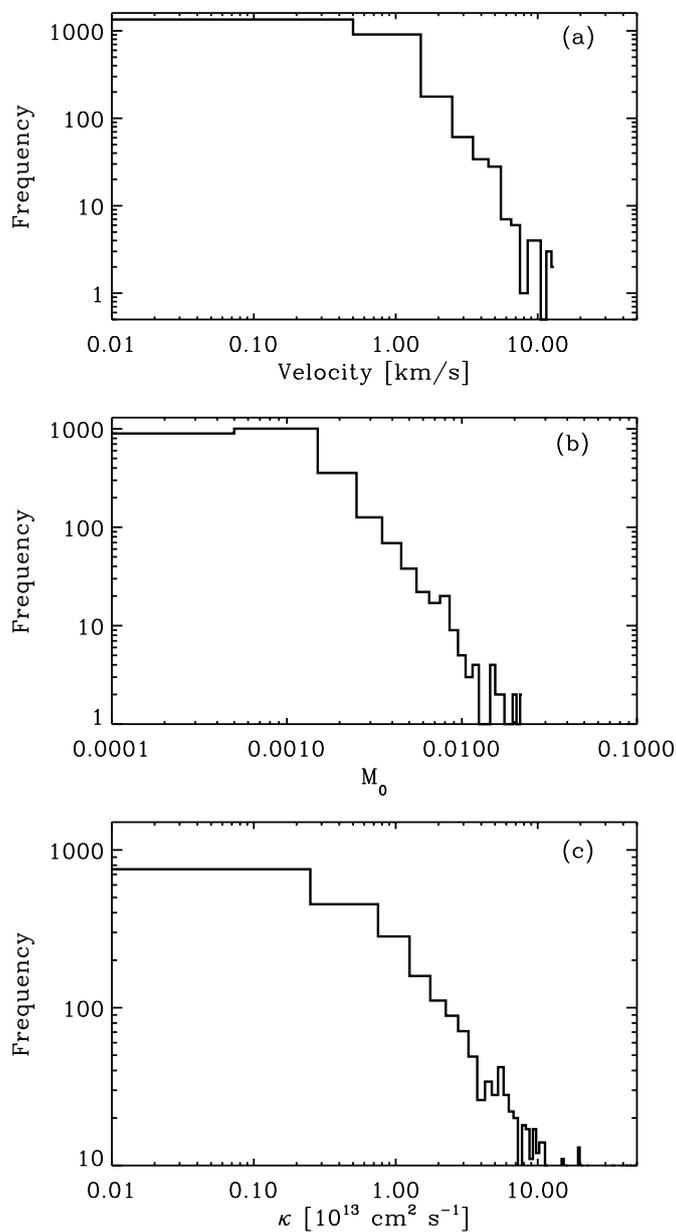} }
\vspace{-0.001\textwidth} 
\caption{Frequency distributions of boundary displacement velocities (a), magnetic reconnection rates (b), and diffusion coefficients (c).}
\label{fig4}
\end{figure}

We calculate the magnetic diffusion coefficient for each $5^{\circ}$ segment of each CH observation using Equation \ref{kappa} by substituting the boundary displacement distances in the segment ($\delta r$) and the time elapsed ($\delta t$) between consecutive observations. The resulting distribution of diffusion coefficients is shown in Figure~\ref{fig4}c, with values typically in the range 0--3~$\times$~10$^{13}~$cm$^{2}~$s$^{-1}$. Using typical loop footpoint distances and characteristic times for reconnection, \cite{Fisk01} have determined the diffusion coefficients inside and outside of a CH to be 3.5~$\times$~10$^{13}$~cm$^{2}$~s$^{-1}$ and 1.6~$\times$~10$^{15}$~cm$^{2}$~s$^{-1}$, respectively. 

\subsection{Magnetic field in coronal holes and the quiet Sun}
\begin{figure}[!t]
\centerline{ 
\includegraphics[scale=0.7, trim=0 20 200 330, clip]{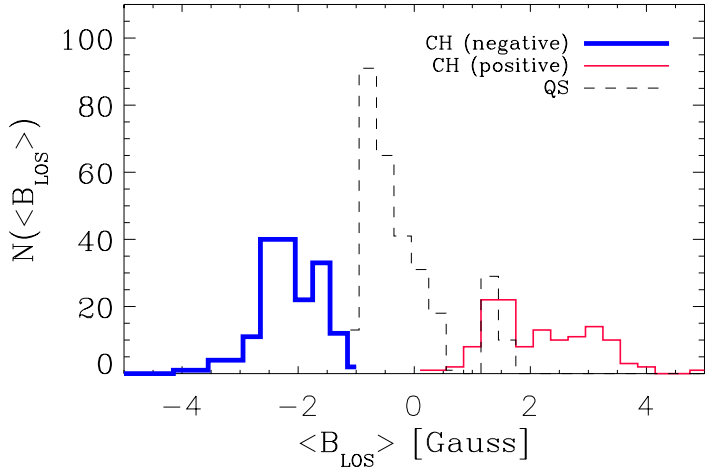} }
\vspace{-0.001\textwidth} 
\caption{The $<$$B_{\mathrm{LOS}}$$>$ histogram of the negative and positive polarity CHs (thick/blue and thin/red line, respectively) and the QS (dashed line).}
\label{fig5}
\end{figure}

\begin{figure}[!t]
\centerline{\hspace*{0.015\textwidth}  
\includegraphics[scale=0.6, trim=10 20 70 40, clip]{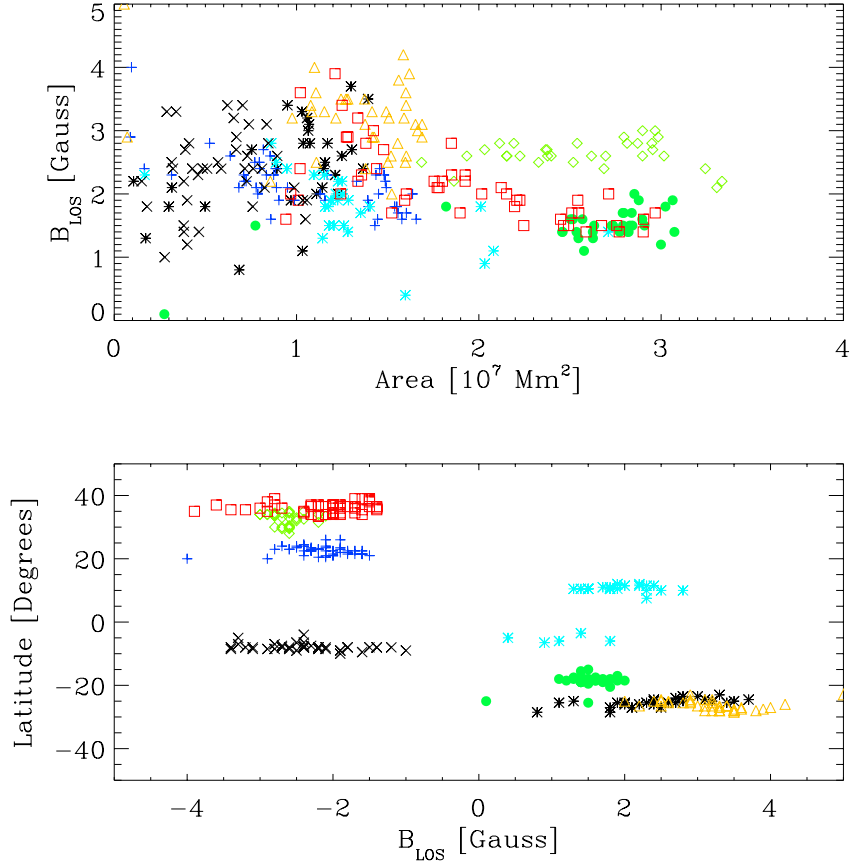} }
\vspace{-0.001\textwidth} 
\caption{Top: The CH mean $B_{\mathrm{LOS}}$ as a function of CH area. Bottom: The latitudinal locations of individual CHs and the CH mean $B_{\mathrm{LOS}}$ measures in the CHs. Different colours/symbols represent different CHs.}
\label{fig6}
\end{figure}

Establishing the CH boundaries also allowed us to determine the average magnetic field strength ($<$$B_{\mathrm{LOS}}$$>$) of the CH and the surrounding QS within the ``box". For the studied CHs, the signed $<$$B_{\mathrm{LOS}}$$>$ was found to be 2.3$\pm$0.9~G and -2.3$\pm$0.5~G, while -0.1$\pm$0.7~G in the QS surrounding the CH. Figure~\ref{fig5} shows the mean $B_{\mathrm{LOS}}$ distributions from all CH observations and the corresponding QS regions. The thick/blue and thin/red lines correspond to the predominantly negative and positive CHs respectively, and the dashed black line corresponds to the QS. The figure shows a clear separation between the negative polarity holes and the QS, while there is some mixing between the positive polarity holes and the QS. 
The top plot in Figure~\ref{fig6} shows the unsigned $<$$B_{\mathrm{LOS}}$$>$ as a function of the CH area. The plot shows that the unsigned $<$$B_{\mathrm{LOS}}$$>$ in larger CHs varies less. However, no correlation has been found between the area and the unsigned $<$$B_{\mathrm{LOS}}$$>$ of the CHs (correlation coefficient $\sim$0.23). The bottom plot in Figure~\ref{fig6} shows the location of the CHs, their polarity, and the range of the corresponding $<$$B_{\mathrm{LOS}}$$>$ values.

\section{Discussion and Conclusions}
\label{chevol_conclusions} 

The relocation of open magnetic field lines through interchange reconnection is the most widely accepted method to explain some of the distinctive characteristics of CHs -- namely their rigid rotation, the lack of loops with heights $>$10~Mm inside CHs, and the CH boundary relocations through interaction with QS loops. In this study CH boundary locations were determined using an intensity thresholding technique described in detail in Chapter~\ref{chapter:detection}. This method determines the CH boundaries based on observations rather than allowing the observer to choose a subjective threshold. A number of CHs were tracked for several days to study the changes in the CH boundaries. The CH boundary tracking was carried out by temporally coupling the boundary contours, which allowed relative distances of boundary movements and hence velocities to be established. In our study the displacement velocities were found to be isotropic with unsigned values typically 0--2~km~s$^{-1}$, which imply reconnection rates of 0--3~$\times$~10$^{-3}$. Studying the distribution of the velocity, magnetic reconnection rate and the diffusion coefficient (Figure~\ref{fig4}) the decay was found to follow a power law, which suggests that these properties are scale-free.

These boundary velocities are in good agreement with the isotropic nature and $\sim$1--2~km~s$^{-1}$ average velocity of random granular motions \citep{Markov91}. Hence, we suggest that the majority of the reconnection at CH boundaries is likely to be initiated through random granular motions. Velocities in excess of $\sim$2~km~s$^{-1}$ were occasionally observed during our analysis. Such surges in the boundary displacement could be explained by consecutive interchange reconnections in the case of favourable loop alignments, whereby a number of loops have opposite magnetic polarity to the CH flux in their nearest footpoint to the CH. In close proximity the CH open field lines could continuously reconnect with the loops and advance into the QS, before being halted by loops that face the CH with the same polarity in their nearest footpoint to the CH.

Interchange reconnection has been previously discussed by \cite{Fisk01} as a mechanism that allows the transfer of magnetic flux from one point to another (see Equation~\ref{eq_fisk}). We determine typical values of diffusion coefficient to be 0--3~$\times$~10$^{13}$~cm$^{2}$~s$^{-1}$ using Equation~\ref{kappa} with every individual CH boundary displacement measurement (i.e., from every 5 degree segment of each CH time pairing). Although the values derived in this study are comparable to those determined by \citeauthor{Fisk01} (3.5~$\times$~10$^{13}$~cm$^{2}$~s$^{-1}$ inside a CH), it is important to note that the results are achieved by different methods. Those presented here are based on CH boundary tracking using observations and reflect upon the diffusion processes happening at the CH boundary, while \citeauthor{Fisk01} provided estimates of the magnetic diffusion in CHs based on typical loop heights, foot-point distances and sizes of supergranular cells.

The lack of systematic relative motion between CHs and the QS on timescales of days indicates that the small CHs studied rotate at approximately the local photospheric QS rotation rate. In the interchange reconnection model, QS loops continuously reconnect with the CH open field and effectively ``pass" through the CH (an argument also used to explain the rigid rotation of CHs). This suggests that over several solar rotations a relative velocity should be observed between the CH and the QS due to the differential rotation of the QS. If we assumed a CH to rotate uniformly at a velocity equivalent to the QS rotation velocity at the latitudinal location of the CH centroid, we would expect the CH to shift eastward at lower latitudes and westward at higher latitudes so as to compensate for the differential rotation (and the shearing effect) of the surrounding QS. Although no evidence for this has been observed in this study, we suggest it is merely due to the fact that the longest observation period was six days, during which only a few degrees shift would occur over a CH latitudinal width of $<$~35$^{\circ}$ (bounding box height). 

Studying the small-scale changes in CH boundaries can greatly expand our understanding of not only CHs but the QS as well. The interaction between a CH and the QS can shed light on the processes controlling the magnetic field evolution in the QS and how open field is accumulated (CH birth) and diffused (CH decay). Although our current understanding of exactly how CHs are created and destroyed is still limited, it is likely that the random granular motions in the QS allow temporary accumulations of open field. Small CH candidates will then grow or decay based on the local loop-emergence rate and the specific orientation of QS loops near the CH boundary. The decay of a CH can be influenced by nearby QS loops but also by processes occurring inside the CH. As well as magnetic bipoles emerging within a CH, small loops admitted into the CH during interchange reconnection could accumulate over a CH lifetime and ultimately lead to the fragmentation and diffusion of the CH. Bipoles and open flux regions would have to be tracked in time to provide observational evidence for the processes governing the birth and the decay of CHs. This could also provide information on how small CHs turn into large, stable and long-lived CHs, and what processes initiate the decay of a long-lived CHs.



\chapter{High-speed solar wind forecasting} 

\label{chapter:forecasting}

\ifpdf
    \graphicspath{{5/figures/PNG/}{5/figures/PDF/}{5/figures/}}
\else
    \graphicspath{{5/figures/EPS/}{5/figures/}}
\fi

\flt{\it Recurring high-speed solar wind streams observed at 1~AU are known to be associated with CHs. However, the relationship between CH properties and in-situ HSSW properties needs to be further investigated to aid space weather forecasting. In this Chapter, we describe the Minor Storm ({\it MIST}) algorithm developed by the author in order to link CHs to HSSW streams detected at Earth. Over 600 CH-HSSW pairs were analysed over nine years which allowed us to determine the expansion factor of open magnetic flux tubes rooted in CHs. The expansion factor was found to be 2--48, which is similar to values previously found using theoretical models. The study also revealed the HSSW stream velocities to have a strong correlation with the proton temperatures and $Kp$ indices. Furthermore, we confirmed the relationship between CH-related HSSW streams and enhanced high-energy electron fluxes measured in the Earth's radiation belt during the descending phase of the solar cycle.
}
\flb

\newpage

\section{CHs and high-speed solar wind}

Unlike flares and CMEs, the prediction of CH related geomagnetic disturbances can be forecasted with high certainty. This is due to three main reasons:
\begin{itemize}
\item
CHs are observable for long periods of time, and thus a forecast can be made several days in advance before the CH related HSSW reaches Earth.
\item
The velocity distribution of the CH-related HSSW streams is known, and can be used to determine the arrival-time range.
\item
CHs are sources of continuous HSSW flows and thus no eruption time needs to be determined as in the case of flares and CMEs.
\end{itemize}
These characteristics are used in the present research to forecast the arrival time of HSSW streams at Earth. The start and end-time range of the HSSW periods are determined based on CH location, however, the end of the HSSW period at Earth is less straightforward to estimate. Since flux tubes rooted in CHs expand super-radially (Section~\ref{expansion}), the HSSW period end-time depends on the expansion factor \citep{Guhathakurta98}. In Section~\ref{manual} we discuss a semi-automated method to determine the expansion factor, which is then used in the {\it MIST} algorithm to determine the time-range for the plasma arrival and end at Earth. The method is run over a long period (2000-2009) to obtain CH and HSSW properties and geomagnetic disturbance indices, which are compared to establish any relationship (Section~\ref{auto}). 

\subsection{Previous forecasting methods}
\label{forecast_methods}

\subsubsection{The Robbins method}
\label{robbins_method}

The empirical forecasting model of \cite{Robbins06} relies on the CH detection method developed by \citeauthor{Harvey02} (described in Section~\ref{harvey_method};\citeauthor{Harvey02} \citeyear{Harvey02}). KPVT He~{\sc i} 10830~\AA\ spectrograms and photospheric magnetograms were used to determine the location, size and magnetic properties of CHs, while the solar wind data was obtained from the OMNIWeb website, provided by the US National Space Science Data Center. 

The heliographic CH maps were first divided into 23 longitudinal sectors in a $\pm$77$^\circ$ longitudinal and $\pm$60$^\circ$ latitudinal range. In each sector the percentage of CH coverage was determined, and cross-correlated with the solar wind speed time series. First, a model solar wind time series was created by determining the CH area percentage in the 14$^\circ$ sectors. These CH area percentages in each sector were then rescaled to agree with the observed solar wind speed, and the resulting linear scaling coefficients were used to forecast HSSW velocities. The model was used on the period 28 May 1992 - 25 Sep 2003, where for the best one-month periods the model was found to be accurate to within 10\% of the observed solar wind measurements. The linear correlation coefficient was determined to be $\sim$0.38 for the 11 years studied.

The KPVT CH maps were also used to predict the IMF polarity at Earth. The pixel values in CHs were set to +1 or -1 depending on a positive or negative magnetic field strength, respectively, while QS pixels were set to zero. The magnetic field was averaged in the sector to determine its overall polarity, and then used as a baseline for the polarity forecasting. Using this method the IMF polarity was correctly forecast 58\% of the time. In their paper, \citeauthor{Robbins06} stated that their model had inherent inaccuracies which were likely to be caused by the oversimplification of the magnetic field structure associated with CHs and the corresponding solar wind streams. They also assumed that the solar wind velocity is related linearly to the CH area, whereas it has been shown that at a certain scale size the wind velocity is independent of CH area \citep{Poletto04}.

Despite its partial success, it is important to note that the \citeauthor{Robbins06} method provided a fast and simple approach to HSSW forecasting by solely using CH maps and magnetograms.

\subsubsection{The Vr\v{s}nak method}
\label{vrsnak_method}

\citeauthor{Vrsnak07a} carried out two studies relating CHs and HSSW streams. The first study used CH properties to forecast solar wind parameters, while the second study forecasted the scale of geomagnetic disturbances based on the previous work.

First, the CHs were identified in soft X-ray images from the Soft X-ray Imager aboard the GOES-12 spacecraft (SXI; \citeauthor{Hill05}, \citeyear{Hill05}). The CH boundaries were determined by setting a fixed intensity threshold of 0.15~DN~s$^{-1}$. For each day the fractional CH area was determined in three 20$^{\circ}$ wide longitudinal sectors. The studied longitudinal ranges were [-40$^{\circ}$, -20$^{\circ}$], [-10$^{\circ}$, 10$^{\circ}$], and [20$^{\circ}$, 40$^{\circ}$] from the central meridian. The latitudinal range of the analysis was $\pm$45$^{\circ}$.
 
\begin{figure}[!t]
\centerline{
\includegraphics[scale=0.5, trim=10 0 10 0, clip ]{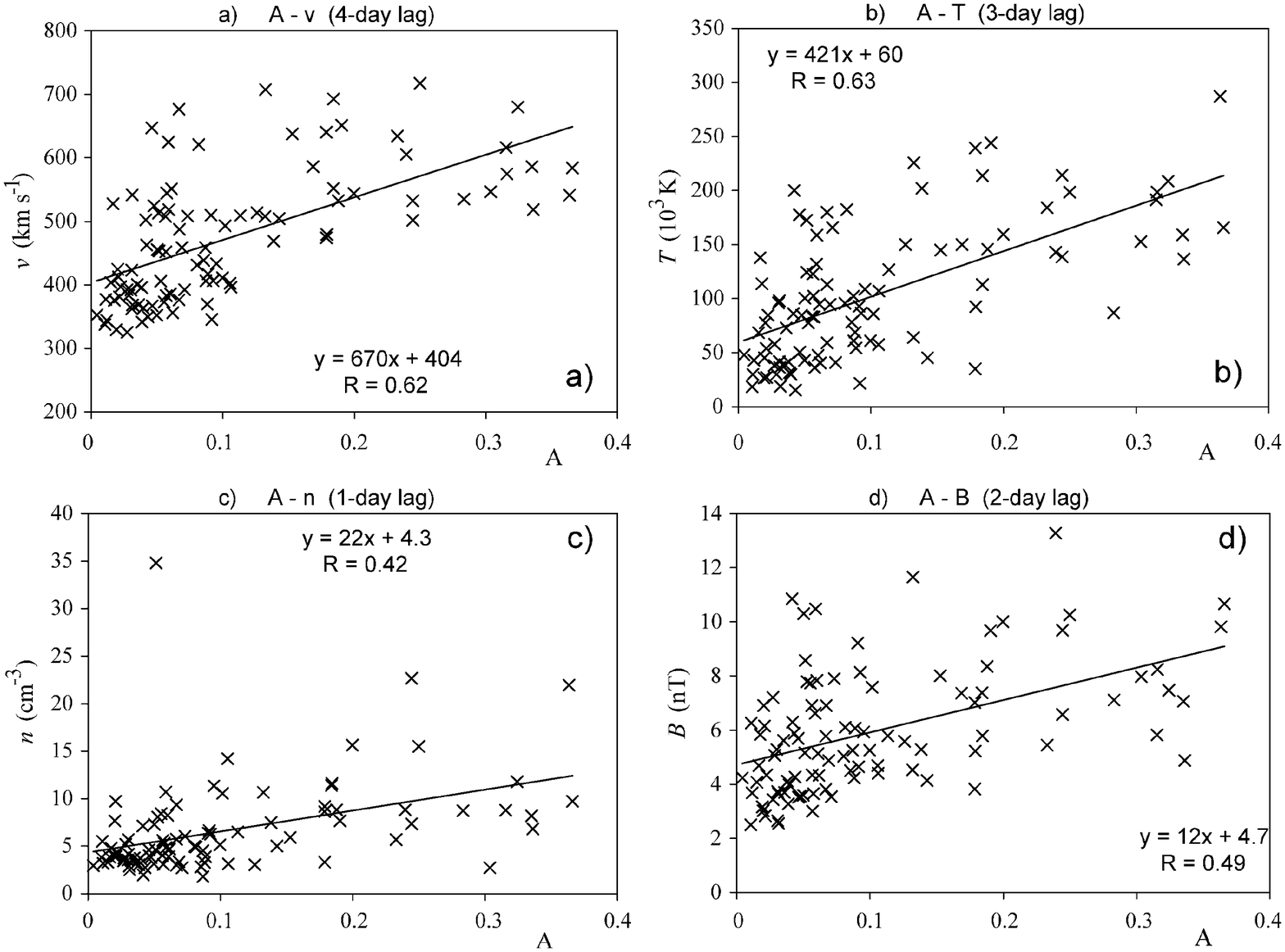}}
\caption{The time-lagged correlations of the CH areas in the central-meridian sector and the peak values of a) solar wind velocity; b) proton temperature; c) density and d) magnetic field, for 25 January -- 5 May 2005. The least-squares fits and the corresponding correlation coefficients are shown in the insets. The time lags used are shown in the plot headers \citep{Vrsnak07a}.} 
\label{vrsnak_plot} 
\end{figure}

The time range of the study was 25 January to 5 May 2005, where eleven distinct peaks in the solar wind properties were identified and linked to CHs. The HSSW properties measured were the velocity ($v$), magnetic field strength ($B$), proton number density ($n$), and the proton temperature ($T$), using data from the ACE/SWEPAM and MAG instruments. The time when a CH had the largest area in the sector was compared to the onset time of the solar wind property enhancements. The time delays were quantified by applying a cross-correlation analysis. This was followed by the time-lagged correlation of the maximum CH areas with the HSSW properties as shown in Figure~\ref{vrsnak_plot}. The obtained coefficients were then used in a fully automated manner to forecast HSSW properties several days in advance.

The most accurate prediction was obtained for the solar wind velocity, for which the average relative difference between the calculated and the observed peak values was 10\%. The forecast reliability for $T$, $B$, and $n$ was found to be 20, 30, and 40\%, respectively. It was also determined that the peak in HSSW properties appear in the sequence: $n$, $B$, $T$, and $v$ with time delays of 1, 2, 3, and 4 days, respectively, relative to the CH transit through the central meridian (CM).

Note that in this method no magnetograms were used, which leaves room for error in determining whether all low intensity regions included in the study were in fact CHs. While a CH linked to HSSW can be accepted as a proof of the CH status, it is the western CH boundary position that should be linked to the start of the HSSW period rather than the time of the maximum area attained by the CH in a CM sector. In addition, several CHs being present on the disk could introduce errors in linking CHs to HSSW streams when using only a CM sector based analysis. The arbitrarily chosen CH boundary intensity may also mean that obtained CH properties could be falsely dependent on the solar cycle, if the method was used on long-term analyses.

The geomagnetic index (Dst) forecasting method was based on the empirical correlation between the CH area/location, and the Dst index \citep{Vrsnak07b}. Studying the same period as mentioned above, it was found that the highest geomagnetic activity occurs 3.2$\pm$1.6~days after a low-latitude CH crosses the central meridian, and the amplitude of the Dst dip (i.e. enhanced geomagnetic activity) is correlated with the CH area and depends on the dominant magnetic polarity of the CH. (Note that the area used here is a fractional CH area in the sector located at the CM.) The Dst dip amplitude forecasts were found to have a rather low accuracy of 30\%, and the time of the Dst minimum could only be forecasted with an error of ±2 days.

\subsection{Parker spiral and the radial plasma propagation}
\label{blob_propagation}

In Section \ref{spiral_structure} we discussed the heliospheric structure of solar wind streams. Here, we readdress the heliospheric propagation of the solar wind plasma in order to determine its arrival time at Earth. 

Let us consider a static coordinate system with the Earth at its origin. In this system the Sun rotates around its axis at the {\it synodic} rotation rate ($\Omega$), relative to Earth. The Sun's rotation speed relative to the stars is called the {\it sidereal} angular velocity, from which the Earth's orbital rotation rate (i.e. $\sim$360$^{\circ}$/365.25~days~$\simeq$ ~1$^{\circ}$/day) is subtracted to get the synodic angular velocity. This is essentially the rotation rate at which the Sun appears to rotate from Earth. The synodic angular velocity of the Sun has been determined to be $\sim$13.3$^{\circ}$/day \citep{Allen73, Wang93, Beck00}.

The equation to determine the arrival time of an ejected plasma blob from the Sun can be arrived at in two different ways. The simplest way is to consider an observer on Earth facing the Sun in the stationary coordinate system. In this system all material leaving the Sun travels radially. Thus, if a region on the Sun located at the Sun-Earth line ejects material (i.e. solar wind streaming from a CH), the plasma will travel radially a distance of 1~AU to reach Earth. The velocity of the plasma can be treated as constant since the acceleration phase occurs under 20~R$_{\odot}$ which is relatively small compared to the 214~R$_{\odot}$ distance between the Sun and Earth \citep{Kojima04}. Hence, the time it takes for the plasma blob to arrive at the target (Earth) from the time of the observation at the CM can be determined from the simple equation

\begin{equation}
\label{traveltime}
dt_{1}=\frac{dr}{v_{sw}}
\end{equation}
where $dr$ is the distance the plasma travels (1~AU), and $v_{sw}$ is the solar wind velocity. Note, that if the source region is not at CM, the time it takes for the source to rotate to CM needs to be calculated using
\begin{equation}
\label{traveltime}
dt_{2}=\frac{d\alpha}{\Omega},
\end{equation}
where $d\alpha$ is the angular distance from the source to the CM. Taking this into consideration, the total time from observation to arrival can be written as
\begin{equation}
\label{totaltravel}
dt=dt_{1}+dt_{2}=\frac{dr}{v_{sw}}+\frac{d\alpha}{\Omega}
\end{equation}
This equation can also be derived using Equation \ref{spiral}:
\begin{equation}
r-r_{0}=-\frac{u_{r}}{\Omega}(\alpha - \alpha_{0}),	
\end{equation}
where $u_{r}=v_{sw}$, $r-r_{0}=dr$ and $\alpha - \alpha_{0}=d\alpha=\Omega dt_{2}$. Substituting these notations leads to Equation~\ref{traveltime}. 

\begin{figure}[!t] 
\centerline{
\includegraphics[scale=0.6, trim= 20 70 20 150, clip]{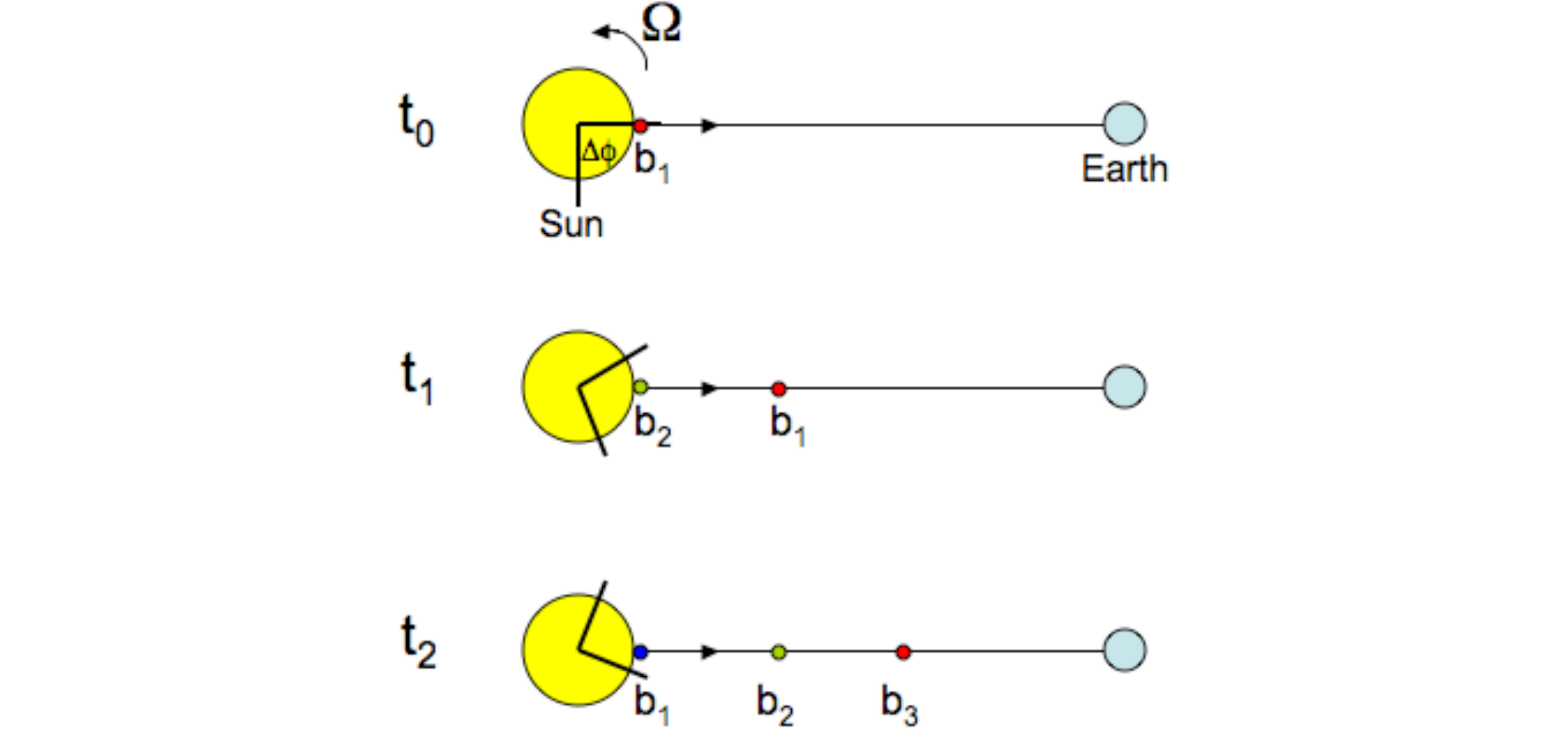} }
\caption{A poleward view of the radial propagation of plasma elements ($b_{1}$, $b_{2}$. $b_{3}$) along the Sun-Earth line. The elements are ejected from different longitudinal positions as the Sun rotates ($t_{0}$, $t_{1}$ and $t_{3}$ represent different times of observation). $\Omega$ is the synodic solar angular velocity.} 
\label{radial} 
\end{figure}

\begin{figure}[!t] 
\centerline{
\includegraphics[scale=0.6, trim= 20 40 20 50, clip]{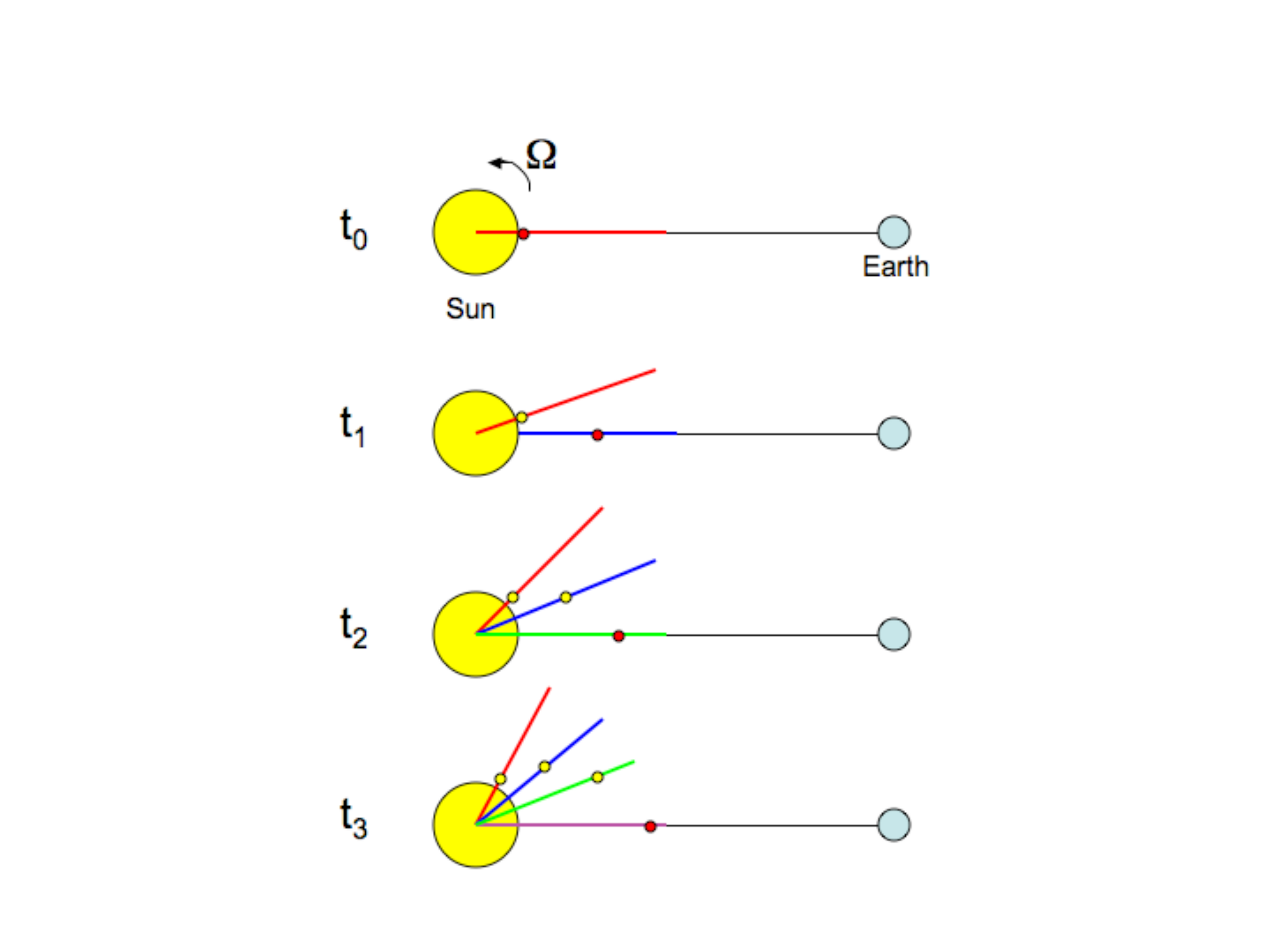} }
\caption{The propagation of a single plasma element in time along the Sun-Earth line. The red blob shows the plasma position at a given observation time. The yellow blobs represent the previous locations of the plasma element relative to the source location on the Sun. The coloured radial lines show the radial lines relative to the source location at given observation times. At $t_{3}$ a spiral structure appears to have formed relative to the source location on the Sun.} 
\label{blobspiral} 
\end{figure}

\begin{figure}[!t] 
\centerline{
\includegraphics[scale=0.6, trim= 20 0 20 15, clip]{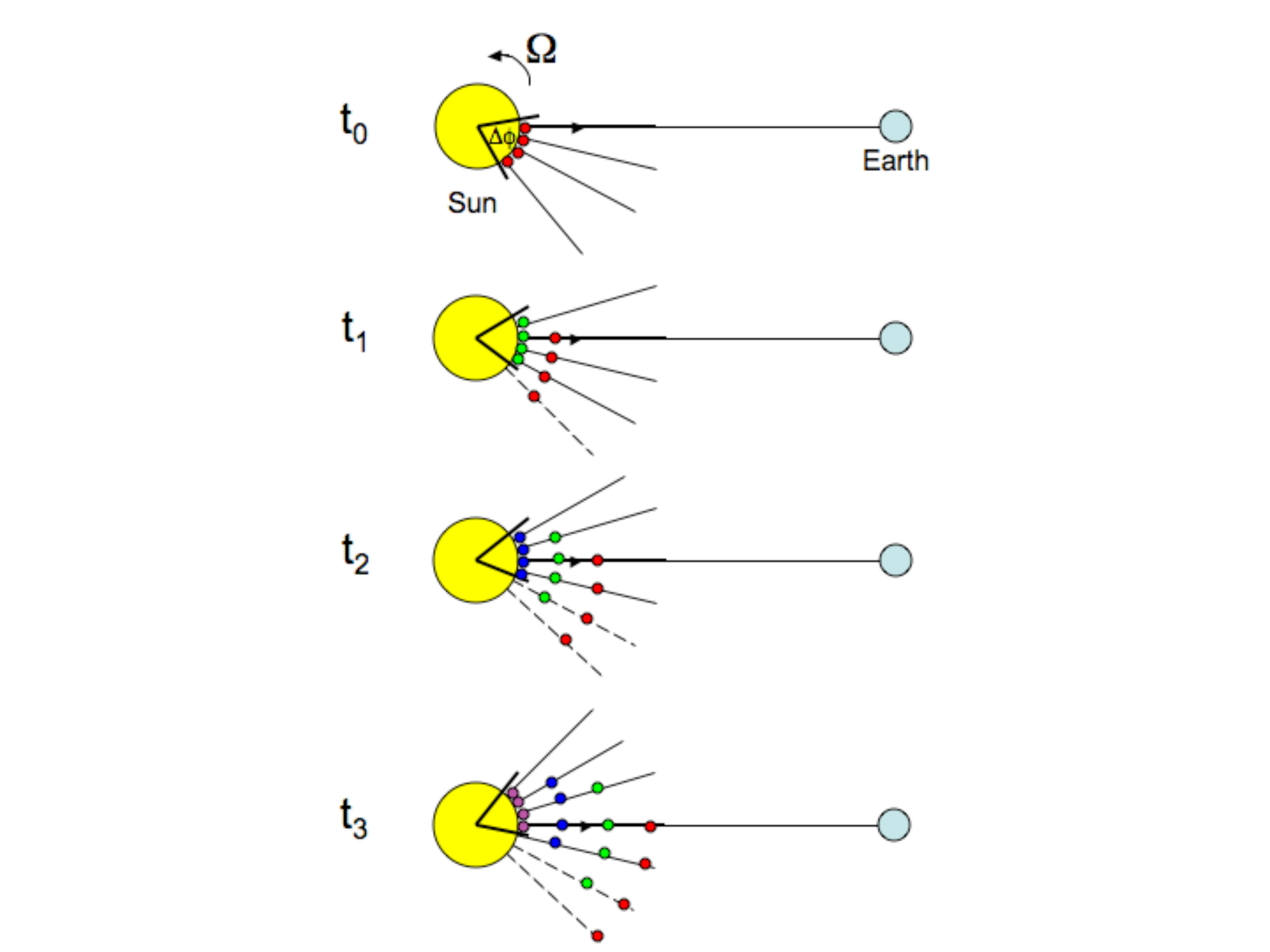} }
\caption{Multiple plasma ejections from non-point source on the Sun. Different colour blobs correspond to different times of ejection. As the source region rotates the relative positions of the ejected plasma elements at consecutive observation times recreate the well-known heliospheric structure of the solar wind.} 
\label{multiblobs} 
\end{figure}

Figures~\ref{radial}, \ref{blobspiral} and \ref{multiblobs} demonstrate the connection between the radially propagating solar wind elements and the spiral structure of the streaming solar wind. In Figure~\ref{radial} plasma elements ($b_{1}$, $b_{2}$ and $b_{3}$) are continuously ejected from a solar point-source along the Sun-Earth line, as seen from a poleward view. The source rotates with the Sun at the synodic rotation rate, hence, the source of the plasma elements propagating towards Earth is different at different observation times ($t_{0}$, $t_{1}$ and $t_{3}$). In Figure~\ref{blobspiral} the location of a single plasma element (red blob) is followed at consecutive observation times relative to the source point. The yellow blobs represent the previous locations of the plasma element relative to the source point. With the rotation of the Sun, the source rotates away from the Sun-Earth line. Hence, if an observer was located in the source point, the location of plasma element at different times would reconstruct a spiral path.

Figure~\ref{multiblobs} shows the plasma elements continuously being ejected from an extended source region on the Sun. The blobs streaming along the Sun-Earth line originate from different longitudinal locations on the Sun, and the colours denote different ejection times. Note the spiral geometry taking shape relative to the source region due to the rotation of the stream source. This figure also demonstrates the relationship between the latitudinal width (i.e. $d\phi$) of the source region and the duration of the plasma stream observed at the target (Earth).

Although understanding the heliospheric propagation of the solar wind plasma is important, we only need to consider the radial propagation of the plasma to provide HSSW forecasts based on CH locations (i.e. Equation~\ref{totaltravel}). 

\subsection{CH locations and the HSSW}
\label{ch_hssw}

As discussed in the previous section, to determine the HSSW arrival time at Earth the location of the source region has to be determined. In this study it is assumed that the magnetic field line of HSSW stream to first connect to Earth is rooted in the west-most point of a CH. Here, we demonstrate the relationship between the CH location on the Sun and the HSSW periods detected at 1~AU.

\begin{figure}[!t] 
\vspace{-0.57\textwidth}   
\centerline{\hspace*{0.1\textwidth} 
\includegraphics[angle=90,clip=, scale=0.6, trim= 0 0 0 20]{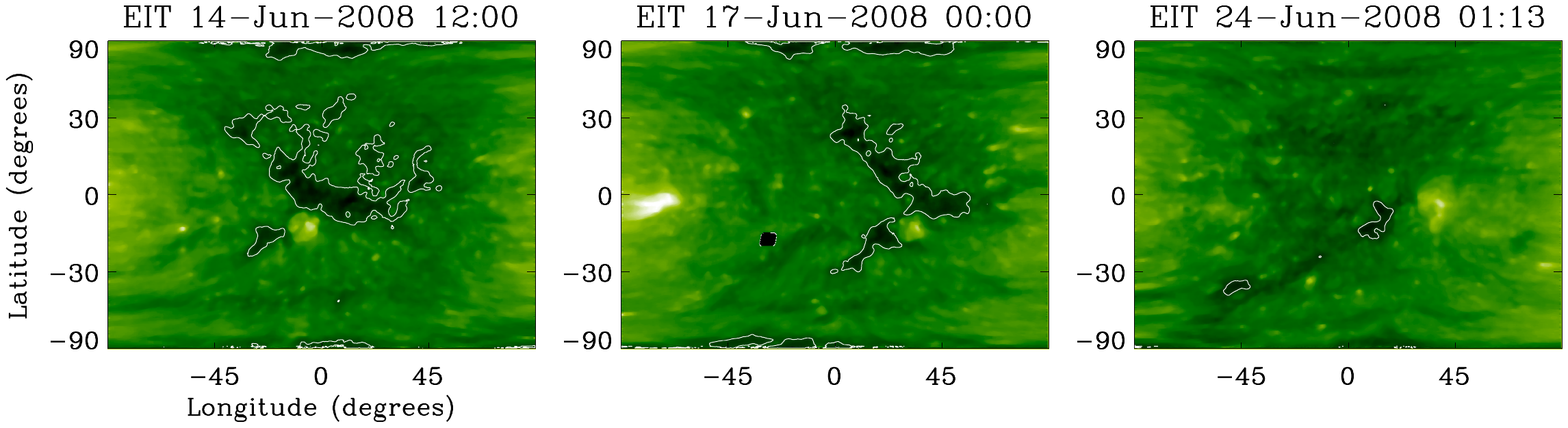} }
\vspace{-0.5\textwidth}   
\centerline{\hspace*{-0.1\textwidth} 
\includegraphics[angle=90, clip=, scale=0.6]{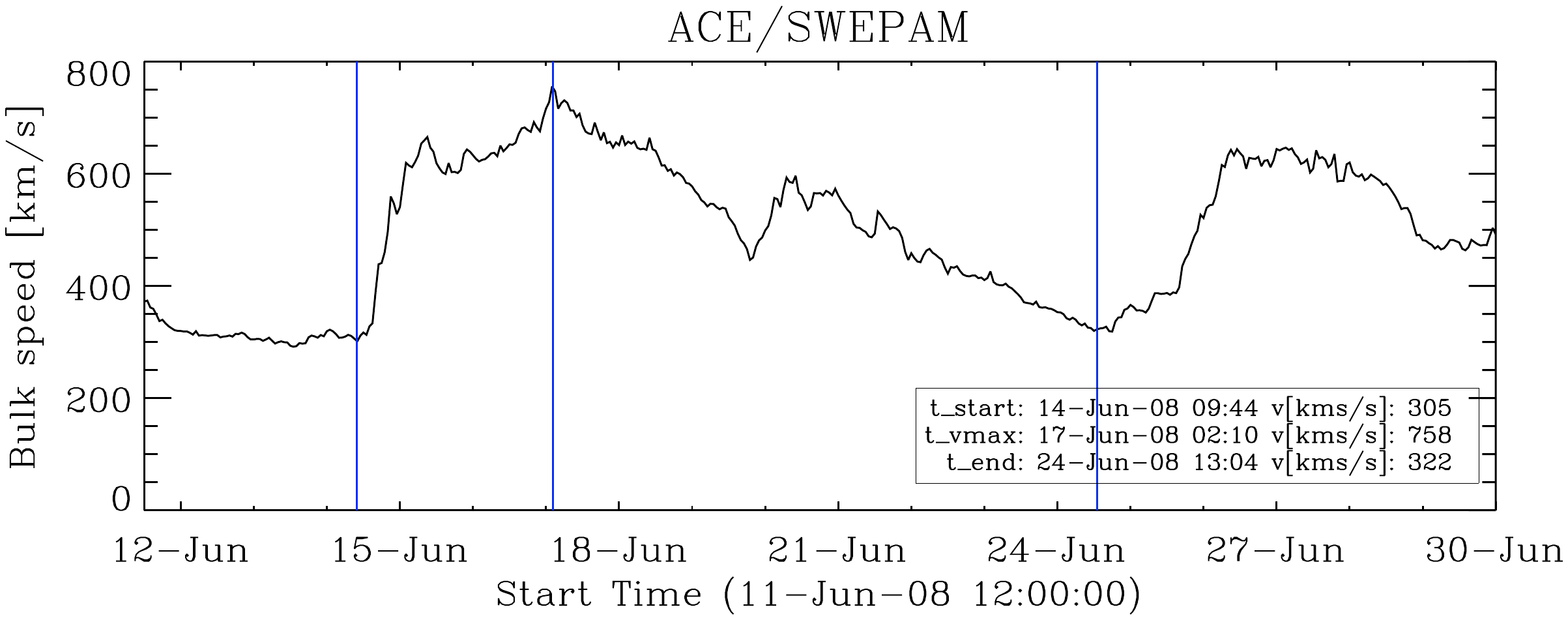} }
\caption{{\it Top}: EIT Lambert maps showing the CH position on the day of the start, the maximum and the end of the HSSW period. {\it Bottom}: The corresponding in-situ ACE solar wind speed measurements for the studied period.} 
\label{ace} 
\end{figure}

Figure~\ref{ace} shows the CH positions at the start, maximum and end of the high-speed solar wind period on 14--24 June 2008. The images show that the west-most CH boundary was at approximately W50$^{\circ}$ when the HSSW stream arrived at the ACE spacecraft. The CH has passed over the west limb by the time the solar wind speed dropped back to normal levels (300--400~km~s$^{-1}$). 

\begin{figure}[!t]
\vspace{-0.02\textwidth}   
\includegraphics[angle=90,scale=0.5, trim= 20 50 400 0, clip]{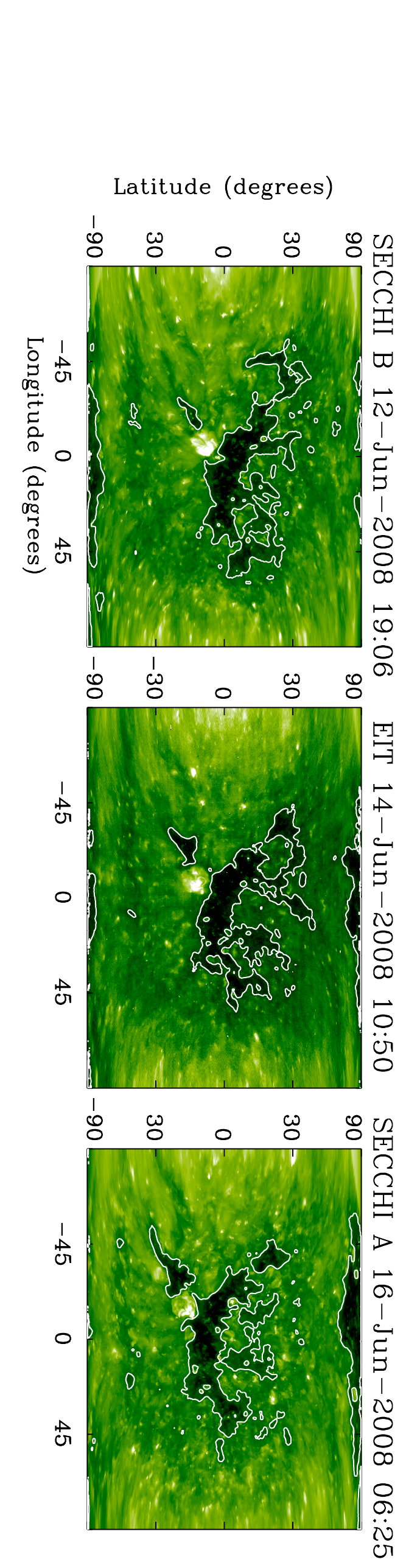}
\includegraphics[angle=90, scale=0.8, trim= 20 200 450 20, clip]{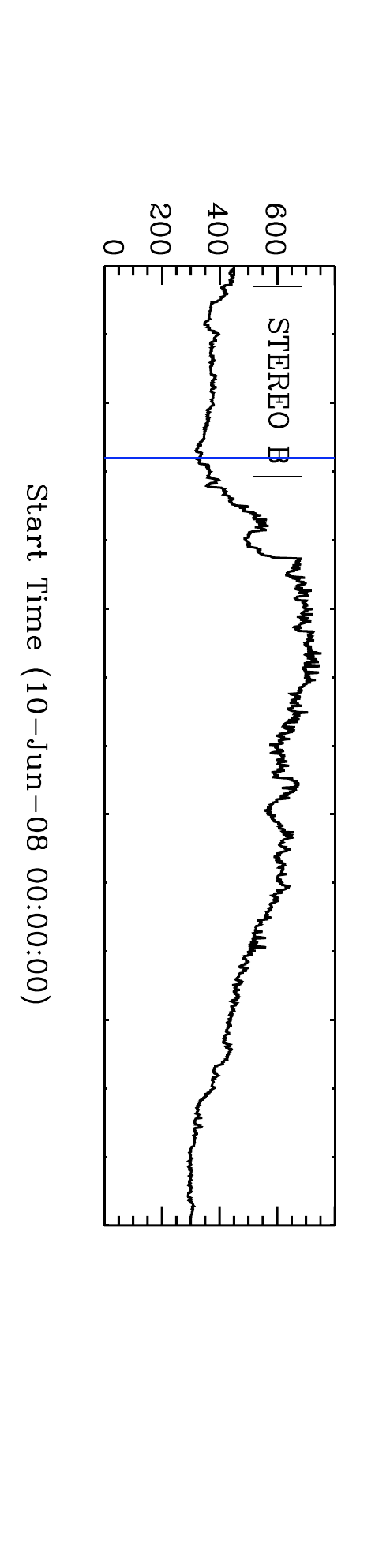} 
\includegraphics[angle=90, scale=0.8 ,trim= 20 100 450 20, clip]{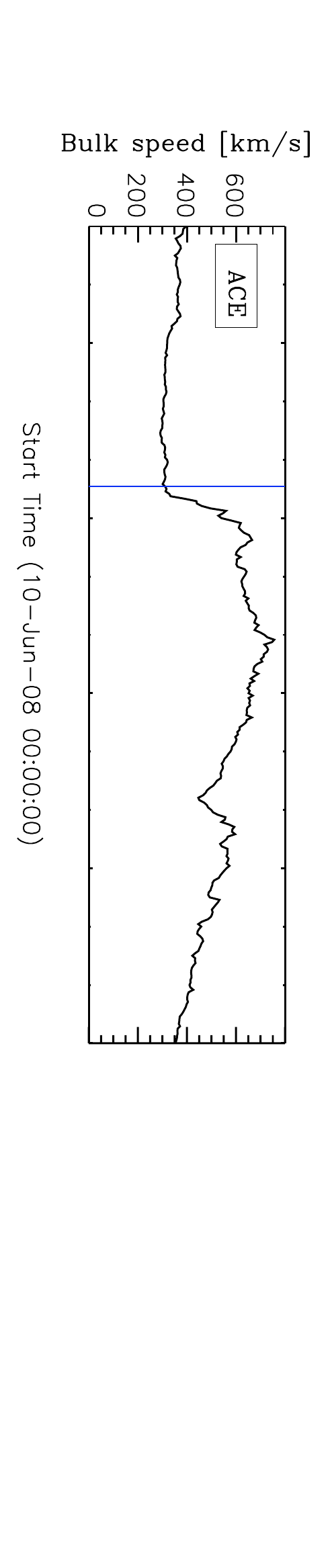}
\includegraphics[angle=90, scale=0.8 ,trim= 20 100 450 20, clip]{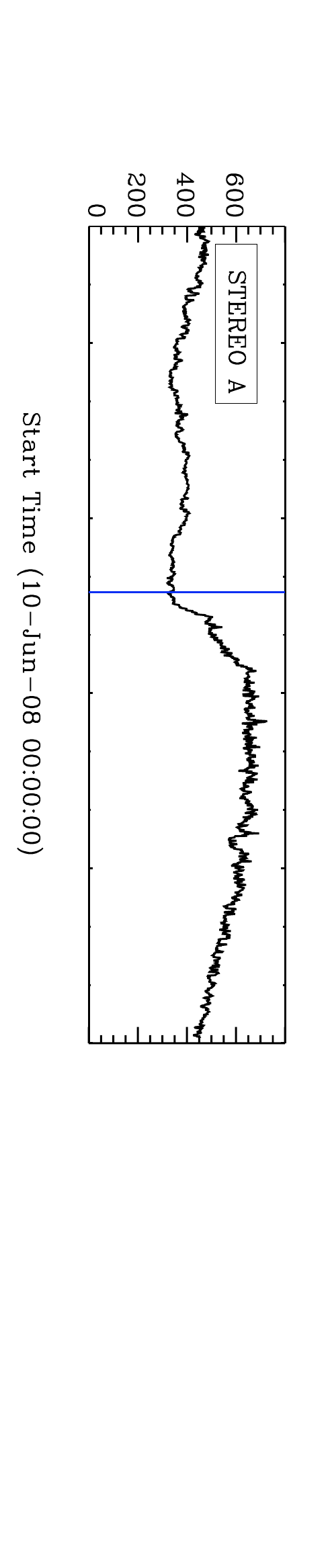} 
\caption{ {\it Top}: EIT Lambert maps showing the CH position on three different days when the high-speed solar wind reached STEREO B, ACE and STEREO A. {\it Bottom}: The solar wind profiles obtained from the three satellites, with the line marking the start of the high-speed solar wind.} 
\label{sw_profiles} 
\end{figure}

The relationship between the CH and its associated solar wind stream measured at three distinct points in the Heliosphere was further explored using imaging and in-situ measurements from STEREO B, ACE and STEREO A. The top row of Figure~\ref{sw_profiles} shows the CH complex on three days: 12, 14 and 16 June 2008. 

The corresponding solar wind velocity measurements, given in the bottom three panels of Figure~\ref{sw_profiles}, show a high-speed solar wind stream sweeping around the inner Heliosphere, from STEREO B to ACE to STEREO A. The HSSW stream was initially detected by STEREO B at 12 Jun 2008 19:10~UT, when the west-most boundary of the CH was at $\sim$W53 from the perspective of STEREO B. Subsequently, the solar wind speed began to rise in ACE in-situ measurements two days later on 14 Jun 2008 11:00~UT. At this point in time, the CH's west-most boundary was at $\sim$W50 as seen by SOHO. The high-speed stream finally traversed the point in the Heliosphere monitored by STEREO A on 16 Jun 2008 6:30~UT, at which time the western boundary of the CH was at $\sim$W43. The complimentary imaging and in-situ capabilities of SOHO, ACE and the two STEREO satellites therefore offer a unique opportunity to study the relationship between CHs rotating across the solar disk and high-speed solar wind streams sweeping through the Heliosphere.

\subsection{Flux tube expansion }
\label{expansion}

It is known that magnetic field lines originating from CHs reach interplanetary distances. The field lines extend from within the confined boundaries of CHs and form a flux tube (or solar wind stream). It has been shown that these flux tubes expand super-radially and the expansion factor can be derived from the mass conservation equation:
\begin{equation}
\frac{dM}{dt}=A\rho v,
\label{m_cons}
\end{equation}
where  $dM/dt$ is the mass loss rate, $A$ is the cross-sectional area of the flux tube, $\rho$ is the density and $v$ is the solar wind velocity. Given that the plasma flow is constant, the mass-loss rate will be constant, and hence
\begin{equation}
A\rho v= const.
\label{masscons}
\end{equation}
This is the one-fluid description of a proton-electron solar wind. To give the cross-sectional area of the flux tube at any point along the tube ($s$), the equation
\begin{equation}
A=A_{0}f(s) \left(\frac{s}{s_{0}}\right)^{2}
\label{expfac}
\end{equation}
is used, where $s_{0}$ and $s$  are the radial distances where the area is measured, and $f$ is the non-radial flux tube expansion factor \citep{Guhathakurta99b}. 

\begin{figure}[!t]
\centerline{
\includegraphics[scale=0.5, trim=40 40 10 30, clip ]{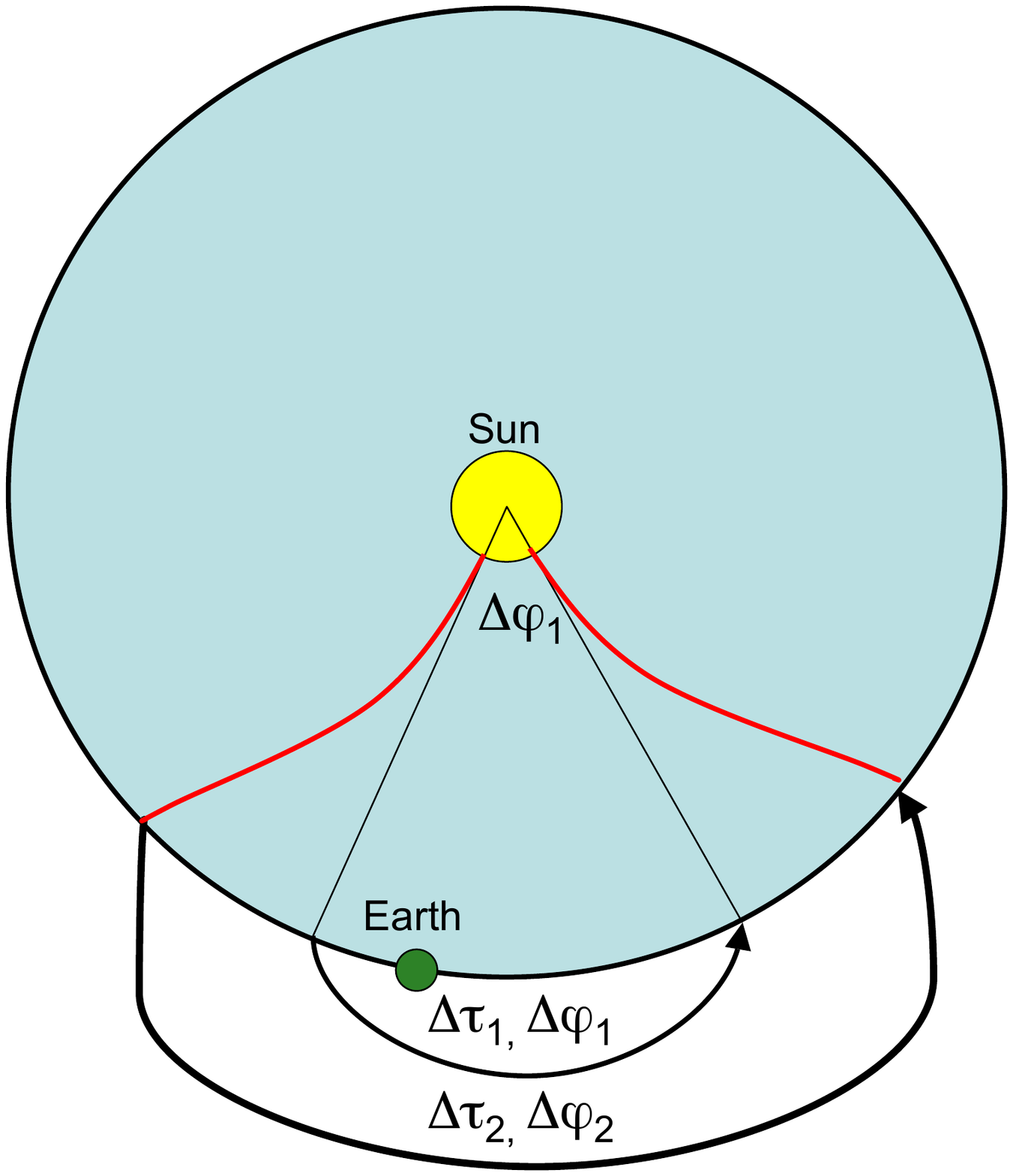}}
\caption{Schematic showing the super-radial expansion of a CH flux tube. The CH has a latitudinal width $\Delta \phi_{1}$, the radial expansion is shown with the inner black lines extending from the CH, the the outer red lines indicate the super-radial expansion.} 
\label{suprad} 
\end{figure}

Equation~\ref{expfac} is modified to contain the observed parameters, namely the latitudinal width of the solar wind source and the HSSW duration measured at Earth. This is done by assuming the flux tube to have a circular cross-section and denoting the area as 
\begin{equation}
A_{1}=\pi r_{1}^{2}
\label{area1}
\end{equation}
for the radially expanded flux tube, and 
\begin{equation}
A_{2}=\pi r_{2}^{2}
\label{area2}
\end{equation}
for the superradially expanded flux tube. Here, $r_{1}$ and $r_{2}$ are the flux tube radii for the radially and superradially expanded flux tubes, while $A_{1}$ and $A_{2}$ are the flux tube cross-sectional areas for the radially and superradially expanded flux tubes. If we assume a radial expansion, with the flux tube having a $\Delta \phi_{1}$ latitudinal width at Earth, the width can be approximated as $s\Delta \phi_{1}=2r_{1}$. Substituting $r_{1}$ into Equation~\ref{area1} hence yields
\begin{equation}
A_{1}=\frac{\pi s^{2} (\Delta\phi_{1})^{2}}{4}. 
\end{equation}
This is substituted into Equation~\ref{expfac}, for the radial expansion ($f=1$) to get
\begin{equation}
\frac{\pi s^{2} (\Delta\phi_{1})^{2}}{4}=\frac{A_{0}{s}^{2}}{{s_{0}}^2}.
\label{exprad}
\end{equation}
Applying the same steps, by substituting Equation~\ref{area2} and $s\Delta \phi_{2}=2r_{2}$ into Equation~\ref{expfac}, for superradial expansion ($f\neq1$) we get
\begin{equation}
\frac{\pi s^{2} (\Delta\phi_{1})^{2}}{4}=f(s)\frac{A_{0} s^{2}}{s_{0}^2}.
\label{expsuprad}
\end{equation}
These latter two equations can be rearranged to yield
\begin{equation}
\Delta\phi_{1}=\sqrt{\frac{4A_{0}}{\pi {s_{0}}^2}},\;\;
\Delta\phi_{2}=\sqrt{f(s)\frac{4A_{0}}{\pi {s_{0}}^2}}
\end{equation}
which can then be related as
\begin{equation}
\Delta\phi_{2}=\sqrt{f(s)}\Delta\phi_{1}.
\label{delphi}
\end{equation}
If $\Delta\tau_{2}$ is the HSSW duration measured at Earth, we can assume $\Delta\phi_{2}=\omega\Delta\tau_{2}$, which we substitute it into Equation~\ref{delphi} to express the expansion factor
\begin{equation}
f=\left(\frac{\omega\Delta\tau_{2}}{\Delta\phi_{1}}\right)^{2}.
\label{expansionfactor}
\end{equation}
As a demonstration, Figure~\ref{suprad} shows the Sun (yellow disk) with a solar wind source of $\Delta \phi_{1}$ latitudinal width, and the expanded flux tube with a $\Delta \phi_{2}$ width at Earth. The solar wind is shown to expand radially and is measured to last for $\Delta \tau_{1}$ at Earth (green disk), while the super radial expansion is shown to give rise to a longer solar wind period of $\Delta \tau_{2}$ at Earth.

\subsection{Expansion factor: semi-automated approach}
\label{manual}

In order to determine the HSSW stream expansion factor, the properties of 47 CHs were studied in the period 2005-2008. The CHs were identified and grouped using the {\it CHARM} algorithm, and the west-most boundary locations were used to determine estimated HSSW arrival times at Earth assuming an average solar wind velocity of 600 km~s$^{-1}$. 

\begin{figure}[!t]
\vspace{-5mm}
\centerline{ 
\includegraphics[scale=0.6, trim=0 30 70 180, clip ]{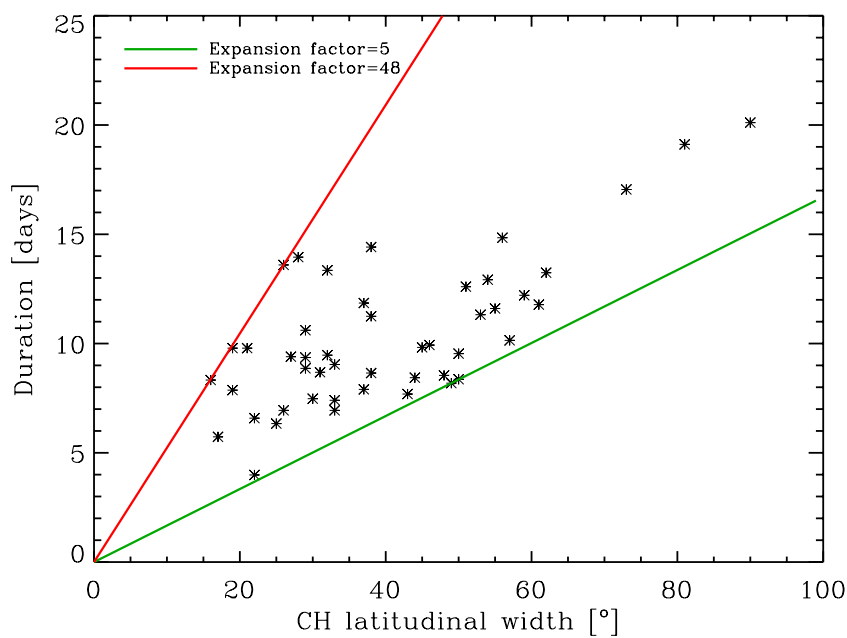} } 
\caption{Scatterplot showing the relationship between latitudinal width of CHs and the HSSW duration.}
\label{expfacman} 
\end{figure}

Figure~\ref{expfacman} shows the correlation between the CH latitudinal widths and the corresponding HSSW durations of the 47 CHs analysed in this study. The green line was fitted to the data point with the largest gradient, while the red line was fitted to data point with the smallest gradient.

An estimated end-time was also determined using the east-most CH boundary location, assuming a simple radial expansion. The estimated arrival and end-time was used to avoid confusion between multiple HSSW periods in the time range. With the help of the estimated arrival and end-times, the HSSW period start and end was manually determined using solar wind velocity data obtained from the ACE/SWEPAM instrument. 
\begin{figure}[!t]
\includegraphics[scale=0.35, trim=-70 0 20 50]{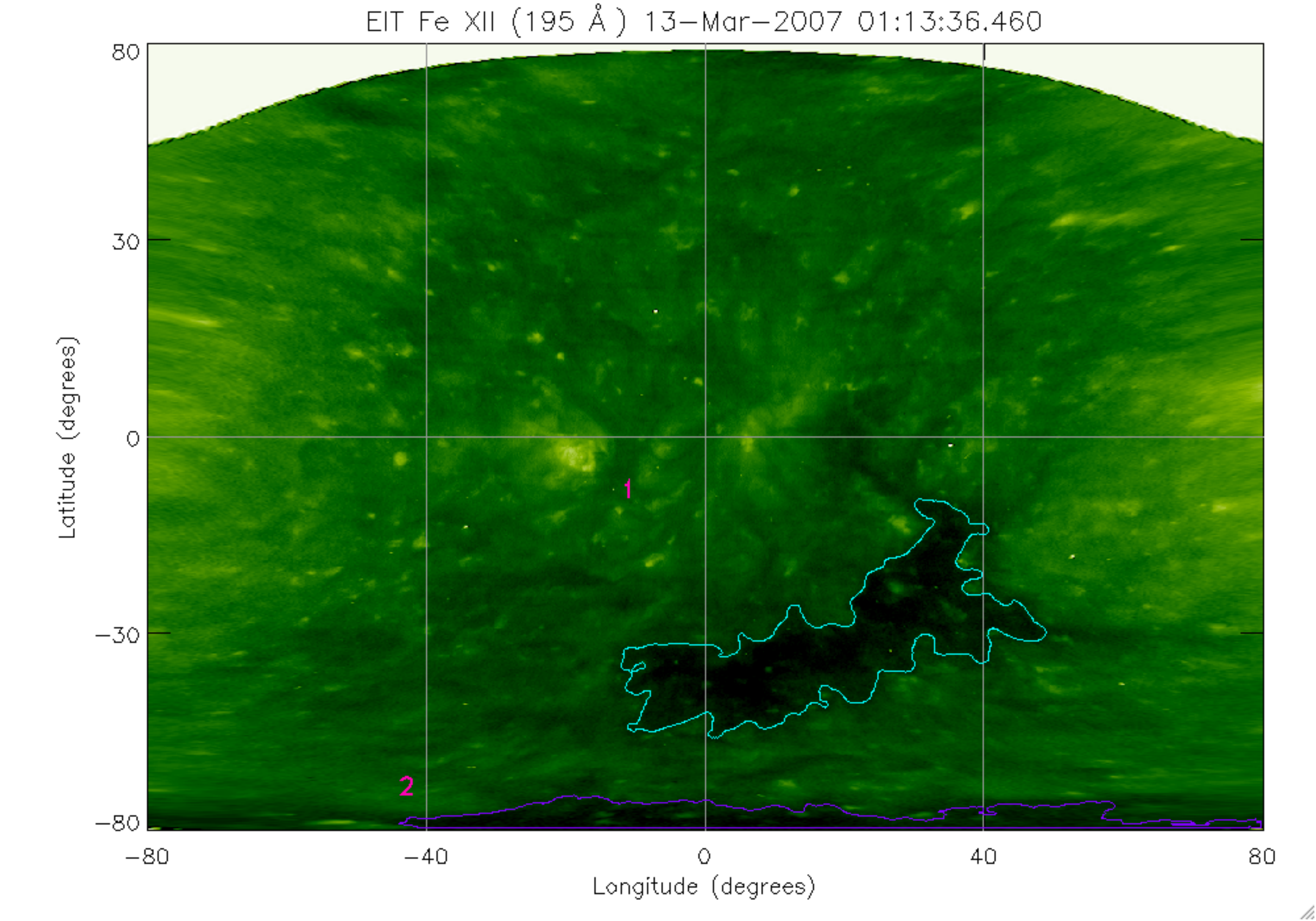}
\includegraphics[scale=0.65, trim=30 20 0 320, clip]{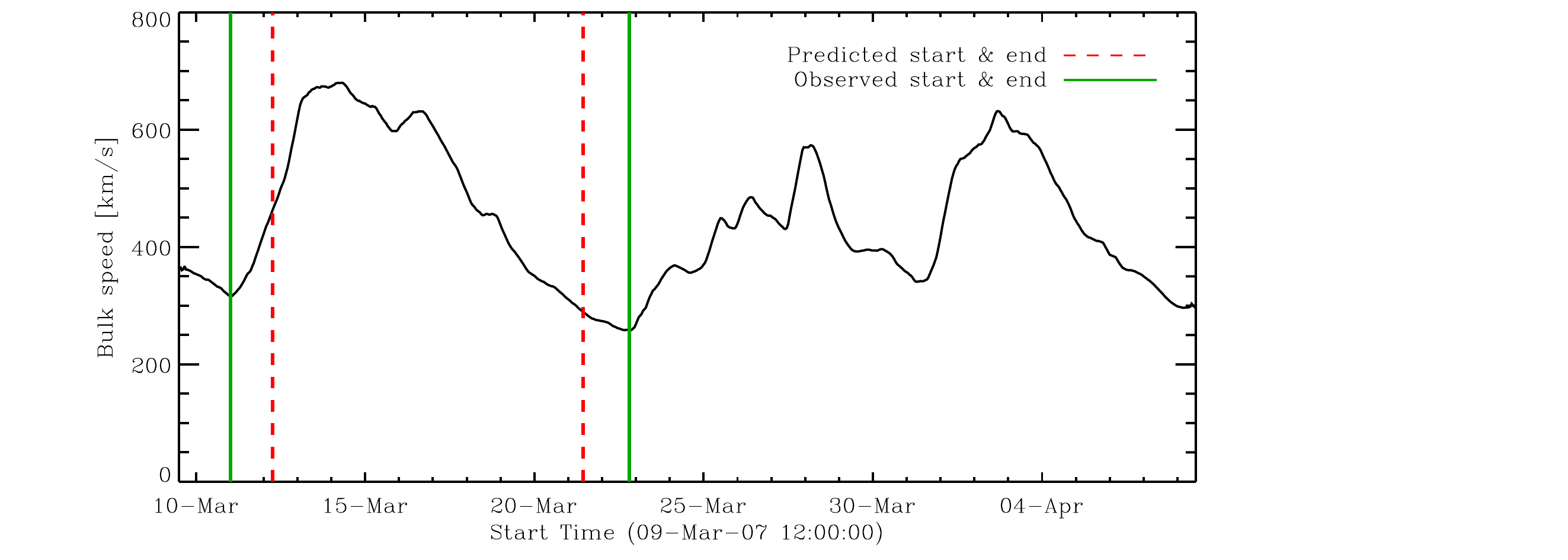}
\caption{The automatically detected CH group (SOHO/EIT 195~\AA\ Lambert map) and the semi-automatically identified corresponding HSSW stream (ACE/SWEPAM).}
\label{hssw_man} 
\end{figure} 
Figure~\ref{hssw_man} shows an example of a CH automatically detected on 13 March 2007, and the corresponding HSSW period identified in the ACE/SWEPAM solar wind data. The dashed red lines show the estimated start and end of the HSSW period (left and right respectively), based on a 600~km~s$^{-1}$ solar wind velocity and a radial expansion. The green lines show the detected HSSW start and end time.

\begin{figure}[!t]
\vspace{-0.03\textwidth} 
\centerline{ 
\includegraphics[scale=0.65, trim=0 10 50 180, clip]{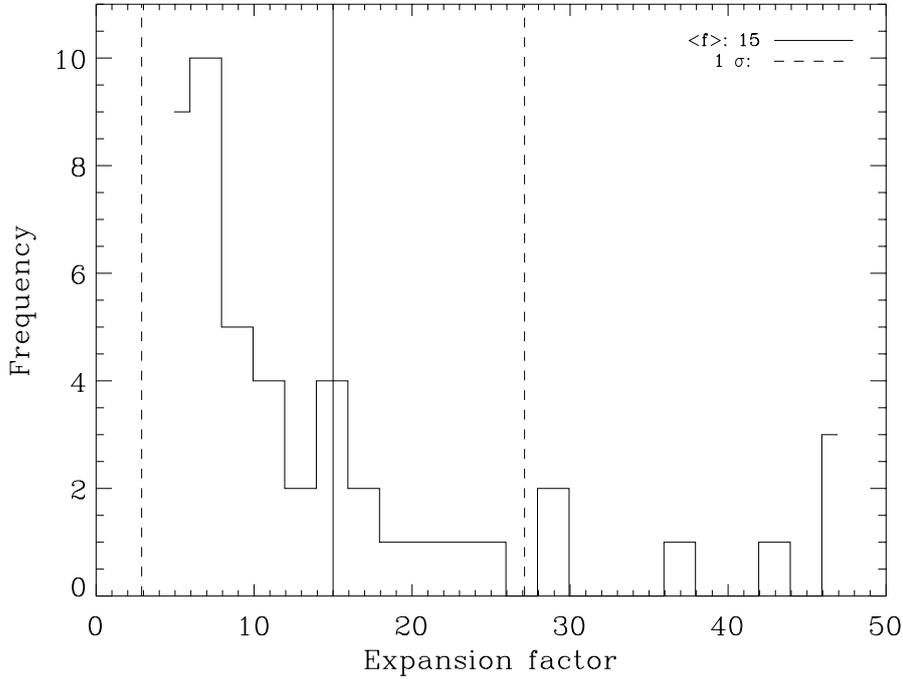} } 
\caption{Histogram showing the frequency distribution of the expansion factors obtained from 47 CHs.}
\label{expfacmanhist} 
\end{figure} 

\begin{figure}[!t]
\centerline{ 
\includegraphics[scale=0.4, trim=50 200 30 200, clip]{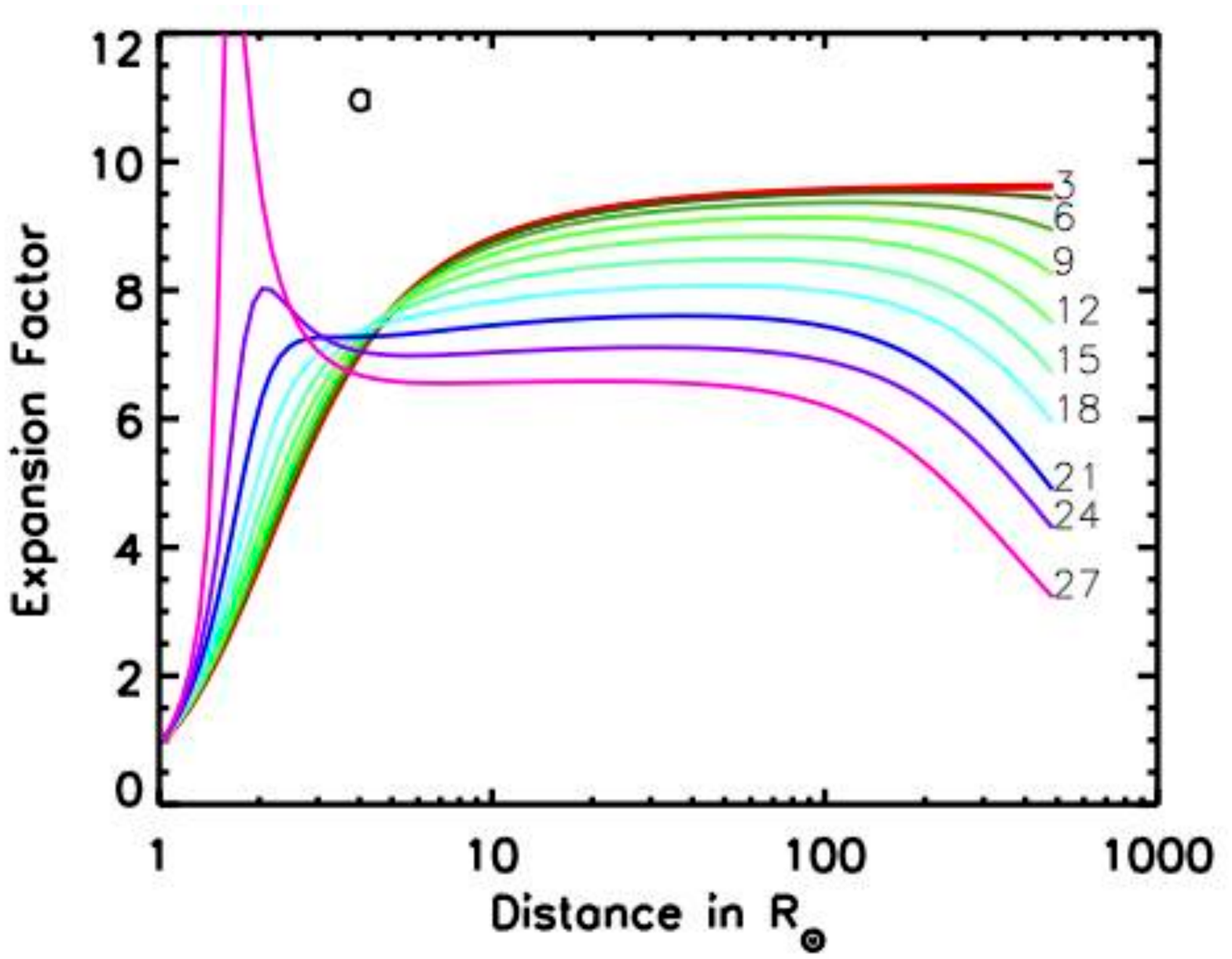} } 
\caption{CH magnetic flux tube expansion factors as a function of distance for ten co-latitudes from 0$^{\circ}$ to 30$^{\circ}$ \citep{Guhathakurta99b}.}
\label{lika_f} 
\end{figure}

Using Equation~\ref{expfac} and the line fit gradients ($\Delta\tau/\Delta\phi$) shown in Figure~\ref{expfacman}, the expansion factor was determined to be in the range 5--48. Figure~\ref{expfacmanhist} shows the frequency distribution of the expansion factors, where the mean expansion factor was found to be 15. However, the most frequent expansion factors were found to be $<$10, while large expansion factors ($>$10) were found to be rare in this study. The expansion factor range determined in our study was found to be similar to previous theoretical and observational results that find the expansion factor to be up to 20 at 1AU \citet{Wang94, Guhathakurta98, Guhathakurta99b, Deforest01}. Figure~\ref{lika_f} shows the expansion factors inferred by \cite{Guhathakurta99b} from Ulysses observations at 1 AU.

\subsection{Expansion factor: automated approach}
\label{auto}

\begin{figure}[!t]
\includegraphics[scale=0.32, trim=-100 0 0 0, clip]{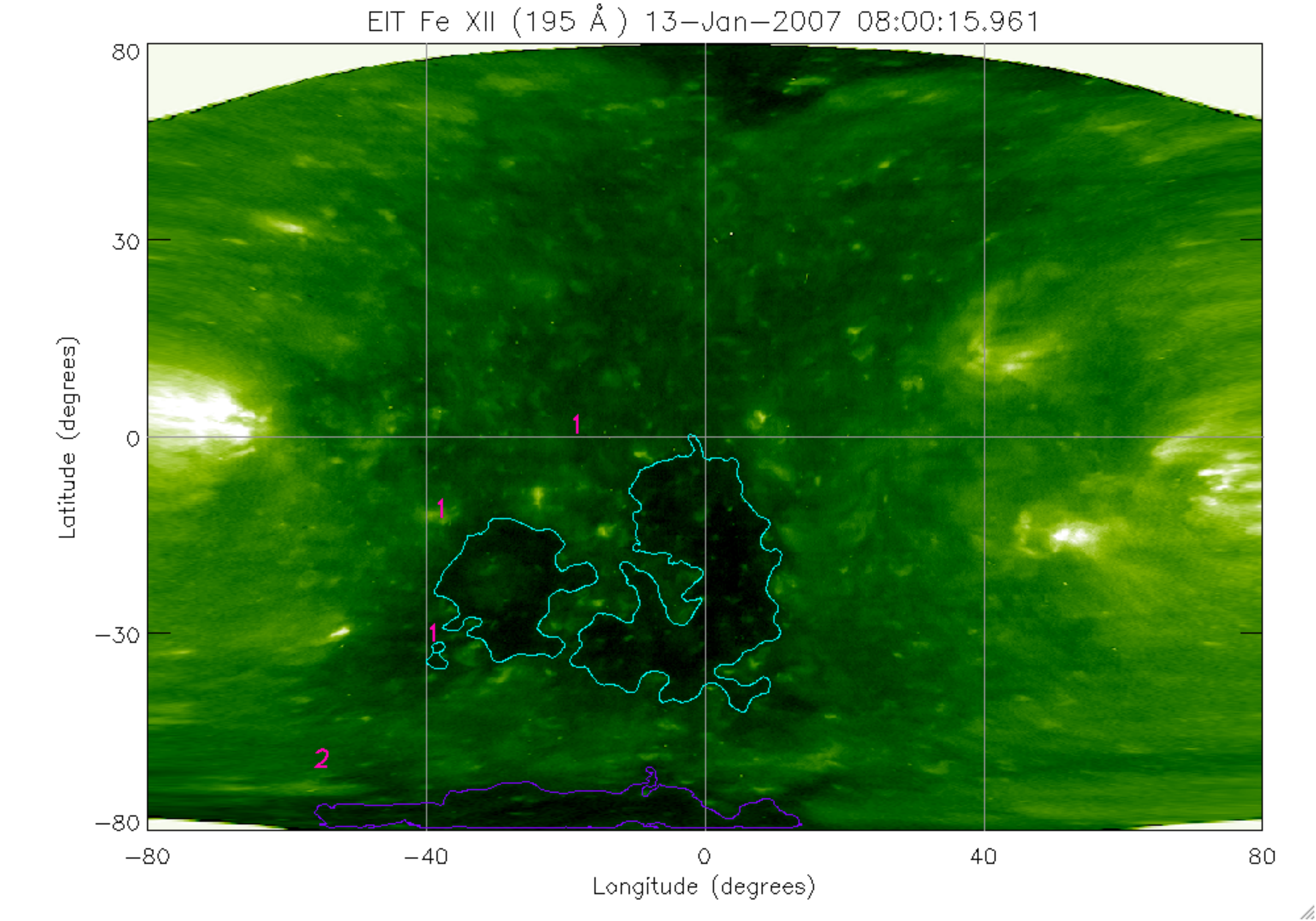}
\includegraphics[scale=0.55, trim=0 35 0 27, clip]{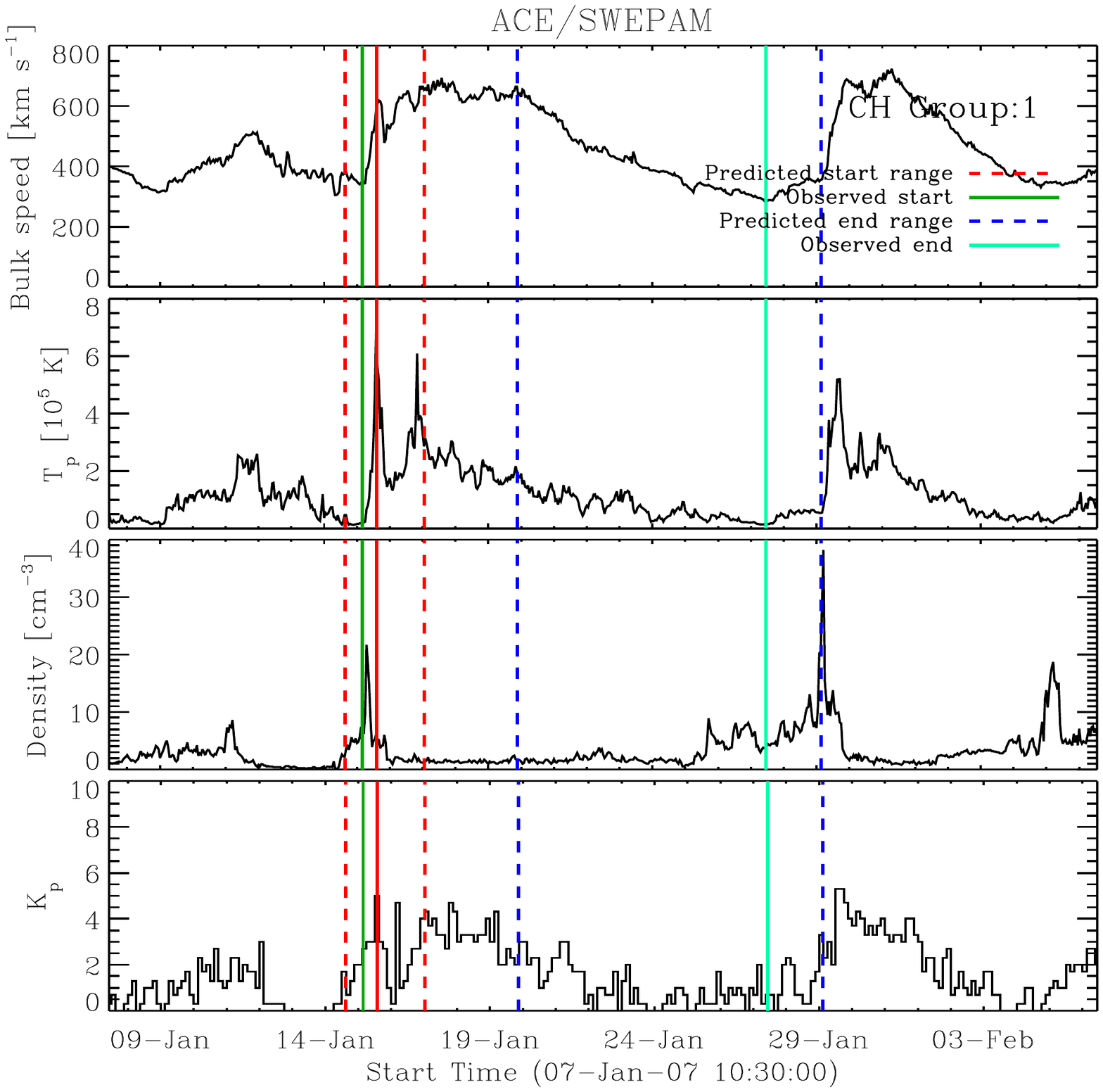}
\caption{The automatically detected CH group (SOHO/EIT 195~\AA\ Lambert map) and the corresponding HSSW stream (ACE/SWEPAM bulk velocity).}
\label{hssw_auto} 
\end{figure}

Using the expansion factor determined from the semi-automated approach, a similar analysis was carried out in an automated manner to obtain the expansion factors using a larger number of CHs and HSSW observations. Using the west-most boundary position of a CH, and a velocity range 400--900~km~s$^{-1}$, an approximate arrival time range was determined, where the rising phase in the HSSW velocity was automatically identified. Similarly, the estimated end-time range was determined from the east-most CH boundary location and previously obtained expansion factors. In this end-time range the HSSW end is determined where the velocity reaches minimum. Figure~\ref{hssw_auto} shows an example CH group which was identified on the 13 January 2007 in an SOHO/EIT 195~\AA\ image, with the corresponding HSSW stream identified in ACE solar wind velocity data. Note the very similar onset times of the $v$, $T$, $\rho$ and $Kp$ enhancement.

In the present analysis the following criteria were used: CH groups have to be of substantial size ($>$1.5$\times$10$^{4}$~Mm), the HSSW rise phase has to occur within $\sim$2-4~days of the CH boundary appearing at the CM (calculated from suggested SW velocities of 400--900~km~s$^{-1}$), the HSSW velocity enhancement has to be at least 150~km~s$^{-1}$, and the maximum velocity has to be above 400~km~s$^{-1}$. An example is shown in Figure~\ref{hssw_auto}, where a large CH group is shown with the corresponding HSSW stream detected in ACE solar wind data. In the solar wind velocity plots the solid red line shows the approximate HSSW arrival time if the solar wind speed is of 600~km~s$^{-1}$. The dashed lines correspond to HSSW arrival times for velocities of 900~km~s$^{-1}$ and 400~km~s$^{-1}$ (left and right, respectively). The green solid line represents the observed (``true") start of the HSSW period. This time was automatically detected based on the previously mentioned criteria. The estimated end-time range are indicated with dashed blue lines and the cyan solid line shows the automatically detected end-time. The label ``CH Group: 1" indicates that the detected HSSW period is linked to CH group 1, shown in the EIT Lambert map above.

\begin{figure}[!t]
\centerline{ 
\includegraphics[scale=0.65, trim=0 30 70 180, clip ]{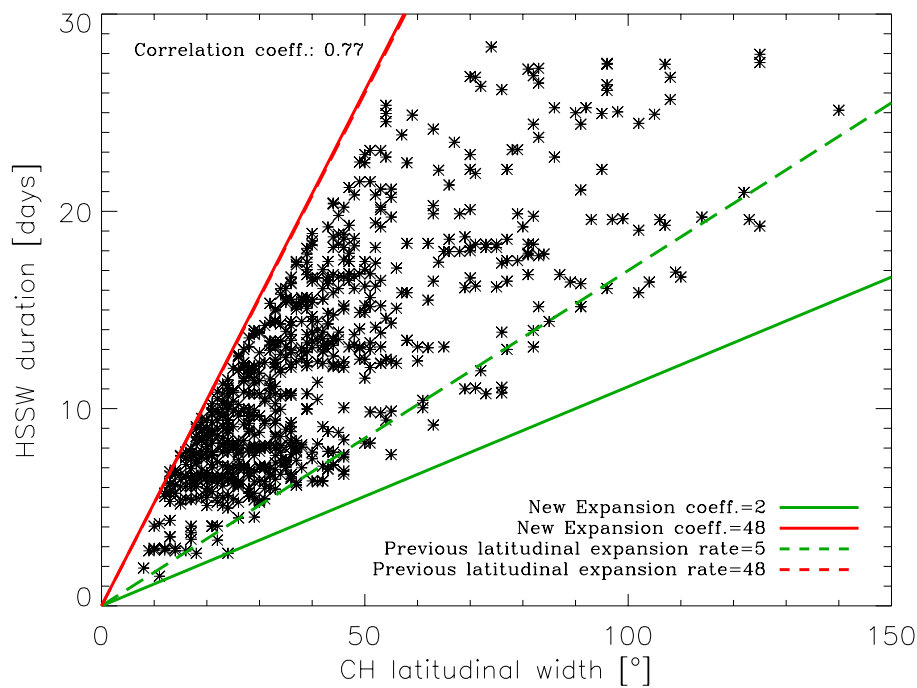} } 
\caption{The automatically obtained CH latitudinal widths as a function of the corresponding HSSW durations. The green and red line fits are obtained from the automated data, the dashed lines corresponding the semi-automated method are shown for comparison.}
\label{expfactor_new} 
\end{figure}

\begin{figure}[!t]
\centerline{ 
\includegraphics[scale=0.5, trim=0 10 50 180, clip]{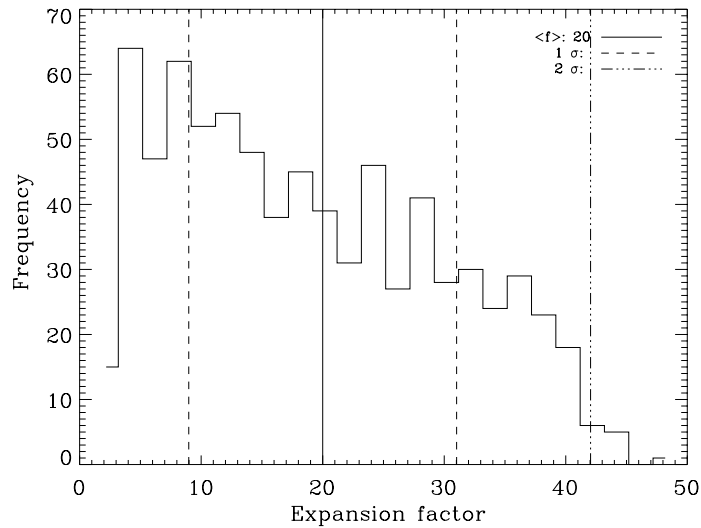} } 
\caption{Histogram showing the frequency distribution of the expansion factors automatically obtained from 616 CHs.}
\label{expfacnewhist} 
\end{figure}

In the period 2000-2009, 616 CHs were automatically linked to HSSW streams. From the automatically obtained CH and HSSW properties, the new expansion factor range was determined to be 2--48. Figure~\ref{expfactor_new} shows the HSSW durations as a function of the CH latitudinal widths of the 616 CHs studied. The line fits from the semi-automated method are shown for comparison in dashed red and green lines, while the line fits for the automated method are shown in solid lines. Figure~\ref{expfacnewhist} shows the frequency distribution of the expansion factors, where the mean expansion factor was found to be 20. Similar to the semi-automated approach, the most frequent expansion factors were found to be $<$10. It is also clear from the histogram that small expansion factors are much more frequent than large expansion factors.

\section{Results}

\subsection{CH area and HSSW properties}
\label{hssw_results}

\begin{landscape}
\begin{figure}[!t]
\centerline{
\includegraphics[scale=0.6, trim=50 20 300 30, clip ]{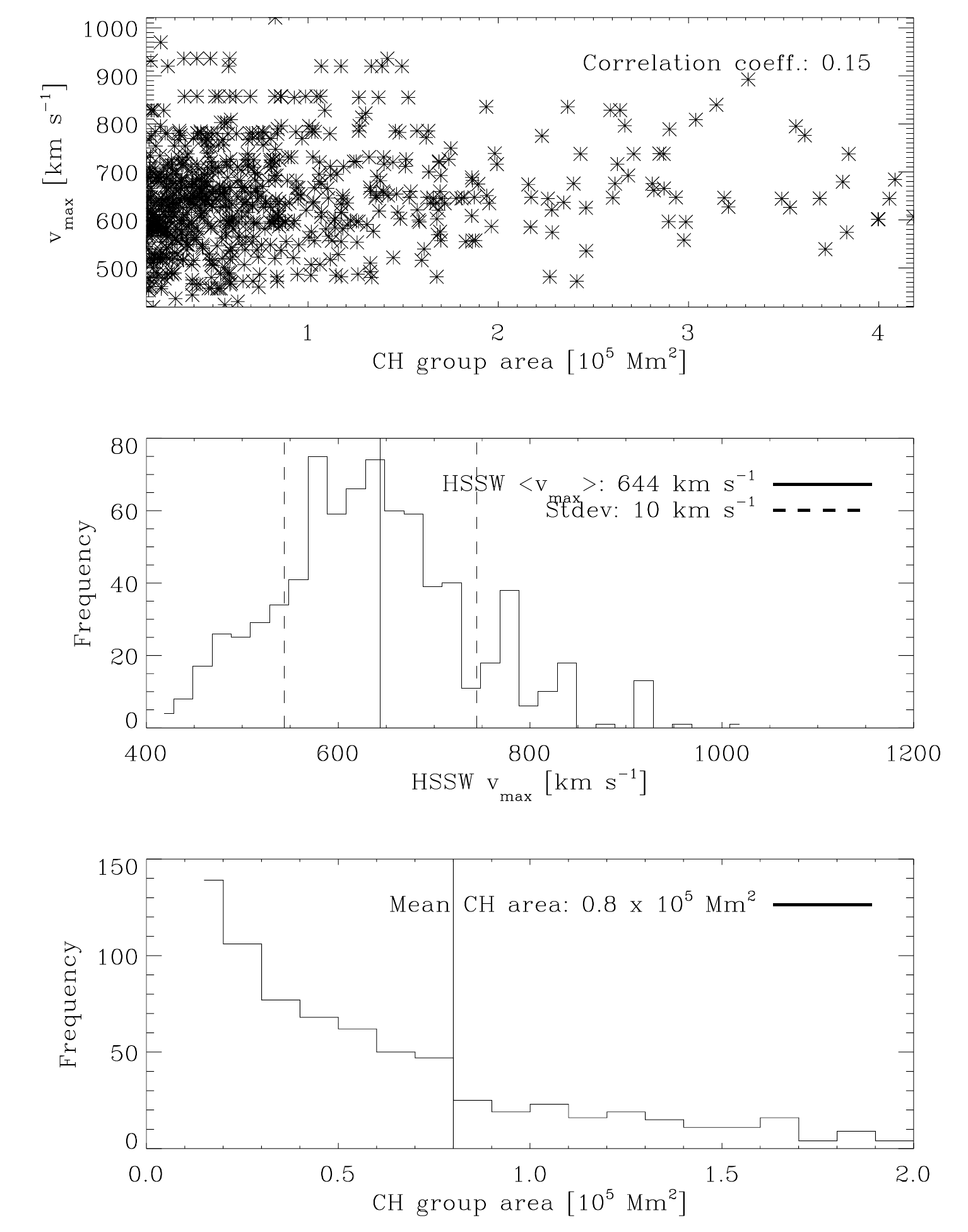}
\includegraphics[scale=0.6, trim=50 20 300 30, clip ]{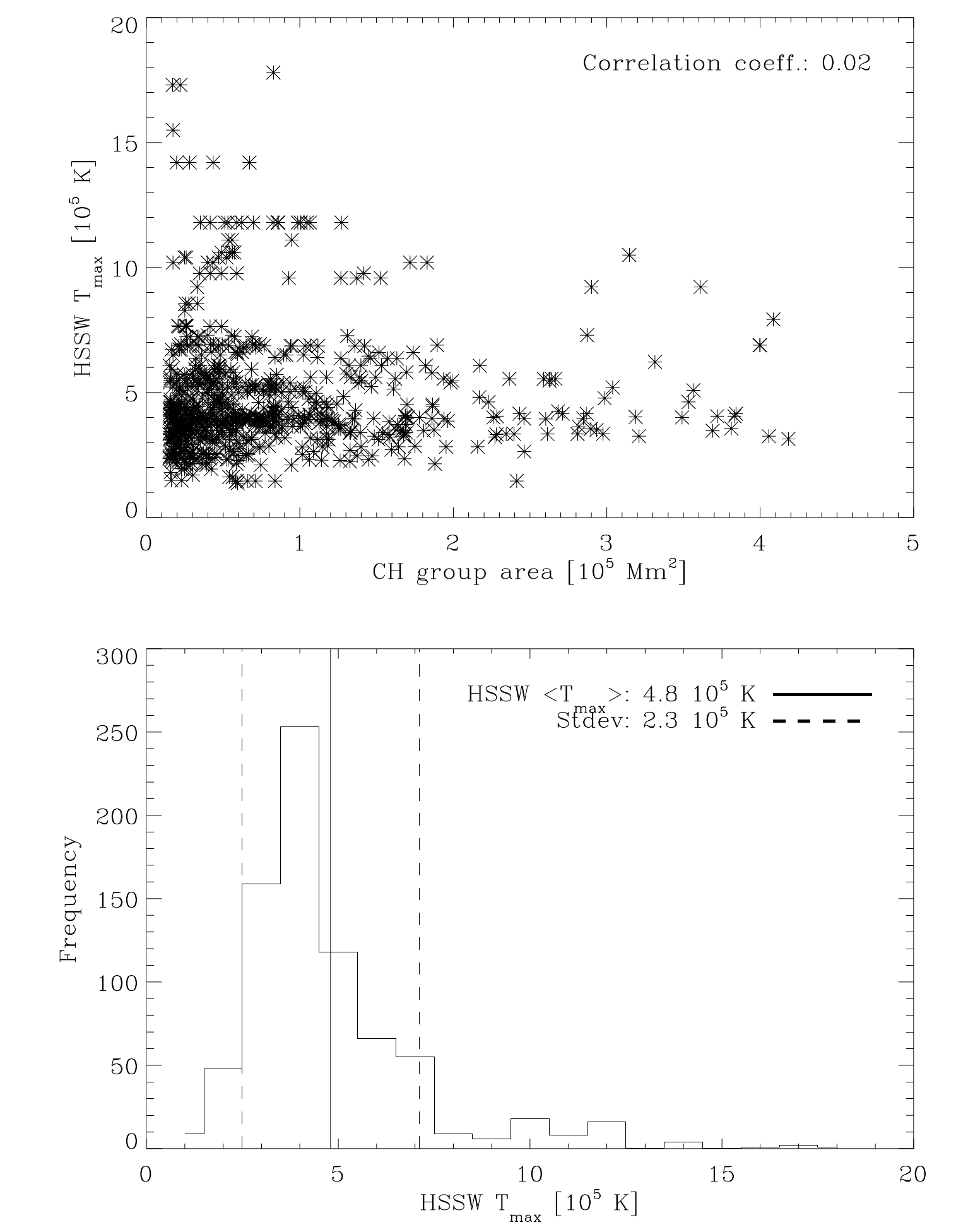} } 
\caption{Top left and right: scatterplot showing the relationship between the CH group area and the maximum velocity, and maximum temperature of the corresponding HSSW stream, respectively. Bottom left: maximum HSSW temperature frequency histogram. Right middle and bottom: CH group area and maximum HSSW velocity frequency histogram, respectively.}
\label{areamax1} 
\end{figure}
\end{landscape}

\begin{landscape}
\begin{figure}[!t]
\centerline{
\includegraphics[scale=0.62, trim=50 20 300 30, clip ]{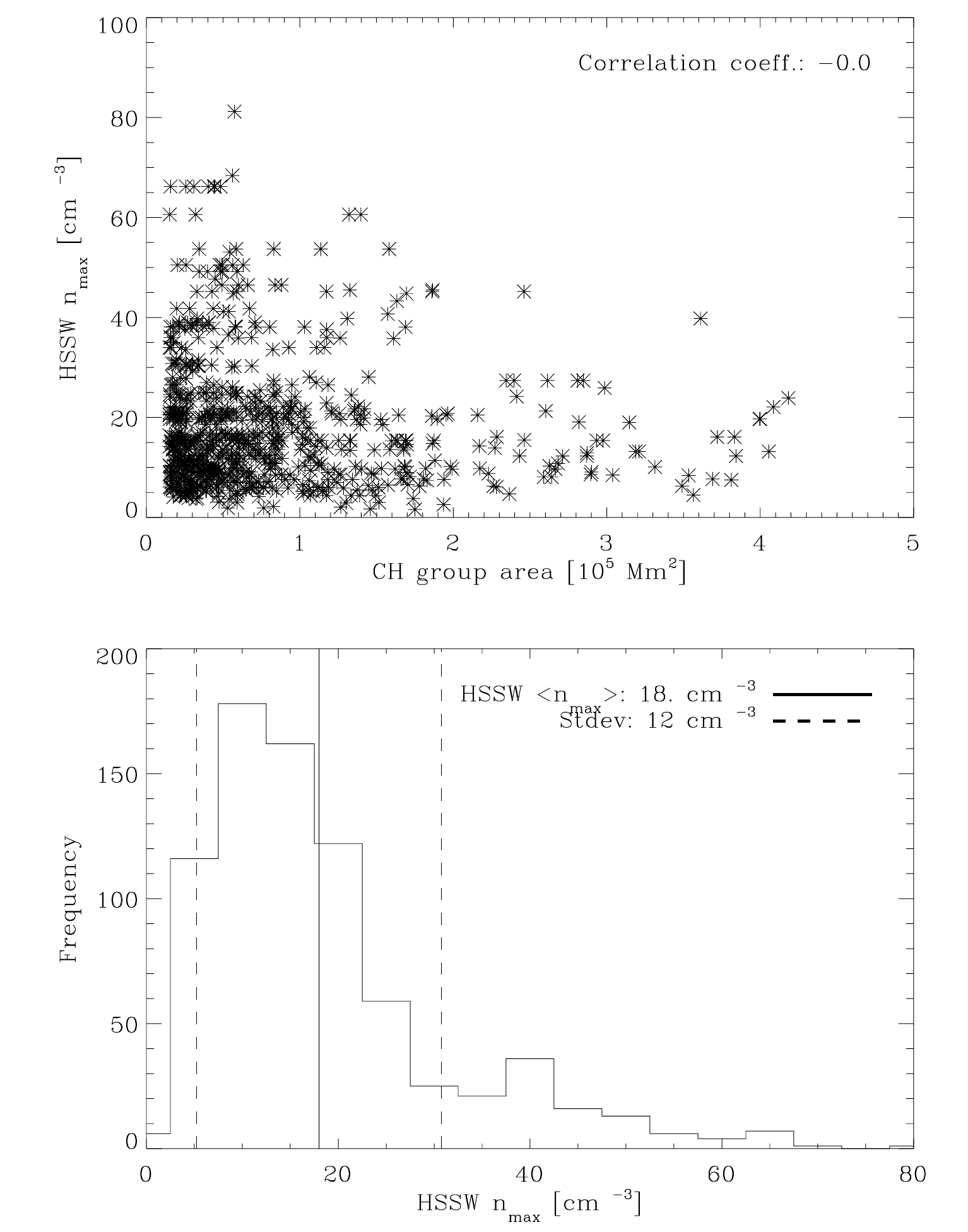}
\includegraphics[scale=0.62, trim=50 20 300 30, clip ]{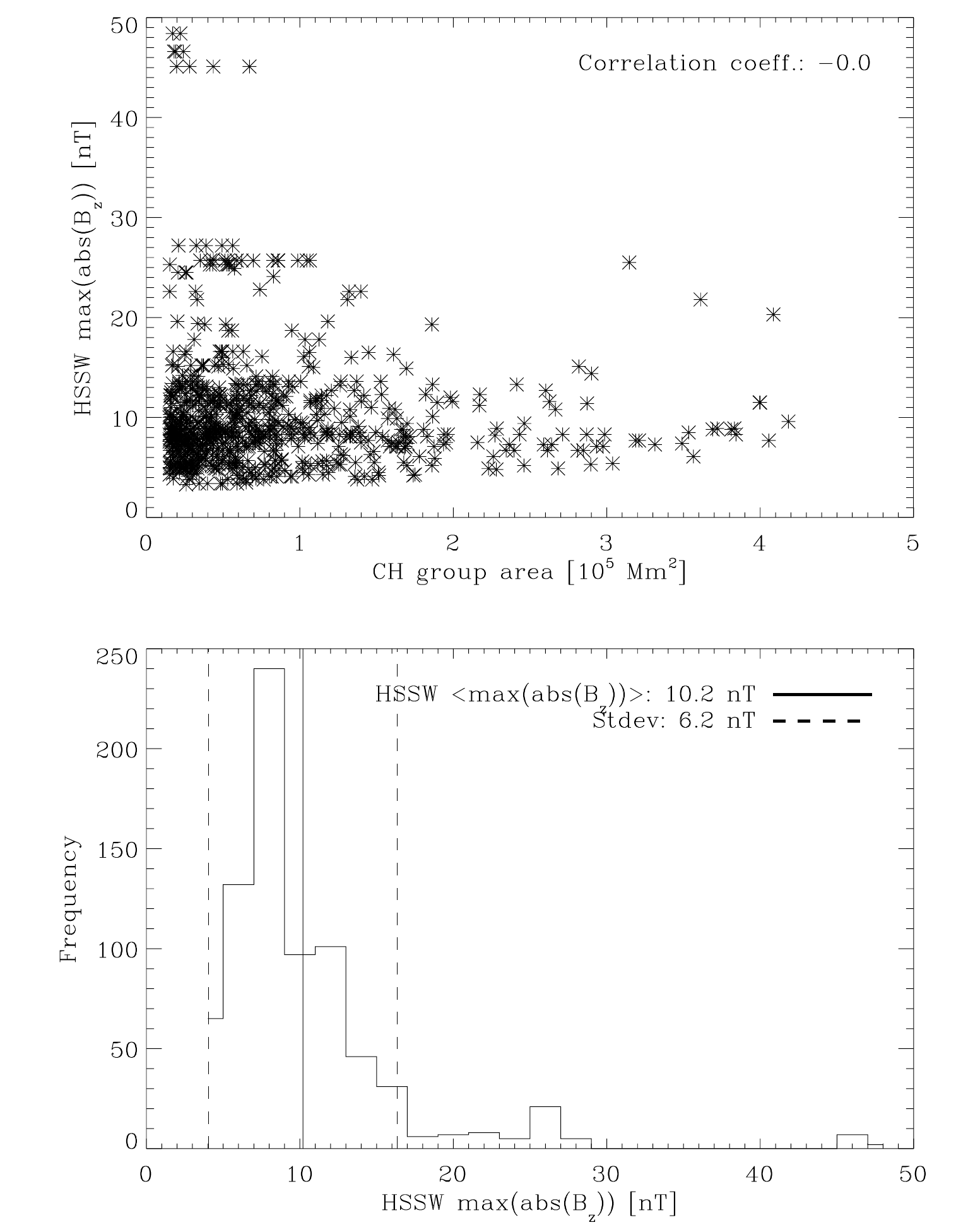} } 
\caption{Top left and right: scatterplot showing the relationship between the CH group area and the maximum density, and maximum B$_{z, max}$ of the corresponding HSSW stream, respectively. Bottom left and right: maximum HSSW density and B$_{z}$ frequency histogram, respectively.}
\label{areamax2} 
\end{figure}
\end{landscape}

\begin{landscape}
\begin{figure}[!t]
\centerline{
\includegraphics[scale=0.62, trim=50 20 300 30, clip ]{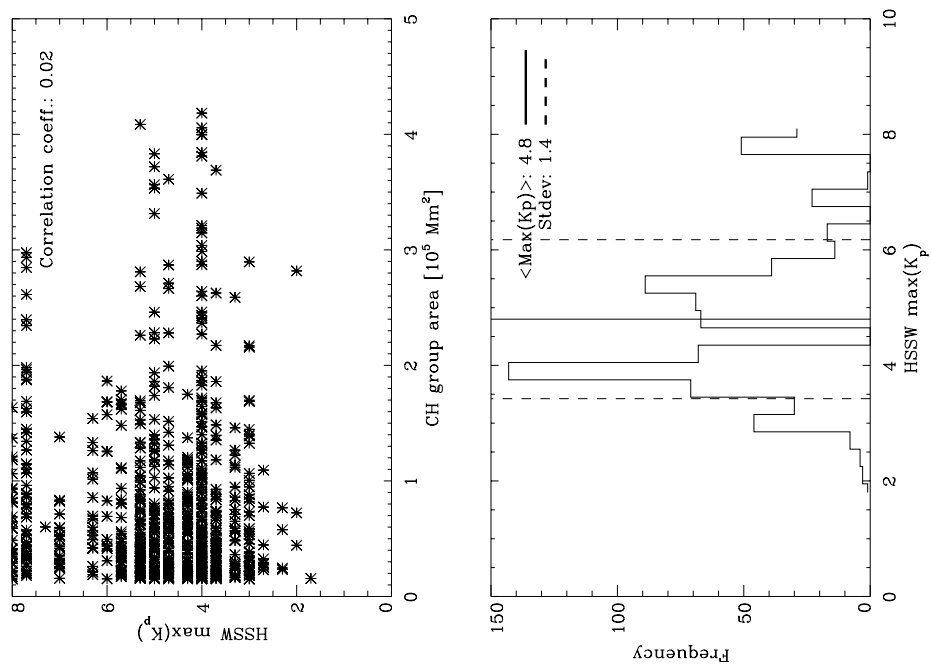} } 
\caption{Scatterplot showing the relationship between the CH group area and the maximum $Kp$ index of the geomagnetic disturbance caused by the corresponding HSSW stream.}
\label{areakpmax} 
\end{figure}
\end{landscape}

Previous work by \citep{Vrsnak07a} has shown HSSW properties to be influenced by the size of the corresponding CH (see Figure~\ref{vrsnak_plot} in Section~\ref{vrsnak_method}). In the study of \citeauthor{Vrsnak07a}, CH areas were determined in a CM sector of 20$^{\circ}$ in the time period 25 January to 5 May 2005 and were compared to peak values of HSSW properties. In our study CH areas were determined for CH groups without sectorial limits in the time period 10 August 2000 to 24 June 2009. We found a weak correlation of $\sim$0.15 between the CH group area and the maximum HSSW velocity. No correlation was found between the CH area and HSSW $n_{max}$, $|B_{z, max}|$, $Kp_{max}$, $T_{max}$. These results are in disagreement with those acquired by \citep{Vrsnak07a}. The scatter plots and the frequency histograms of the HSSW properties are shown in  Figures~\ref{areamax1}, \ref{areamax2}, and \ref{areakpmax}.

\begin{landscape}
\begin{figure}[!t]
\centerline{
\includegraphics[scale=0.62, trim=50 20 300 30, clip ]{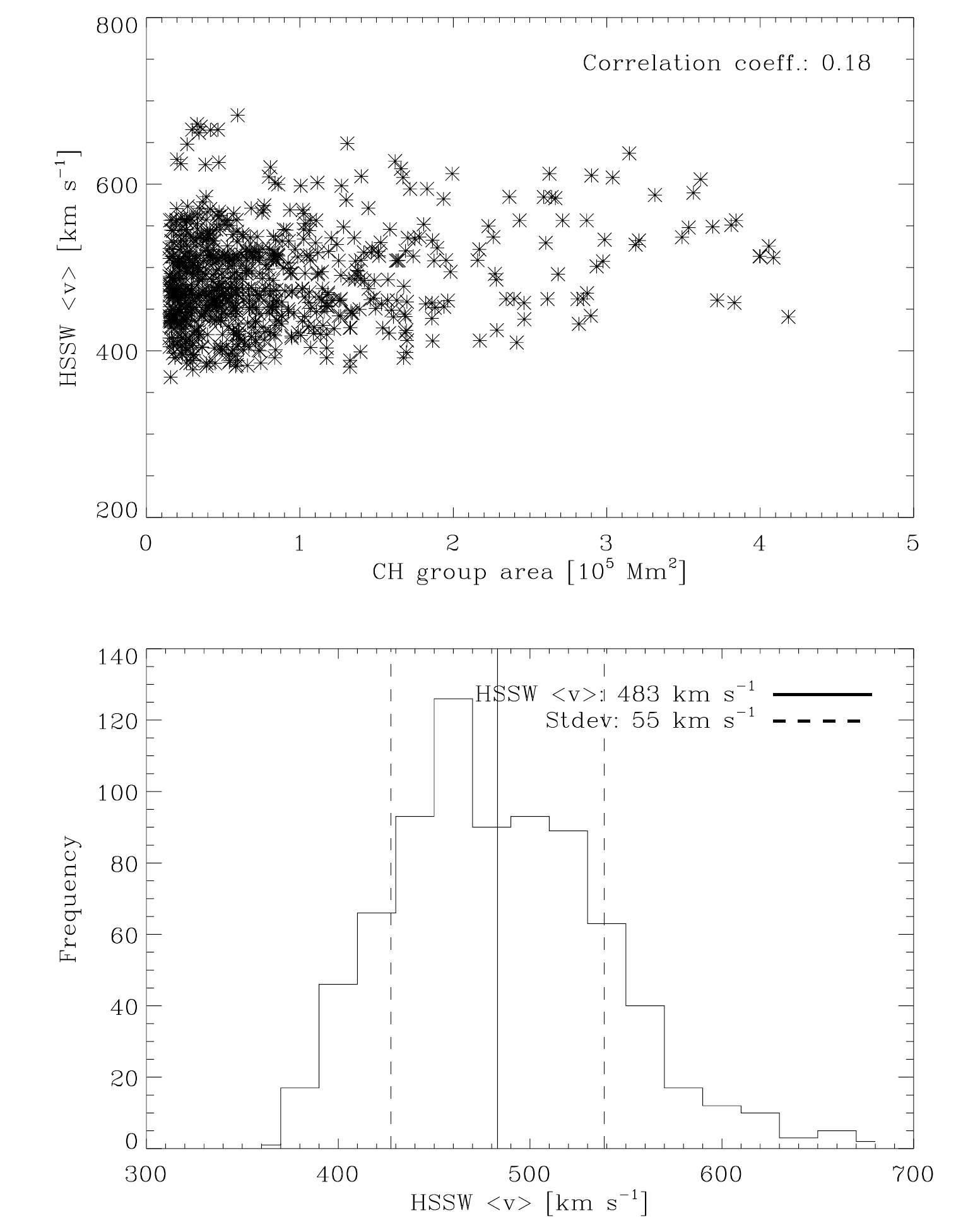} 
\includegraphics[scale=0.62, trim=50 20 300 30, clip ]{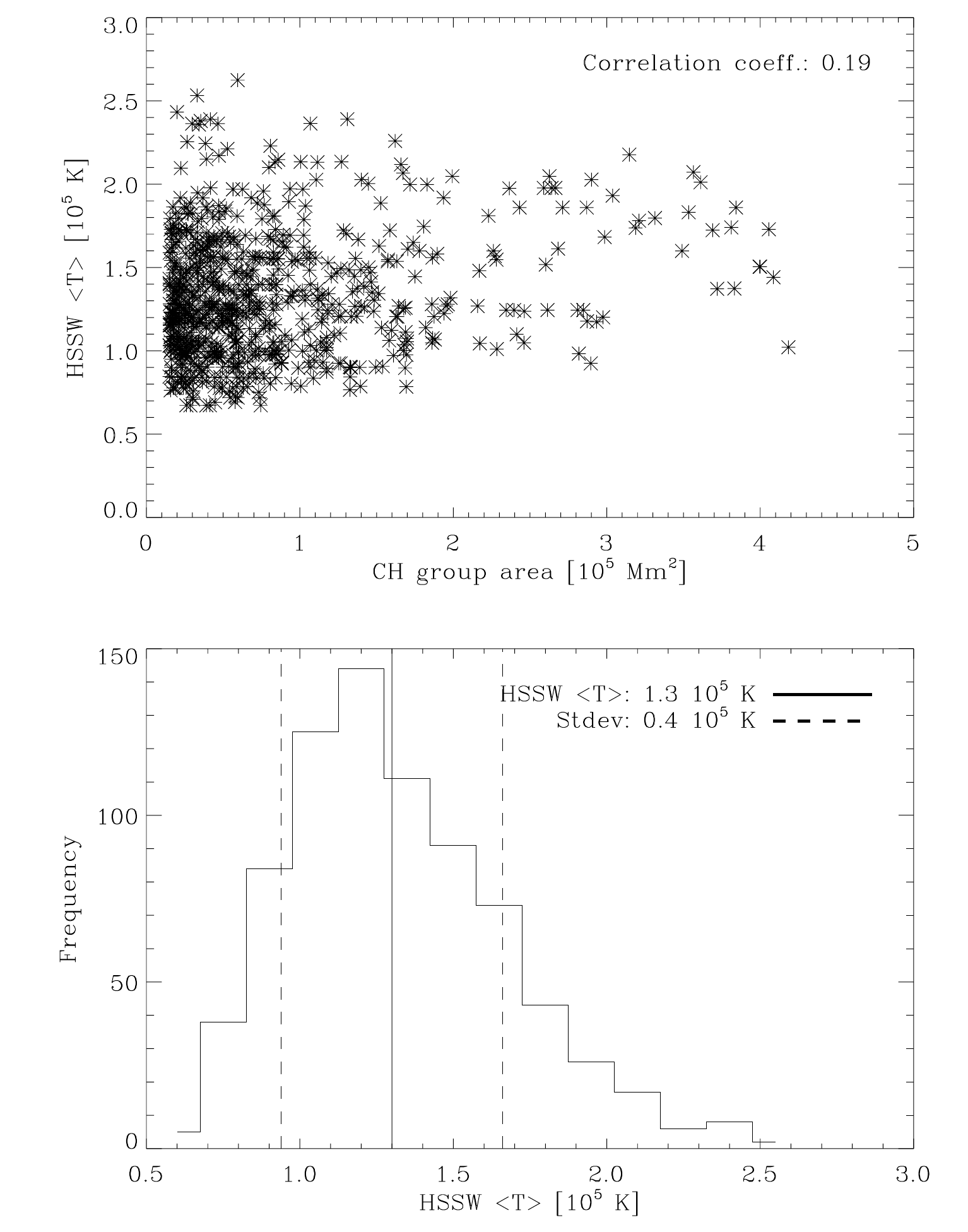} } 
\caption{Top left and right: scatterplot showing the relationship between the CH group area and the mean velocity, and temperature of the corresponding HSSW stream, respectively. Bottom left and right: mean HSSW velocity, and temperature frequency histogram, respectively.}
\label{areamean1} 
\end{figure}
\end{landscape}

\begin{landscape}
\begin{figure}[!t]
\centerline{
\includegraphics[scale=0.62, trim=50 20 300 30, clip ]{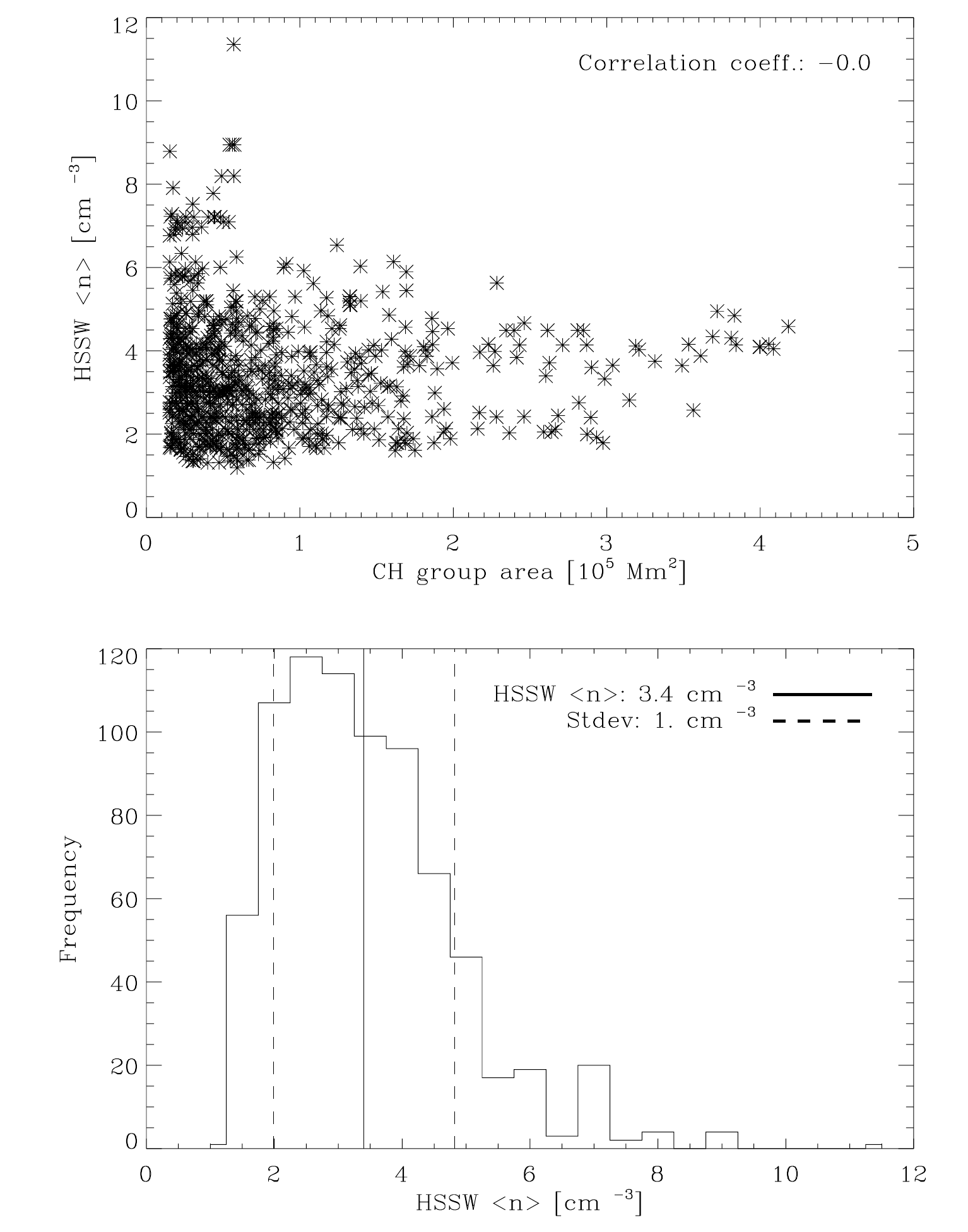} 
\includegraphics[scale=0.62, trim=50 20 300 30, clip ]{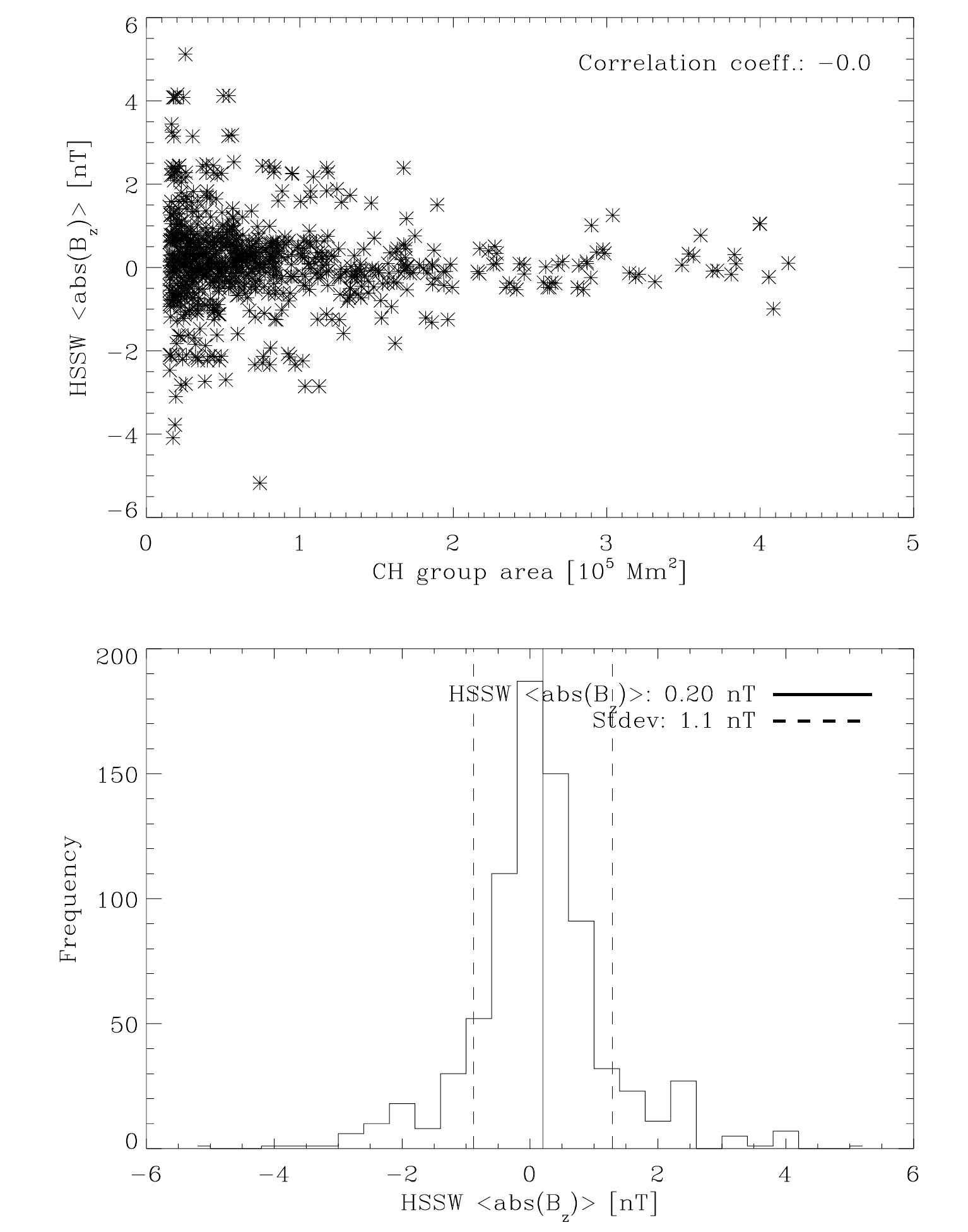} } 
\caption{Top left and right: scatterplot showing the relationship between the CH group area and the mean density, and B$_{z}$ of the corresponding HSSW stream, respectively. Bottom left and right: mean HSSW density, and B$_{z}$ frequency histogram, respectively.}
\label{areamean2} 
\end{figure}
\end{landscape}

\begin{landscape}
\begin{figure}[!t]
\centerline{
\includegraphics[scale=0.62, trim=50 20 300 30, clip ]{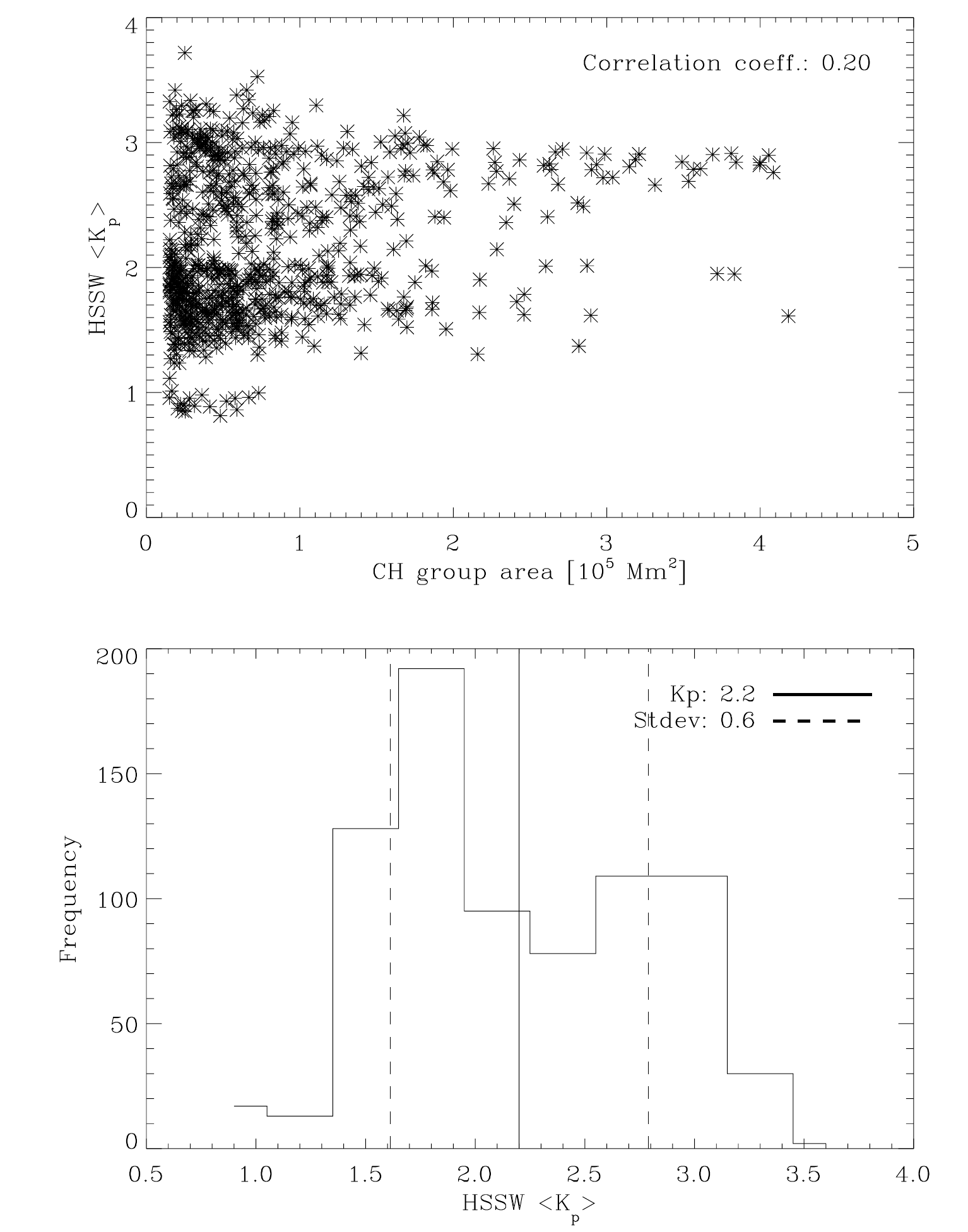} 
\includegraphics[scale=0.62, trim=50 20 300 30, clip ]{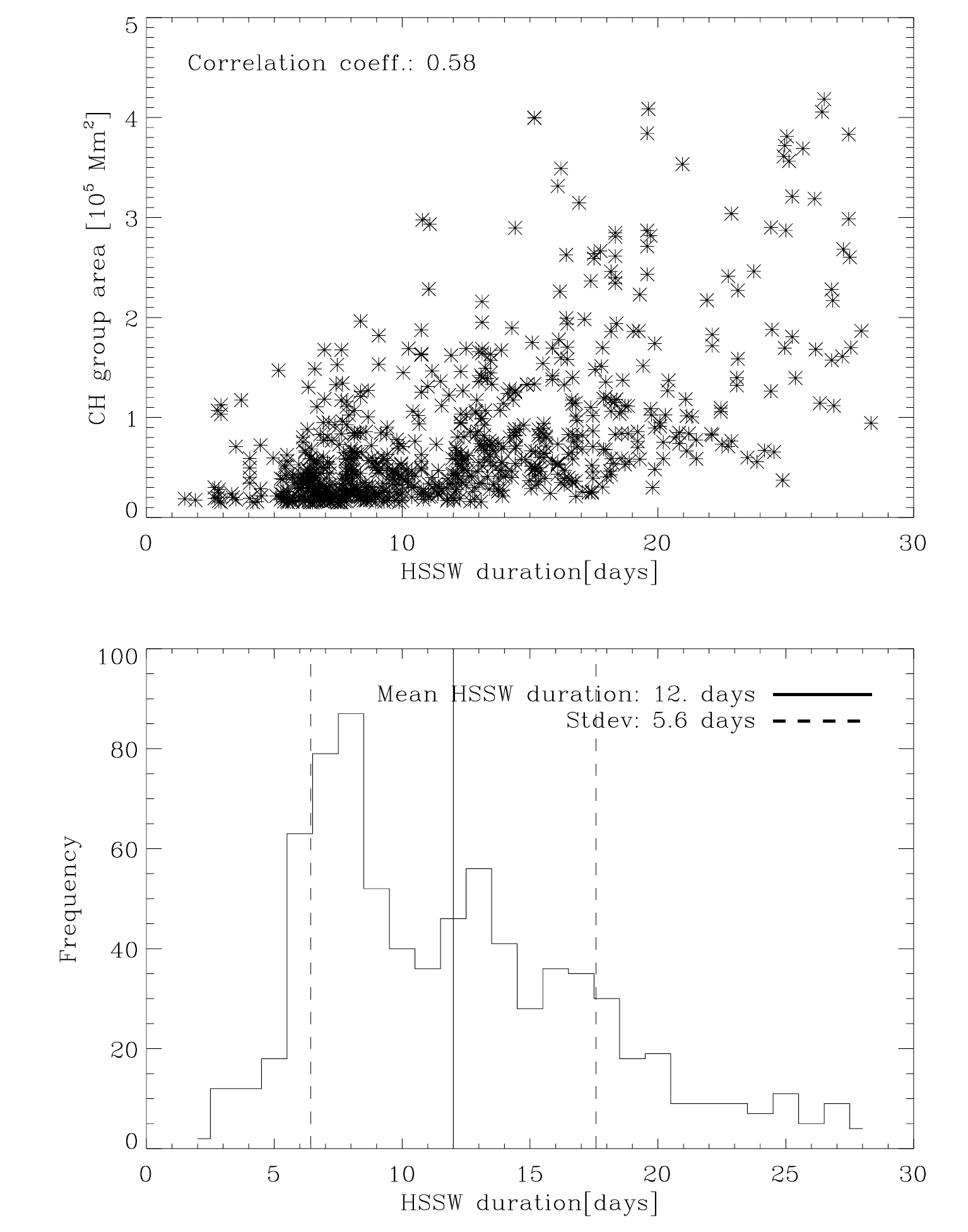} } 
\caption{Top left and right: scatterplot showing the relationship between the CH group area and the mean HSSW $Kp$ index, and duration, respectively. Bottom left and right: mean HSSW $Kp$ index, and duration frequency histogram, respectively.}
\label{areakpdur} 
\end{figure}
\end{landscape}

The correlation between the CH area and the mean HSSW properties was also studied. A weak correlation of $\sim$0.2 was found between the area and $<$$T$$>$, $<$$Kp$$>$ and $<$$v$$>$. No correlation was found between the area and $<$$n$$>$, $<$$|{B}_{z}|$$>$ (Figures~\ref{areamean1}, \ref{areamean2}, and \ref{areakpdur}).
 
When inspecting the relationship between the CH area and HSSW duration, a strong correlation was found (correlation coefficient of $\sim$0.58); this is in agreement with the super-radial flux tube expansion model (Figure~\ref{areakpdur}).

When relating the maximum HSSW velocity to the maximum temperature a strong correlation of $\sim$0.61 was found, and and even stronger correlation was shown between the mean HSSW velocity and mean temperature (correlation coefficient of $\sim$0.89). This correlation may indicate that the mechanism accelerating the solar wind in CHs is also responsible for the heating of the plasma (e.g. Alfv\'en wave heating).

\begin{landscape}
\begin{figure}[!t]
\centerline{
\includegraphics[scale=0.62, trim=50 30 300 30 ]{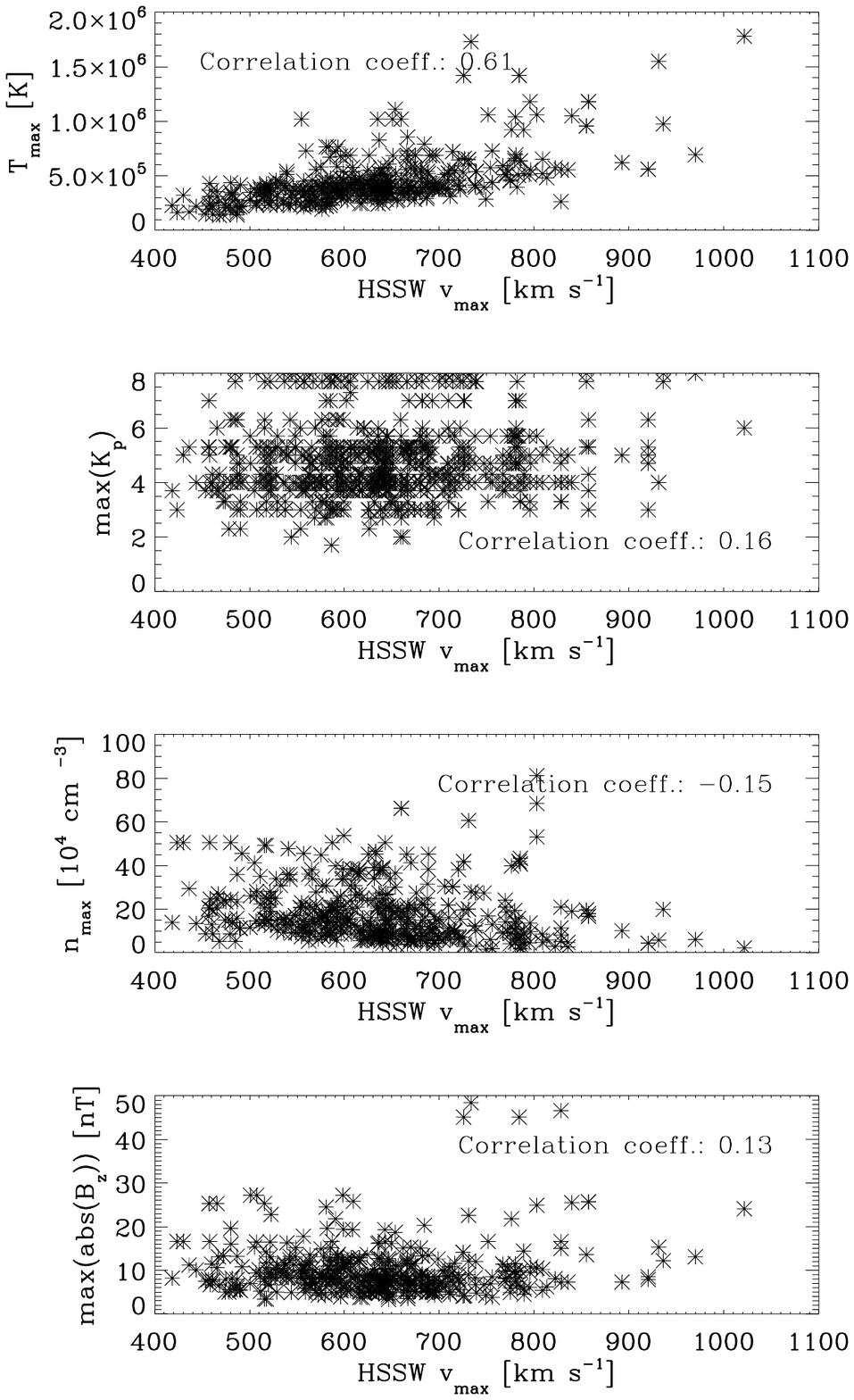}
\includegraphics[scale=0.62, trim=50 30 300 30 ]{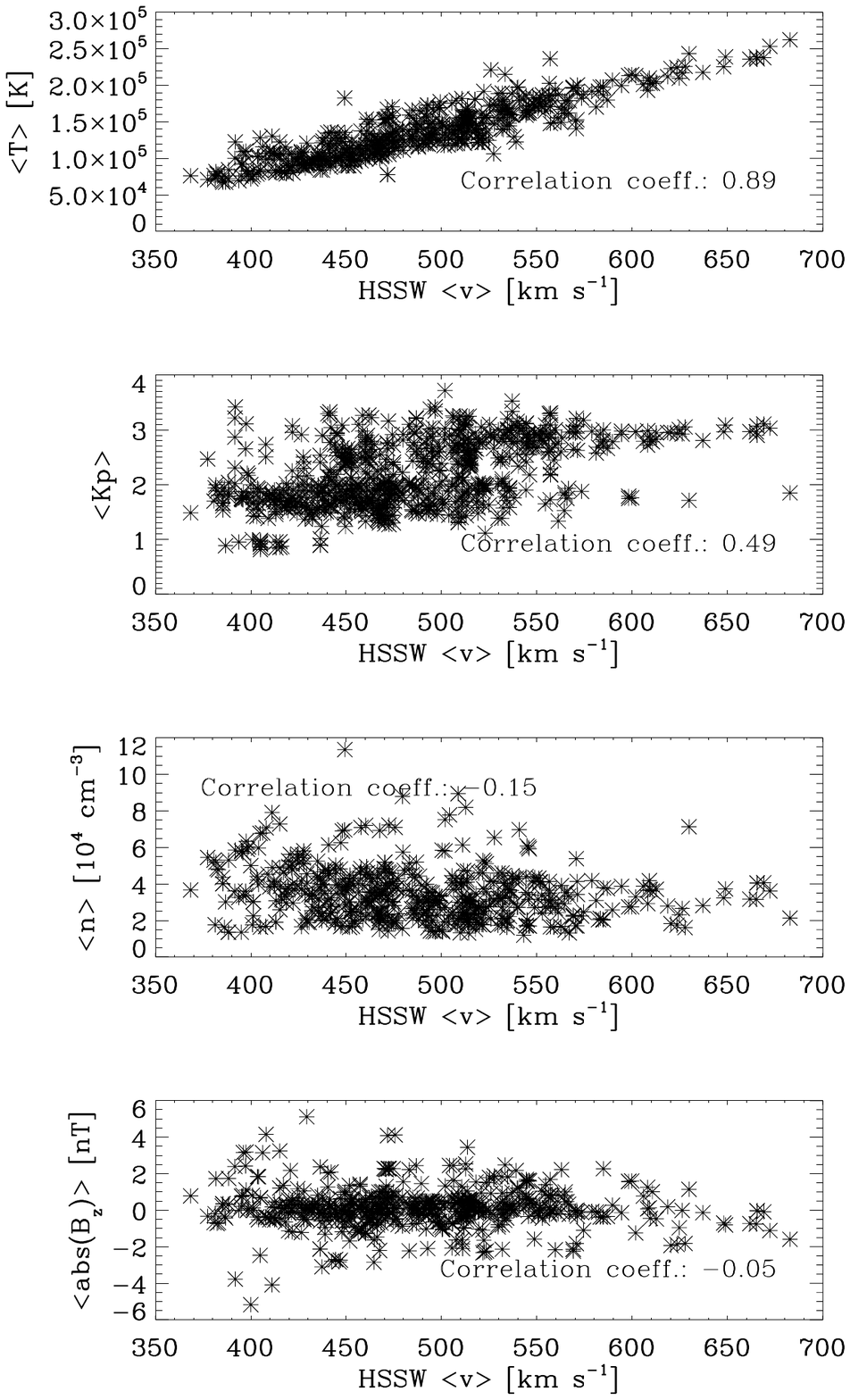} } 
\caption{Left: scatterplots showing the relationship between the maximum HSSW velocity and the maximum temperature, $Kp$ index, density, and $|\mathrm{B}_{z}|$. Right: scatterplots showing the relationship between the mean HSSW velocity and the mean temperature, $Kp$ index, density, and $|\mathrm{B}_{z}|$.}
\label{corr} 
\end{figure}
\end{landscape}

The mean HSSW velocity and the mean $Kp$ index have a correlation of $\sim$0.49. The rest of the HSSW properties show little or no correlation. The described scatter plots are shown in Figure~\ref{corr}. The frequency histograms of the discussed properties are shown in the corresponding figures.

\subsection{Geomagnetic effects of HSSW streams}
\label{hssw_var}

\begin{figure}[!t]
\centerline{
\includegraphics[scale=0.5, trim=0 50 0 50, clip ]{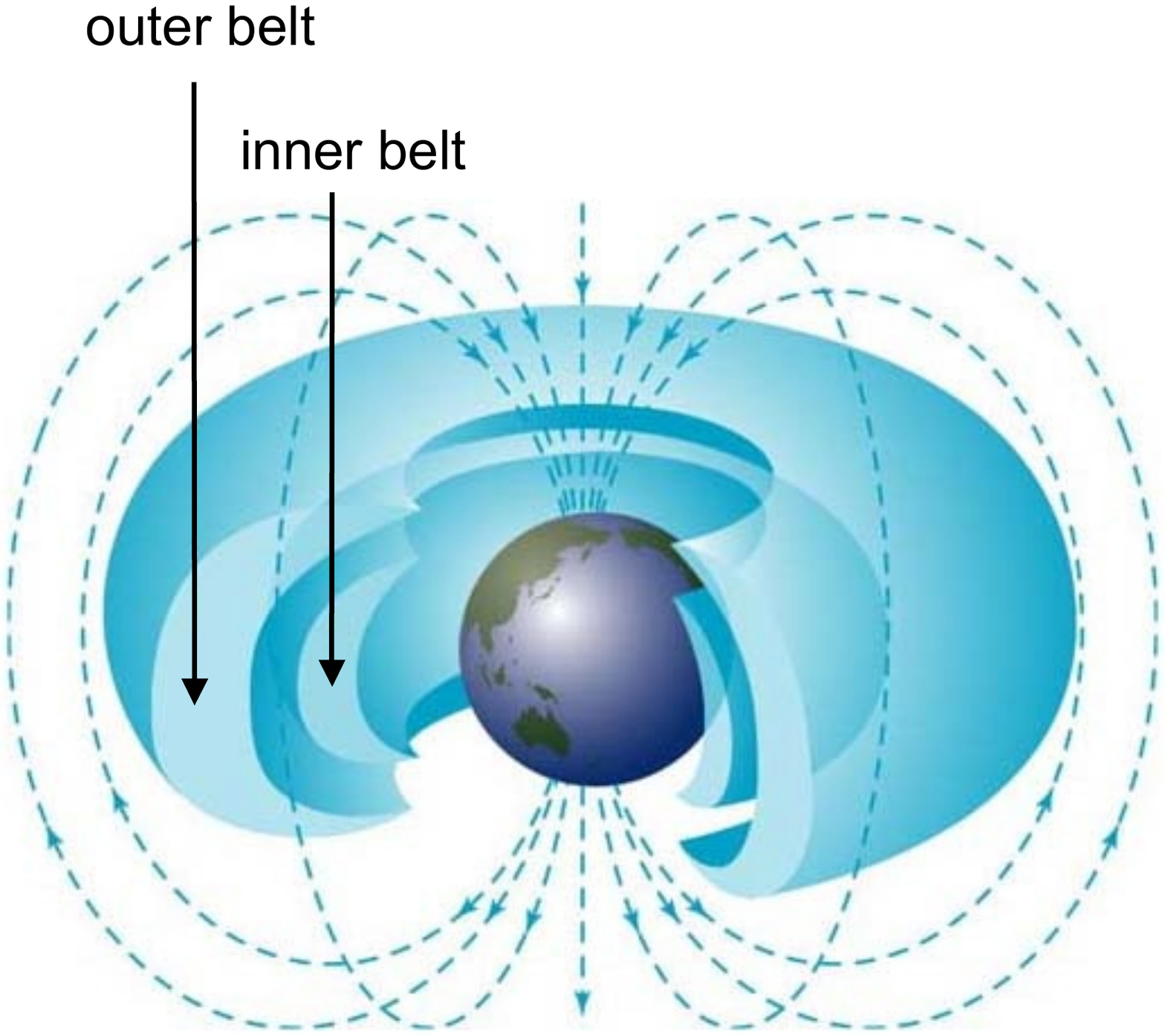}}
\caption{The inner and outer Van Allen radiation belts.}
\label{vanallen} 
\end{figure}

The CH location, area and magnetic flux are known to follow the solar cycle \citep{Wang96}. Hence, it is safe to assume that the properties of HSSW streams originating from CHs should consequently show some variation over the solar cycle. The space weather effects of CHs and the corresponding HSSW streams is particularly important in this study. A key aspect of the analysis presented in this section is the variation in the Van Allen radiation belts, which have been linked to solar storms. Relativistic ions and electrons trapped in the Earth's magnetosphere make up the Van Allen belts, which have been shown to change under the influence of magnetic and electric fields. Figure~\ref{vanallen} shows the radiation belts, which are made up of two layers - the inner ($\sim$1.5~R$_{Earth}$) and outer ($\sim$4--7~R$_{Earth}$) belt. Geomagnetic storms caused by CMEs and CIRs cause enhancements in the radiation belt relativistic electron flux, which can cause serious damage to electrical devices on satellites \citep{Miyoshi11, Pilipenko06}. It has been shown that during solar maximum activity the anomalies experienced on satellites are mostly caused by energetic solar protons originating in CMEs, while in the declining and growing phase of the cycle it is the magnetospheric electrons that cause satellite disturbances. The outer radiation belt has been found to be influenced mainly by CIRs, while the inner belt by CMEs.

\begin{landscape}
\begin{figure}[!ht]
\centerline{
\includegraphics[scale=0.4, trim=50 20 100 150, clip ]{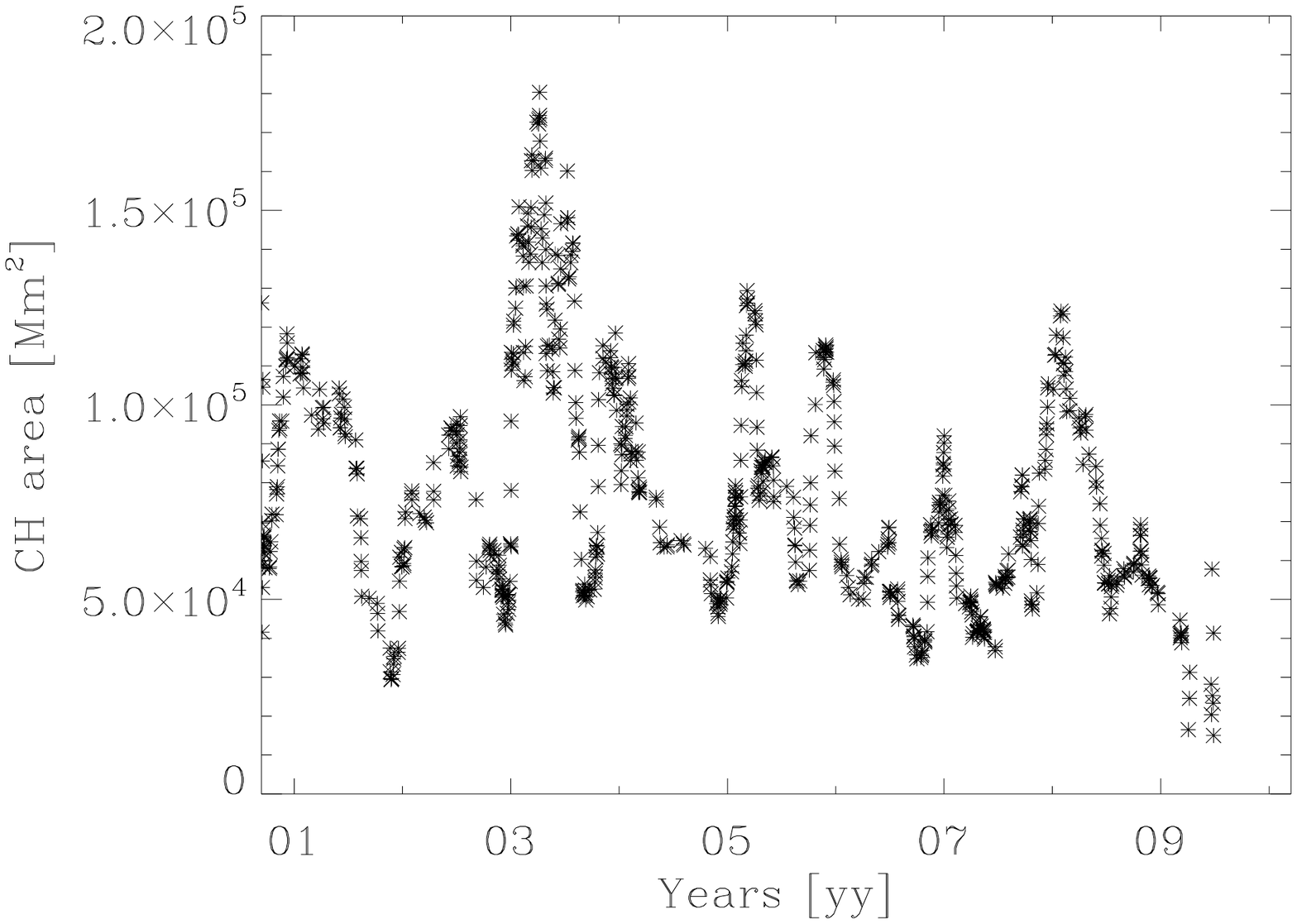}
\includegraphics[scale=0.4, trim=50 20 100 150, clip ]{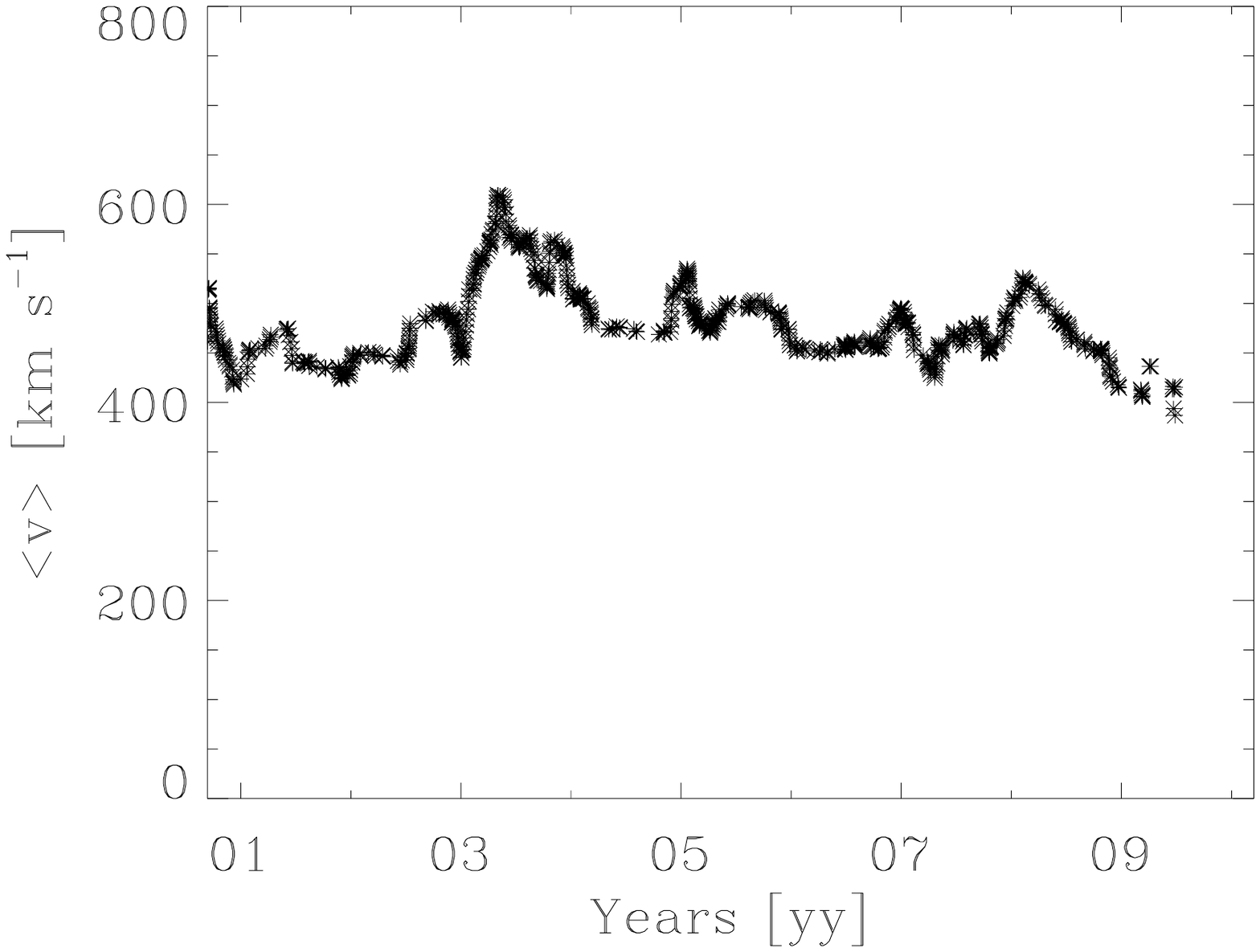} } 
\centerline{
\includegraphics[scale=0.4, trim=50 20 100 150, clip ]{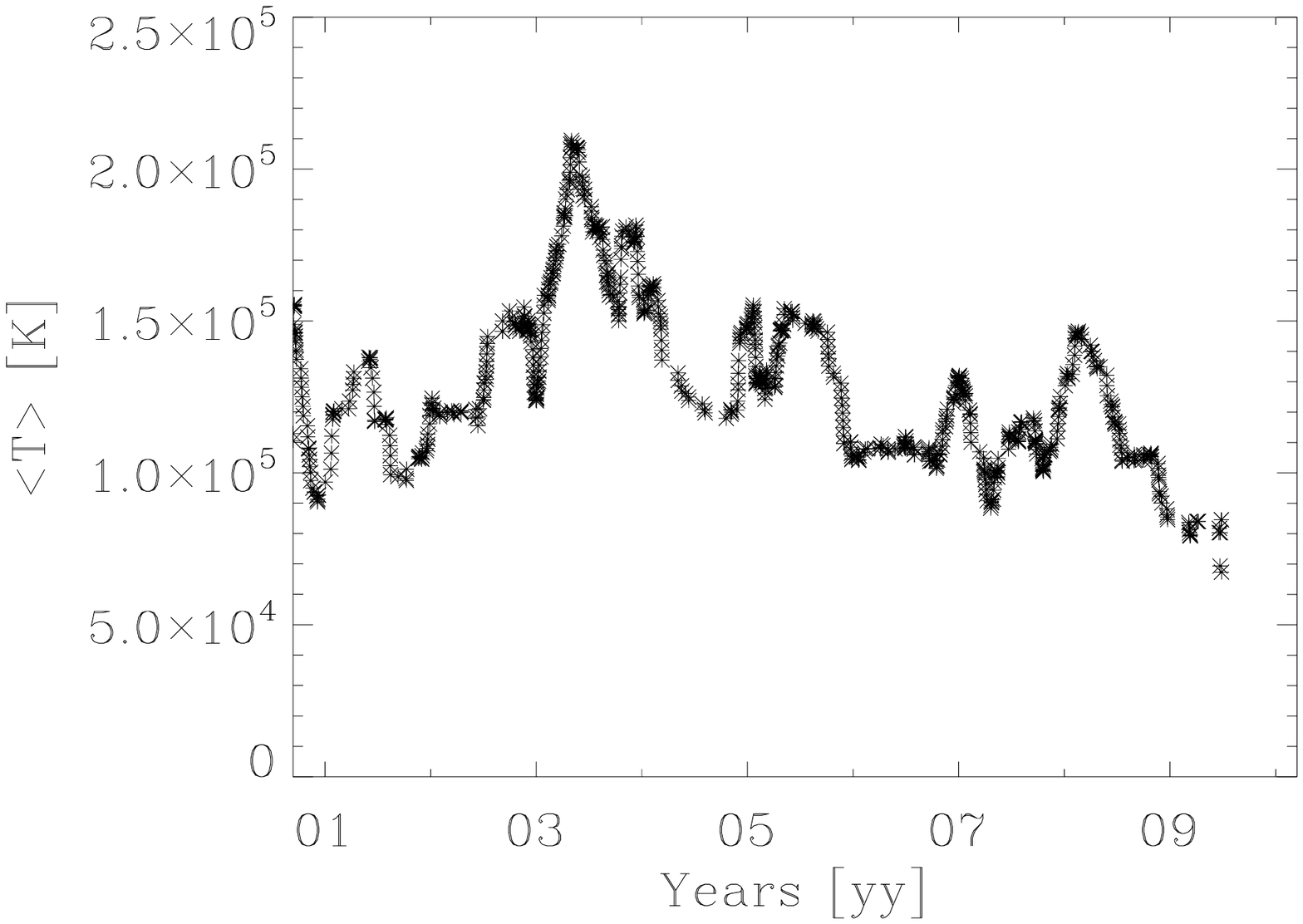}
\includegraphics[scale=0.4, trim=50 20 100 150, clip ]{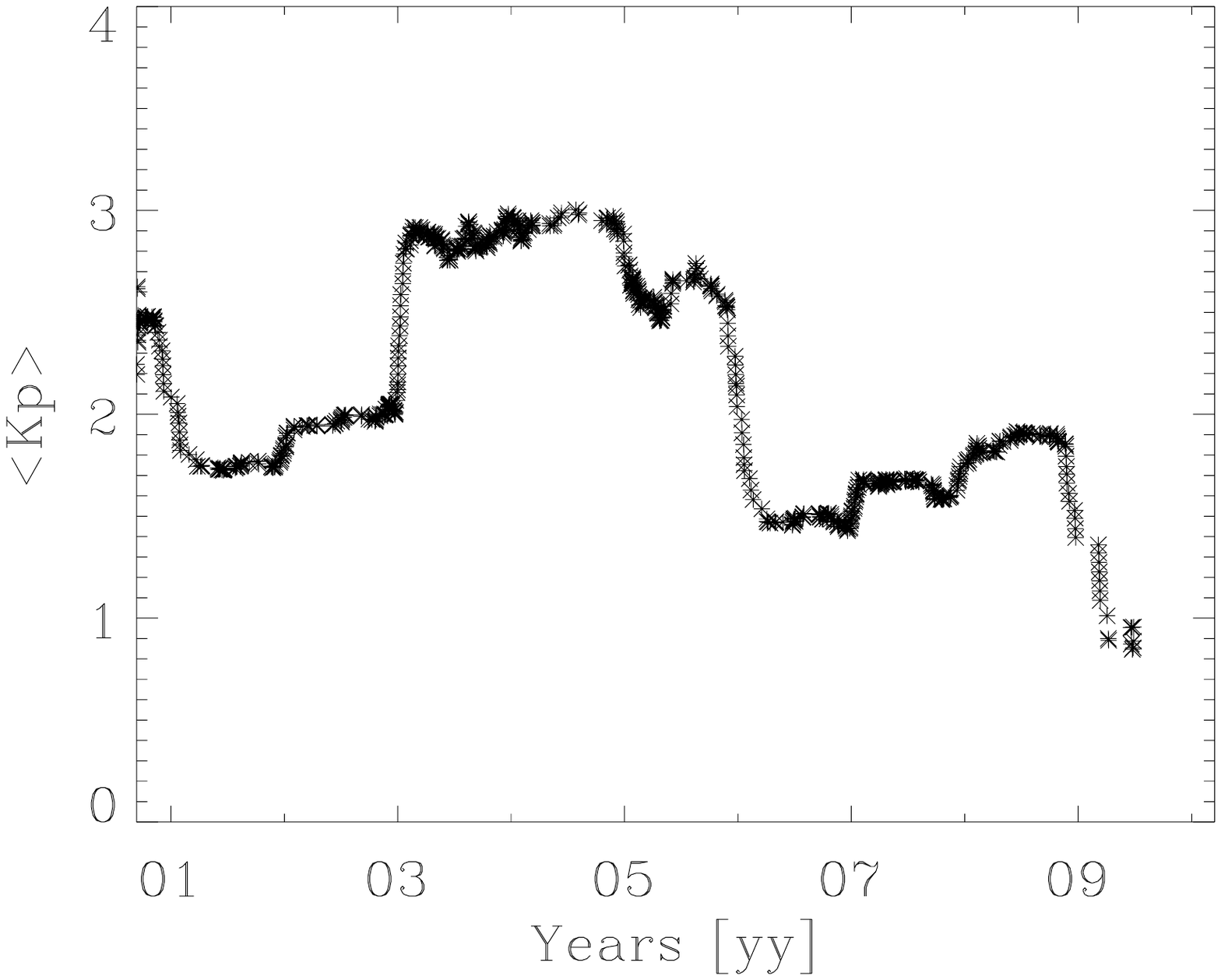} } 
\caption{Top left and right: the variation of the CH group area and the mean HSSW velocity, respectively. Bottom left and right: the variation of the mean HSSW temperature and $Kp$ index, respectively.}
\label{maxvar1} 
\end{figure}
\end{landscape}

\begin{landscape}
\begin{figure}[!ht]
\centerline{
\includegraphics[scale=0.4, trim=50 20 100 150, clip ]{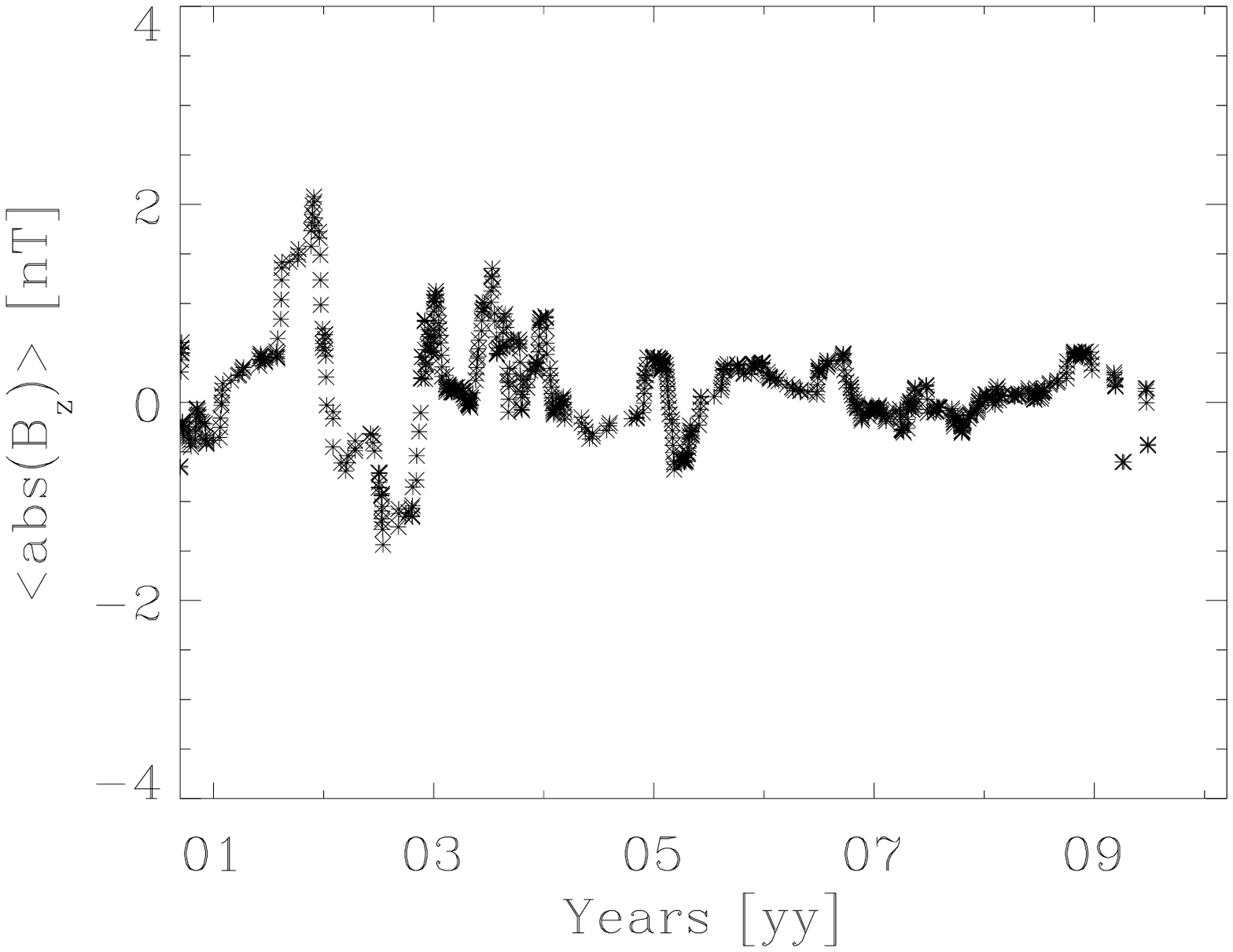}
\includegraphics[scale=0.4, trim=50 20 100 150, clip ]{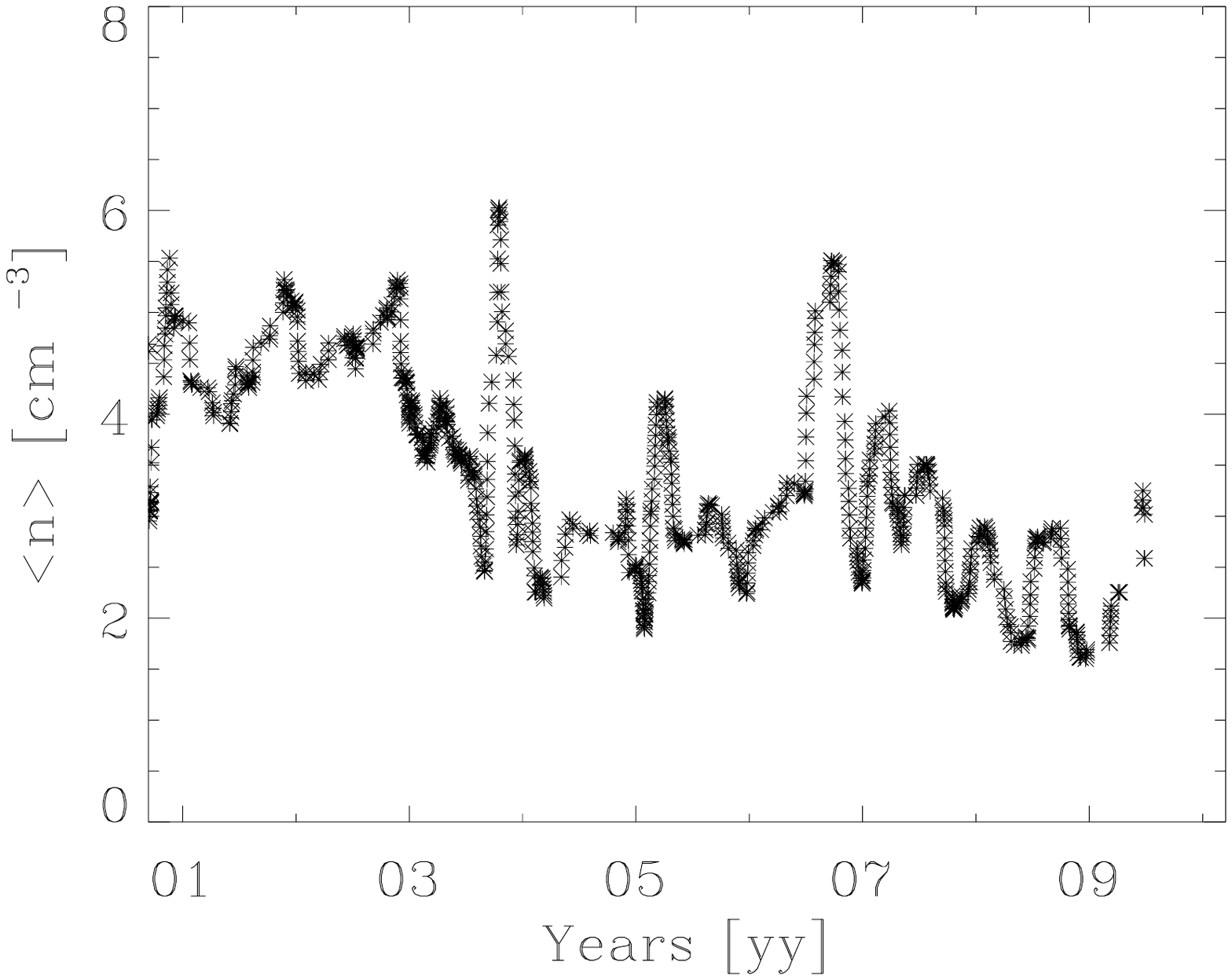} } 
\includegraphics[scale=0.4, trim=-40 0 0 150, clip ]{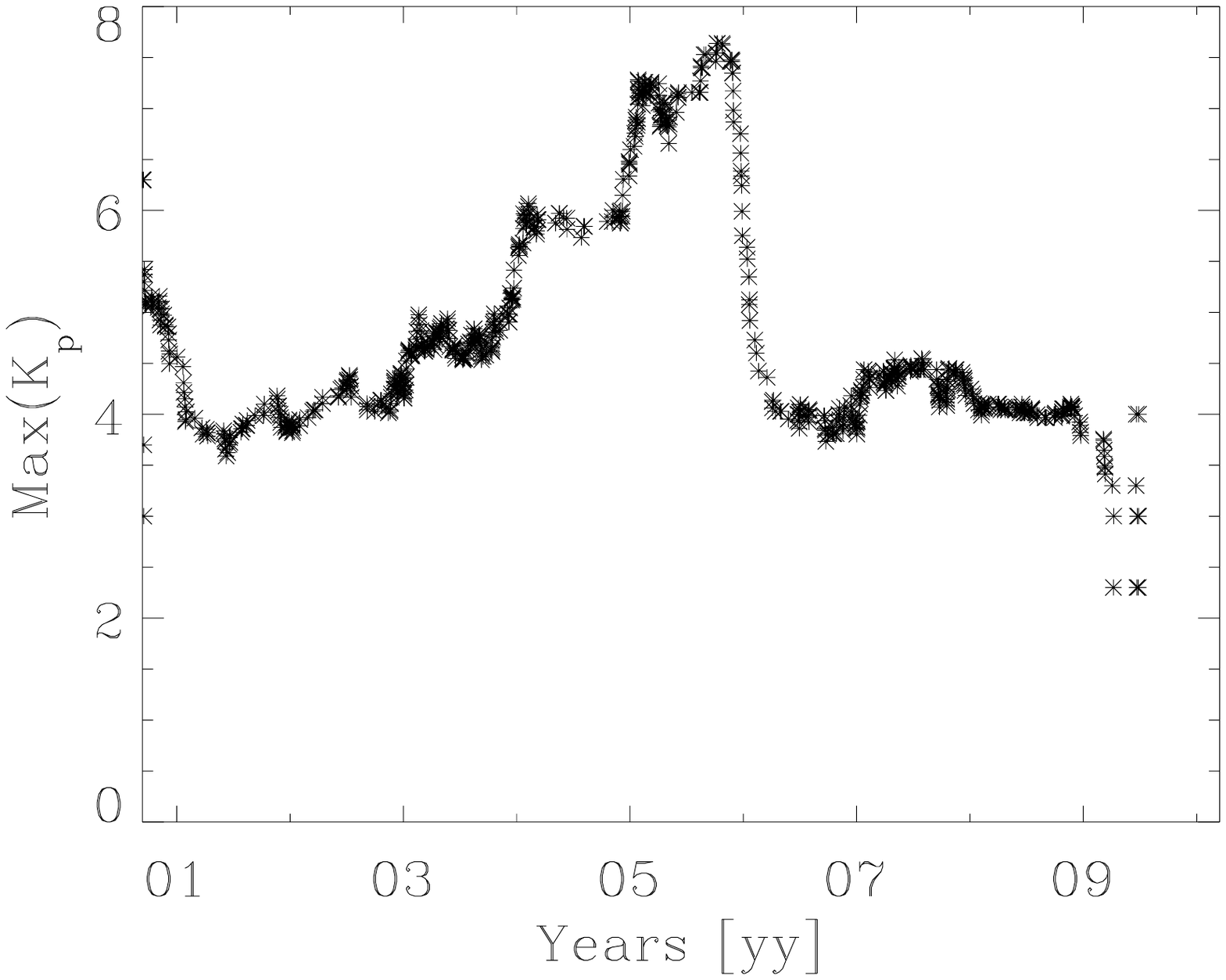}
\caption{Top left and right: the variation of the mean $|\mathrm{B}_{z}|$ and density, respectively. Bottom: The variation of the CH related maximum geomagnetic disturbance indices (Kp) over 2000-2009.}
\label{maxvar2} 
\end{figure}
\end{landscape}

In this study we show the variation in the HSSW properties over nine years from 2000 to 2009, which has shown an interesting trend in the CH area, the mean HSSW velocity, mean temperature and mean $Kp$ index. Near the start of 2003 all four properties show a remarkable rise (hence the relatively strong correlation described in the previous section). This is shown in Figures~\ref{maxvar1} and \ref{maxvar2}.

\begin{figure}[!ht]
\centerline{
\includegraphics[scale=1.3, trim=0 0 0 0, clip ]{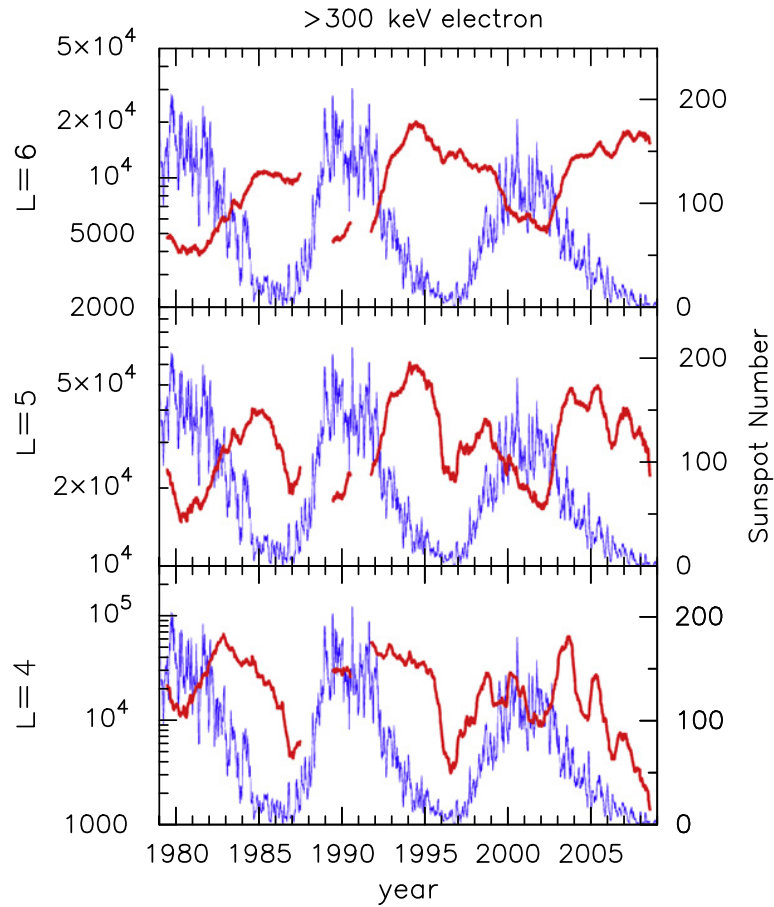}}
\caption{The yearly average $>$~300~keV electron flux (red line) from 1979--2008 for different layers of the Earth's radiation belt (L=4, 5 and 6). The sunspot numbers are shown with a blue line for comparison \citep{Miyoshi11}.}
\label{miyoshi_flux} 
\end{figure}

Similar enhancements have been shown in the yearly averaged $>$~300~keV electron flux measured in the Earth's Van Allen radiation belt \citep{Miyoshi11}. Figure~\ref{miyoshi_flux} compares the variation in the sunspot number (blue line) during the solar cycle and the electron flux (red flux) measured in the outer portions of the outer radiation belt by the MEPED instrument aboard the NOAA/POES satellites. The radiation belt shells are described with ``L", which corresponds to a set of magnetic field lines which cross the Earth's magnetic equator at a distance equal to the L-value measured in Earth-radii. In all outer radiation belt shells (especially L=5 and 6) the high energy electron flux rises during the declining phase of the solar cycle. The maximum value is reached between 2003-2004 which is in perfect agreement with the enhancements shown in the CH area, the corresponding HSSW stream velocity, temperature and the $Kp$ index of the resulting geomagnetic storms (Figure~\ref{maxvar1}). Although the rise is not as significant, it can still be recognised in the HSSW density and $|B_{z}|$ (Figure~\ref{maxvar2}). The CH area, the mean HSSW velocity and the temperature show a significant rise which is followed by a similarly steep decline to the end of 2005. The average $Kp$ index however stays at $\sim$3 from 2003--2006.

The peak values of the HSSW streams were also studied over the nine years, but no significant rise was identified in the year 2003. However, the maximum $Kp$ index shows a significant enhancement from 2003 to 2005, reaching a peak value of $Kp$=7.5 in 2005-2006, which is followed by a steep decline at the start of 2006, after which it levels off at $Kp\simeq$4 (Figure~\ref{maxvar2}).

\begin{figure}[!ht]
\centerline{
\includegraphics[scale=1, trim= 0 0 0 0, clip ]{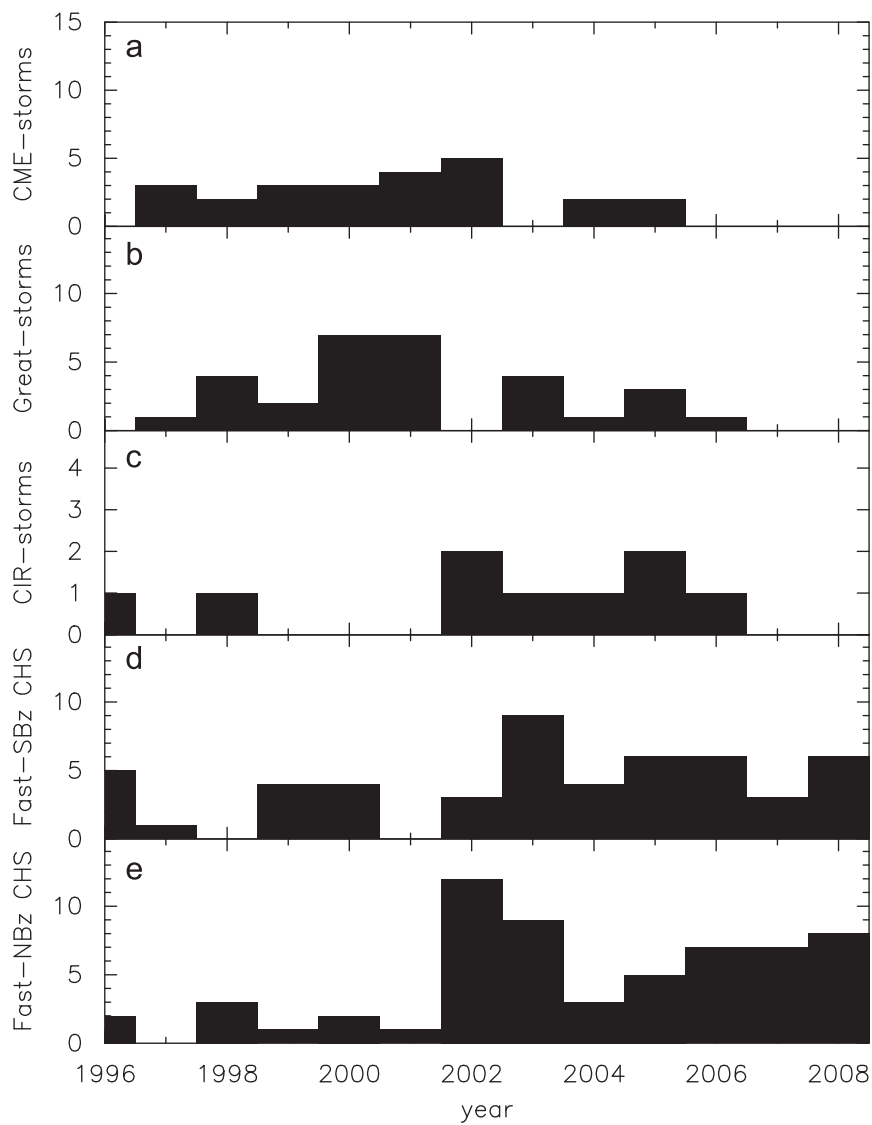}}
\caption{The frequency of geomagnetic storms corresponding to different sources: CMEs (a $\&$ b), CIRs (c), and CHs (d $\&$ e, for north and south directed magnetic fields, respectively). Figure courtesy of \citep{Miyoshi11}.}
\label{miyoshi_kp} 
\end{figure}

Figure~\ref{miyoshi_kp} of \cite{Miyoshi11} shows the geomagnetic storm $Kp$ indices corresponding to different causes from 1996--2008. Panel (a) shows the frequency of CME related storms, with a small peak in 2002, which coincides with the solar cycle maximum. Larger CME related storms appear to peak between 2000--2001, also the time of the maximum solar activity - as shown in panel (b). Panel (c) shows the CIR related storm frequency enhancement from 2002--2006, which is in agreement with the CH related maximum $Kp$ indices found in our study. Panel (d) and (e) show the CH related geomagnetic storm frequency with significant enhancements in 2002 and 2003, which were also found to be in agreement with our results.

\section{Discussion and conclusions}
\label{hssw_conclusions}

CIRs created by the HSSW streams originating in CHs are well known to cause long-lasting geomagnetic storms due to the high plasma velocity and magnetic field fluctuations \citep{Borovsky06}. While the geomagnetic storms associated with CHs are normally of a minor scale ($Kp$=4--5), the number of occurrences and duration of these minor storms make them comparable to CMEs in their contribution to the overall geomagnetic activity. Minor storms are especially important to consider during the declining phase of the solar cycle when the number of their occurrences and the scale of the disturbances rise. 

In order to automatically detect and forecast CH related HSSW streams, the author of the present thesis developed the {\it MIST} algorithm, using which the expansion of HSSW flux tubes was estimated. The CH latitudinal widths were used in conjunction with the HSSW durations to calculate the super-radial expansion factor which was found to be 2--48. Although this result is similar to the values of 2-20 determined previously using theoretical models \citep{Wang94}, it is yet to be investigated why the observational values give a much higher upper limit. However, it is likely that the circular flux-tube approximation based on the maximum latitudinal width of a CH will introduce some variation in the results, since CHs are known to often have complicated morphologies. Furthermore, in our analysis HSSW streams are linked to CH groups rather than individual CHs, and hence the latitudinal width of the stream is related to the width of the CH group.

The relationship between CHs and the HSSW streams and geomagnetic disturbances was also investigated. For the first time, a large-scale study involving nine years of data from 2000-2009 linked CH occurrences to geomagnetic storms and radiation belt electron flux enhancements. An enhancement was found in the CH area, mean HSSW velocity, temperature and $Kp$ index in the year 2003, the same time when the high-energy electron flux measured in the Earth's outer radiation belt rose significantly. It was also shown that while the average $Kp$ indices reached a maximum value in 2003 and plateaued until 2006, the most geoeffective storms reaching $Kp=7.5$ occurred in 2005, and kept occurring for approximately one year. It is yet to be investigated what causes the enhanced geoeffectivity associated with CHs during the declining phase of the solar cycle, and which CHs are likely to cause such strong storms.  

Forecasting HSSW streams related to CHs is also beneficial for the calculation of the interplanetary CME (or ICME) arrival times at Earth. It is well known that CMEs are slowed down as they pass through the ambient slow solar wind (also known as the ``snow-plough" effect), while this deceleration reduces when a CME passes through the high-speed, low density plasma originating from CHs \cite{Tappin06, Odstrcil05}. The morphology of the CME front could also be modified due to interaction with the CIR. For these reasons, it is very important to continue research on CH related HSSW streams.
\newline
\newline
The author would like to thank Edward Erwin (NOAA/NGDC) for providing the $Kp$ data used in this study.



\chapter{Conclusions and Future Work} 
\label{chapter:conclusions}

\ifpdf
    \graphicspath{{6/figures/PNG/}{6/figures/PDF/}{6/figures/}}
\else
    \graphicspath{{6/figures/EPS/}{6/figures/}}
\fi

\flt
{\it The goal of the research presented in this thesis was to study the mechanisms controlling coronal hole evolution, and to determine the space weather effects of the high-speed solar wind streams originating from coronal holes. In order to carry out the research, the author developed an automated detection method to standardise coronal hole identification throughout the solar cycle and allow the use of different instruments and wavelengths. It consequently allowed the investigation of the interaction between coronal holes and the quiet Sun, which reflects on the mechanisms controlling the short-term evolution of coronal hole boundaries. The study of CHs was extended to analyse the long-term changes in coronal hole and corresponding high-speed solar wind stream properties. In addition, a method was developed to provide forecasts of coronal hole related geomagnetic storms several days in advance of their arrival at Earth. In this Chapter, the primary results of the analyses are summarised and discussed. Suggestions for improvements and future work are addressed at the end of the Chapter.
 }
\flb

\newpage
\section{Thesis summary}

The goal of the research presented in this thesis was to investigate the behaviour of CHs from their small-scale boundary motions seen in solar coronal images to the geomagnetic effects experienced at Earth due to the HSSW emanating from them. 

It was shown that the fast and robust detection method ({\it CHARM}) developed by the author allows the objective identification of CHs throughout the solar cycle. Using an additional package ({\it CHEVOL}) we were able to study the boundary displacements in CHs and determine the diffusion coefficient ($\kappa$=0--3$\times10^{13}~\mathrm{cm}^{2}~\mathrm{s}^{-1}$), as well as the magnetic reconnection rate ($M_{0}$=0--3$\times10^{-3}$). The results indicated that the interchange reconnection between CHs and QS loops is likely to be accommodated by random photospheric granular motions. Our results support the interchange reconnection model at CH boundaries.

The {\it MIST} algorithm allowed CHs to be linked to HSSW streams in order to study their geomagnetic effects. An extensive study was carried out, which included over 600 CH and HSSW stream pairs. Using the CH latitudinal widths and HSSW durations measured at Earth we were able to determine the open flux tube expansion factor ($f$=2--48), which was found to be similar to those derived from theoretical studies. The CH and HSSW properties were compared to determine any relationship, however, the CH area was only found to correlate with the duration of the HSSW streams. Nevertheless, a strong correlation was found between the solar wind velocity and temperature, which could support the theory of Alfv\'en wave heating in CHs. 

Furthermore, long-term trends were discovered in the CH related HSSW stream properties, which were linked to enhanced electron fluxes measured in the outer Van Allen radiation belt. This confirms that the the Earth's radiation belt is strongly affected by the HSSW streams originating from CHs, especially during the descending phase of the solar cycle when the frequency and scale of CH related disturbances increases.

\subsection{Principal results}

The aim of the work presented in this thesis was to further our understanding of the short-term and long-term evolution of CHs. The small-scale changes were investigated to reflect on the mechanisms controlling CH growth and decay, while the long-term study showed the relationship between CHs and the corresponding HSSW streams and geomagnetic effects. Here, we summarise the results obtained from these analyses.

\begin{itemize}
	\item
The {\it CHARM} algorithm was developed by the author to provide an objective approach to CH detection throughout the solar cycle \citep{Krista09}. As opposed to previous methods where arbitrary thresholds were chosen, here, a local intensity thresholding method was used to determine the CH boundary intensity. Additionally, the flux imbalance was determined using magnetograms to establish whether the detected region was a CH. The detected CHs were then grouped to aid the forecasting algorithm and to help the future tracking of CHs. As part of the algorithm the CH properties (location, boundary positions, area, line-of-sight magnetic field strength and intensity) were stored for analysis. The developed algorithm is fast and robust, and allows the use of different instruments and wavelengths.

\item 
The {\it CHARM} algorithm was used in conjunction with the {\it CHEVOL} package to study the short term changes in CH boundaries. The study aimed to compare the observational results to those derived from the interchange reconnection model. The model describes the ``relocation" of open magnetic field lines through reconnection between open and closed field lines. This mechanism can explain the observed changes occurring in the CH boundaries on timescales of hours. 

\item
The results show that the CH boundary displacements were isotropic and had velocities in the range 0--2~km~s$^{-1}$. This is in agreement with the velocity and isotropic nature of photospheric granular motions. The author suggests that the interchange reconnection is aided by the random granular motions and allows small-scale loops get closer to open field lines, and accommodate interchange magnetic reconnection. This mechanism could also play an important role in the birth, evolution and decay of CHs. It is still unclear what initiates the formation of CHs and what determines the location of CH formation, however, it is possible that random concentrations of open field lines are aided by granular motions. Similarly, random granular motions could also be responsible for the fragmentation and diffusion of open field in decaying CHs.

\item CH boundary relocation speeds in excess of 2~km~s$^{-1}$ were also observed which may indicate that reconnection may be accommodated by mechanisms other than the granular motions. A bipole emergence near a CH boundary may cause continuous reconnection between the open and closed field lines at a faster rate, while favourable loop alignments could also lead to a faster rate of boundary displacements. However, to confirm these theories it is necessary to analyse images of higher-temporal cadence than the 12~minute EIT images.

\item
Using the interchange reconnection model the diffusion coefficient was calculated to be 0--3$\times10^{13}~\mathrm{cm}^{2}~\mathrm{s}^{-1}$ at CH boundaries. This was found to be in agreement with the theoretical results derived by \cite{Fisk05}, who determined the diffusion coefficients inside and outside of a CH to be 3.5~$\times$~10$^{13}$~cm$^{2}$~s$^{-1}$ and 1.6~$\times$~10$^{15}$~cm$^{2}$~s$^{-1}$, respectively.

\item
Using the CH boundary displacement velocities and the Alf\'en speed typical for the lower corona, the magnetic reconnection rate (or magnetic Mach number) at CH boundaries was determined to be in the range 0--3~$\times$~10$^{-3}$. In comparison, a reconnection rate of 0.001-0.01 was found by \cite{Isobe02} in the case of a long duration event flare in a cusp-shaped arcade. As shown in Figure~\ref{flare_reconn}, the estimated coronal magnetic field strength is similar to that used in our study ($\sim$10~G), however, the inflow and Alf\'en speeds are significantly higher in the case of the flare.

\begin{figure}[!t]
\centerline{
\includegraphics[scale=0.5, trim=20 0 20 0, clip]{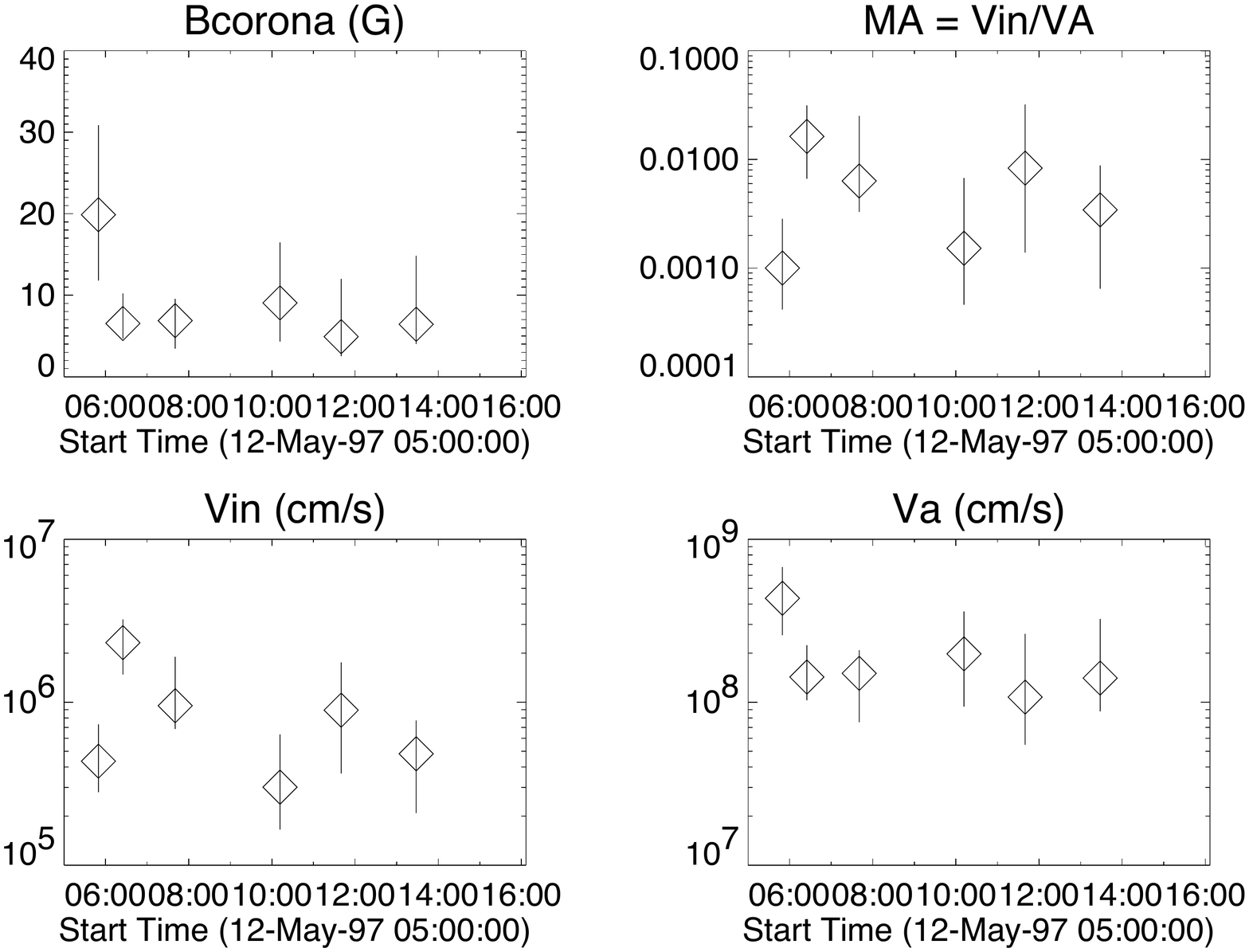}}
\caption{The coronal magnetic field strength, the inflow and Alf\'en speeds, and the magnetic reconnection rates determined for the long duration event flare of 12 May 1997 \citep{Isobe02}.}
\label{flare_reconn} 
\end{figure}

\item
The {\it MIST} algorithm was developed to build on the {\it CHARM} algorithm. It allows CHs to be linked to HSSW streams, and it can also be used to provide HSSW forecasts several days in advance. Using this algorithm over 600 CHs were linked with HSSW periods between 2000--2009. The determined CH latitudinal widths and the HSSW period durations were used to calculate the expansion factor of the open flux tubes rooted in CHs. This was determined by the author to be 2--48, while previous theoretical results indicated a range of 2--20. It is not yet clear what causes the difference between the results, but it may be due to the approximations used in our study (assuming a circular morphology for CHs and the related magnetic flux tubes).

\item
The properties of the CHs and the corresponding HSSW periods identified between 2000--2009 were compared to establish any relationship in order to aid future forecasting. However, no correlation was found between the CH area and the HSSW properties, except for the HSSW duration. Nevertheless, a very strong correlation was found between the mean HSSW velocity and temperature. This could confirm that the mechanism accelerating the solar wind in CHs is also responsible for the heating of the plasma (e.g. Alfv\'en wave heating). It has also been shown that HSSW streams of higher mean velocities are more likely to cause geomagnetic storms with higher $Kp$ indices (correlation coefficient $\sim$0.5). The correlation found between the CH area and the duration of the HSSW stream experienced at Earth can also aid future forecasting purposes (at present the HSSW duration at Earth is determined based on the latitudinal CH width and the expansion factor).

\item
The long-term analysis of CHs and HSSW properties allowed the determination of solar cycle trends in the properties of the CH related HSSW streams. The CH area, mean HSSW velocity, mean temperature and mean $Kp$ index were all found to rise in the descending phase of the solar activity cycle. The peaks were reached within the start and middle of 2003, and the values reduced significantly in the consecutive two years. 

\item
The maximum $Kp$ index showed a significant rise in 2003, reaching a peak value of $Kp$=7.5 between 2006-2007. This was followed by a steep decline in 2006--2007 where is levelled off at $Kp\simeq$4. Note that a geomagnetic storm of $Kp$=7.5 is considered ``strong", causing power fluctuations and long lasting storms (typical of CHs) may result in transformer damage on electrical grids. Spacecraft operations are also likely to be affected during such storms. 

\item
The time of elevated HSSW properties were found to coincide with the time of enhanced high-energy electron flux measured in the outer region of the Earth's Van Allen radiation belt. \cite{Miyoshi11} previously suggested that CHs may be responsible for this effect, but no systematic long-term analysis has been carried out before the study presented in this thesis. Based on the author's analysis it is clear that during the descending phase of the solar cycle CHs are responsible for the electron surges in the radiation belt, which can also cause serious damage to electrical devices on satellites.

\end{itemize}

\section{Future work and outstanding questions}

\subsection{CH detection}

One of the main goals of this thesis is to improve previous CH detection methods by determining the CH boundary threshold using local intensity histograms. However, it is important to note that similarly to previous methods a global intensity threshold is used in a single image, while CH boundary intensities could be different from each other. Furthermore, different parts of a single CH boundary could be of different intensities too. Thus, individual CH boundary thresholds could be an improvement to the present detection method.

For further improvement of the algorithm, it would also be beneficial to test other segmentation methods which could potentially improve or supplement the local intensity thresholding, i.e., the watershed operation can determine boundaries between areas of similar intensities based on points where single pixels separate similar areas \citep{Gonzalez02}. This method proved to be successful on single CH study cases, but needs improvement in order to detect CHs through the solar cycle. Optimal thresholding could be another approach, where a multimodal intensity distribution of an image is fitted with Gaussian distributions to determine thresholds between areas of different intensity \citep{harra00}. Using this method it is important to {\it a priori} determine the number of distributions to be fitted and the approximate locations of the distributions. These properties will have to depend on the time of the solar cycle and the number of different intensity phenomena present on the Sun that correspond to different intensity distributions. Laplacian filtering could also be a useful approach to detect CH boundaries. By computing the second derivatives of the image it helps to identify whether the change in adjacent pixel values is due to an edge or a continuous progression. Applying this approach to the intensity histograms rather then an image could also prove useful to determine boundary intensities \citep{Oshea00}. Multiscale analysis of solar images could also be used in order to detect CH boundaries, by identifying edges at different scales and excluding those corresponding to noise. 

CHs are known to be dominated by a single polarity and for this reason determining the CH status of a low intensity region is dependent on the flux imbalance. At present the {\it CHARM} algorithm uses an arbitrarily chosen value as a threshold. Further analysis needs to be carried out to justify this value and to possibly correct it. The CH grouping algorithm could also be further developed, as at the present the grouping is based solely on the distance between individual CHs. It may be advisable to group CHs based on their polarity as well as their distance, since regions of similar polarity may be interconnected \citep{Antiochos07} and could be treated as a group from a physical perspective. Another addition to the CH detection and grouping algorithm could be the identification of single CHs (or groups) in order to study their evolution over several rotations. The study of individual CHs could reveal what locations CHs are more likely to form at, and what determines the maximum size and the lifetime of a CH. It could also prove useful in determining the long term geomagnetic effects of single CHs (or CH groups) over several rotations.

It would also be very beneficial to carry out a long-term CH study using X-ray observations and compare the obtained CH boundaries with those determined from EUV observations. With the advances of solar imaging, it will be very important to study CHs using narrow-band filters that correspond to different distinct temperatures, as this could reflect on the three-dimensional morphology of CHs in the inner corona. The 3D morphology could also be investigated by observing CHs from different spatial angles, i.e. using SOHO as well as STEREO A and B observations. In addition, it would be very useful to carry out a long-term study on the temperature, density and and element abundances of individual CHs using spectral data and compare the CH properties to corresponding in-situ HSSW measurements.

\subsection{Short and long-term CH evolution}

 A further improvement of the {\it CHEVOL} algorithm could be the use of high-cadence EUV (or X-ray) imagery. The recently launched SDO mission provides high temporal and spatial cadence EUV images that will be very useful in the future not only for short-term studies but also for long-term studies of CHs, once a large database is built up.

The short-term CH evolution study raised the question of bipole evolution. Since bipoles influence CH evolution through interchange reconnection, it would be useful to track the emergence and the displacement of individual bipoles and to determine their effect on the CH boundary. It would also be important to determine whether bipoles ``pass through'' the rigid rotating CH, as they move with the differentially rotating photosphere, or whether they remain inside the CH once admitted through interchange reconnection. This could also reflect on the role of bipoles in the decay of CHs.

\begin{figure}[!t]
\centerline{
\includegraphics[scale=0.6, trim=20 0 15 0, clip]{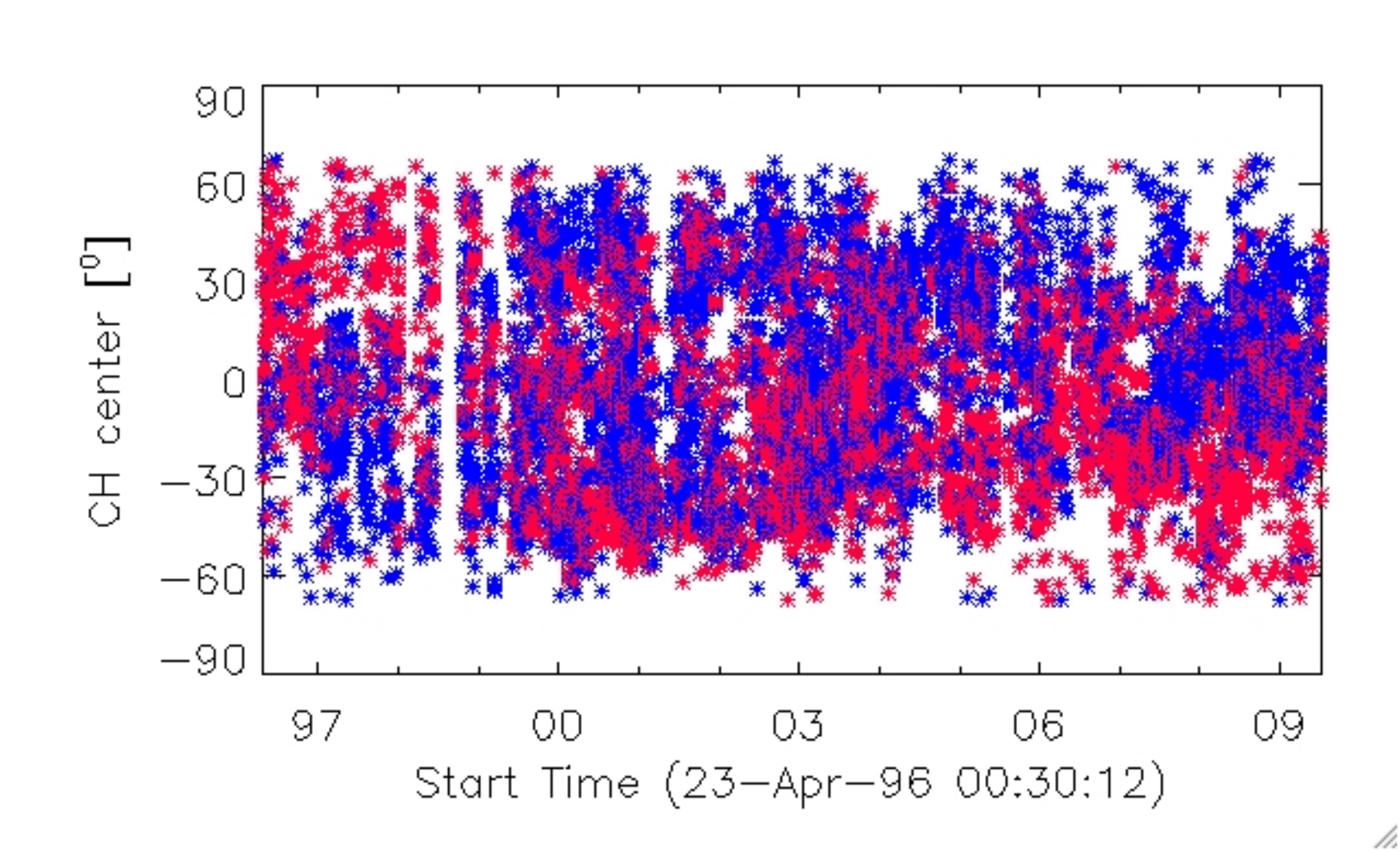}}
\caption{The polarity-wise separation of CHs over a solar cycle. The latitudinal location of CH centerpoints are shown as a function of time between 2006--2009. Red symbols correspond to positive polarity, while blue corresponds to negative polarity CHs.}
\label{pol_sep} 
\end{figure}

As part of the long-term analysis, the author investigated the latitudinal distribution and polarity of CHs from 1996--2009. Figure~\ref{pol_sep} shows a very clear polarity-wise separation in CHs from $\sim$1996--1999 and $\sim$2005--2009, the ascending and descending phase of the solar cycle, respectively. During these phases the solar hemispheres appear to be dominated by CHs of the same polarity as the polarity of the corresponding pole. This clearly dipolar configuration of the global solar magnetic field breaks down during the solar cycle maximum, when the global magnetic field becomes multipolar \citep{Bilenko02}. This can be seen in Figure~\ref{pol_sep} between 1999--2005, where the latitudinal location of CHs is mixed in both hemispheres.

It is also interesting that while CHs can form in the vicinity of decaying ARs with diffusing unipolar magnetic fields, ARs are not necessarily the reason for CH occurrence \citep{Bilenko04}. The dynamics of non-AR CHs is likely to be determined by the global magnetic field. We suggest further analysis to be carried out on the polarity-wise separation of CHs, which may reflect on the mechanisms involved in the dipole reversal.

\subsection{Space weather}

The {\it MIST} algorithm uses the simplified approach of considering the maximum latitudinal width of CHs to determine the HSSW duration, without considering the latitudinal location of CHs. It is a challenge to determine the shape of the open flux tube at Earth and to forecast at what part of the flux tube the Earth is likely to pass through. However, it is commonly known that open field lines expand towards the ecliptic, and thus the method used is a good approximation. However, it is important to note, that the location of the equatorial current sheet can significantly modify the location of the flux tube at 1~AU and hence cause altered HSSW durations.

The equation used to calculate the expansion factor is based on the CH area (Equation~\ref{expfac}). This equation was rewritten to include the latitudinal width of CHs and the duration of HSSW streams at Earth (Equation~\ref{expansionfactor}). In doing this we approximated the CHs to have a circular shape and their maximum latitudinal width to be related to a flux tube of a similarly circular cross-section. However, it is clear from observations that CHs can have elongated shapes and an intricate morphology. For this reason it would be useful to further develop this calculation to account for the non-circular shape of CHs, which may lead to more reliable expansion factors.

The forecasting algorithm at present uses several criteria (i.e. CH area, HSSW velocity difference from start to peak value) that limits the number of CHs and HSSW streams to be linked. With further testing these criteria could perhaps be reduced or even relaxed to allow a larger number of CHs to be included in future studies. 

It would also be very useful to extend the analysis on an even longer time period and determine whether the CH related HSSW properties show the same enhancements in the descending phase of the solar cycle as shown in 2003.

The forecasting accuracy will also need to be compared to those given by theoretical models. The Community Coordinated Modeling Center\footnote[4]{\href{http://ccmc.gsfc.nasa.gov/index.php}{http://ccmc.gsfc.nasa.gov/index.php}} (CCMC) is a multi-agency partnership that allows access to solar, heliospheric, ionospheric and magnetospheric models. The heliospheric models may also be used for forecasting purposes. Model requests can be submitted to CCMC to map the ambient solar corona and inner heliosphere for a Carrington Rotation selected by the user. These models provide information on the solar wind velocity, temperature, density, and magnetic field strength in the heliosphere, from a poleward view as well as a cone-view. 

\begin{figure}[!t]
\centerline{
\includegraphics[scale=0.4, trim=10 0 6 6, clip]{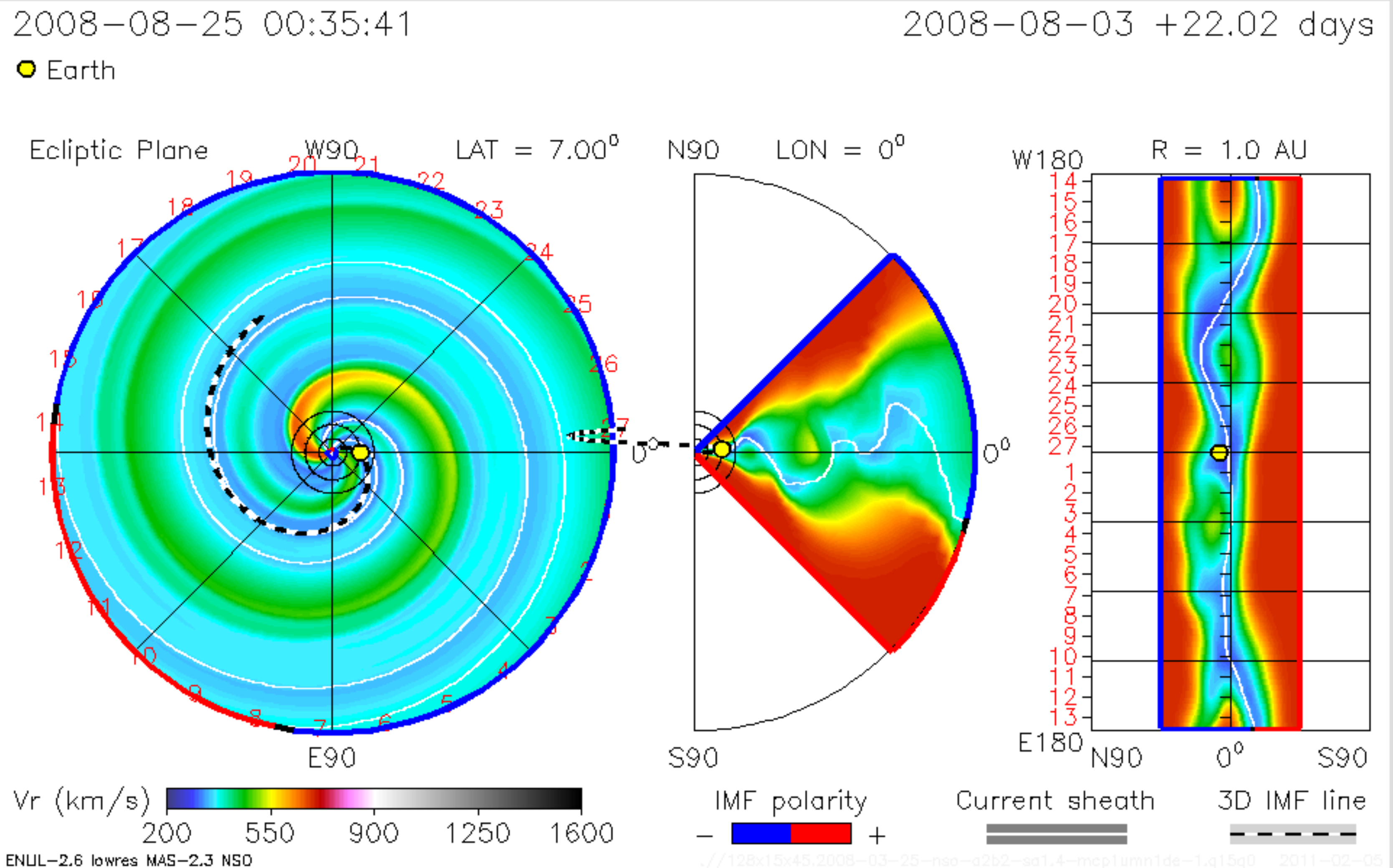}}
\caption{The ENLIL-WSA model of the solar wind velocity in the heliosphere from a poleward view (left), ecliptic cone-view of $\pm45^{\circ}$ from the Sun (middle), and the full ecliptic view (right). The white line represents the heliospheric current sheet location, and the Earth location is shown with a yellow circle. The date on the right is the start time of the model run, the date on the left is the time stamp of the diagrams shown.} 
\label{enlil} 
\end{figure}

Figure~\ref{enlil} shows the result of the ENLIL-cone model run request. In the left, the radial solar wind velocity is shown in the heliosphere from a poleward view. In the middle, the cone-model shows the heliosphere in the Sun-Earth direction in an ecliptic view of $\pm45^{\circ}$ from the Sun. The diagram in the left shows the full ecliptic view of the heliosphere (i.e. a 360$^{\circ}$ extension of the middle figure). In all diagrams the white line represents the location of the heliospheric current sheet, and the location of Earth is shown with a yellow circle. 

ENLIL uses the Wang-Sheeley-Arge (WSA) model which extends a potential field source surface (PFSS) magnetic field to 21.5~$R_{\odot}$, past the critical point where the plasma velocity exceeds the sound speed. It uses a heliospheric current sheet model and a slow and high speed solar wind distribution depending on CH locations which are determined from magnetograms. ENLIL results map the heliosphere from 21.5~$R_{\odot}$ -- 1.6 AU, and from -58 to +58~$^{\circ}$ heliographic latitude.
 
\begin{figure}[!t]
\centerline{
\includegraphics[scale=0.5]{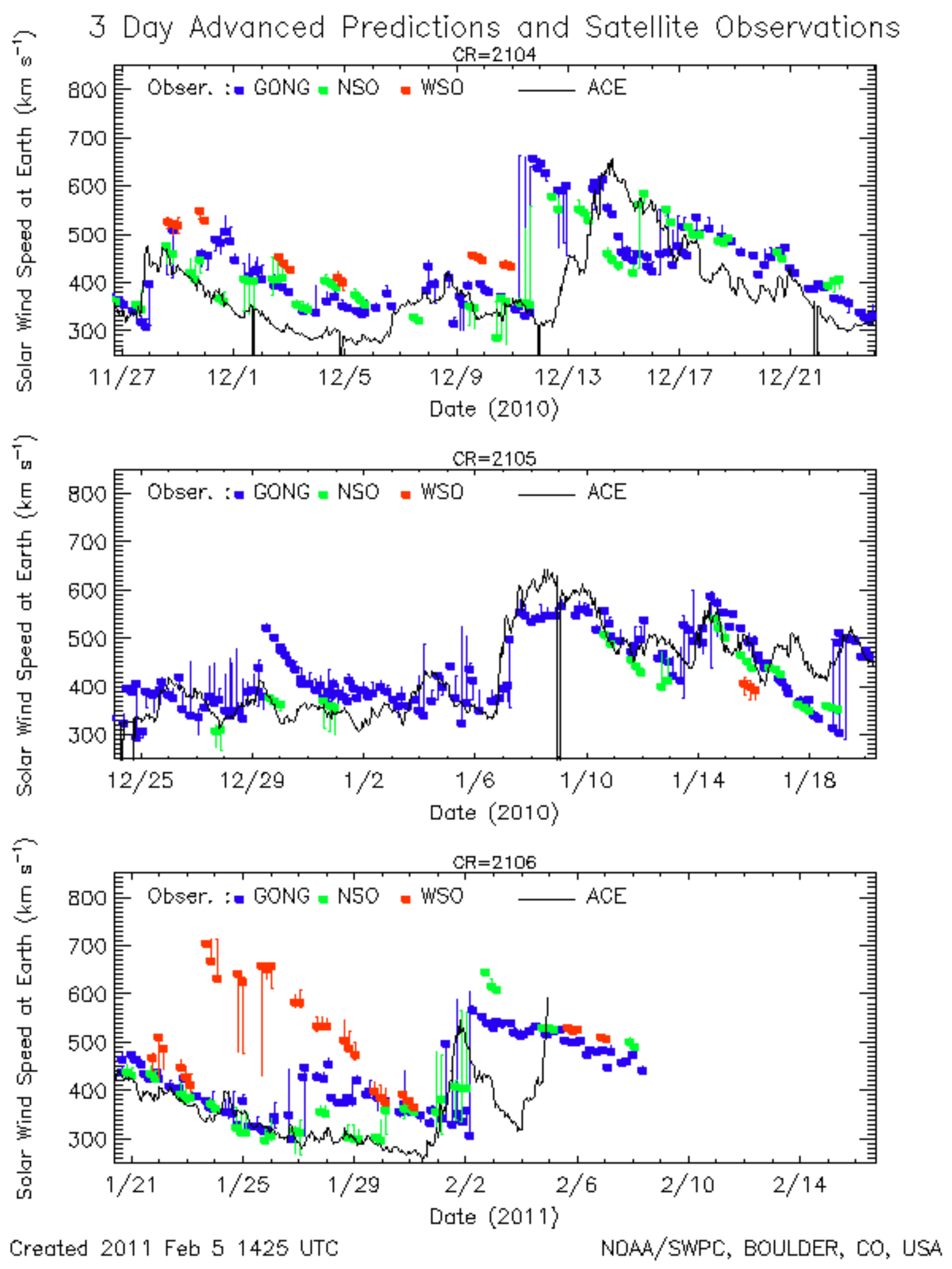}}
\caption{The WSA solar wind velocity forecast. ACE solar wind speed data is shown with solid black line, while the colored dots show the WSA model predictions. The red, green, and blue symbols correspond to predictions based on magnetograms obtained from Wilcox (WSO), Mount Wilson (MWO) and SOLIS (NSO) solar observatories, respectively. The uncertainty estimates are shown with vertical bars.} 
\label{wsa_swpc} 
\end{figure}

The CCMC heliospheric models all use Carrington rotation magnetograms to identify open and closed field regions on the Sun in order to carry out the magnetic field extrapolation. This however means that it is an approximation of the heliospheric magnetic field, since it uses data which is time delayed (the Carrington maps are created over $\sim$27 days). Emerging regions and considerable changes in the global magnetic field may lead to errors in the resulting solar wind properties. However, these models provide a remarkable insight into the heliospheric morphology of the solar wind. The WSA model is currently used at the NOAA Space Weather Prediction Center and the forecast are available online\footnote[5]{\href{http://www.swpc.noaa.gov/ws/}{http://www.swpc.noaa.gov/ws/}}. The author suggests that these models could be aided by EUV or X-ray based CH maps, as it would increase the accuracy of the size and location of open magnetic fields, and thus may increase the accuracy of the model-based forecasts. A result of a recent solar wind forecast is shown in Figure~\ref{wsa_swpc}.

A very exciting new development is the HELIO\footnote[6]{\href{http://www.helio-vo.eu/aboutus/helio_consortium.php}{http://www.helio-vo.eu/aboutus/helio\_consortium.php}} project, which will provide access to data produced by different domains involved in solar and heliospheric physics. It will also allow the use of solar and heliospheric models in order to study solar events and phenomena in solar observations and link them to heliospheric effects, such as geomagnetic disturbances occurring on planets. The algorithms presented in this thesis will all be available through the HELIO development in the future. The CH Lambert maps and data products are already in the progress of being integrated into the HELIO database, while the HSSW profiles corresponding to CHs will also be publicly available in the coming months. These products will also be available on \href{www.SolarMonitor.org}{www.SolarMonitor.org}.







\bibliographystyle{Latex/Classes/jmb} 
\renewcommand{\bibname}{References} 

\bibliography{9_backmatter/references} 

\begin{thebibliography}{151}
\bibitem[{{Krimigis} {\em et~al.\/}(2009){Krimigis}, {Mitchell}, {Roelof},
  {Hsieh} \& {McComas}}]{Krimigis09}
{\sc {Krimigis}, S.M., {Mitchell}, D.G., {Roelof}, E.C., {Hsieh}, K.C. \&
  {McComas}, D.J.} (2009). {Imaging the Interaction of the Heliosphere with the
  Interstellar Medium from Saturn with Cassini}. {\em Science\/}, {\bf 326},
  971--.
\end{thebibliography}








\end{document}